\documentclass[11pt,BCOR18mm]{scrreprt}
\usepackage[latin1]{inputenc}
\usepackage[T1]{fontenc}
\usepackage{ae}
\usepackage{dsfont}
\usepackage{setspace}
\usepackage{hyperref}
\usepackage{epsfig,graphicx}

\usepackage{amsmath,amsthm,amsfonts,amssymb}

\theoremstyle{plain}
\newtheorem{thm}{Theorem}
\newtheorem{lem}[thm]{Lemma}
\newtheorem{prop}[thm]{Proposition}
\newtheorem{cor}{Corollary}
\theoremstyle{definition}
\newtheorem{defn}{Definition}

\begin{document}
\onehalfspacing

 \begin{titlepage}
  \begin{center}

   \vspace*{90pt}
    {\huge Computational Perspectives on Bell Inequalities and Many-body Quantum Correlations}\\
    \vspace*{50pt}

   {\Large Matthew Joseph Hoban}\\\vspace*{70pt}
   {\it A thesis submitted to}\\\vspace*{10pt}
    {\large University College London}\\\vspace*{10pt}
    {\it for the degree of}\\\vspace*{10pt}
    {\large Doctor of Philosophy}\\\vspace*{120pt}
    {\large Department of Physics and Astronomy}\\
    {\large University College London}\\
    \today\\
  \end{center}

 \end{titlepage}
\begin{quote}
 I, Matthew Joseph Hoban confirm that the work presented in this thesis is my own. Where information has been derived from other sources, I confirm that this has been indicated in the thesis.
\end{quote}
\newpage
\chapter*{Abstract}
\noindent
The predictions of quantum mechanics cannot be resolved with a completely classical view of the world. In particular, the statistics of space-like separated measurements on entangled quantum systems violate a Bell inequality \cite{bellineq}.

We put forward a computational perspective on a broad class of Bell tests that study correlators, or the statistics of joint measurement outcomes. We associate particular maps, or functions to particular theories. The violation of a Bell inequality then implies the ability to perform some functions, or computations that classical, or more generally, local hidden variable (LHV) theories cannot.

We derive an infinite class of Bell inequalities that establish a link to so-called ``non-local games'' \cite{nonlocal}. We then make the connection between Raussendorf and Briegel's formulation of Measurement-based Quantum Computing (MBQC) \cite{mbqc1}, and these non-local games. Not only can we show that a quantum violation implies a computational advantage in this model, we show that adaptive measurements are required to perform all quantum computations.

Finally, we explore post-selection of data in Bell tests from both a practical and conceptual point-of-view, with particular consideration to so-called ``loopholes''. Loopholes allow LHV theories to simulate quantum correlations through post-selection. We give a computational description of how loopholes can emerge in different post-selection scenarios.  This motivates us to find a form of post-selection that does not lead to loopholes. Central again to this discussion is the description of LHV theories in terms of computations.

Interestingly, quantum correlators can be made more ``non-classical'' with this loophole-free post-selection. This method of post-selection also can simulate information processing tasks, such as MBQC, that have time-like separated components. This opens up new avenues for the study of time-like tasks studied within the space-like separated scenario of the Bell test.

\newpage
\chapter*{Publications}
The majority of the work in this thesis is based on the following publications:
\newline

\noindent
M. J. Hoban, E. T. Campbell, K. Loukopoulos, and D. E. Browne, \emph{Non-adaptive Measurement-based Quantum Computation and Multi-party Bell Inequalities}, New J. Phys. \textbf{13} 023014 (2011).
\newline

\noindent
M. J. Hoban and D. E. Browne, \emph{Stronger Quantum Correlations with Loophole-Free Postselection}, Phys. Rev. Lett. \textbf{107}, 120402 (2011).
\newline

\noindent
M. J. Hoban, J. J. Wallman, and D. E. Browne, \emph{Generalized Bell-inequality experiments and computation}, Phys. Rev. A \textbf{84}, 062107 (2011).
\newline

\newpage
\chapter*{Acknowledgements}
I am indebted to many different people for many things throughout the process of completing my PhD. First of all, I desperately need to thank my supervisor, Dan Browne. To paraphrase Winston Churchill, ``Never was so much owed by one PhD student to one tireless supervisor.'' I thank him for his patience, enthusiasm, insight and for sharing his great ideas. Not only has he been a great supervisor, he has been a good friend.

My examiners Jon Barrett and Sougato Bose need to be thanked for their thorough reading of this thesis. Their experience and insight has only improved this document. It remains to be said that any remaining inaccuracies result from me.

A lot of the work contained in this tome would never have seen the light of day if I had not had such good collaborators. Earl Campbell was instrumental in the first two years of my PhD, sharing his great insights and discussing his new ideas. I am very proud of the paper we wrote together with Klearchos Loukopoulos, whose contribution to many discussions I am also grateful for. Joel Wallman has also been a great collaborator for discussing very many ideas in our time-zone defying email discourses. 

I also thank Bob Coecke and Samson Abramsky for allowing me to join their group at Oxford. They have given me an independence that has allowed the continuation of some of the ideas resulting from the work contained herein. 

I would also like to thank the Quantum Information group at UCL for producing a delightful environment in which to do research. In particular, I am indebted to Janet Anders for showing me around on my first day and always being around for good conversation, regardless of the topic. Also, my office mates at any one particular time including Sai-Yun Ye (for long discussions on short topics), Hussain Anwar, Brad Augstein, Tahir Sharaan, Hulya Yadsan-Appleby, and Peter Burns.

Outside of my office, along the corridor are the Public Engagement Unit. I would like to thank them for being just brilliant friends, especially Hilary Jackson and Gemma Moore. They put up with my semi-coherent ramblings for far longer than they needed to. I will, and do already miss them.

Outside of my corridor, in big old London, I have to thank my friends for distracting me from everything. In particular James Millen (who became integrated into my office friends), Andy Sykes (who became integrated into my corridor friends), Philippa Stanger, Andy Webster, Keira Poland and countless others. They make it easy to miss London.

Outside of London, in the United Kingdom at large, I want to thank my family. My mum and dad, Christine and Chris Hoban, for everything they have done to allow me to get this far. I doubt anyone loves their parents as much I do and defy anyone to say otherwise. Along with my parents, I want to thank my brothers Kieran, Niall and Dominic for their support and love, and for testing the water before my arrival. Kieran especially needs to be thanked for the countless lunches, excellent conversation and emotional support.

Finally, and most importantly, I want to thank Francesca Richards for her kindness, intelligence, love and humour. Cesca (and Mwg) kept me sane throughout the whole of my PhD and I cannot begin to thank her enough.

\newpage
\tableofcontents
\newpage
\begin{center}
To the memory of Margaret Ellen Hoban.
\end{center}
\newpage

\chapter{Introductions}\label{chap1}
\textit{``I tell you, we are here on Earth to fart around, and don't let anybody tell you different.''}\newline

-Kurt Vonnegut
\newline
\newline

\noindent
If this thesis has one central motivation it is this: to explore the interplay between the foundations and applications of quantum physics. The emergence of quantum information (the application of computer science ideas to quantum physics \cite{nielsen}) has motivated new insights into quantum mechanics. Indeed, new interpretations of quantum physics have been influenced by information theoretic concepts (e.g. \cite{fuchs}). In turn, ideas in quantum foundations have inspired new technological ideas and applications (e.g. \cite{ekert,wootters}). The hope is that this work contributes to this fertile area of research by considering quantum mechanical correlations from a computational point-of-view.

In discussing the interplay between computation and correlations (in particular correlations of measurement statistics), we will discuss issues central to both computer science and quantum theory. Before we can address these issues we need to introduce basic concepts in quantum mechanics and quantum information. We will also mention how ideas in the foundations of quantum mechanics have inspired new applications of quantum theory, with a particular focus on the Bell inequality \cite{bellineq}.

First, we introduce quantum mechanics and discuss the concept of entanglement \cite{schroedinger}. Einstein, Podolsky and Rosen used entanglement to argue that quantum mechanics is an incomplete theory \cite{epr}. This leads us to discuss Bell's argument that quantum mechanics is incompatible with ``local realism'' \cite{bellineq}. This incompatibility is epitomised by a violation of a Bell inequality \cite{bellineq,chsh}. 

After the above discussion, we give a brief overview of quantum information science. We indicate that entanglement has been shown to be a resource in quantum information \cite{nielsen}. The incompatibility of quantum mechanics with local realism is also a resource for certain tasks: device-independent quantum information protocols \cite{mayers,acin3,pironio}. We finish by discussing possible connections between Bell inequalities and Measurement-based Quantum Computing \cite{mbqc1}; the latter utilises entangled states to perform computational tasks. All of the work in this chapter is introductory and does not consist of new results produced by the author of this thesis.

\section{Quantum Mechanics and Entanglement}\label{sec11}

\noindent
In this section, we give a brief overview of the postulates of quantum mechanics. We also look at one of the consequences of these postulates: entanglement. There are very many clear and excellent pedagogical introductions to the quantum formalism (e.g. \cite{peres2,nielsen}). We base our introduction on that of Nielsen and Chuang \cite{nielsen}. The more relevant aspects of quantum theory will be emphasized, especially with regards to measurements. 

Quantum mechanics is a mathematical framework for making predictions of outcomes of experiments. The problem of how this framework relates to a picture of physical reality is still open. An interesting research direction is to recover the quantum formalism from a set of axioms rooted in less mathematical, or more physical principles (e.g. \cite{hardy,chiribella}). This subject will not be addressed in this thesis as it would be too much of a diversion from our discussion. Although, the issue of realism in a limited form will be encountered in section \ref{sec21}.

\subsection{Postulates of Quantum Mechanics} 

In this subsection, we assume familiarity with linear algebra, complex vector spaces and Dirac notation (\cite{nielsen} is an excellent reference for these subjects). Physical systems described by quantum mechanics are associated with a complex inner product vector space, or Hilbert space $\mathcal{H}$. This idea can be formalised in the following postulate taken verbatim from \cite{nielsen}.
\newline

\noindent
\textbf{Postulate 1} \cite{nielsen}. \textit{Associated to any isolated physical system is a complex vector space with inner product (that is, a Hilbert space) known as the state space of the system. The system is completely described by its state vector, which is a unit vector in the system's state space.}
\newline

A unit vector $|\psi\rangle$ in this Hilbert space $\mathcal{H}$ must satisfy $\langle\psi|\psi\rangle=1$, where $\langle\psi|$ is the dual vector to $|\psi\rangle$ in the dual Hilbert space $\mathcal{H}^{*}$. For two-dimensional Hilbert spaces, all unit vectors are called ``qubits'' (quantum bits) and can be written as $|\psi\rangle=\alpha|0\rangle+\beta|1\rangle$, where $\alpha$ and $\beta$ are complex numbers satisfying $|\alpha|^{2}+|\beta|^2=1$. By convention we choose the basis states in a $d$-dimensional Hilbert space $\mathcal{H}$ (where $d$ is finite) to be $|j\rangle$ where $j\in\{0,1,...,(d-1)\}$\footnote{The set of integers $\{0,1,...,(d-1)\}$ can be described in terms of the cyclic group $\mathbb{Z}_{d}$.}.

More generally, quantum states can be associated with ``density matrices'' $\rho$, or an element of the space $L(\mathcal{H})$ of linear operators on $\mathcal{H}$. We may need to consider density matrices for physical systems that are not isolated or when an experimenter is not sure which state $|\psi\rangle$ a system is in; they assign probabilities to the possibilities. These density matrices represent statistical ensembles of the unit vectors described by Postulate 1. The unit vectors $|\psi\rangle$ are associated with ``pure states'' that are the density matrices $\rho=|\psi\rangle\langle\psi|$. If $\rho$ is a statistical ensemble of pure states $|\psi_{j}\rangle$ then it can be represented as
\begin{equation}
\rho=\sum_{j}p_{j}|\psi_{j}\rangle\langle\psi_{j}|
\end{equation}
where $j$ labels all possible pure states in an ensemble. The probabilities $p_{j}$ are associated with each pure state $|\psi_{j}\rangle$ where $\sum_{j}p_{j}=1$ and all $p_{j}\geq 0$.

For density matrices, the inner product is generalised to the operator trace $\textrm{Tr}(...)$ such that $\textrm{Tr}(\rho)=\sum_{j}p_{j}\textrm{Tr}(|\psi_{j}\rangle\langle\psi_{j}|)=1$, due to the cyclicity of trace. This is one of the conditions that a density matrix must satisfy along with the positivity condition $\rho\geq 0$. This second condition is satisfied for any arbitrary state $|\phi\rangle\in\mathcal{H}$ as $\langle\phi|\rho|\phi\rangle=\sum_{j}p_{j}|\langle\phi|\psi_{j}\rangle|^{2}\geq 0$ due to $|\langle\phi|\psi_{j}\rangle|^{2}=\langle\phi|\psi_{j}\rangle\langle\psi_{j}|\phi\rangle$.

The second postulate describes how quantum states can be transformed over time. Again this and all postulates are reproduced verbatim from \cite{nielsen}.
\newline

\noindent
\textbf{Postulate 2} \cite{nielsen}. \textit{The evolution of a closed system is described by a unitary transformation. That is, the state} $|\psi\rangle$ \textit{of the system at time} $t_{1}$ \textit{is related to the state} $|\psi'\rangle$ \textit{of the system at time} $t_{2}$ \textit{by a unitary operator} $U$ \textit{which depends only on the times} $t_{1}$ \textit{and} $t_{2}$,
\begin{equation}
|\psi'\rangle=U|\psi\rangle.
\end{equation}
\newline

We immediately see that a unitary operator preserves normalisation of a state as $\langle\psi'|\psi'\rangle=\langle\psi|U^{\dagger}U|\psi\rangle=\langle\psi|\psi\rangle=1$ where $U^{\dagger}$ is the adjoint of $U$ so that $U^{\dagger}U=\mathbb{I}$, the identity matrix. Unitary operators can also be applied to a density matrix as
\begin{equation}
U\rho U^{\dagger}=\sum_{j}p_{j} U|\psi_{j}\rangle\langle\psi_{j}|U^{\dagger}=\sum_{j}p_{j}|\psi'_{j}\rangle\langle\psi'_{j}|.
\end{equation}
For open systems (i.e. systems that are not closed) we can generalise the unitary operator to a linear operator that must be completely positive and not increase the trace of $\rho$. The next postulate of quantum mechanics relates to measurements which are a form of completely positive and non-trace-increasing linear operator. 
\newline

\noindent
\textbf{Postulate 3} \cite{nielsen}. \textit{Quantum measurements are described by the collection} $\{M_{m}\}$ \textit{of measurement operators. These are operators acting on the state space of the system being measured. The index} $m$ \textit{refers to the measurement outcomes that may occur in the experiment. If the state of the quantum system is} $|\psi\rangle$ \textit{immediately before the measurement then the probability that result $m$ occurs is given by}
\begin{equation}
p(m)=\langle\psi|M^{\dagger}_{m}M_{m}|\psi\rangle,
\end{equation}
\textit{and the state of the system after the measurement is}
\begin{equation}
\frac{M_{m}|\psi\rangle}{\sqrt{\langle\psi|M^{\dagger}_{m}M_{m}|\psi\rangle}}.
\end{equation}
\textit{The measurement operators satisfy the completeness equation,}
\begin{equation}
\sum_{m}M_{m}^{\dagger}M_{m}=\mathbb{I}.
\end{equation}
Again, the above postulate can be extended to density matrices $\rho$ where $p(m)$ becomes $p(m)=\textrm{Tr}(\rho M^{\dagger}_{m}M_{m})$ and the state of the system after measurement is now
\begin{equation}
\frac{M_{m}\rho M^{\dagger}_{m}}{\textrm{Tr}(\rho M^{\dagger}_{m}M_{m})}.
\end{equation}
Therefore measurement operators $M_{m}$ act on density matrices in an analogous fashion to unitary operators. In calculating the probabilities of particular outcomes and satisfying the completeness relation, $M_{m}^{\dagger}$ and $M_{m}$ always appear together. For probabilities of measurement outcomes we rewrite $M_{m}^{\dagger}M_{m}$ as an operator $E_{m}$ associated with a measurement outcome $m$. The operator $E_{m}$ is called an ``element'' of a Positive Operator-Valued Measure (POVM) and is a positive operator\footnote{For all choices of states $|\phi\rangle$, $\langle\phi|E_{m}|\phi\rangle$ is a probability by definition, so $E_{m}$ is a positive operator.} such that $\sum_{m}E_{m}=\mathbb{I}$ with probabilities $p(m)=\textrm{Tr}(\rho E_{m})$. The set of operators $\{E_{m}\}$ is then a POVM. 

A special case of all possible measurements is the von Neumann projective measurement (PVM). This is the set $\{P_{m}\}$ where each element $P_{m}$ is a projector associated with a measurement outcome $m$. These projectors satisfy an orthogonality constraint $P_{m}P_{m'}=\delta^{m}_{m'}P_{m}$ and for an arbitrary pure state $|\psi\rangle$, the state after a PVM is $|\nu_{m}\rangle\propto P_{m}|\psi\rangle$. Then, to satisfy the orthogonality constraint, we need $N$ orthogonal vectors $|\nu_{m}\rangle$ to describe the projectors $P_{m}=|\nu_{m}\rangle\langle\nu_{m}|$, where $N$ is the number of possible outcomes $m$ of a measurement. 


A PVM can be associated with an ``observable'' which matches each projector $P_{m}$ of a PVM with a real eigenvalue $\lambda_{m}$. This observable $\hat{O}$ can be written as $\hat{O}=\sum_{m}\lambda_{m}P_{m}$ where $\lambda_{m}$ is an observed outcome. The eigenvalue $\lambda_{m}$ corresponds to a system being projected into the eigenstate $|\mu_{m}\rangle$ associated with $P_{m}$\footnote{These are eigenvalues and eigenstates as $\hat{O}|\mu_{m}\rangle=\lambda_{m}|\mu_{m}\rangle$.}. For example, for a two-dimensional Hilbert space, we can have observables with eigenvalues $\lambda_{m}=\pm 1$ associated with two-dimensional vectors $|\mu_{m}\rangle$ where $m$ takes two possible values.

There is a beautiful result due to Naimark that shows that any POVM on a quantum state can be associated with a PVM \cite{naimark}. That is, every POVM acting on a Hilbert space $\mathcal{H}$ can be implemented with a PVM on a larger Hilbert space $\mathcal{K}$. We can obtain some auxiliary (often referred to as an ancilla) system and take the composite of this system and our original Hilbert space $\mathcal{H}$ and perform a PVM on this new space. In order to consider composite systems we need to introduce the next postulate.
\newline

\noindent
\textbf{Postulate 4} \cite{nielsen}. \textit{The state space of a composite physical system is the tensor product of the state spaces of the component physical systems. Moreover, if we have systems numbered $1$ through $n$, and system number $i$ is prepared in the state $|\psi_{i}\rangle$, then the joint state of the total system is $|\psi_{1}\rangle\otimes|\psi_{2}\rangle\otimes...\otimes|\psi_{n}\rangle$.}
\newline

We can replace pure states $|\psi_{i}\rangle$  in this postulate with density matrices $\rho_{i}$. Composite systems can be represented by density matrices as linear operators on a tensor product Hilbert space, i.e. $\rho\in L(\mathcal{H}_{1}\otimes\mathcal{H}_{2}\otimes...\otimes\mathcal{H}_{n})$ where $\mathcal{H}_{i}$ is the Hilbert space of each $i$th system. While in the postulate, we mention one pure state, $\bigotimes_{i=1}^{n}|\psi_{i}\rangle=|\psi_{1}\rangle\otimes|\psi_{2}\rangle\otimes...\otimes|\psi_{n}\rangle$, in particular, this is not the most general pure state in a composite Hilbert space $\bigotimes_{i=1}^{n}\mathcal{H}_{i}=\mathcal{H}_{1}\otimes\mathcal{H}_{2}\otimes...\otimes\mathcal{H}_{n}$. The state $\bigotimes_{i=1}^{n}|\psi_{i}\rangle$ is a ``product state'', but pure states that cannot be expressed in this form are said to be ``entangled''. This property will be discussed in the next subsection.

We have given a brief overview of the mathematical construction of quantum mechanics. In this thesis, we will be utilising the definition of a measurement and the description of composite systems. If a composite system consists of two space-like separated systems $\mathcal{H}_{1}$ and $\mathcal{H}_{2}$, then experimenters in each of these space-like separated systems can perform measurements on each of their respective subsystems. This way measurements can be written as a tensor product of these localised measurements, i.e. $\mathcal{M}^{1}_{m}\otimes \mathcal{M}^{2}_{m}$ where $\mathcal{M}^{i}_{m}\in L(\mathcal{H}_{i})$, a linear operator on $\mathcal{H}_{i}$. Assume that one can prepare all possible states (by whatever means) on the composite system, i.e. $\rho\in L(\mathcal{H}_{1}\otimes\mathcal{H}_{2})$. If the measurements are performed on \emph{entangled} states then the statistics produced by this total system do not always factorise, i.e.
\begin{equation}
p(m,m')=\textrm{Tr}(\rho \mathcal{M}^{1}_{m}\otimes \mathcal{M}^{2}_{m'})\neq\sum_{j}p_j\textrm{Tr}(\rho_{1,j}\mathcal{M}^{1}_{m})\textrm{Tr}(\rho_{2,j}\mathcal{M}^{2}_{m'})
\end{equation}
as $\rho$ is not necessarily equal to $\sum_{j}p_j\rho_{1,j}\otimes\rho_{2,j}$ where $\rho_{i,j}$ corresponds to a pure state of the $i$th system in the $j$th term of the decomposition of $\rho$.

This inability for the statistics of space-like separated measurements to be factorised will be central to the discussion of quantum correlations in this thesis. Entanglement is central to this subject. In the next subsection we will briefly discuss entanglement and how it can be quantified. 

\subsection{Entanglement}

Schr\"{o}dinger first introduced the term ``entanglement'' \cite{schroedinger}. This concept has become formalised for all possible density matrices $\rho$. First we describe systems in a bipartite scenario, that is where the Hilbert space of the system in question is the tensor product of two Hilbert spaces. An entangled state represented by a density matrix $\rho$ \textbf{cannot} be expressed as
\begin{equation}\label{separable}
\rho=\sum_{i}p_{i}|\psi^{1}_{i}\rangle\langle\psi^{1}_{i}|\otimes|\psi^{2}_{i}\rangle\langle\psi^{2}_{i}|.
\end{equation}
The pure state $|\psi^{j}_{i}\rangle\langle\psi^{j}_{i}|$ is the $j$th party's state for the $i$th pure state in the probabilistic ensemble of $\rho$. There may be multiple, even infinite possible decompositions of $\rho$ into a convex combination of pure states $|\psi^{1}_{i}\rangle\langle\psi^{1}_{i}|\otimes|\psi^{2}_{i}\rangle\langle\psi^{2}_{i}|$. For example, the density matrix $\rho=\frac{1}{4}\mathbb{I}$ in a composite Hilbert space of two, two-dimensional Hilbert spaces can be written as
\begin{eqnarray}
\rho&=&\frac{1}{4}\left(|00\rangle\langle 00|+|01\rangle\langle 01|+|10\rangle\langle 10|+|11\rangle\langle 11|\right)\nonumber \\
&=&\frac{1}{4}\left(|++\rangle\langle ++|+|+-\rangle\langle +-|+|-+\rangle\langle -+|+|--\rangle\langle --|\right),\nonumber \\
\end{eqnarray}
where $|+\rangle=\frac{1}{\sqrt{2}}(|0\rangle+|1\rangle)$ and $|-\rangle=\frac{1}{\sqrt{2}}(|0\rangle-|1\rangle)$\footnote{We make the standard abbreviation of omitting the tensor product for composite pure states, e.g. $|0\rangle\otimes|0\rangle$ becomes $|00\rangle$.}. This multiplicity of decomposition makes it difficult to ascertain whether an arbitrary density matrix is entangled or otherwise.

If a density matrix is a bipartite pure state, then there is a definite method to detect whether this state is entangled or not \cite{popescu,plenio}. This method of detection also can quantify the \emph{amount of entanglement}. For mixed states, this detection is a hard problem to compute \cite{gurvits}.

The method of detecting entanglement for bipartite pure states involves finding the ``Entropy of Entanglement'' \cite{popescu}. To calculate this quantity, first one needs to find the reduced density matrix of $\rho_{1}$ and $\rho_{2}$ corresponding to party $1$ and $2$. The reduced density matrix is calculated from the partial trace of $\rho$, where we only take a trace over one party's system instead of the whole composite system. The partial trace of $\rho$ over system $1$ of two systems is written as $\textrm{Tr}_{1}(\rho)$ and is calculated as
\begin{equation}
\textrm{Tr}_{1}(\rho)=\sum_{i}\langle i_{1}|\rho|i_{1}\rangle,
\end{equation}
where $|i_{1}\rangle$ are basis states on system $1$. Without loss of generality, we assume that all subsystems have the same dimensional Hilbert space. We then calculate the von Neumann entropy $S(\rho_{1})$ \cite{vonneumann} of this reduced density matrix\footnote{The von Neumann entropy is the same for either sub-system \cite{nielsen}.} $\rho_{1}$:
\begin{equation}
S(\rho_{1})=-\textrm{Tr}(\rho_{1}\log_{2}{(\rho_{1})}).
\end{equation}
If $S(\rho_{1})=0$, then the reduced state $\rho_{1}$ is a pure state and so $\rho=\rho_{1}\otimes\rho_{2}$ with both $\rho_{j}$ being pure states. Importantly if $S(\rho_{1})>0$ then the pure state $\rho$ is entangled. For $S(\rho_{1})=1$, then $\rho_{1}=\frac{1}{2}\mathbb{I}$. The state $\rho$ that results in $S(\rho_{1})=1$ is the ``maximally entangled state'' of two qubits, as it gives the maximum value of $S(\rho_{1})$ for two qubits. The maximally entangled state $|\Psi\rangle$ of two $d$-dimensional systems can be written as 
\begin{equation}
|\Psi\rangle=\frac{1}{\sqrt{d}}\sum_{j=0}^{(d-1)}|jj\rangle.
\end{equation}
If we take the partial trace over system $1$, then
\begin{eqnarray}
\textrm{Tr}_{1}(|\Psi\rangle\langle\Psi|)&=&\sum_{i=0}^{(d-1)}\frac{1}{d}\left(\sum_{j=0}^{(d-1)}\langle i|jj\rangle\right)\left(\sum_{j=0}^{(d-1)}\langle jj|i\rangle\right)\nonumber \\
&=&\frac{1}{d}\sum_{j=0}^{(d-1)}|j\rangle\langle j|=\frac{1}{d}\mathbb{I}.
\end{eqnarray}
In the case of two qubits we retrieve the value of entropy mentioned above, but in general, for these states $S(\rho_{2})=S(\rho_{1})=\log_{2}(d)$. 

We have only discussed the bipartite case. In this thesis, we will also be interested in multipartite quantum systems. The definition of an entangled multipartite state is now where an entangled state cannot be written as \eqref{separable} but with $|\psi^{1}_{i}\rangle\langle\psi^{1}_{i}|\otimes|\psi^{2}_{i}\rangle\langle\psi^{2}_{i}|$ now replaced with $\bigotimes_{j=1}^{n}|\psi^{j}_{i}\rangle\langle\psi^{j}_{i}|$. Entanglement of multipartite systems is relatively less well-studied but there do exist measures of entanglement in this scenario \cite{plenio}. There is also not one particular maximally entangled state for the multipartite setting like there is for the bipartite setting.

So far entanglement has been discussed as a mathematical construct and we have not discussed its physical consequences. In the next section we will discuss the impact of entanglement upon the foundations of quantum mechanics. That is, it causes a tension between quantum physics and a classical physics view of the world \cite{bell}. If quantum mechanics describes what is actually happening in the world then we need to accept some behaviour that is potentially incompatible with everyday intuition. We will make these issues more rigorous in the next section.

\section{EPR Paradox and Bell Inequalities}\label{sec12}

\noindent 
Albert Einstein played a crucial role in the development of quantum theory \cite{einstein}. However, upon being developed formally, he was famously dissatisfied with it. At its core, quantum mechanics predicts probabilities, and does not always make deterministic predictions\footnote{Einstein's dissatisfaction can be summarised with one of his famous playful quotes: ``... He[God] does not throw dice.'' \cite{einstein2}}. It could be argued that this probabilistic feature convinced Einstein that quantum mechanics was a statistical theory akin to classical statistical, or Liouvillian mechanics \cite{liouville}. In Liouvillian mechanics objects have defined positions and momentum, but we may not have complete knowledge of these properties . Therefore, a probability distribution is assigned over a space of potential properties of a system. The state $|\psi\rangle$ could also resemble a probability distribution over some underlying reality describing a system. For more discussion of Einstein's potential view of quantum physics, see work by Harrigan and Spekkens \cite{harrigan}.

A particular focus for Einstein's criticism of quantum mechanics became the issue of ``locality''. Locality has many different guises but we heuristically use it here in the sense that events in space-time can only ``affect'' each other if they are within each other's light-cone. It has been suggested by Bacciagaluppi and Valentini that Einstein had an argument against quantum theory based on a violation of locality at the 1927 Solvay Conference \cite{valentini}. This discussion is beyond the scope of this thesis but we only mention it as a prelude to the argument presented by Einstein, Podolsky and Rosen (EPR) \cite{epr}, often called the ``EPR paradox''\footnote{The paradox being that if one accepts a particular picture of reality, then quantum mechanics contradicts this picture. It is not a paradox in the sense of demonstrating that quantum mechanics is inconsistent.}.

\subsection{Realism and ``Incompleteness'' of Quantum Mechanics}

In the original EPR paper, they argued that if one can predict a physical property, or quantity, with certainty then we associate that quantity with an ``element of reality'' \cite{epr}. If by the definition of EPR, a theory is ``complete'' then the properties that are found with certainty must be incorporated into the theory describing the system. Take two observables $\hat{O}_{1}$ and $\hat{O}_{2}$ that do not commute, i.e. $[\hat{O}_{1},\hat{O}_{2}]\neq 0$, and a state $|\psi\rangle$ being an eigenstate of $\hat{O}_{1}$ (with eigenvalue $\lambda$). If we make the constraint that the two observables do not share eigenstates nor are any of the eigenstates of one observable orthogonal to eigenstates of the other. We can predict the outcome $\lambda$ of observable $\hat{O}_{1}$ with certainty, but cannot predict the outcome of $\hat{O}_{2}$ with certainty\footnote{If we make a measurement of the observable $\hat{O}_{1}$ on $|\psi\rangle$, we obtain $\lambda$ so that the projection $P_{m}= |\psi\rangle\langle\psi|$ has been performed on $|\psi\rangle$, giving the probability $p(m)= \langle\psi|\psi\rangle\langle\psi |\psi\rangle=1$. However, $\hat{O}_{2}$ consists of projectors $P_{m}= |\phi\rangle\langle\phi|$ where $|\psi\rangle\neq |\phi\rangle$ so for $\hat{O}_{2}$ $p(m)=\langle\psi|\phi\rangle\langle\phi|\psi\rangle\neq 1$.}. This means that we can only associate the observable $\hat{O}_{1}$ with an element of reality but not both observables. The following contradiction emerges if one asserts that elements of reality can only be associated with commuting observables. We follow Bohm's version of the EPR argument \cite{bohm}.

Imagine that two parties share the entangled state (that is equivalent to the maximally entangled state\footnote{One applies the unitary $\mathbb{I}\otimes U$ such that $U=|0\rangle\langle 1|-|1\rangle\langle 0|$ to both qubits.}):
\begin{equation}
|\Psi\rangle_{\textrm{EPR}}=\frac{1}{\sqrt{2}}\left(|01\rangle-|10\rangle\right),
\end{equation}
such that one party has access to one of the two-dimensional subsystems, or qubit, and the other party has access to the other qubit. We have put no constraint on the distance between the two parties, and in fact we make them space-like separated. The first party makes measurements of the observables $\hat{X}=\left(|+\rangle\langle +|-|-\rangle\langle -|\right)$ or $\hat{Z}=\left(|0\rangle\langle 0|-|1\rangle\langle 1|\right)$\footnote{These are the Pauli-X and Pauli-Z measurements respectively.}. Each observable is associated with outcomes, or eigenvalues $\pm 1$ and projectors $P_{m}$ associated with this eigenvalue. According to EPR because the parties are space-like separated they can no longer ``interact'', and regardless of the observable performed by the first party, we must assign the same elements of reality to the second party \cite{epr}. 

If party $1$ measures $\hat{X}$ and gets $+1$ or $-1$ then the second party's state will be $|-\rangle$ or $|+\rangle$ respectively with certainty (upto a global phase). Since we can predict the second party's state with certainty we must assign this property with an element of reality. If, on the other hand, party $1$ measures $\hat{Z}$ then for outcomes $+1$ or $-1$ the second party's state will be $|1\rangle$ or $|0\rangle$ respectively (upto a global phase). Again, we can assign an element of reality since after the measurement, the first party knows the second party's state with certainty.

To summarise, if party $1$ measures $\hat{X}$, then we can assign an element of reality with the second party's observable $\hat{X}$. When party $1$ measures $\hat{Z}$, we assign an element of reality with the second party's observable $\hat{Z}$. Since the measurements performed by party $1$ are space-like separated from party $2$, the elements of reality for party $2$ should not be affected by the first party's measurements. This is the locality argument in the EPR paradox. However, $\hat{X}$ and $\hat{Z}$ do not commute, so we cannot assign an element of reality to each observable arriving at a contradiction. EPR reasoned that this contradiction means that quantum mechanics does not result in a complete picture of reality \cite{epr}.

John Bell formalised the language of the EPR paradox away from the discussion of ``incompleteness'' and ``elements of reality'' into more mathematically rigorous concepts \cite{bellineq,bell}. He showed that the assumption upon which the EPR paradox is based is that all physical systems obey ``local realism'' \cite{bell}. Local realism combines two separate assumptions invoking locality and realism and can be seen to limit the statistics of space-like separated measurements. In the following subsection we will briefly review local realism and show that it puts constraints on these statistics.

\subsection{CHSH Inequality}

We will describe local realism mathematically in section \ref{sec21} of the next chapter but for now, we review the work of Clauser-Horne-Shimony-Holt (CHSH) \cite{chsh}. The seminal work of Bell \cite{bellineq} led to the formulation of the Bell inequality. This work was developed by CHSH into a mathematical expression that can be experimentally testable: the CHSH inequality.

We now describe the Bell-CHSH scenario, or ``test'' \cite{chsh}. There are two parties and each party chooses between two measurements. The choice of measurement is a completely random, free choice of the parties. This is a key assumption in the construction of Bell inequalities \cite{bell} (for consequences of dropping this assumption see \cite{barrett2,hall}). Each measurement has two possible outcomes $\pm 1$. The measurements that the $j$th party chooses from are $\mathcal{M}_{j}^{0}$ and $\mathcal{M}_{j}^{1}$. These measurements can be described by an arbitrary theory and not just quantum theory. The statistics in this experiment that will be of interest to us are the correlations of the form 
\begin{equation}
\mathbb{E}(\mathcal{M}_{1}^{k}\mathcal{M}_{2}^{k'})=p(\mathcal{M}_{1}^{k}\mathcal{M}_{2}^{k'}=1)-p(\mathcal{M}_{1}^{k}\mathcal{M}_{2}^{k'}=-1), 
\end{equation}
the expectation values of the joint outcome of both parties' measurements for choices $k$, $k'\in\{0,1\}$ where $p(\mathcal{M}_{1}^{k}\mathcal{M}_{2}^{k'}=\pm 1)$ is the probability of getting the joint measurement outcome $\pm 1$.

There is actually a class of CHSH inequalities for this scenario \cite{fine}, but we just pick out one particular expression
\begin{equation}\label{exchsh}
\mathbb{E}(\mathcal{M}_{1}^{0}\mathcal{M}_{2}^{0})+\mathbb{E}(\mathcal{M}_{1}^{0}\mathcal{M}_{2}^{1})+\mathbb{E}(\mathcal{M}_{1}^{1}\mathcal{M}_{2}^{0})-\mathbb{E}(\mathcal{M}_{1}^{1}\mathcal{M}_{2}^{1})\leq 2,
\end{equation}
where the upper bound of $2$ is satisfied for all physical systems that satisfy local realism \cite{chsh}. Local realism means outcomes of $\mathcal{M}_{j}^{k}$ are only dependent on some set of objective properties of each party's system. Secondly, these properties (which can be seen as elements of reality) are localised to each space-like separated region. Whilst they may have been shared properties when parties were not separated in the past, they are not affected by anything outside of their region. A locally realistic property for measurement $M_{j}^{k}$ is then $\chi_{j}^{k}\in\{\pm1\}$, so we can write the left-hand-side of \eqref{exchsh} as
\begin{equation}\label{exchsh2}
\mathbb{E}(\chi_{1}^{0}(\chi_{2}^{0}+\chi_{2}^{1})+\chi_{1}^{1}(\chi_{2}^{0}-\chi_{2}^{1}))=\sum_{\chi}p_{\chi}\left(\chi_{1}^{0}(\chi_{2}^{0}+\chi_{2}^{1})+\chi_{1}^{1}(\chi_{2}^{0}-\chi_{2}^{1})\right)
\end{equation}
where $p_{\chi}$ is a probability distribution over all possible assignments of $\chi_{j}^{k}$ to measurements such that $\sum_{\chi}p_{\chi}=1$ and $p_{\chi}\geq 0$. By convexity we can upper bound the right-hand-side of \eqref{exchsh2} by just considering the maximum value of $\chi_{1}^{0}(\chi_{2}^{0}+\chi_{2}^{1})+\chi_{1}^{1}(\chi_{2}^{0}-\chi_{2}^{1})$. If $(\chi_{2}^{0}-\chi_{2}^{1})$ is non-zero then $(\chi_{2}^{0}+\chi_{2}^{1})$ will be zero, resulting in
\begin{equation}\label{exchsh3}
\sum_{\chi}p_{\chi}\left(\chi_{1}^{0}(\chi_{2}^{0}+\chi_{2}^{1})+\chi_{1}^{1}(\chi_{2}^{0}-\chi_{2}^{1})\right)\leq 2.
\end{equation}
This expression then gives exactly the same right-hand-side of \eqref{exchsh}.

This result is interesting as we can derive a consequence of a theory with very few prior assumptions. More importantly though, in the following theorem, we can actually say something about quantum theory using the expression in \eqref{exchsh}.
\newline

\noindent
\textbf{Bell's Theorem} \cite{bellineq}: \textit{The predictions of quantum mechanics are not compatible with a locally realistic theory.}

\textit{Proof}: To prove this theorem, we just need to show that the inequality \eqref{exchsh} is not satisfied for all predicted values of $\mathbb{E}(\mathcal{M}_{1}^{k}\mathcal{M}_{2}^{k'})$ in quantum theory. We prove this by example. If two space-like separated parties share the state $|\Psi\rangle_{\textrm{EPR}}$ and make the measurements $\mathcal{M}_{1}^{0}=\hat{X}$, $\mathcal{M}_{1}^{1}=\hat{Z}$, $\mathcal{M}_{2}^{0}=\frac{1}{\sqrt{2}}(-\hat{Z}-\hat{X})$, and $\mathcal{M}_{2}^{1}=\frac{1}{\sqrt{2}}(\hat{Z}-\hat{X})$, then the left-hand-side of \eqref{exchsh} is
\begin{eqnarray}\label{quantchsh}
& &\mathbb{E}(\mathcal{M}_{1}^{0}\mathcal{M}_{2}^{0})+\mathbb{E}(\mathcal{M}_{1}^{0}\mathcal{M}_{2}^{1})+\mathbb{E}(\mathcal{M}_{1}^{1}\mathcal{M}_{2}^{0})-\mathbb{E}(\mathcal{M}_{1}^{1}\mathcal{M}_{2}^{1})\nonumber \\
&=&\langle\left(\frac{\hat{X}\otimes(-\hat{Z}-\hat{X})}{\sqrt{2}}+\frac{\hat{X}\otimes(\hat{Z}-\hat{X})}{\sqrt{2}}+\frac{\hat{Z}\otimes(-\hat{Z}-\hat{X})}{\sqrt{2}}-\frac{\hat{Z}\otimes(\hat{Z}-\hat{X})}{\sqrt{2}}\right)\rangle.\nonumber\\
\end{eqnarray}
We have used the short-hand notation $\langle(...)\rangle=\langle\psi|(...)|\psi\rangle$ where $|\psi\rangle=|\Psi\rangle_{\textrm{EPR}}$. Calculation of the right-hand-side of \eqref{quantchsh} yields a value of $2\sqrt{2}$, which is greater than $2$, thus violating the CHSH inequality\footnote{This value of $2\sqrt{2}$ is known as Tsirelson's bound \cite{tsirelson} as it is the largest possible quantum value of the left-hand-side of \eqref{quantchsh}.}. Therefore quantum mechanics is incompatible with a theory satisfying local realism. $\square$
\newline

The simplicity of the theorem and its proof has remarkable implications for the foundations of quantum mechanics. It means we must abandon the intuition of local realism, a constraint satisfied by classical physical systems. If the predictions of quantum theory are experimentally verified then if measurement outcomes result from elements of reality, then this reality does not satisfy locality. Or we could just abandon realism all together and not have to worry about locality. 

\subsection{Geometric Construction of Bell Inequalities}

Beginning with the work of Froissart \cite{froissart}, then developments by Fine \cite{fine}, Pitowsky \cite{pitowski} and Peres \cite{peres}, the geometric picture of Bell inequalities has been well-developed. Correlations of space-like separated measurements are now elements of a vector in some real space. The space of correlations satisfying local realism is a convex polytope which can be described in terms of linear inequalities \cite{grunbaum}. These linear inequalities are examples of Bell inequalities. Finding these inequalities is then a problem in convex geometry. 

This polytope approach to Bell inequalities is now an effective way of understanding the consequences of local realism. We will elaborate on and describe this approach in section \ref{sec21c} of the next chapter. Also we will comment on the hardness of finding the Bell inequalities that define the polytope of locally realistic correlations. The convex geometric approach has also been extended to the study of correlations that satisfy only a form of locality: space-like separated measurements that do not allow instantaneous communication \cite{barrett3, pironio3}. These issues will be discussed in section \ref{sec23}.

\subsection{Experimental implementations for testing local realism}

Testing whether quantum mechanics violates a Bell inequality in the laboratory is a difficult task. Firstly, measurements have to be space-like separated but transporting fragile quantum states over large distances can be hard. States may interact with the environment and become mixed states that are no longer entangled. Secondly, apparatus in the lab is not perfect and detectors may not always perfectly detect a measurement outcome. These difficulties can lead to ``loopholes'' (as we shall discuss in section \ref{sec41} of chapter \ref{chap4}) whereby locally realistic theories are no longer constrained by the Bell inequality being tested \cite{pearle,garg}. 

If we do not have space-like separated measurements then the local aspect of locally realistic theories is not constrained and we have the ``locality loophole''. For imperfect detection, the associated ``detection loophole'' is more subtle as it allows the possibility that the objective properties of a system can describe the statistics of detection \cite{pearle}. If we make the extra assumption that properties of the system we are observing are independent of the detection system, often called the ``fair-sampling assumption'' \cite{clauser}, then violations of a Bell inequality have been observed in photonic systems \cite{aspect,weihs}. Without this extra assumption, then ion-based systems have got around the detection loophole but suffer from the locality loophole \cite{rowe}. At the time of writing this thesis, completely loophole-free Bell inequality violations have not been observed. Although, there are promising avenues for future experimental work \cite{monroe,vertesi}. In chapter \ref{chap4}, we will give a more thorough discussion of loopholes in Bell tests.

\subsection{The GHZ Paradox}

Bell's theorem can be proven using the now-famous Bell inequality. Did we need to construct this expression? There have been several arguments which have shown that quantum mechanics is incompatible with local realism but without use of a Bell inequality. For example, in 1983, Heywood and Redhead \cite{heywood} developed a proof that local realism cannot be compatible with the statistics of two space-like separated, yet entangled spin-$1$ systems. This proof relied on an argument of determinism, in the spirit of the original EPR argument \cite{epr}. Later in 1989, Greenberger, Horne, and Zeilinger (GHZ) developed a proof of Bell's theorem without inequalities for three space-like separated parties \cite{ghz}. The GHZ argument has subsequently been developed by Mermin\footnote{This argument was a development of a proof that quantum mechanics is ``contextual'' by Asher Peres developed into a proof of Bell's theorem.} \cite{mermin,mermin2}. Another notable example of Bell's theorem without the inequality is ``Hardy's Paradox'' which can be seen as a ``possibilistic'' proof, i.e. some things are possible in quantum mechanics that are not possible with locally realistic theories \cite{hardy3}. We now present the GHZ argument, or ``GHZ paradox'' to which it is often referred, as a simple and beautiful proof of Bell's theorem.

We have three, space-like separated parties who (like in the CHSH construction) each have a completely free choice of measurement from a set of two measurements. We label the two measurements for the $j$th site $\mathcal{M}_{j}^{0}$ and $\mathcal{M}_{j}^{1}$ and each measurement takes one of two possible outcomes $\pm 1$. As with the CHSH construction, each outcome is then a result of some objective property of each party's local system (which may have been shared in the past). Therefore, each measurement $\mathcal{M}_{j}^{k}$ again is assigned the value $\chi_{j}^{k}\in\{\pm 1\}$. Again we are interested in the correlations $\mathbb{E}(\mathcal{M}_{1}^{k}\mathcal{M}_{2}^{l}\mathcal{M}_{3}^{m})$ where $k$, $l$, $m\in\{0,1\}$. If we now obtain the following deterministic correlations for a particular set of measurements
\begin{eqnarray}\label{ghzassign1}
\mathbb{E}(\mathcal{M}_{1}^{0}\mathcal{M}_{2}^{0}\mathcal{M}_{3}^{0})=-1\\\label{ghzassign2}
\mathbb{E}(\mathcal{M}_{1}^{0}\mathcal{M}_{2}^{1}\mathcal{M}_{3}^{1})=-1\\\label{ghzassign3}
\mathbb{E}(\mathcal{M}_{1}^{1}\mathcal{M}_{2}^{0}\mathcal{M}_{3}^{1})=-1
\end{eqnarray}
then we can assign values of $\chi_{j}^{k}$ deterministically to $\mathbb{E}(\mathcal{M}_{1}^{k}\mathcal{M}_{2}^{l}\mathcal{M}_{3}^{m})=\chi_{1}^{k}\chi_{2}^{l}\chi_{3}^{m}$. If we multiply rows \eqref{ghzassign1}, \eqref{ghzassign2}, and \eqref{ghzassign3} together after they have been assigned values of $\chi_{j}^{k}$ and observe that $(\chi_{j}^{k})^{2}=1$, then in a locally realistic theory, we \emph{must} obtain
\begin{equation}\label{contra}
\mathbb{E}(\mathcal{M}_{1}^{1}\mathcal{M}_{2}^{1}\mathcal{M}_{3}^{0})=\chi_{1}^{1}\chi_{2}^{1}\chi_{3}^{0}=-1.
\end{equation}
However, measurements on an entangled quantum state can satisfy \eqref{ghzassign1}, \eqref{ghzassign2}, and \eqref{ghzassign3} but contradict \eqref{contra}. The entangled state consists of three qubits
\begin{equation}
|\Psi\rangle_{\textrm{GHZ}}=\frac{1}{2}\left(-|00+\rangle+|01-\rangle+|10-\rangle+|11+\rangle\right), 
\end{equation}
where each $j$th site has one of these qubits and performs the measurements $\mathcal{M}_{j}^{0}=\hat{X}$ or $\mathcal{M}_{j}^{1}=\hat{Z}$. Calculating all expectation values, the statistics from these measurements on the state $|\Psi\rangle_{\textrm{GHZ}}$ satisfy correlations in \eqref{ghzassign1}, \eqref{ghzassign2}, and \eqref{ghzassign3}. However, 
\begin{equation}\label{contra2}
\mathbb{E}(\mathcal{M}_{1}^{1}\mathcal{M}_{2}^{1}\mathcal{M}_{3}^{0})=\langle\hat{Z}\otimes\hat{Z}\otimes\hat{X}\rangle=+1,
\end{equation}
thus contradicting \eqref{contra}. These quantum correlations have deterministically shown that local realism is inconsistent with quantum mechanics.

Mermin showed that we can still construct a Bell inequality from the correlations of the GHZ argument \cite{mermin}. We construct the following inequality 
\begin{equation}\label{ghzineq}
-\mathbb{E}(\mathcal{M}_{1}^{0}\mathcal{M}_{2}^{0}\mathcal{M}_{3}^{0})-\mathbb{E}(\mathcal{M}_{1}^{0}\mathcal{M}_{2}^{1}\mathcal{M}_{3}^{1})-\mathbb{E}(\mathcal{M}_{1}^{1}\mathcal{M}_{2}^{0}\mathcal{M}_{3}^{1})+\mathbb{E}(\mathcal{M}_{1}^{1}\mathcal{M}_{2}^{1}\mathcal{M}_{3}^{0})\leq 2,
\end{equation}
which the correlations in \eqref{ghzassign1}, \eqref{ghzassign2}, \eqref{ghzassign3} and \eqref{contra} satisfy. Mermin showed that this inequality is satisfied for all locally realistic theories \cite{mermin}, but the quantum mechanical correlations described above give a value of $4$ for the left-hand-side of \eqref{ghzineq}. We will show in section \ref{sec14} that the GHZ-Mermin argument against local realism in quantum physics will be relevant to discussion about quantum information. 

The CHSH inequality and the GHZ argument are ways of putting constraints on what is possible in a classical, or more generally, a locally realistic theory. The fact that quantum mechanics predicts contradictions to both constraints gives a remarkable departure from a classical view of the world. It indicates that when we are utilising the quantum mechanical formalism we can produce non-classical phenomena. One of the most enticing prospects for quantum mechanics is to use non-classical behaviour to perform some useful task that we could not achieve with classical resources. This motivation has led to the relatively nascent field of ``quantum information science'' \cite{nielsen}. One of the goals of this field is to process information via computation or communication and use quantum mechanical systems to do this ``better'' than with classical resources. In the next section we will give a broad overview of the field and how quantum systems could out-perform classical systems.

\section{Quantum Information Processing}\label{sec13}

\noindent
We have seen how quantum physics can be seen as non-classical in some concrete sense. Quantum information science has been developed to answer whether the non-classicality of quantum physics can be used to perform information processing tasks thought difficult or intractable with classical physical systems \cite{nielsen}. We now give a broad, and incomplete, overview of the field of quantum information in order to show that quantum resources can be useful for information processing. 

The history of quantum information is itself an interesting topic for discussion. Stephen Wiesner developed the idea of ``conjugate coding'' circa 1970 but the result was not published until the 1980s \cite{wiesner}; this idea went on to influence the field of quantum cryptography. Alexander Holevo published his famous theorem in 1973 limiting the classical information in, say, a qubit to being at most one classical bit \cite{holevo}. Holevo's theorem is one of the most significant results in information theory applied to quantum systems and quantum ``channels''\footnote{Quantum channels consist of the positive linear operators on some ``input'' quantum state, mapping this state to another state. Perhaps this channel is a perfect communication channel for qubits and so would be the identity operator $\mathbb{I}$.}. The idea of a ``quantum computer'', or some quantum system capable of performing computations was first suggested by Richard Feynman in 1982 \cite{feynman}; the work of David Deutsch later formalised this concept \cite{deutsch}. Wootters and Zurek showed that unknown quantum information cannot be copied, called the ``no-cloning'' theorem \cite{wootters}. In the light of all of this work, we begin our discussion in the next subsection in 1984, with the seminal work by Bennett and Brassard (BB) on quantum cryptography \cite{bb84}. This work by BB brought together the ideas of the no-cloning theorem and conjugate cloning in a simple yet powerful way.

\subsection{Quantum Cryptography}

Two parties, referred to as Alice and Bob\footnote{These two characters have a long and auspicious career in computer science. Such is their success that the quantum information community talk often of Alice and Bob in quantum information procedures.} want to communicate to each other without fear of eavesdroppers intercepting their messages. Alice encodes her message into another message or ``ciphertext'' with a ``key'' that Bob knows but no-one else does. Bob can use the ``key'' to unlock Alice's message from the ciphertext. An eavesdropper can try and guess or calculate the key, but if it is random and Alice applies the ``one-time pad'', then a message can be made perfectly secure as defined by Shannon \cite{shannon}. The one-time pad consists of one bit of a message $x\in\{0,1\}$ being added (modulo $2$) to a random bit $r\in\{0,1\}$ giving $y$, i.e. $y=x\oplus r$ where $\oplus$ represents modulo $2$ addition. We need at least as many random bits as there are bits in the message, but as long as Bob knows every one of these random bits he can recover $x$ by adding (modulo $2$) $r$ to $y$ as $y\oplus r=x \oplus r\oplus r=x$. Shannon showed that this makes the ciphertext secure if an eavesdropper cannot obtain all values of $r$ \cite{shannon}.

How does Alice share the key consisting of the values of $r$ to Bob? Since their goal was to communicate securely in the first place, they must find a secure way so that each party can communicate the random key. In 1984, BB showed that the combination of publicly communicating quantum states $|\psi\rangle$ from Alice to Bob and publicly communicating classical information about these states between Alice and Bob, secure values of $r$ can be generated \cite{bb84}. An eavesdropper cannot perfectly copy the state that is publicly communicated by the no-cloning theorem, so must make a measurement to learn $|\psi\rangle$. The security is partly based on the fact that when an eavesdropper makes a measurement on the quantum state that is sent from Alice to Bob, they project the state into another state which may be different from $|\psi\rangle$. If the eavesdropper projects into a different state, Alice and Bob can compare measurement outcomes on the state to detect this. If Alice and Bob proceed with a particular protocol, with public quantum and classical communication, they can generate a secure random key. We then describe this as a method of ``quantum key distribution'' (QKD).

In 1991, Artur Ekert developed another method of QKD that utilised entanglement \cite{ekert}. This result alongside the discovery of ``quantum teleportation'' \cite{bennett} based upon sharing entanglement and classical communication led, in earnest, to entanglement being investigated as a resource for quantum information processing. Ekert based his protocol on a modified version of the CHSH Bell inequality test where Alice and Bob each receive one-half of the bipartite entangled state $|\Psi\rangle_{\textrm{EPR}}$. The intuition behind the protocol is that a key is revealed by the act of space-like separated measurements on this entangled state; if a key existed before measurement it would be an ``element of reality'' and so incompatible with an entangled state. 

For a given choice of measurements as discussed in the EPR paradox, outcomes are perfectly correlated generating a shared random bit. Alice and Bob randomly choose measurements and announce the choice after receiving measurement outcomes. An eavesdropper can intercept the quantum state before it reaches either Alice or Bob and make a measurement, but this interception leaves the state in a separable state. They use the CHSH inequality to confirm that the state is entangled when they make measurements on it. Therefore, the protocol requires that the state is entangled and the CHSH inequality just confirms this, the security of the original 1991 protocol does not hinge directly on the incompatibility with local realism. Remarkably, in the spirit of Ekert's intuition, Barrett, Hardy and Kent designed a protocol whereby security was guaranteed by a Bell inequality violation \cite{barrett7}. Ac\'{i}n et al then made the connection to the original CHSH inequality that Ekert used (without assuming the quantum state shared), to confirm the security of a key \cite{acin3}. 

\subsection{Quantum Computing}

If one does not use quantum cryptographic means to establish secure communication, then what means are there to establish a secure key? One of most commonly used tools is the Rivest-Shamir-Adleman (RSA) algorithm which is based upon a computational premise \cite{rsa}. It is believed that it is hard for computers to find the prime factors of a large number. The RSA algorithm involves a public and a private key, where Alice makes the product of two large primes public and keeps these factors private. Bob receives the public key, encodes his message using it and sends his ciphertext to Alice in such a way that it can only be decrypted using Alice's private data. Therefore, if one can find the two prime factors of the public key efficiently, one can decode the message. However, as we mentioned, it is believed that this cannot be done efficiently with current computers and Alice receives the information from Bob securely. The RSA algorithm, as a result, is used quite successfully in many internet-based financial transactions.

Remarkably, if one could build a computer that works on quantum mechanics, a quantum computer\footnote{Current desktop PCs rely on quantum theory to describe their workings. A quantum computer full exploits the quantum formalism and is based on the postulates of the theory.}, one could find the prime factors of a large number efficiently, thus breaking the RSA algorithm. The algorithm for finding these prime factors was invented by Peter Shor in 1994 \cite{shor} and became a key motivator for building a quantum computer. 

In 1985, David Deutsch described a universal quantum computer which can perform any possible quantum computation \cite{deutsch}. A computation can be described in the ``circuit model'' of quantum computation where a quantum state consisting of $n$ qubits is prepared in the product state $|\textbf{0}\rangle=\bigotimes_{j=1}^{n}|0\rangle$ \cite{nielsen}. A computation then consists of a sequence of unitary operations performed on these qubits. Each unitary is considered $1$ computational step, or ``gate''. After the requisite number of unitary operators is performed, some, or all of the $n$ qubits can be measured. 

Various algorithms have been designed for quantum computers indicating a potential improvement in computational time over classical computers \cite{dj,shor,grover,harrow}. This improvement is conjectured in computational complexity terms as we currently do not even know the power of classical computers \cite{pap}. If quantum computers are more powerful than classical computers, then it would be of interest to know what aspect of quantum mechanics gives this improvement. It might even be the case that this property of quantum mechanics can assert the assumed separation between quantum and classical computers. Jozsa and Linden showed that in quantum computations on pure states, unbounded entanglement is necessary if there is to be a computational speed-up \cite{jozsa}. This does not mean that if there is entanglement in pure state quantum computation, the circuit cannot be simulated efficiently on a classical computer. A ``Clifford circuit'' is an example of a such a circuit that can be simulated efficiently with a classical computer \cite{nielsen,aaronson}.  It has also been shown by Vidal that if entanglement is bounded, then the quantum computation can be simulated efficiently classically \cite{vidal}. For quantum computations on mixed states, which will be those that are performed in the laboratory, the role of entanglement is unknown or possibly not even relevant \cite{jozsa,datta}.

\subsection{Measurement-based Quantum Computing}

There are several models of quantum computing that are equivalent to the circuit model of quantum computing\footnote{Equivalence means that every computation in one model can be efficiently simulated in another model.} \cite{mbqc1,zanardi,aharonov,kitaev,leung}. In one particular class of models, the presence of entanglement is by construction a key ingredient in performing a computation. This is the class of models of Measurement-based Quantum Computing (MBQC) \cite{mbqc1,mbqc2,jozsa}. One of the origins of this model can be seen in the teleportation-based quantum gate model developed by Gottesman and Chuang \cite{gottesman,nielsen2,leung}. In teleportation-based quantum computing, $n$ parties share bipartite maximally entangled states with their nearest neighbours, and make measurements at each site. Richard Jozsa has shown that this model is equivalent to a model proposed by Raussendorf and Briegel (RB) in 2001 \cite{jozsa,mbqc1}.

The model of MBQC proposed by RB consists of a multipartite entangled state, or ``resource state'' shared by $n$ parties \cite{mbqc1}. This state is the ``cluster state'' consisting of $n$ qubits \cite{mbqc1,mbqc2}. However, in our discussion, we allow any possible state to be shared by these parties (see section \ref{sec34}) and do not restrict this aspect of MBQC. Also, from now on when we mention MBQC, we make it synonymous with the original RB construction but with any possible resource state \cite{anders}. The computation proceeds by each site performing a measurement with two outcomes $\pm 1$ on their respective system. Measurements on, say, cluster states can have completely random outcomes, but a set of gates in the quantum circuit model corresponds to a set of unitary operators \cite{mbqc1}. The unitary evolution of a state is a deterministic operation. Remarkably, one can achieve determinism in MBQC by applying corrections at the end of the measurements and making measurements adaptive \cite{jozsa}. That is, the choice of measurement during the computation must be dependent on previous measurement outcomes. 

To take into account this correction and adaptivity, a crucial component of MBQC is needed: the ``classical control computer'' \cite{mbqc2,mbqc3,anders}. This computer is a classical processor and processes bits corresponding to a choice of measurement and its respective outcome at each site. In the model of RB, one just needs a choice between two measurements at each site to get a universal quantum computer labelled by bit-values. The outcomes of a measurement are $\pm 1=(-1)^{x}$ as described above where $x$ is now a bit-value. In MBQC as formulated by RB, the control computer does not require all possible operations, or gates in classical computing. In fact, as pointed out by Anders and Browne, all that is needed is modulo $2$ addition between classical data \cite{anders}. Using only these operations, a computer cannot perform all logical, or Boolean operations, and is therefore not \emph{functionally complete} for all classical computations.

\subsection{Entanglement as a Resource}

In MBQC, entanglement can be seen as a resource that is ``consumed'' via single-qubit measurements \cite{mbqc3}. This model also provides a nice distinction between the quantum and classical parts of computation; the quantum part being the measurements on a quantum state and the classical control computer providing some, albeit limited processing to utilise this measurement data. 

The idea of entanglement being a resource for information processing that is consumed can be seen in many aspects of quantum information \cite{horodecki}. Historically, beginning with entanglement as a resource for producing secure keys, then used as a channel for communicating quantum states via teleportation \cite{ekert,bennett}. The interplay between quantum gates and teleportation as highlighted by Gottesman and Chuang, also highlights the role of entanglement with respect to computation \cite{gottesman,jozsa}. Also relevant to quantum computation and communication, entanglement has been utilised as a resource for correcting errors \cite{brun}. 

Inspired by these information processing tasks, the resource theory of entanglement has been developed \cite{horodecki}. If parties are restricted to being only able to perform local operations on their respective subsystem and communicating classical information (LOCC), then they cannot produce an entangled quantum state \cite{horodecki}. Therefore, if parties have an entangled state, they can do tasks that they otherwise could not do with only LOCC. This theory has become well-developed and we refer the reader to \cite{horodecki} for a review of entanglement in quantum information. 

\section{Bell Inequalities and Quantum Information}\label{sec14}

\noindent
Since entanglement is a resource for information processing, and entanglement was used to show an incompatibility of quantum physics with local realism, can this incompatibility also be used as a resource? In recent years, the answer to this question has been answered in the affirmative. The intuition behind Ekert's 1991 QKD protocol that if a key is some element of reality held by each party then an eavesdropper can threaten security and learn this data \cite{ekert}. As mentioned, Barrett, Hardy and Kent developed this intuition \cite{barrett7} and then Ac\'{i}n et al made the connection between security and Bell's theorem concrete \cite{acin3}. They showed that if we put no constraint on the devices that Alice and Bob use (these devices can even be produced by the eavesdropper), then the security of a key can be established \emph{directly} by the violation of a Bell inequality. This is an example of ``device-independent'' quantum information processing \cite{mayers,acin3,pironio}, and we now review this nascent field very briefly.

\subsection{Device-independent Quantum Information}

A violation of a Bell inequality indicates that if we assume our system is quantum mechanical, then the shared quantum state was entangled. Therefore, it is natural to say that a violation must detect entanglement without making any assumption on the system. Indeed this idea has been developed both in the bipartite and multipartite setting where a Bell inequality is used as a ``witness'' of entanglement \cite{liang,rabelo}. In calculating entanglement of a state directly, one calculates this quantity directly from the state. However, if we do not know the state and we observe a violation, then it must be entangled\footnote{As well as this device-independent approach to entanglement, Bell inequalities can be used to gain information about the dimension of a quantum sytem \cite{gallego}.}.

There are two aspects of bipartite entanglement that make it useful for QKD. The fact that random, yet completely correlated outcomes can be generated for the shared key, and if measured, the system will no longer be entangled. The second fact ensures that the randomly generated key is securely generated. But randomness is in of itself a useful resource for many tasks \cite{knuth}, including secure key distribution and cryptography in general \cite{shannon}. For example, in a Monte Carlo simulation of complicated systems, a random source is required to pick a data point at random on which to calculate something \cite{metropolis}. Also, randomly sampling from a probability distribution to perform statistical analysis is useful for ruling out statistical bias in this analysis. 

Genuinely random processes are difficult to come by as classical physical systems are seemingly random due to lack of knowledge about all parameters of the systems. The underlying parameters of the system have deterministic properties but our inability to access all of them leads to the assignment of probabilities. This form of randomness can be seen as not true randomness due to the underlying determinism, but ``pseudorandomness'' \cite{knuth}. However, if we assume that locality must be respected then the random outcomes of observables on either side of a bipartite, space-like separated maximally entangled state cannot be due to some underlying real parameters. The randomness of the maximally entangled state is a good source of randomness.

If we do not assume that we have a maximally entangled state shared between two parties, Pironio et al showed that true randomness can be generated from the violation of a Bell inequality \cite{pironio}. This randomness from a violation can then be used as a ``seed'' to generate something more random. Therefore, randomness can be generated without assuming anything about the underlying system that can possibly generate it, and so is device-independent. The generation of random numbers \cite{pironio,colbeck} and cryptography \cite{mayers,acin3,pironio2,silman} are two main current implementations of device-independent protocols. The motivation behind both of these tasks comes from cryptography, but in the next section we give an example of computing based on a violation of a Bell inequality in the form of the GHZ paradox.

\subsection{GHZ Paradox and Measurement-based Quantum Computing}

Models of computing have been related to Bell inequalities. Communication complexity is a model where we have several parties and each party has unbounded computational power \cite{kushilevitz}. Each party has some data and the goal is to compute some function on all of this data. The question is whether all of this data needs to be sent between parties in order for the function to be computed? Communication complexity studies the minimum amount of communication needed to calculate a particular function. If a system violates a particular Bell inequality, then it can exhibit an advantage in a communication complexity task over a system that does not violate a Bell inequality \cite{brukner,buhrman}. Another example of a computational model related to Bell inequalities is a ``non-local game'' \cite{nonlocal}. We will discuss these models in sub-section \ref{sec33b} of chapter \ref{chap3} and so postpone discussion of the model until then. 

In MBQC, the classical control computer can only perform addition modulo $2$, or ``XOR gates'' as they are called in the Boolean circuit model of classical computing \cite{anders}. In order to have a full power classical computer, we require another gate: the ``NAND gate'' \cite{pap}. The XOR gate on two bits $x_{1}$ and $x_{2}$ is the function $f(x_{1},x_{2})=x_{1}\oplus x_{2}$ but the NAND gate is $f(x_{1},x_{2})=1\oplus x_{1}x_{2}$ where this function is $0$ for $x_{1}=x_{2}=1$ and $1$ otherwise. Anders and Browne (AB) showed that in MBQC a NAND gate can be performed with three measurement sites and a single round of measurements \cite{anders}. This three-party system is also the minimal resource in MBQC that can produce this function. AB used the GHZ paradox to demonstrate this result and we now review this result \cite{anders}. 

If we inspect the correlations in \eqref{ghzassign1}, \eqref{ghzassign2}, \eqref{ghzassign3} and \eqref{contra2}, the choice of measurements can be labeled by bit-values $s_{j}\in\{0,1\}$ at each $j$th site. We relabel the values $s_{1}$ and $s_{2}$ to be some bit-values $x_{1}$ and $x_{2}$ respectively. Therefore, for the specific correlations in \eqref{ghzassign1}, \eqref{ghzassign2}, \eqref{ghzassign3} and \eqref{contra2}, the choice of third measurement $s_{3}$ is equal to $x_{1}\oplus x_{2}$. This can be modeled as a computation in MBQC where the classical control computer sets the choice of measurement on the first two sites to be $x_{1}$ and $x_{2}$, and $x_{1}\oplus x_{2}$ for the third site; the classical control computer calculates this third choice. Then if the measurements and quantum state are those in the GHZ paradox then we observe the correlations are
\begin{equation}
\mathbb{E}(\mathcal{M}_{1}^{x_{1}}\mathcal{M}_{2}^{x_{2}}\mathcal{M}_{3}^{x_{1}\oplus x_{2}})=(-1)^{x_{1}x_{2}\oplus 1}.
\end{equation}
The function corresponding to the NAND gate then appears on the right-hand-side. We obtain the measurement outcome from the $j$th site as $(-1)^{m_{j}}$ where $m_{j}\in\{0,1\}$ and then the joint outcome of all three parties is $(-1)^{m_{1}\oplus m_{2}\oplus m_{3}}$. Then every instance of $\mathcal{M}_{1}^{x_{1}}\mathcal{M}_{2}^{x_{2}}\mathcal{M}_{3}^{x_{1}\oplus x_{2}}$ must deterministically produce an outcome $(-1)^{m_{1}\oplus m_{2}\oplus m_{3}}$ equal to $(-1)^{x_{1}x_{2}\oplus 1}$. If each site sends the value $m_{j}$ to the classical control computer then it can calculate $m_{1}\oplus m_{2}\oplus m_{3}$ thus obtaining $x_{1}x_{2}\oplus 1$ deterministically. 

We have shown that classical correlations cannot reproduce the above quantum correlations. In order that we produce a NAND gate with classical correlations communication in the form of adaptivity is required in the measurement-based circuit. This does not minimise the resources required for a full classical computer and shows that correlations that are incompatible with local realism are useful in MBQC. These ideas will be developed further in section \ref{sec34} of chapter \ref{chap3}.

\section{Chapter Summary}

\noindent
We have introduced the quantum formalism and shown that it has an interesting mathematical consequence: entanglement. Not only is entanglement a mathematical curiosity, it has consequences for our understanding of quantum theory. In particular, it challenges the notion of local realism that is satisfied in classical physical systems \cite{bell}. The advent of quantum information placed entanglement in yet another context, that as an information theoretic resource \cite{horodecki}. It then has become an interesting research avenue to link the incompatibility with local realism with a potential information theoretic advantage. This has led to the development of device-independent quantum information.

Finally we have an indication of the possible applications of Bell inequalities to some computational models. When implementing a test of a Bell inequality, or Bell test, measurements must be space-like separated and classical communication is ruled out. Computation often involves time ordering of operations, or gates, and this time-ordering allows the possibility of communication between parties. However, when we process statistics from Bell tests, we are performing a computation on this data, and in the example of the GHZ paradox we can use this processing to obtain something ``useful'' from quantum correlations \cite{anders}. Correlations then are computations in this example after this processing. Throughout this thesis, this picture will become central to our understanding of correlations. That is, correlations can be used to compute particular functions on an ``input'' corresponding to the choice of measurement settings at all sites.

This computational insight on Bell tests will be used to give a new perspective on established ideas in Bell tests as well as new ideas for Bell tests. We will review the issue of loopholes in Bell tests \cite{pearle} and indicate that they have a computational interpretation that makes this subject more amenable pedagogically. Motivated by these issues, we describe a way of expanding Bell tests to include processing of statistical data but without introducing loopholes. We give the notion of a loophole a more technical grounding and present these results in chapter \ref{chap4}. 

We have hinted at a connection between Bell tests and MBQC. We extend this connection and make it more concrete in chapter \ref{chap3} by discussing MBQC without adaptivity. Using Bell tests we can actually say something about the power of MBQC without adaptivity as well as showing that quantum physics can do something that classical physics cannot. In chapter \ref{chap4} we will discuss whether MBQC with adaptivity can be framed in terms of a Bell test. We give some indication that this is possible using a method of loophole-free data processing. Before we discuss applications of the Bell test to computation and vice versa, in the next chapter we introduce our framework for Bell tests.

\chapter{Correlators and Bell Tests}\label{chap2}

\noindent
In this chapter, we will lay the foundations for our study of Bell tests \cite{bellineq}. More precisely we will motivate the study of what we call correlators: the statistics of the joint outcome of many parties. We have already discussed (in section \ref{sec12}) correlators in terms of the expectation value of measurements made in the Bell-CHSH test \cite{chsh}. We now describe correlators in terms of conditional probabilities of a joint outcome given some measurement settings. We will make the connection to the Bell-CHSH test concrete and show that considering these conditional probabilities allows for greater scope when considering a broad class of Bell tests.

One prominent tool utilised in this chapter is to describe the correlators, or conditional probabilities as \emph{stochastic maps} from a set of inputs (describing the measurement settings) to a set of outputs (describing corresponding measurement outcomes). These maps are then probabilistic maps from an input to an output. We will define particular classes of functions and show how they relate to correlators resulting from particular physical theories, more specifically locally realistic and quantum theories. 

The famous Bell inequality emerges from a discussion on the geometry of stochastic maps. Correlators can be represented as vectors in a real vector space; every vector is a list of conditional probabilities for each joint outcome for every choice of measurement settings. In this real space, the space of LHV correlators can be defined as a \emph{convex polytope}, an object which is the convex hull of a finite number of correlators (called extreme points) \cite{grunbaum}. The boundary, or surface of a convex polytope is made up of objects called faces. If the dimension of the space a polytope lives in is $\Delta$, then a $(\Delta-1)$-dimensional face is called a \emph{facet} and facets are defined by particular linear inequalities. These inequalities define half-spaces in the $\Delta$-dimensional real space, and the intersection of these half-spaces also define a convex polytope \cite{grunbaum}. For the convex polytope of locally realistic correlators, the linear inequalities that define its facets are the facet Bell inequalities \cite{pitowski,froissart,peres,fine}. 

We combine this geometric picture of correlators with the discussion of stochastic maps or functions and show that we can describe Bell tests in terms of computations. This computational aspect allows us to capture locally realistic correlators in terms of computational expressiveness. Not only is this method used to describe correlators, it is used to say something about the full probability of distribution for all possible measurement outcomes and settings. In particular, correlators single out particular probability distributions that only satisfy special relativity (non-signalling) and no other physical constraints \cite{pr}. Finally, we also characterise correlators, and correlations in general that appear in a model constructed originally by George Svetlichny \cite{svet}. This model allows a sub-set of parties to share unconstrained correlations but satisfy local realism with respect to others.

This chapter in the main motivates the study of correlators as a simplification from studying the full statistics of a Bell test. Despite the simplification, the study of correlators yields significant insights into the study of the full probability distribution. We also establish the framework upon which results in later chapters are built. Section \ref{sec21} consists of review material, and the work in section \ref{sec22} introduces a new computational framework for correlators. Section \ref{sec23} consists of new results describing non-signalling correlations and section \ref{sec24} recasts Svetlichny correlations in terms of a computational description. The original work in sections \ref{sec22}, \ref{sec23} and \ref{sec24} was completed in collaboration with Joel Wallman and Dan Browne and published as \cite{hoban3}.

\section{A General Framework for Bell tests}\label{sec21}
\noindent
Bell tests are carried out by space-like separated parties that each make a choice from a set of measurements and each measurement produces an outcome from a set of possible outcomes \cite{bellineq}. From this starting point, it has been insightful to think of Bell tests, and other physical processes from an operational point-of-view \cite{hardy,hardy2,barrett1}. In an operational framework, each measurement site is an abstract object, often referred to as a ``box'', that takes an ``input'' as the choice of measurement setting and returns an ``output'' in the form of a measurement outcome. Operationally then, we only concern ourselves with the statistics resulting from these boxes and not necessarily their ``inner-workings''. We only want to infer the properties of these boxes from their statistics making minimal assumptions.

\begin{figure}
	\centering
		\includegraphics[width=0.70\textwidth]{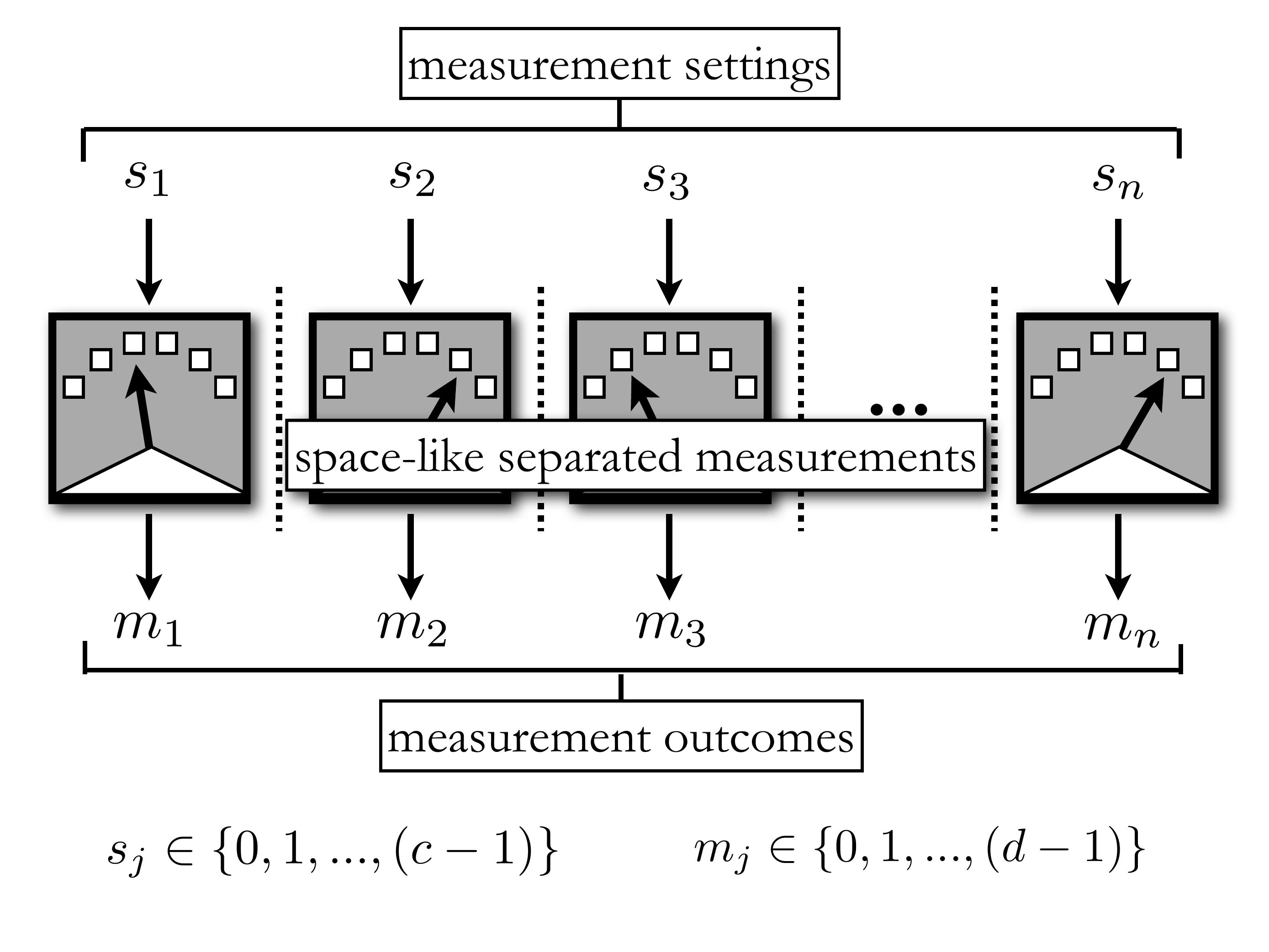}
			\caption{In a Bell test, $n$ parties each make a measurement from $c$ possible choices, where each measurement has $d$ possible outcomes. Labelling the $j$th party's measurement choice and outcome by $s_j$ and $m_j$ respectively, we can describe each run of the experiment with $n$-digit strings $\textbf{m}$ and $\textbf{s}$. (Copyright: American Physical Society, 2011).}
			\label{fig:fig1}
\end{figure}

We then consider $n$ space-like separated parties, or boxes. Each $j$th site for $j\in\{1,2,...,n\}$ makes a measurement $\mathcal{M}_{s_{j}}$ from a choice of $c_{j}$ measurements where $s_{j}\in\{0,1,...,(c_{j}-1)\}$ labels the choice of measurement and is expressed in terms of an integer, or digit in $\mathbb{Z}_{c_{j}}$, the cyclic group of $c_{j}$ elements. Each measurement $\mathcal{M}_{s_{j}}$ has $d(s_{j})$ possible outcomes $\mathcal{O}_{m(s_{j})}$, where $m(s_{j})\in\{0,1,...,(d(s_{j})-1)\}$ is an element of $\mathbb{Z}_{d(s_{j})}$, the cyclic group of $d(s_{j})$ elements. Therefore, in operational terms, each $j$th box takes an input $s_{j}$ and returns an output $m(s_{j})$ for each input. From now on, we assume that $d(s_{j})=d_{j}$ is constant for all measurements labelled by $s_{j}$. We include a schematic of the Bell test in Figure \ref{fig:fig1}.

Inputs into all $n$ boxes are represented by $n$-length digit-strings $\textbf{s}\in\bigoplus_{j=1}^{n}\mathbb{Z}_{c_{j}}$, the Cartesian product of all sites' inputs $\textbf{s}=\{s_{1},s_{2},...,s_{n}\}$ \footnote{We are using this perhaps unconventional notation for the Cartesian product for the sake of brevity. Typically the Cartesian product between sets $A$ and $B$ is represented as $A \times B$ and $A_1 \times ... \times A_n$ for an $n$-fold Cartesian product of sets $A_j$. In this non-standard notation $A_1 \times ... \times A_n=\bigoplus_{j=1}^{n}A_j$.}. All digit-strings will be expressed in bold typeface, with $j$th elements not in bold, but with sub-script $j$. Outputs are then expressed as $n$-length digit-strings $\textbf{m}=\{m_{1},m_{2},...,m_{n}\}$ where we have not explicitly written the dependence on $\textbf{s}$ for brevity, but this dependence is implied. That is, every output is always a particular output $\textbf{m}$ for a given input $\textbf{s}$. Given non-empty sub-sets $\mathcal{J}\subseteq\{1,2,...,n\}$ of all $n$ with $|\mathcal{J}|$ being the number of parties in the sub-set, the outputs of this sub-set is written as $\textbf{m}^{\{j|j\in\mathcal{J}\}}=\{m_{j}|j\in\mathcal{J}\}$ such that singleton sub-sets are the elements $m_{j}$. The same notation is applied also to the inputs $\textbf{s}$ with $\textbf{s}^{\{j|j\in\mathcal{J}\}}=\{s_{j}|j\in\mathcal{J}\}$ being the inputs on a sub-set $\mathcal{J}$. When $\mathcal{J}$ includes all $n$ parties then we recover $\textbf{m}$ as the output again. There are then $(2^{n}-1)$ of these non-empty sub-sets $\mathcal{J}$.

Central to the standard construction of Bell tests is that choice of measurement setting is independent of anything else in the experiment \cite{bellfree}. In other words, the choice of measurement is completely random, i.e. $p(\textbf{s})=\prod_{j=1}^{n}\frac{1}{c_{j}}$. The consequences of relaxing the constraint of measurement independence have been shown to be detrimental to Bell tests \cite{barrett2,hall}.

As mentioned, in Bell tests, statistics are calculated from the data obtained from the boxes. The statistics are the conditional probabilities $p(\mathbf{m}^{\{j|j\in\mathcal{J}\}}|\textbf{s})$, the probability of obtaining outputs $\mathbf{m}^{\{j|j\in\mathcal{J}\}}$ given the input $\mathbf{s}$ for all sub-sets $\mathcal{J}$ of $n$ parties. Crucially though we can obtain every probability $p(\mathbf{m}^{\{j|j\in\mathcal{J}\}}|\textbf{s})$ for a proper sub-set $\mathcal{J}$ from the full distribution $p(\mathbf{m}|\textbf{s})$ by taking a sum of outcomes on the complement sub-set $\mathcal{J}^{c}=\{j|\{1,2,....,n\}\setminus\mathcal{J}\}$ to $\mathcal{J}$ of all $n$ parties, i.e. $p(\mathbf{m}^{\{j|j\in\mathcal{J}\}}|\textbf{s})=\sum_{j\in\mathcal{J}^{c}}p(\textbf{m}|\textbf{s})$. Therefore from now on we only need to consider the full probability distribution $p(\textbf{m}|\textbf{s})$. In the following sub-section, we discuss the basic geometric objects that will dominate our discussion of correlations: the convex polytope.

\subsection{Convex Polytopes and Stochastic Maps}\label{sec21a}

The conditional probabilities $p(\textbf{m}|\textbf{s})$ are stochastic maps producing the map, or function $f: \bigoplus_{j=1}^{n}\mathbb{Z}_{c_{j}}\rightarrow \bigoplus_{j=1}^{n}\mathbb{Z}_{d_{j}}$ with some probability. Throughout this thesis, we will use a geometric picture to consider these (and other forms of) stochastic maps. These conditional probabilities $p(\textbf{m}|\textbf{s})$ are elements of a vector $\vec{p}$ in a real vector space. We can reduce the number of probabilities we need to consider by the normalisation condition that $\sum_{\textbf{m}}p(\textbf{m}|\textbf{s})=1$ where $p(\textbf{0}|\textbf{s})=1-\sum_{\textbf{m}\neq\textbf{0}}p(\textbf{m}|\textbf{s})$. Vectors $\vec{p}$ have length $D=(\prod_{j=1}^{n}d_{j}-1)\prod_{j=1}^{n}c_{j}$ in $\mathbb{R}^{D}$ real space \footnote{This can be seen from the fact that we have $\prod_{j=1}^{n}c_{j}$ normalisation conditions (one for each input string) and $\prod_{j=1}^{n}d_{j}\prod_{j=1}^{n}c_{j}$ original probabilities $p(\textbf{m}|\textbf{s})$.}. Since the elements of $\vec{p}$ are probabilities, they will live in a bounded sub-space in $\mathbb{R}^{D}$ satisfying the constraints that all $p(\textbf{m}|\textbf{s})\geq 0$ and $\sum_{\textbf{m}\neq\textbf{0}}p(\textbf{m}|\textbf{s})\leq 1$; the positivity and normalisation constraints respectively. These inequalities essentially describe a \emph{convex polytope} in $\mathbb{R}^{D}$ which we call $\mathcal{F}$. Convex polytopes will be a central part of this thesis for all manner of different real spaces and so we shall define them for all possible real spaces now. 

A convex polytope $\mathcal{C}$ in a real space $\mathbb{R}^{\Delta}$ of dimension $\Delta$ can be defined in two ways: first is the \emph{half-space representation} and the second is the \emph{vertex representation} \cite{grunbaum}. We will now formally define $\mathcal{C}$ in terms of each representation:

\begin{defn}\label{defn211}
\textit{(Half-space representation)}: A convex polytope $\mathcal{C}$ in a real space $\mathbb{R}^{\Delta}$ of dimension $\Delta$ is the intersection of closed half-spaces. These closed half-spaces are defined by linear inequalities of the form $\sum_{j}^{\Delta}a_{j}v_{j}\leq b$ for real values $a_{j}$ and elements $v_{j}$ of a vector $\vec{v}\in\mathbb{R}^{\Delta}$.
\end{defn}

This definition is general enough to encompass \emph{unbounded} polytopes. We say a convex polytope is bounded if it can be contained in a ball of finite radius and unbounded otherwise. We impose the extra constraints that there are a finite number of inequalities that form a bounded polytope \cite{grunbaum}.

The linear inequalities in the above definition are ``facet-defining'' which we define formally later on but can be informally seen as the boundary of the convex polytope $\mathcal{C}$. If we return to the example of $\mathcal{F}$ as the space of all possible conditional probabilities $p(\mathbf{m}|\mathbf{s})$ then the linear inequalities defining $\mathcal{F}$ are the positivity and normalisation constraints. As mentioned, dual to the half-space representation, the vertex representation of a convex polytope describes the polytope in terms of all points, or vectors in the polytope:

\begin{defn}\label{defn212}
\textit{(Vertex representation)}: \textbf{convex polytope} $\mathcal{C}$ in a real space $\mathbb{R}^{\Delta}$ of dimension $\Delta$ is the convex hull of $E$ extreme points, or vectors $\vec{v}_{e}\in\mathbb{R}^{\Delta}$ for $e\in\{1,2,...,E\}$.
\end{defn}

The convex polytope $\mathcal{C}$ then is the set of vectors that can be written as a convex combination of $E$ vectors in the $\Delta$-dimensional real space. For example, the polytope $\mathcal{F}$ can then be written in terms of the convex combination of $E$ vectors $\vec{p}$ which we call $\vec{p}_{e}$ for $e\in\{1,2,...,E\}$ with probability distribution $p(E)$ over each $\vec{p}_{e}$. These vectors have the elements  $\vec{p}_{e}$ that are the \emph{deterministic} probabilities $p(\textbf{m}|\textbf{s})\in\{0,1\}$. For these deterministic probabilities we associate values $\{0,1\}$ with each map $f:\textbf{s}\rightarrow\textbf{m}$. This way the probabilities can be written as $p(\textbf{m}|\textbf{s})=\sum_{E}p(E)\vec{p}_{E}=\sum_{f}p_{f}\delta^{f(\textbf{s})}_{\textbf{m}}$ where $f(\textbf{s})$ is the image of $\textbf{s}$ under $f$ and $p_{f}$ is a probability distribution over all maps $f$.

\subsection{The Non-signalling Polytope}\label{sec21b}

One can make extra assumptions upon the statistical data obtained from space-like separated sites: each measurement site cannot communicate with each other outside each other's light-cone. This assumption is expressed in terms of the \emph{no-signalling condition} which can be formally stated as:
\begin{equation}
\sum_{\textbf{m}^{\{j|j\in\mathcal{J}\}}}p(\textbf{m}|\textbf{s})=\sum_{\textbf{m}^{\{j|j\in\mathcal{J}\}}}p(\textbf{m}|\textbf{s}')=p(\textbf{m}^{\{j|j\in\mathcal{J}^{c}\}}|\textbf{s}^{\{j|j\in\mathcal{J}^{c}\}}),
\end{equation}
where $\mathcal{J}$ is any sub-set of all $n$ parties and $\mathcal{J}^{c}=\{j\in\{1,2,...,n\}|j\notin\mathcal{J}\}$ is the complement of this sub-set and $\textbf{s}\neq\textbf{s}'$ such that the inputs differ in elements $s_{j}$ with $j\in\mathcal{J}$ \cite{pr}. Each of these conditions forms a hyperplane in $\mathbb{R}^{D}$ and then the intersection of hyperplanes is the space of correlations that satisfies the no-signalling condition. Or just as before, one can reduce the dimensionality of the space of statistics by imposing these equalities and then define inequalities on the reduced space. 

Therefore one can construct another convex polytope called $\mathcal{NS}$ which is the intersection of half-spaces defined by inequalities resulting from the normalisation, positivity and no-signalling conditions \cite{barrett3}. We shall discuss the polytope $\mathcal{NS}$ in section \ref{sec23} of this chapter. Now we consider the correlations that satisfy local realism, or local hidden variable theories.

\subsection{Local Hidden Variable Theories and Bell Inequalities}\label{sec21c}

A Bell test is an experiment that aims to test whether the statistics produced by boxes can be satisfied by a theory that obeys \emph{local realism}. Systems that satisfy local realism satisfy two conditions (covered thoroughly in \cite{bell}):
\begin{enumerate}
\item Realism: There are objective properties of a system that are elements, or ``hidden'' variables $\lambda\in\Lambda$ in a (generally continuously defined) space of hidden variables $\Lambda$. These variables have a pre-existing value before the measurement is made and can influence measurement outcomes;
\item Locality: The variables $\lambda$ possessed by a party at any site are not affected by events that occur outside of the light-cone of the measurement made at this site. These variables are called \emph{Local Hidden Variables} (LHV).
\end{enumerate}
Each party's measurement is influenced by $\lambda$ and the measurement choice made at that party's site. In an LHV theory, space-like separated parties cannot communicate their measurement information to each other via the LHV, or any other means, due to locality. Measurement outcomes are then influenced by $s_{j}$ and $\lambda$ alone.

To be more precise, there is a probability distribution $p(\lambda)d\lambda$ over $\Lambda$ such that $p(\lambda)\geq 0$ and $\int_{\Lambda}p(\lambda)d\lambda=1$. This can occur, for example, if the parties have some shared source of randomness over the variables $\lambda$. Therefore each set of measurement outcomes conditioned upon measurement settings can be written in the following form \cite{bellineq},
\begin{equation}\label{lhv1}
p(\textbf{m}|\textbf{s})=\int_{\Lambda}p(\lambda)d\lambda\prod_{j=1}^{n}p(m_{j}|s_{j},\lambda).
\end{equation}
This expression can be written in terms of a convex combination of deterministic maps $g_{j}:\mathbb{Z}_{c_{j}}\rightarrow\mathbb{Z}_{d_{j}}$ at each site. The single site probabilities are then written as a convex combination over all deterministic maps, i.e. $p(m_{j}|s_{j},\lambda)=\sum_{g_{j}}p_{g_{j}}\delta^{m_{j}}_{g_{j}}$ where $g_{j}$ are the single site maps with $p_{g_{j}}\geq 0$ and $\sum_{g_{j}}p_{g_{j}}=1$. If one considers all $n$ deterministic single-site maps, then we can deterministically obtain the resulting output digit-string from all parties $\textbf{m}=\{g_{1}(s_{1}),g_{2}(s_{2}),g_{3}(s_{3}),...,g_{n}(s_{n})\}$ where $g_{j}(s_{j})$ is the image of $s_{j}$ under the single-site map $g_{j}$. As a result, equation \eqref{lhv1} can be rewritten as:
\begin{equation}\label{lhv2}
p(\textbf{m}|\textbf{s})=\sum_{g_{1},g_{2},...,g_{n}}p_{g_{1},g_{2},...,g_{n}}\prod_{j}^{n}\delta^{m_{j}}_{g_{j}(s_{j})},
\end{equation}
taking a convex combination over all combination of single site maps $g_{j}$ so that $p_{g_{1},g_{2},...,g_{n}}\geq 0$ and $\sum_{g_{1},g_{2},...,g_{n}}p_{g_{1},g_{2},...,g_{n}}=1$ is satisfied. Note that the decomposition in \eqref{lhv2} is not unique; uniqueness is only guaranteed when $p_{g_{1},g_{2},...,g_{n}}=1$ for a particular choice of single site maps.

We see immediately from \eqref{lhv2} that the space $\mathcal{L}_{\mathcal{F}}\subseteq\mathcal{F}$ of LHV correlations $p(\textbf{m}|\textbf{s})$ is also a convex polytope as defined in terms of a vertex representation. The vertices of $\mathcal{L}_{\mathcal{F}}$ are the deterministic probabilities $p(\textbf{m}|\textbf{s})=\prod_{j}^{n}\delta^{m_{j}}_{g_{j}(s)}$ corresponding to each combination of single site maps $g_{j}$. There is also the facet representation of the polytope $\mathcal{L}_{\mathcal{F}}$ in terms of facet-defining linear inequalities. These linear inequalities are the facet-defining Bell inequalities, which we abbreviate to \textbf{facet Bell inequalities}, that constrain and define the consequences of LHV theories \cite{collins, froissart, pitowski, peres}. We now formally define what is means for a linear inequality to be facet-defining. 

\begin{defn}\label{defn213}
 A linear inequality is \textbf{facet-defining} for a convex polytope $\mathcal{C}$ in a real space $\mathbb{R}^{\Delta}$ of dimension $\Delta$ when at least $\Delta$ affinely independent extreme points of $\mathcal{C}$ saturate the inequality (i.e. satisfy the equality of the linear inequality).
\end{defn}

\begin{figure}
	\centering
		\includegraphics[width=0.80\textwidth]{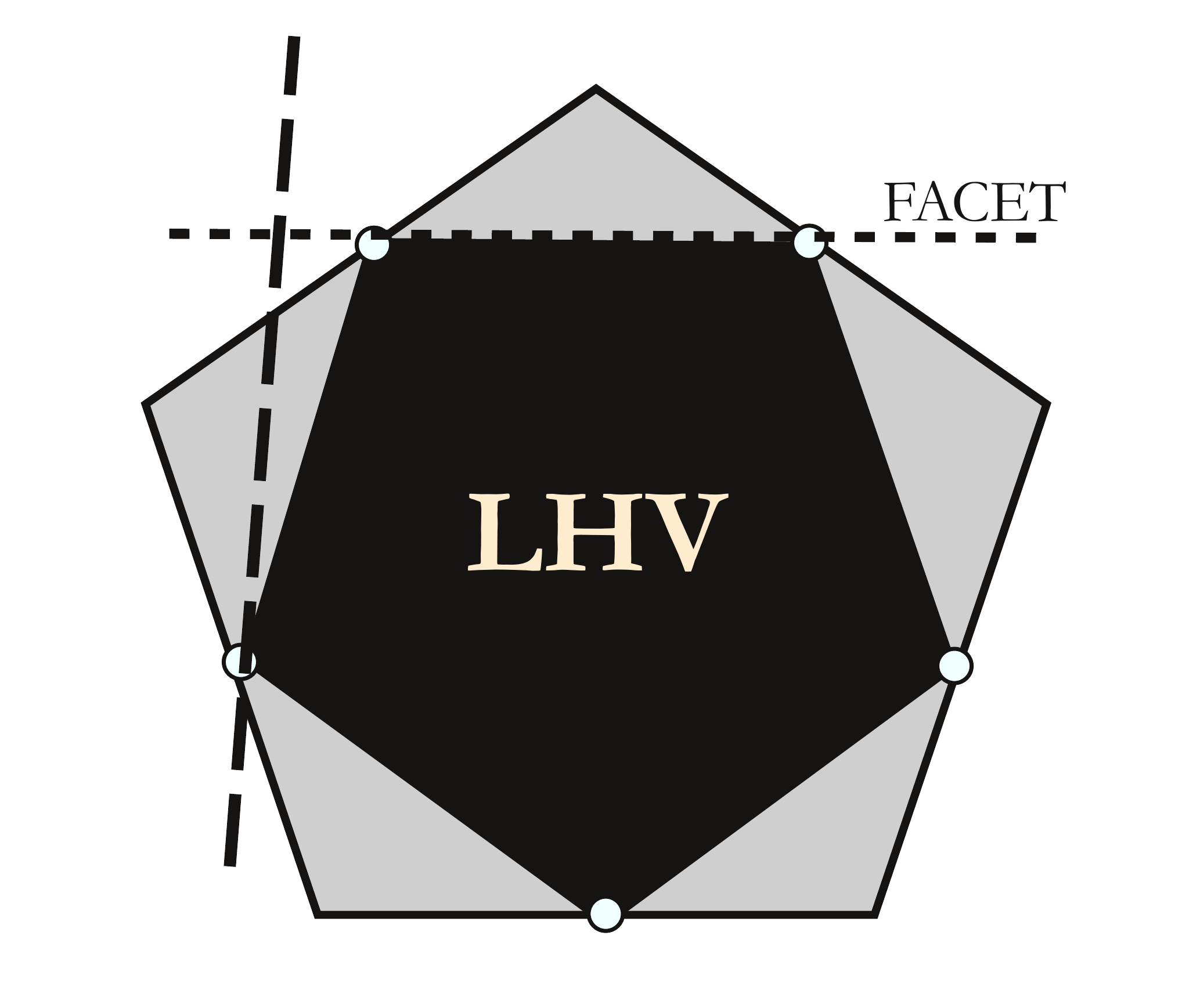}
			\caption{The LHV polytope can be defined in terms of the facet Bell inequalities. These inequalities intersect the surface of this polytope; whereas an arbitrary Bell inequality might only intersect one vertex as shown in this schematic.}
			\label{fig:space}
\end{figure}

A set $\mathcal{S}$ of $K$ vectors $\vec{p}_{i}$, $\mathcal{S}=\{\vec{p}_{0},\vec{p}_{1},...,\vec{p}_{(K-1)}\}$ is \emph{affinely independent} if for every $\vec{p}_{k}\in\mathcal{S}$, the $(K-1)$ vectors in the set $\{\vec{p}_{i}-\vec{p}_{k}|\vec{p}_{i}\neq\vec{p}_{k}\}$ are linearly independent. A linear inequality for the space of correlations is of the form:
\begin{equation}\label{bellineq1}
\sum_{\textbf{m},\textbf{s}}\beta_{\textbf{m},\textbf{s}}p(\textbf{m}|\textbf{s})\leq \gamma_{\mathcal{L}},
\end{equation}
where $\beta_{\textbf{m},\textbf{s}}$ are real pre-factors depending on $\textbf{m}$ and $\textbf{s}$ and $\gamma_{\mathcal{L}}\in\mathbb{R}$ as the upper bound resulting from LHV correlations in \eqref{lhv2}. All LHV correlations satisfy \eqref{bellineq1} whether the inequality is facet-defining or otherwise. For the inequalities to be facet Bell Inequalities the following conditions must be satisfied:
\begin{equation}\label{facet1}
\sum_{\textbf{m},\textbf{s}}\beta_{\textbf{m},\textbf{s}} \prod_{j}^{n}\delta^{m_{j}}_{g_{j}(s)}= \gamma_{\mathcal{L}},
\end{equation}
for at least $(\prod_{j=1}^{n}d_{j}-1)\prod_{j=1}^{n}c_{j}$ affinely independent vectors $\vec{p}$ such that elements are $p(\textbf{m}|\textbf{s})= \prod_{j}^{n}\delta^{m_{j}}_{g_{j}(s)}$. We can demonstrate this schematically in Figure \ref{fig:space} where we show that a facet Bell inequality picks out the surface of the LHV polytope, whereas the inequalities in \eqref{bellineq1} might just bound the LHV polytope. We will make these ideas concrete in chapter \ref{chap3}.

The problem of finding the facets of a polytope given the vertices is known as the \emph{facet enumeration problem} \cite{collins} and software does exist that performs this task (e.g. \cite{polymake}). However, it is currently in general both theoretically and practically hard to find these inequalities as we shall discuss in the subsequent sub-section. The hardness of this problem will motivate us to think about simplified Bell inequality settings, and then relate these simplified settings to a more general setting. 

\subsection{Facet Bell Inequalities and Computational Complexity}\label{sec21d}

\noindent
Given our abstract setting for $n$ parties each with $c_{j}$ possible inputs and $d_{j}$ possible outputs, it is immediately natural to ask how hard is it to obtain Facet Bell Inequalities? Pitowsky notably studied this question by studying the intimate link between convex polytopes and propositional logic \cite{pitowski}. The latter then has a deep connection to computational complexity, the branch of theoretical computer science devoted to the hardness of computational problems \cite{pap}. 

Whether a problem is easy or not can be defined in terms of time efficiency of finding a solution on a Turing Machine (an abstract computer that can simulate all other computers \cite{turing}) with respect to the size of the input. Time is defined in terms of computational time, or the number of computational steps in an algorithm. The computational time as a function of input size is then the indicator of computational efficiency, or hardness. If this function is a polynomial in the size of the input, then it is efficient, if super-polynomial (e.g. exponential) then it is inefficient \cite{pap}. 

The problems that are given to a Turing Machine are decision problems. That is, given an input the machine outputs `yes', represented as the bit $0$ or `no', represented as bit $1$; the decision problems are questions with binary potential outcomes. If the algorithm performed by the machine to make this decision operates in a number of steps that is polynomial in the size of the input, then it is in the complexity class called $\textrm{P}$. It is key to note that the algorithm must be polynomial in the input size for all possible inputs, as there may be inputs that are easier to compute than others. If problems in $\textrm{P}$ are efficiently solvable, then there is another class of problems where solutions can be checked (for veracity) in an efficient amount of time. This complexity class is called $\textrm{NP}$. 

The question of whether $\textrm{P}$ is equal to $\textrm{NP}$ is one of the greatest mathematical unsolved puzzles. Discussion of this problem is outside of the discussion of this thesis. However, if $\textrm{P}=\textrm{NP}$ then in loose terms it would be as easy to solve a problem as to check the validity of the solution; this may seem intuitively incorrect to both a casual reader and a computer scientist. The hardest decision problems in $\textrm{NP}$ are called  $\textrm{NP-complete}$ and so if $\textrm{P}=\textrm{NP}$ then these problems have a polynomial time solution. At the current time, no polynomial time solution is known for $\textrm{NP-complete}$ problems.

Pitowsky has shown that finding the facet Bell inequalities is at least as hard as any of the $\textrm{NP-complete}$ problems, if not necessarily in $\textrm{NP}$ \cite{pitowski,pitowski2}. In the terminology of computatational complexity, this problem is $\textrm{NP-hard}$. Heuristically, Pitowsky showed this by relating the problem of finding a facet Bell inequality to a Boolean satisfiability problem \cite{pitowski2}. These problems ask whether there exist variables that result in a Boolean function being `true' and are $\textrm{NP-complete}$ \cite{pap}. The vertices of the LHV polytope $\mathcal{L}_{\mathcal{F}}$ consist of vectors with elements being $0$ or $1$, which are truth value assignments. This relationship between vertices of the polytope and truth assignments allowed Pitowsky to say that finding facet Bell inequalities is at least as hard as a Boolean satisfiability problem.

We have given an overview of the geometric construction of Bell inequalities and the space of correlations. The implications of local realism are connected to the study of convex polytopes. Polytopes have also been used to study the space of non-signalling correlations; we shall return to this subject in section \ref{sec23} of this chapter. Finally we have shown that finding the Bell inequalities that define the LHV polytope is a computationally hard problem. This motivates our study of correlators, the probability of a joint outcome between all $n$ parties instead of the full probability distribution. This simplifies a hard problem by decreasing its dimensionality. Correlators also have a computational perspective that will become crucial to our study of Bell inequalities.

\section{Correlators}\label{sec22}

\noindent
Finding facet Bell inequalities is difficult. This difficulty scales with the size of the problem such as the number of possible inputs and outputs for $n$ parties. Firstly, we assume that $d_{j}=d_{k}=d$ (for $j\neq k$) is the same for all parties and is prime. We also simplify the type of probabilities that we need to consider from the correlations $p(\textbf{m}|\textbf{s})$ to a \emph{correlator} which takes the form:
\begin{equation}
p(k|\textbf{s})=\sum_{\textbf{m}}\delta^{\left[\sum_{j=1}^{n}m_{j}\right]_{d}}_{k}p(\textbf{m}|\textbf{s})=p([\sum_{j=1}^{n}m_{j}]_{d}={k}|\textbf{s}),
\end{equation}
where throughout this thesis (unless otherwise stated) we write all modulo $x$ arithmetic in brackets with a sub-script $[...]_{x}$. From this perspective, the Bell test now consists of inputs $\textbf{s}\in\bigoplus_{j=1}^{n}\mathbb{Z}_{c_{j}}$ and a single value output $k=\left[\sum_{j=1}^{n}m_{j}\right]_{d}$ is returned.

The correlator $p(k|\textbf{s})$ is a stochastic map $f:\bigoplus_{j=1}^{n}\mathbb{Z}_{c_{j}}\rightarrow\mathbb{Z}_{d}$ and due to the normalisation $\sum_{k=0}^{d-1}p(k|\textbf{s})=1$ for all $\textbf{s}$, we only need to consider $(d-1)\prod_{j=1}^{n}c_{j}$ correlators. We do not consider, then, the correlator $p(0|\textbf{s})$ for all $\textbf{s}$ as it can be recovered by normalisation. These correlators are now elements of a real vector $\vec{k}\in\mathbb{R}^{(d-1)\prod_{j=1}^{n}d_{j}}$ which we call a \emph{correlator vector} \footnote{If there is no conflict in meaning, we may shorten correlator vector to just correlator. For example, if we refer to correlators being in some space, this means the resulting correlator vectors are in some space.}.

We can describe the space of all possible correlator vectors as another convex polytope $\mathcal{P}$. This object, analogous to $\mathcal{F}$, has a simple description in terms of vertices and linear inequalities. First, $\mathcal{P}$ has $E$ vertices, or extreme points, $\vec{k}_{e}$ for $e\in\{1,2,...,E\}$ that have the elements $p_{e}(k|\textbf{s})\in\{0,1\}$ for all $k$ and $\textbf{s}$. Therefore, these vectors correspond to deterministic maps where a single value of $k$ is the output given the input $\textbf{s}$ so elements are $p_{e}(k|\textbf{s})=\delta^{k}_{f(\textbf{s})}$ where $f(\textbf{s})$ is the image of $\textbf{s}$ under the map $f:\bigoplus_{j=1}^{n}\mathbb{Z}_{c_{j}}\rightarrow\mathbb{Z}_{d}$. Any correlator vector $\vec{k}\in\mathbb{R}^{(d-1)\prod_{j=1}^{n}d_{j}}$ can be written as a convex combination of these extreme points:
\begin{equation}
\vec{k}=\sum_{e}p_{e}\vec{k}_{e}=\sum_{f}p_{f}\vec{k}_{f}
\end{equation}
where every extreme point $\vec{k}_{e}$ corresponds one-to-one with a vector $\vec{k}_{f}$ resulting from a function $f:\bigoplus_{j=1}^{n}\mathbb{Z}_{c_{j}}\rightarrow\mathbb{Z}_{d}$ and $p_{e}\geq 0$, $p_{f}\geq 0$ with $\sum_{e}p_{e}=1$ and $\sum_{f}p_{f}=1$ for all functions $f$.

Equivalently we can describe $\mathcal{P}$ in terms of the linear inequalities corresponding to positivity and normalisation: $p(k|\textbf{s})\geq 0$ for all $k$ and $\textbf{s}$ and $\sum_{k\neq 0}p(k|\textbf{s})\leq 1$. This is analogous to the way we defined $\mathcal{F}$ but interestingly, every vector in $\mathcal{P}$ can be produced by at least one probability distribution in $\mathcal{NS}$, the non-signalling polytope. If we allow all probability distributions that satisfy only the no-signalling condition we can completely saturate $\mathcal{P}$. As an example, every vertex of $\mathcal{P}$ corresponding to the map $f:\bigoplus_{j=1}^{n}\mathbb{Z}_{c_{j}}\rightarrow\mathbb{Z}_{d}$, we can always write this probability distribution:
\begin{equation}\label{genbox}
p(\textbf{m}|\textbf{s}) = 
\begin{cases} d^{1-n} & \text{if $[\sum_{j=1}^{n}m_{j}]_{d}=f(\textbf{s})$,}
\\
0 &\text{otherwise.}
\end{cases}
\end{equation}
for the function $f$ as above. All reductions of this probability distribution are $p(\textbf{m}^{\{j|j\in\mathcal{J}\}}|\textbf{s}^{\{j|j\in\mathcal{J}\}}) = d^{-|\mathcal{J}|}$ if $|\mathcal{J}|\neq n$ for all $\textbf{m}^{\{j|j\in\mathcal{J}\}}$ and $\textbf{s}^{\{j|j\in\mathcal{J}\}}$. Since this distribution is uniformly random for all sub-sets of parties, it satisfies the no-signalling condition. We shall elaborate on the connections between $\mathcal{NS}$ and $\mathcal{P}$ in a subsequent section \ref{sec23}.

Another motivation for these correlators is that they are a generalisation of the well-studied CHSH Bell Inequality setting for many parties \cite{chsh,ww,zb}. This generalisation also coincides with the Collins-Gisin-Linden-Massar-Popescu (CGLMP) setting again generalised to many parties \cite{cglmp}. An example of work in a many-setting CGLMP framework includes that by Ac\'{i}n et al \cite{acin}. 

Also, in the literature, correlators can be considered to result from the expectation value of the outcome of joint measurements if the outcomes of measurements are complex numbers of unit modulus \cite{general, slk}. More specifically, every $j$th party's measurement $\mathcal{M}_{s_{j}}$ has the outcome values $e^{i2\pi\frac{k}{d}}$ for $k\in\mathbb{Z}_{d}$, then the expectation value of the joint measurement $\mathbb{E}(\textbf{s})=\mathbb{E}(\prod_{j=1}^{n}\mathcal{M}_{s_{j}})$ is:
\begin{eqnarray}
\mathbb{E}(\textbf{s})&=&\sum_{\textbf{m}}\prod_{j=1}^{n}e^{i2\pi\frac{m_{j}}{d}}p(\textbf{m}|\textbf{s}) \nonumber \\
&=&\sum_{k=0}^{(d-1)}e^{i2\pi\frac{k}{d}}p(k|\textbf{s})\nonumber \\
&=&1+\sum_{k=1}^{(d-1)}\left[e^{i2\pi\frac{k}{d}}-1\right]p(k|\textbf{s}).
\end{eqnarray}
These expectation values $\mathbb{E}(\textbf{s})$ can be written in terms of correlators. Every measurement that has two possible outcomes $\{+1,-1\}$ results in expectation values of measurements being $\mathbb{E}(\textbf{s})=1-2p(1|\textbf{s})$; the expectation values are equivalent to a single correlator $p(1|\textbf{s})$. This is the many-party generalisation of the CHSH setting for two parties. This equivalence has allowed research in the past to interchangeably use expectation values as well as conditional probabilities.

One can coarse-grain research into generalized Bell inequality setting as either obtaining statistics in terms of correlators (e.g. \cite{chsh,cglmp,acin}) or the full probability distribution (e.g. \cite{ch,collins}). The latter can be reduced to the former but much literature has been devoted to the study of correlators. As well as being able to infer structure of $\mathcal{NS}$ from $\mathcal{P}$ (see section \ref{sec23}), these correlators are at the centre of much research into Bell inequalities. We will now try and formalise the structure of the space of correlator vectors by considering the maps performed by all possible theories.

\subsection{Correlators as Computations}\label{sec22a}

\noindent
Throughout this thesis we argue for a computational approach to Bell inequality experiments by considering in what sense correlations are computing a function $f:\bigoplus_{j=1}^{n}\mathbb{Z}_{c_{j}}\rightarrow\mathbb{Z}_{d}$ on the inputs $\textbf{s}$. In this section we now want to introduce some of the tools associated with these functions so that we can be more specific about the computational power of correlations from physical (or non-physical) theories. 

Every function $f:\bigoplus_{j=1}^{n}\mathbb{Z}_{c_{j}}\rightarrow\mathbb{Z}_{d}$ can be written as a list (a single column table) with each row representing $f(\textbf{s})$, the image of $\textbf{s}$ under $f$. In turn this list is an element of the module $\mathbb{M}$ over the ring $\mathbb{Z}_{d}$. The module $\mathbb{M}$ consists of the abelian group $\mathbb{Z}_{d}^{D}$ for $D=\prod_{j=1}^{n}c_{j}$ with the group multiplication being modulo $d$ addition of these elements, the module also has (left or right) scalar multiplication $\mathbb{Z_{d}}\times\mathbb{Z}_{d}^{D}\rightarrow\mathbb{Z}_{d}^{D}$ of elements in the group. In order to satisfy $\mathbb{M}$ being a module then for all $\textbf{x}$, $\textbf{y}$ in $\mathbb{Z}_{d}^{D}$, and all $a$, $b$ in $\mathbb{Z}_{d}$ then:

\begin{enumerate}
\item $1\textbf{x}=\textbf{x}$ (existence of the identity)
\item $a(b\textbf{x})=(ab)\textbf{x}$ (associativity)
\item $a(\textbf{x}+\textbf{y})=a\textbf{x}+a\textbf{y}$ (distributivity over $\mathbb{Z}_{d}^{D}$)
\item $(a+b)\textbf{x}=a\textbf{x}+b\textbf{x}$ (distributivity over $\mathbb{Z}_{d}$),
\end{enumerate}

where we could have written the scalar multiplication in terms of left or right multiplication \cite{anderson}. All arithmetic is modulo $d$ but we have suppressed the notation $[...]_{d}$ for clarity. 

Every $f\in\mathbb{M}$ can be written in terms of Kronecker delta functions with elements $f(\textbf{s})=\delta^{\textbf{s}}_{\textbf{y}}$ which is $1$ for only one input $\textbf{s}=\textbf{y}\in\bigoplus_{j=1}^{n}\mathbb{Z}_{c_{j}}$ and $0$ otherwise. Therefore every element $f(\textbf{s})$ of any function $f$ can be written as
\begin{equation}
f(\textbf{s})=\sum_{\textbf{y}\in\bigoplus_{j=1}^{n}\mathbb{Z}_{c_{j}}}f(\textbf{y})\delta^{\textbf{s}}_{\textbf{y}}
\end{equation}
with $f(\textbf{y})\in\mathbb{Z}_{d}$. The delta functions then form something analogous to the basis vectors for a vector space and we can replace one of the delta functions with the constant, all-ones function with elements $f(\textbf{s})=1$. The delta function we choose to replace is $f(\textbf{s})=\delta^{\textbf{s}}_{\textbf{y}}$ with $\textbf{y}=\textbf{0}$, the all-zeroes digit-string.

For every $s_{j}\in\mathbb{Z}_{c_{j}}$ we can choose to represent $\mathbb{Z}_{c_{j}}$ as a Cartesian product of cyclic groups of dimension being the prime factors of $c_{j}$. The set of prime factors of $c_{j}$ are written as $\{^{1}c_{j},^{2}c_{j},...,^{q_{j}}c_{j}\}$ for $^{k}c_{j}$ as the $k$th prime factor and $q_{j}$ being the number of prime factors, therefore $s_{j}=\{^{1}s_{j},^{2}s_{j},...,^{q_{j}}s_{j}\}\in\bigoplus_{k=1}^{q_{j}}\mathbb{Z}_{^{k}c_{j}}$ with $^{k}s_{j}\in\mathbb{Z}_{^{k}c_{j}}$. The delta functions $\delta^{\textbf{s}}_{\textbf{y}}$ can be written now in terms of inputs $\textbf{s}\in\bigoplus_{j=1}^{n}\left(\bigoplus_{k=1}^{q_{j}}\mathbb{Z}_{^{k}c_{j}}\right)$ and $\textbf{y}\in\bigoplus_{j=1}^{n}\left(\bigoplus_{k=1}^{q_{j}}\mathbb{Z}_{^{k}c_{j}}\right)$ giving
\begin{eqnarray}
\delta^{\textbf{s}}_{\textbf{y}}&=&\prod_{j=1}^{n}\prod_{k=1}^{q_{j}}\delta^{^{k}s_{j}}_{^{k}y_{j}}\nonumber\\
&=&\prod_{j=1}^{n}\prod_{k=1}^{q_{j}}\left[1-({^{k}s_{j}}-{^{k}y_{j}})^{^{k}c_{j}-1}\right]_{^{k}c_{j}}\nonumber\\
&=&\prod_{j=1}^{n}\prod_{k=1}^{q_{j}}\left[1-\sum_{l=0}^{^{k}c_{j}-1}(-1)^{l}{{^{k}c_{j}-1}\choose{l}}(^{k}y_{j})^{l}(^{k}s_{j})^{{^{k}c_{j}-(l+1)}}\right]_{^{k}c_{j}}
\end{eqnarray}
where the second line is guaranteed by Fermat's little theorem. That is, the modular arithmetic expression is $\left[({^{k}s_{j}}-{^{k}y_{j}})^{^{k}c_{j}-1}\right]_{^{k}c_{j}}=1$ for ${^{k}s_{j}}\neq{^{k}y_{j}}$ and coprime with ${^{k}c_{j}}$. The third line above just results from the binomial theorem. 

In the instance where $c_{j}=c_{k}=d$ being prime for all $j\neq k$, the delta functions just simplify to being a polynomial over the field $\mathbb{Z}_{d}$ as indicated by the third line above. For example, for $d=2$ the delta functions are Boolean functions $f:\mathbb{Z}_{2}^{n}\rightarrow\mathbb{Z}_{2}$ expressed as polynomials over $\mathbb{Z}_{2}$. We will use these properties more explicitly in subsequent chapters of this thesis.

We now introduce classes of functions that will be used to characterise the correlators resulting from particular theories. The first class of functions we now describe as ``$n$-partite linear functions'' have connections to LHV theories.

\begin{defn}\label{defn221}
An $n$-\textbf{partite linear function} is a function $g:\bigoplus_{j=1}^{n}\mathbb{Z}_{c_{j}}\rightarrow\mathbb{Z}_{d}$ where the image of $\textbf{s}$ under $g$ can be written as
\begin{equation}
g(\textbf{s})=\left[\sum_{j=1}^{n}g_{j}(s_{j})\right]_{d}
\end{equation}
with $g_{j}(s_{j})$ the image of $s_{j}$ under the single-site map $g_{j}:\mathbb{Z}_{c_{j}}\rightarrow\mathbb{Z}_{d}$.
\end{defn}

These functions are not strictly linear as the single-site maps $g_{j}$ are not always linear in $s_{j}$, but for $c_{j}=d=2$, then these maps are linear. We use the nomenclature of linearity \emph{only} to highlight the fact that there is addition modulo $d$ between single-site maps and not multiplication. If a map cannot be expressed as an $n$-partite linear function then we say it is a $\textbf{non-}n\textbf{-partite linear function}$.

Any function $f:\bigoplus_{j=1}^{n}\mathbb{Z}_{c_{j}}\rightarrow\mathbb{Z}_{d}$ can be described in terms of a sum of an $n$-partite linear function and non-$n$-partite linear function, i.e. $f(\textbf{s})=[g(\textbf{s})+h(\textbf{s})]_{d}$ where $g(\textbf{s})$ and $h(\textbf{s})$ are $n$-partite and non-$n$-partite linear functions respectively. First we write an $n$-partite linear function in terms of the delta functions $\delta^{s_{j}}_{y_{j}}$ for single-site maps to obtain:
\begin{equation}
g(\textbf{s})=\left[\alpha+\sum_{j=1}^{n}\sum_{k=1}^{c_{j}-1}\beta_{j,k}\delta^{s_{j}}_{k}\right]_{d},
\end{equation}
with $\alpha$, $\beta_{j,k}\in\mathbb{Z}_{d}$. The constant $\alpha$ emerges from taking the sum modulo $d$ of the constant function $\alpha_{j}\in\mathbb{Z}_{d}$ for each site, that replaces the delta function $\delta^{s_{j}}_{0}$, as discussed. 

For inputs $\textbf{s}$ with only one non-zero element, there is only a single delta function $\delta^{s_{j}}_{k}$ (for $s_{j}$ being the non-zero element) that describes the value of $g(\textbf{s})=\left[\alpha+\beta_{j,k}\delta^{s_{j}}_{k}\right]_{d}$. For the all-zeroes digit-string $\textbf{s}=\textbf{0}$, then the only function describing $g(\textbf{s})$ is the constant function $\alpha$. We call the set of digit-strings $\textbf{s}$ with at most one non-zero element $\mathcal{T}$. 

Now we briefly consider a column list of the images of $s$ under $f$ for only these digit-strings $\textbf{s}\in\mathcal{T}$. Then for this restricted list  delta functions $\delta^{s_{j}}_{k}$ (and constant $\alpha$), similar to before, form a basis for any function $f(\textbf{s})$ with $\textbf{s}\in\mathcal{T}$. This is because they are equivalent to the delta functions $\delta^{\textbf{s}}_{\textbf{y}}=\delta^{s_{j}}_{k}\prod_{l\neq j}^{n}\delta^{s_{l}}_{0}$ for these particular input digit-strings. A basis is formed in the sense that these functions are linearly independent over $\mathbb{Z}_{d}$.

In order to achieve any function $f(\textbf{s})$ we need a basis for the functions for all possible input strings $\textbf{s}$ including $\textbf{s}\in\mathcal{T}$. We do this by supplementing the function $g(\textbf{s})$ above with the delta functions $\delta^{\textbf{s}}_{\textbf{y}}$ with $\textbf{y}\notin\mathcal{T}$. Therefore, any function can be written as a sum of an $n$-partite linear function $g(\textbf{s})$ and a non-$n$-partite linear function
\begin{eqnarray}
f(\textbf{s})&=&\left[\alpha+\sum_{j=1}^{n}\sum_{k=1}^{c_{j}-1}\beta_{j,k}\delta^{s_{j}}_{k}+\sum_{\textbf{y}\notin\mathcal{T}}\gamma_{\textbf{y}}\delta^{\textbf{s}}_{\textbf{y}}\right]_{d}\nonumber\\
&=&\left[g(\textbf{s})+h(\textbf{s})\right]_{d},
\end{eqnarray}
with $\gamma_{\textbf{y}}\in\mathbb{Z}_{d}$ and $h(\textbf{s})=\sum_{\textbf{y}\notin\mathcal{T}}\gamma_{\textbf{y}}\delta^{\textbf{s}}_{\textbf{y}}$ as a non-$n$-partite linear function by construction. We describe this form of $f(\textbf{s})$ as the \emph{decomposition} of the function into $n$-partite linear and non-$n$-partite linear functions. If $h(\textbf{s})=0$ for all $\textbf{s}$, then $f(\textbf{s})$ is necessarily an $n$-partite linear function, otherwise it is necessarily a non-$n$-partite linear function. If, on the other hand, $g(\textbf{s})=0$ we have the following result:

\begin{lem}\label{lem221}
If a function $f:\bigoplus_{j=1}^{n}\mathbb{Z}_{c_{j}}\rightarrow\mathbb{Z}_{d}$ has no $n$-partite linear function in its decomposition, then $f(\textbf{s})=0$ for all $\textbf{s}\in\mathcal{T}$.
\end{lem}

\textit{Proof} - If a function $f(\textbf{s})=[g(\textbf{s})+h(\textbf{s})]_{d}$ has no $n$-partite linear function part, i.e. $g(\textbf{s})=0$ for all $\textbf{s}$ but some non-zero non-$n$-partite linear part, i.e. $h(\textbf{s})\neq 0$ for some inputs $\textbf{s}$, then it can be written as,
\begin{equation}
f(\textbf{s})=\left[\sum_{\textbf{y}\notin\mathcal{T}}\gamma_{\textbf{y}}\delta^{\textbf{s}}_{\textbf{y}}\right]_{d}.
\end{equation}
Then $f(\textbf{s})$ must be zero for all $\textbf{s}\in\mathcal{T}$. $\square$
\newline

We will use this lemma in the proof of Lemma \ref{lem231} in section \ref{sec23a} and is a useful consequence of choosing this decomposition of functions. We shall also show in the following subsection that this decomposition is physically motivated and not just mathematically convenient.

Another class of functions will now be introduced and shown to be useful in later sections \ref{sec23} and \ref{sec24} of this chapter. They can be seen to be a relaxation of the constraint of $n$-partite linear functions where instead of taking a sum of single-site maps, we take a sum of maps produced by sub-sets of all parties. In particular, we consider all the ways in which $n$ parties can be partitioned into a non-empty sub-set $\mathcal{J}$ and its complement $\mathcal{J}^{c}$ as introduced in section \ref{sec21}. The class of functions called ``bipartite linear functions" are then a generalization of $n$-partite linear functions defined for these partitions.

\begin{defn}\label{defn222}
A \textbf{bipartite linear function} is a function $f:\bigoplus_{j=1}^{n}\mathbb{Z}_{c_{j}}\rightarrow\mathbb{Z}_{d}$ where the image of $\textbf{s}$ under $f$ can be written as
\begin{equation}
f(\textbf{s})=\left[f^{1}(\textbf{s}^{\{j|j\in\mathcal{J}\}})+f^{2}(\textbf{s}^{\{j|j\in\mathcal{J}^{c}\}}) \right]_{d}
\end{equation}
with $f^{1}:\bigoplus_{j\in\mathcal{J}}\mathbb{Z}_{c_{j}}\rightarrow\mathbb{Z}_{d}$ and $f^{2}:\bigoplus_{j\in\mathcal{J}^{c}}\mathbb{Z}_{c_{j}}\rightarrow\mathbb{Z}_{d}$ being functions mapping inputs for each partition into $\mathcal{J}$ and $\mathcal{J}^{c}$ to a single output.
\end{defn}

These functions are equivalent to an $n$-partite linear function for $n=2$ as the partition can be seen as a coarse-graining of $n$ parties into two sub-sets, where each sub-set can be considered a party in its own right. Then the input string $\textbf{s}^{\{j|j\in\mathcal{J}\}}$ is now a single input to one `collective' party and $\textbf{s}^{\{j|j\in\mathcal{J}^{c}\}}$ the input to the other collective party. Each subset's collective output is just then the sum modulo $d$ of all of their outputs $m_{j}$ for $j\in\mathcal{J}$ and $j\in\mathcal{J}^{c}$ for each respective subset. As a result, for a given partition, any function $f(\textbf{s})$ can be written as a sum of a bipartite linear function and a \textbf{non-bipartite linear function}. 

We have described classes of functions, and every function describes a vertex of $\mathcal{P}$. To recapitulate, a vertex of $\mathcal{P}$ has the elements $p(k|\textbf{s})=\delta^{k}_{f(\textbf{s})}$ for every $\textbf{s}$. A correlator captures a computation whereby given some input, an output is produced with some probability. The region of $\mathcal{P}$ that is subsumed by a particular theory can then have a computational interpretation in terms of how `close' the region of a particular theory gets to vertices of $\mathcal{P}$. In the following subsection we will discuss the region of correlators achievable in an LHV or quantum theory. 

\subsection{Correlators from Physical Theories}\label{sec22b}

\noindent
Bell tests aim to expose statistics that do not result from a particular class of theories viz. LHV theories. We will now describe the space of correlators $\mathcal{L}\subseteq\mathcal{P}$ resulting from LHV theories. This space can be defined in terms of the language of stochastic maps, and in particular, the functions defined in the subsection \ref{sec22a}. We now present the following theorem which defines $\mathcal{L}$ in terms of a sub-class of all possible functions $f:\bigoplus_{j=1}^{n}\mathbb{Z}_{c_{j}}\rightarrow\mathbb{Z}_{d}$.

\begin{thm}\label{thm221}
The space of LHV correlators $\mathcal{L}$ is the convex hull of deterministic correlators $p(k|\textbf{s})=\delta^{k}_{f(\textbf{s})}$ for $f(\textbf{s})$ being all of the $n$-partite linear functions, for all $\textbf{s}$.
\end{thm}

\textit{Proof:} The proof follows simply from how the probabilities $p(\textbf{m}|\textbf{s})$ are defined in \eqref{lhv2} to obtain correlators:
\begin{eqnarray}
p(k|\textbf{s})&=&\sum_{\textbf{m}}\delta^{[\sum_{j=1}^{n}m_{j}]_{d}}_{k}\sum_{g_{1},g_{2},...,g_{n}}p_{g_{1},g_{2},...,g_{n}}\prod_{j}^{n}\delta^{m_{j}}_{g_{j}(s_{j})}\nonumber \\
&=&\sum_{g(\textbf{s})}p_{g(\textbf{s})}\delta^{k}_{g(\textbf{s})},
\end{eqnarray}
where $g(\textbf{s})=[\sum_{j=1}^{n}g_{j}(s_{j})]_{d}$ is an $n$-partite linear function by definition and $p_{g(\textbf{s})}\geq 0$ and $\sum_{g(\textbf{s})}p_{g(\textbf{s})}=1$. Therefore all LHV correlators are contained in the convex hull of $n$-partite linear functions. $\square$
\newline

The consequence of this theorem then is that correlators resulting from LHV theories have a limited \emph{computational expressiveness}. That is, no correlator resulting from an LHV theory can deterministically perform a non-$n$-partite linear function. This is one of the main computational perspectives that we employ in this thesis, and we will return to this result throughout. 

The CHSH inequality is a facet-defining Bell inequality for the LHV polytope $\mathcal{L}$ for $n=2$ and $c_{1}=c_{2}=d=2$ \cite{fine}. We have shown previously that this inequality can be violated by quantum correlators, therefore they cannot in general be confined to the polytope $\mathcal{L}$ for all possible values of $n$, $c_{j}$ and $d$. Tsirelson showed that there is an equivalent CHSH inequality for quantum correlators denoted:
\begin{equation}
p(1|00)+p(1|01)+p(1|10)-p(1|11)\leq 1+\sqrt{2}\approx 2.41,
\end{equation}
whereas the upper bound for LHV correlators is $2$ \cite{tsirelson}. However, the vertex of $\mathcal{P}$ described as $p(1|\textbf{s})=\delta^{1}_{f(\textbf{s})}$ with $f(\textbf{s})=[s_{1}s_{2}+1]_{2}$ gives a value of $3$ for the CHSH inequality. Therefore, there is a hierarchy of spaces of correlators such that $\mathcal{L}\subseteq\mathcal{Q}\subseteq\mathcal{P}$ with $\mathcal{Q}$ as the space of quantum correlators.

Defining the space $\mathcal{Q}$ of quantum correlators (and correlations in general) is still a major open question but we can indicate some general properties of  $\mathcal{Q}$. As Pitowsky has previously shown, $\mathcal{Q}$ is convex, but not a polytope \cite{pitowski}. Quantum correlators can be written in terms of the probabilities $p(\textbf{m}|\textbf{s})$ which result from measurements on a quantum state $\rho$, i.e.
\begin{eqnarray}
p(k|\textbf{s})&=&\sum_{\textbf{m}}\delta^{[\sum_{j=1}^{n}m_{j}]_{d}}_{k}p(\textbf{m}|\textbf{s})\nonumber \\
&=&\sum_{\textbf{m}}\delta^{[\sum_{j=1}^{n}m_{j}]_{d}}_{k}\textrm{Tr}(\rho\bigotimes_{j=1}^{n}P^{s_{j}}_{m_{j}})
\end{eqnarray}
where $P^{s_{j}}_{m_{j}}$ is a single-site POVM corresponding to an outcome $m_{j}$ given the choice of measurement $s_{j}$ so that $\sum_{m_{j}}P^{s_{j}}_{m_{j}}=\mathbb{I}$, the identity matrix. If each measurement site has access to a Hilbert space $\mathcal{H}_{j}$, then the state $\rho$ is in general, a density matrix acting over the tensor-product of these $n$ Hilbert spaces $\bigotimes_{j=1}^{n}\mathcal{H}_{j}$. The dimension of each Hilbert space is arbitrary (and possibly infinite).

Naimark's theorem indicates that any POVM is equivalent to a PVM on an ancilla Hilbert space (it also applies for infinite dimensional systems) \cite{naimark}. Therefore every correlator can be written in terms of a state $\rho'$ and projectors $Q^{m_{j}}_{s_{j}}$ on the Hilbert space $\bigotimes_{j=1}^{n}\mathcal{H}'_{j}$ where $\mathcal{H}'_{j}$ is each $j$th site's enlarged Hilbert space. Projectors $Q^{m_{j}}_{s_{j}}$ are expressed in terms of each site's orthogonal basis $|m_{j}\rangle_{s_{j}}$ for the $s_{j}$ choice of basis, i.e. $Q^{m_{j}}_{s_{j}}=|m_{j}\rangle_{s_{j}}\langle m_{j}|_{s_{j}}$ such that $Q^{m_{j}}_{s_{j}}Q^{m'_{j}}_{s_{j}}=Q^{m_{j}}_{s_{j}}\delta^{m_{j}}_{m'_{j}}$. A density matrix can be constructed from a convex combination of pure states $\rho'=\sum_{l}p_{l}|\psi_{l}\rangle\langle\psi_{l}|$ so that,
\begin{eqnarray}
p(k|\textbf{s})&=&\sum_{l}p_{l}\sum_{\textbf{m}}\delta^{[\sum_{j=1}^{n}m_{j}]_{d}}_{k}\left|\langle\psi_{l}|\bigotimes_{j=1}^{n}|m_{j}\rangle_{s_{j}}\right|^{2}.
\end{eqnarray}
Since $p(k|\textbf{s})=\sum_{\textbf{m}}\delta^{[\sum_{j=1}^{n}m_{j}]_{d}}_{k}\left|\langle\psi_{l}|\bigotimes_{j=1}^{n}|m_{j}\rangle_{s_{j}}\right|^{2}$ is itself a quantum correlator, $\mathcal{Q}$ is convex and the extreme points of $\mathcal{Q}$ will be defined by particular measurements on particular pure states, i.e. $\rho'=|\psi_{l}\rangle\langle\psi_{l}|$.  $\mathcal{Q}$ is not a polytope with a finite number of extreme points as the inner product $\left|\langle\psi_{l}|\bigotimes_{j=1}^{n}|m_{j}\rangle_{s_{j}}\right|^{2}$ is continuously defined over the reals for all pure states and bases $|m_{j}\rangle_{s_{j}}$. Heuristically, if an extreme point of $\mathcal{Q}$ is outside of $\mathcal{L}$ then there is a correlator that is arbitrarily close to this point resulting from a pure state that may also be extreme (see a far more rigorous analysis in \cite{pitowski}).

As mentioned, actually finding the extreme points for all settings is a major open problem in current research. However, there do exist instances where the extreme points can be defined, particularly with $n$ parties where each site has two inputs and two outputs (see section \ref{sec32c}). Numerical methods exist for finding the boundary of $\mathcal{Q}$ using semi-definite programming \cite{navascues} and optimization over measurement bases given a particular state (e.g. the maximally entangled state for $n=2$) \cite{durt,acin2}. We will elaborate on this point further on in section \ref{sec32} of chapter \ref{chap3}.

There has recently been a different tack to defining $\mathcal{Q}$; is there some physical principle that captures the boundary of $\mathcal{Q}$? Instead of being a difficult calculation, is there is an underlying reason why the extreme points are the way they are? There is no definite answer to this, only indications of an answer (e.g. work presented in \cite{mac} and \cite{unc}). Interestingly, this approach has been extended to finding information theoretic principles that define extreme points. For example, if an extreme point were further from $\mathcal{L}$ then parties would be able to accumulate more information than is communicated to them \cite{ic} or perform calculations with a ``trivial'' amount of communication \cite{cc}.

Popescu and Rohrlich began this exploration of finding what defines the quantum region \cite{pr}. They originally asked whether it was special relativity that limits the region of $\mathcal{Q}$ but the answer to this is negative. The vertex of $\mathcal{P}$ corresponding to $p(1|\textbf{s})=\delta^{1}_{f(\textbf{s})}$ with $f(\textbf{s})=[s_{1}s_{2}+1]_{2}$ can be produced by the following distribution:
\begin{equation}\label{popescurohrlichnonlocalbox}
p(m_{1},m_{2}|s_{1},s_{2}) = 
\begin{cases} \frac{1}{2} & \text{if $[m_{1}+m_{2}]_{2}=[s_{1}s_{2}+1]_{2}$,}
\\
0 &\text{otherwise,}
\end{cases}
\end{equation}
which is an example of a non-signalling probability distribution, as described in section \ref{sec22}, more specifically it is a form of ``Popescu-Rohrlich Non-local Box'' (PR box) \cite{pr,barrett3}. However, it violates the Tsirelson-CHSH inequality above and so cannot result from quantum theory. On the other hand, it shows that there is a connection between the structure of $\mathcal{P}$ and $\mathcal{NS}$, the non-signalling polytope. In fact, this PR box is the \emph{only} non-signalling distribution that can produce the corresponding vertex of $\mathcal{P}$. In the following section we make this unique connection more concrete.

\section{Non-signalling Correlations}\label{sec23}

\noindent
In this section, we will elaborate on the connections between the polytopes $\mathcal{P}$ and $\mathcal{NS}$. We have mentioned that the PR box in \eqref{popescurohrlichnonlocalbox} is the only non-signalling correlation that can be associated with achieving a particular vertex of $\mathcal{P}$. If we assume our resources are non-signalling and we achieve a vertex of $\mathcal{P}$ associated with the function $[s_{1}s_{2}+1]_{2}$ with only one possible probability distribution. This is no coincidence, but one example of an infinite number of non-signalling probability distributions of the form \eqref{genbox} that can be uniquely associated with a vertex of $\mathcal{P}$. 

We suggest that a vertex of $\mathcal{NS}$ corresponding uniquely to a vertex of $\mathcal{P}$ is one possible way to generalise a PR box to more scenarios. We introduce another possible generalisation of a PR box in the next chapter in section \ref{sec34b}. First we discuss the $n=2$ situation and show that a vertex of $\mathcal{P}$ that is not in $\mathcal{L}$ can be uniquely associated with a vertex of $\mathcal{NS}$. We use the results we obtained for these bipartite PR boxes to consider the $n>2$ case. In this multipartite scenario, again we can uniquely associate vertices of $\mathcal{P}$ with $\mathcal{NS}$. In all of the discussion in this section, we assume that $d$ is prime.

\subsection{Generalised bipartite PR boxes}\label{sec23a}

The following lemma shows that there is a uniqueness relation between vertices of $\mathcal{P}$ and a distribution in $\mathcal{NS}$ for $n=2$, or bipartite Bell tests. This result gives us many new ways of immediately generalising the PR box.

\begin{lem}\label{lem231}
For every function $f:\mathbb{Z}_{c_{1}}\times\mathbb{Z}_{c_{2}}\rightarrow\mathbb{Z}_{d}$ that is non-$n$-partite linear for $n=2$, the only non-signalling distribution compatible with the corresponding vertex $p(k|\textbf{s})=\delta^{k}_{f(\textbf{s})}$ in $\mathcal{P}$ is
\begin{equation}
p(m_{1},m_{2}|s_{1},s_{2}) = 
\begin{cases} d^{-1} & \text{if $\left[m_{1}+m_{2}\right]_{d}=f(\textbf{s})$,}
\\
0 &\text{otherwise.}
\end{cases}
\end{equation}
\end{lem}

\textit{Proof}: The condition $p(k|\textbf{s})=\delta^{k}_{f(\textbf{s})}$ for all $\textbf{s}=\{s_{1},s_{2}\}$ implies that for every value of $m_{1}$ in $p(m_{1},m_{2}|s_{1},s_{2})$, there exists a unique value of $m_{2}=[f(\textbf{s})-m_{1}]_{d}$. This immediately implies the equality for the following conditional distributions:
\begin{eqnarray}
p(m_{1}=x,m_{2}=[f(\textbf{s})-x]_{d}|\textbf{s})&=&\sum_{m_{2}}p(m_{1}=x,m_{2}=[f(\textbf{s})-x]_{d}|\textbf{s})\nonumber \\
&=&p(m_{1}=x|\textbf{s})\nonumber \\
&=&\sum_{m_{1}}p(m_{1}=x,m_{2}=[f(\textbf{s})-x]_{d}|\textbf{s})\nonumber \\
&=&p(m_{2}=[f(\textbf{s})-x]_{d}|\textbf{s}), \nonumber \\
\end{eqnarray}
for all $x\in\mathbb{Z}_{d}$. The non-signalling condition further implies that $p(m_{1}=x|\textbf{s})$ is equal to $p(m_{1}=x|s_{1})$ and
\begin{equation}\label{marginal}
p(m_{1}=x|s_{1})=p(m_{2}=[f(\textbf{s})-x]_{d}|s_{2}),
\end{equation}
which must be satisfied for all $\textbf{s}$ and all $x$. We will show that repeated application of \eqref{marginal} for varying $\textbf{s}$ allows us to prove that all non-marginal probabilities are equal provided that $f(\textbf{s})$ has a non-$n$-partite linear element. 

A function $f(\textbf{s})$ can be decomposed into a non-$n$-partite linear and $n$-partite linear part, i.e.$f(\textbf{s})=[$g(\textbf{s})$+$h(\textbf{s})$]_{d}$ with $h(\textbf{s})$ as a non-$n$-partite linear function. For every function, the $n$-partite linear part $g(\textbf{s})=[g_{1}(s_{1})+g_{2}(s_{2})]_{d}$ can be removed by local operations performed by each party; $g_{j}(s_{j})$ is a single-site map that can be deleted from each party's outcome. Therefore, we only need to consider functions $f(\textbf{s})$ without an $n$-partite linear part. By lemma \ref{lem221}, we know for functions $f(\textbf{s})$ without an $n$-partite linear part, $f(0,0)=f(0,s_{2})=f(s_{1},0)=0$ for all $s_1$ and $s_2$. So repeatedly applying \eqref{marginal} gives
\begin{align}
p(m_2 = [- x]_{d}|s_2) &= p(m_1 = x|s_{1}=0) \nonumber\\
&= p(m_2 = [- x]_{d}|s_{2}=0) \nonumber\\
&= p(m_1 = x|s_1) \nonumber\\
&= p(m_2 = [f(s_1,s_2) - x]_{d}|s_2)
\end{align}
for all $x$. Repeated iteration implies for the $\alpha$th iteration,
\begin{equation}
p(m_2 = [- x]_{d}|s_2) = p(m_2 = [\alpha f(s_1,s_2) - x]_{d}|s_2)
\end{equation}
for all $\alpha\in\mathbb{Z}_d$. The function $f(\textbf{s})$ is non-$n$-partite linear so there must be at least one value of $\{s_{1},s_{2}\}$ where $f(s_{1},s_{2})$ is non-zero. Since $d$ is prime, $\alpha f(s_1,s_2)$ takes on all values in $\mathbb{Z}_d$, therefore the marginals are $p(m_2|s_2)=d^{-1}$ for all $m_2$. If the marginals are uniformly random for one particular input $s_2$, because $p(m_{2}=[-x]_d|s_{2})=p(m_{1}=x|s_{1}=0)=p(m_{2}=[-x]_d|s'_{2})$ for $s_{2}\neq s'_{2}$, they will be uniformly random for all inputs.

Therefore, by the non-signalling conditions $p(m_{1}|s_{1},s_{2})=p(m_{1}|s_{1},s_{2}')=d^{-1}$ and $p(m_{2}|s_{1},s_{2})=p(m_{2}|s_{1}',s_{2})=d^{-1}$ implying $p(m_1|0,s_2)=p(m_1|0,0)=d^{-1}$ and $p(m_2|s_1,0)=p(m_2|0,0)=d^{-1}$; the marginals for all $\textbf{s}$ must be completely random. Applying equation \eqref{marginal} implies that $p(m_{1},m_{2}|s_{1},s_{2})=d^{-1}$ for all $\textbf{m}$ such that $[m_{1}+m_{2}]_{d}=f(\textbf{s})$. $\square$
\newline

\subsection{Multipartite Generalisations of the PR box}\label{sec23b}

Lemma \ref{lem231} shows that for every vertex of $\mathcal{P}$ outside of $\mathcal{L}$ for $n=2$ and $d$ being prime, there is only one non-signalling probability distribution compatible with this vertex. As a corollary, $\mathcal{P}$ captures a lot of the structure of $\mathcal{NS}$ but with the space of statistics considered being smaller. We now go further and show that this one-to-one correspondence exists for $n>2$. 

Previous work has explicitly found the vertices of $\mathcal{NS}$ for $n=3$, $c_{j}=2$ for all $j$ and $d=2$ \cite{pironio3}. This work revealed that multipartite non-signalling probability distributions can have an extremely complicated and unintuitive structure. For more general scenarios, very little is understood or been investigated. Our approach, culminating in the following result,  shows that correlators can give an insight into the multipartite structure of $\mathcal{NS}$.

\begin{thm}\label{thm231}
For every function $f:\bigoplus_{j=1}^{n}\mathbb{Z}_{c_{j}}\rightarrow\mathbb{Z}_{d}$ that is non-bipartite linear, the only non-signalling distribution compatible with the corresponding vertex $p(k|\textbf{s})=\delta^{k}_{f(\textbf{s})}$ in $\mathcal{P}$ is
\begin{equation}
p(\textbf{m}|\textbf{s}) = 
\begin{cases} d^{1-n} & \text{if $\left[\sum_{j=1}^{n}m_{j}\right]_{d}=f(\textbf{s})$,}
\\
0 &\text{otherwise.}
\end{cases}
\end{equation}
\end{thm}
\textit{Proof}: As well as the above distribution of the form of \eqref{genbox} but for bipartite linear functions $f(\textbf{s})$, we can explicitly construct another non-signalling probability distribution other than \eqref{genbox}. This distribution can produce the corresponding vertex $p(k|\textbf{s})=\delta^{k}_{f(\textbf{s})}$ of $\mathcal{P}$ for a bipartite linear function $f(\textbf{s})$ and is
\begin{equation}\label{bigenbox}
p(\textbf{m}|\textbf{s}) = 
\begin{cases} d^{2-|\mathcal{J}|-|\mathcal{J}^{c}|}=d^{2-n} & \text{if $[\sum_{j\in \mathcal{J}}m_{j}]_{d}=f_{1}(\textbf{s}^{\{j|j\in\mathcal{J}\}})$}
\\
& \text{and $[\sum_{j\in \mathcal{J}^{c}}m_{j}]_{d}=f_{2}(\textbf{s}^{\{j|j\in\mathcal{J}^{c}\}})$,}
\\
0 &\text{otherwise,}
\end{cases}
\end{equation}
since a bipartite linear function can be written as 
\begin{equation}
f(\textbf{s})=[f_{1}(\textbf{s}^{\{j|j\in\mathcal{J}\}})+f_{2}(\textbf{s}^{\{j|j\in\mathcal{J}^{c}\}})]_{d}
\end{equation}
for all functions $f_{1}:\bigoplus_{j\in\mathcal{J}}\mathbb{Z}_{c_{j}}\rightarrow\mathbb{Z}_{d}$ and $f_{2}:\bigoplus_{j\in\mathcal{J}^{c}}\mathbb{Z}_{c_{j}}\rightarrow\mathbb{Z}_{d}$ for strict sub-set $\mathcal{J}$ and complement $\mathcal{J}^{c}$. The distribution is non-signalling across the partition as well as amongst the parties in the sub-set since in the sub-set it has the form \eqref{genbox}. 

This, therefore, leaves non-bipartite linear functions and their corresponding non-signalling probability distributions. As mentioned, every partition into $\mathcal{J}$ and $\mathcal{J}^{c}$ can be seen as a situation with two parties, where each side of the partition makes a choice from $c_{\mathcal{J}}=\prod_{j\in\mathcal{J}}c_{j}$ and $c_{\mathcal{J}^{c}}=\prod_{j\in\mathcal{J}^c}c_{j}$ inputs respectively; each partition also adds all their outputs togethers modulo $d$ to obtain collective outputs $m_{\mathcal{J}}=[\sum_{j\in\mathcal{J}}m_{j}]_{d}$ and $m_{\mathcal{J}^{c}}=[\sum_{j\in\mathcal{J}^{c}}m_{j}]_{d}$ respectively. As a result, Lemma \ref{lem231} now applies and if the resource produces $p(k|\textbf{s})=\delta^{k}_{f(\textbf{s})}$ for $f(\textbf{s})$ being a non-bipartite linear function, for all partitions into $\mathcal{J}$ and $\mathcal{J}^{c}$, then we obtain $p(m_{\mathcal{J}}|\textbf{s}^{\{j|j\in\mathcal{J}\}})=p(m_{\mathcal{J}^{c}}|\textbf{s}^{\{j|j\in\mathcal{J}^{c}\}})=d^{-1}$. This means all output strings $\textbf{m}^{\{j|j\in\mathcal{J}\}}$ and $\textbf{m}^{\{j|j\in\mathcal{J}^{c}\}}$ for all strict sub-sets occur with equal probability, unlike the distribution in \eqref{bigenbox}. This necessarily results in the distribution of the form \eqref{genbox} thus proving the theorem. $\square$
\newline

The uniqueness relation between vertex of $\mathcal{P}$ and a distribution in $\mathcal{NS}$ says something about the vertices of $\mathcal{NS}$. This results from the extremality of vertices of $\mathcal{P}$ and the following result that says all non-signalling probability distributions that produce a vertex of $\mathcal{P}$ must form a face of $\mathcal{NS}$. The uniqueness result of Theorem \ref{thm231} then collapses the face to a single vertex. 

\begin{prop}\label{prop231}
Every non-signalling probability distribution that produces a vertex of $\mathcal{P}$ forms a face of $\mathcal{NS}$.
\end{prop}

\textit{Proof}: Every non-signalling probability distribution $p(\textbf{m}|\textbf{s})$ can be written as a convex combination of the set $E$ of extreme points of $\mathcal{NS}$, i.e. 
\begin{equation}
p(\textbf{m}|\textbf{s})=\sum_{E}p(E)p_{E}(\textbf{m}|\textbf{s})
\end{equation}
where $p_{E}(\textbf{m}|\textbf{s})$ is a vertex distribution of $\mathcal{NS}$ and $p(E)\geq 0$ and $\sum_{E}p(E)=1$. Of the set $E$, a sub-set of extreme points $E'$ will each result in the same vertex of $\mathcal{P}$, and their convex combination will always result in a vertex $\vec{k}_{E}$ of $\mathcal{P}$. The region of distributions in $\mathcal{NS}$ which is formed by the convex hull of extreme points in $E'$ is called $\mathcal{E}$.

First, we will point out that $\mathcal{E}$ has no points in the interior of $\mathcal{NS}$ and elements of $\mathcal{E}$ are only on the boundary (i.e. surface) of $\mathcal{NS}$. If we take the convex combination of an extreme point $p_{E'}(\textbf{m}|\textbf{s})$ in $E'$ and an extreme point $p_{\not E'}(\textbf{m}|\textbf{s})$ in the set of extreme points not in $E'$, then we have the convex line:
\begin{equation}\label{convexline}
qp_{E'}(\textbf{m}|\textbf{s})+(1-q)p_{\not E'}(\textbf{m}|\textbf{s}),
\end{equation}
for $0 \leq q \leq 1$. If $\mathcal{E}$ has any elements in the interior of $\mathcal{NS}$ then $\mathcal{E}$ is intersected by at least one of the convex lines of \eqref{convexline} for $q\neq 0$ or $q\neq 1$. However, if $q\neq 1$ then this means that a probability distribution cannot result in the deterministic correlator $\vec{k}_{E}$ in $\mathcal{P}$, thereby leading to a contradiction. Therefore $\mathcal{E}$ must lie in at most a facet of $\mathcal{NS}$ because if it lies on one or more facets, then there will necessary be interior points of $\mathcal{NS}$ in $\mathcal{E}$.

Finally, we now show that if $\mathcal{E}$ is a $\Delta$-dimensional sub-space of $\mathcal{NS}$, it does not lie in $\mathcal{X}$, a $\Delta'$-dimensional sub-space (or $\Delta'$-face) of $\mathcal{NS}$ where $\Delta'>\Delta$. As a result, $\mathcal{E}$ must be a face of $\mathcal{NS}$. If $\mathcal{E}$ lies in a larger space $\mathcal{X}$, then $\mathcal{X}$ has at least one more extreme point than $\mathcal{E}$; this would mean that points in $\mathcal{E}$ can be written as a convex combination of extreme points in $E'$ \emph{and} not in $E'$. A contradiction again emerges as we would not obtain a deterministic correlator $\vec{k}_{E}$ in $\mathcal{P}$. $\square$

\begin{prop}\label{prop232}
A vertex $p(k|\textbf{s})=\delta^{k}_{f(\textbf{s})}$ of $\mathcal{P}$ corresponding to $f(\textbf{s})$ being a non-bipartite linear function results from a single vertex of $\mathcal{NS}$.
\end{prop}

\textit{Proof}: Since there is only a single non-signalling probability distribution resulting in the vertex $p(k|\textbf{s})=\delta^{k}_{f(\textbf{s})}$ of $\mathcal{P}$ for $f(\textbf{s})$ being a non-bipartite linear function, the region $\mathcal{E}$ from the proof of Proposition 1 will necessarily consist of one extreme point. Therefore, $\mathcal{E}$ becomes a $1$-face or vertex. $\square$
\newline

The space of all possible correlators $\mathcal{P}$, uniquely captures properties of a full probability distribution that only satisfies special relativity. The study of $\mathcal{NS}$ has been motivated recently by foundational issues of what distinguishes quantum physics from something unphysical (e.g. \cite{ic}). Vertices of $\mathcal{NS}$ have also been studied in the context of being an information theoretic resource \cite{barrett3}. Possession of particular resources that produce a vertex of $\mathcal{NS}$ not achievable with LHV or quantum resources (e.g. PR boxes) can lead to an information processing advantage in certain tasks (e.g. communication complexity \cite{cc}). It has also been suggested that PR boxes can be seen as a unit of non-LHV correlations (often abbreviated as ``non-locality''), though there is evidence both for and against this suggestion \cite{barrett4}. The fact that the space of correlators captures generalisations of the PR box (with respect to extremality of $\mathcal{NS}$) motivates the study of correlators as a smaller-dimensional problem revealing more general structures. 

The bipartite linear functions are not only of relevance to Proposition \ref{prop232} but also of relevance to the next section. In the next section, we discuss a generalisation of correlations discussed by George Svetlichny \cite{svet}; these are correlations that exceed LHV correlations but do involve the space of all possible correlations. Interestingly, as Svetlichny has shown, these correlations do not fully capture all quantum correlations.

\section{Svetlichny Correlations}\label{sec24}

\noindent
George Svetlichny suggested an extension to the standard model of local hidden variables in the many-party scenario. More specifically, Svetlichny introduced the scenario where there are three parties, and two parties are allowed to share whatever correlations they wish, but they are restricted to sharing only an LHV with the third party \cite{svet}. Therefore if parties $1$ and $2$ can share whatever correlation they wish (it could even not respect special relativity), then party $3$ only shares some local hidden variable $\lambda\in\Lambda$ with $1$ and $2$, to obtain the following distribution:
\begin{equation}\label{svet1}
p(\textbf{m}|\textbf{s})=\int_{\Lambda}p(\lambda)d\lambda p(m_{1},m_{2}|s_{1},s_{2},\lambda)p(m_{3}|s_{3},\lambda),
\end{equation}
with the probability distribution $p(\lambda)d\lambda$ over $\Lambda$ with $\int_{\Lambda}p(\lambda)d\lambda=1$. There is no reason to privilege some parties over others and we allow permutations of parties so labels can be swapped, i.e. $\{1,2,3\}\rightarrow \sigma(\{1,2,3\})$ and $\sigma$ is just a member of the permutation group.

In full generality, we can allow probabilistic combinations of distributions of the form \eqref{svet1} but with permutations of parties to give
\begin{eqnarray}\label{svet2}
p(\textbf{m}|\textbf{s})&=&p_{1,2}\int_{\Lambda}p_{1,2}(\lambda)d\lambda p(m_{1},m_{2}|s_{1},s_{2},\lambda)p(m_{3}|s_{3},\lambda)\nonumber \\
& &+p_{1,3}\int_{\Lambda}p_{1,3}(\lambda)d\lambda p(m_{1},m_{3}|s_{1},s_{3},\lambda)p(m_{2}|s_{2},\lambda)\nonumber \\
& &+p_{2,3}\int_{\Lambda}p_{2,3}(\lambda)d\lambda p(m_{2},m_{3}|s_{2},s_{3},\lambda)p(m_{1}|s_{1},\lambda),
\end{eqnarray}
with $p_{i,j}$ and $p_{i,j}(\lambda)$ for $i$, $j\in\{1,2,3\}$ and $i\neq j$ being probabilities for a particular permutation such that $p_{1,2}+p_{1,3}+p_{2,3}=1$. Therefore, all Svetlichny-type correlations in the form of \eqref{svet2} are in a sub-region of $\mathcal{F}$ that is a convex polytope $\mathcal{S}_{\mathcal{F}}$; the extreme points of $\mathcal{S}_{\mathcal{F}}$ are distributions of the form \eqref{svet1} but with both probabilities $p(m_{j},m_{k}|s_{j},s_{k})$ and $p(m_{l}|s_{l},\lambda)$ being deterministic for $j\neq k\neq l$.

\subsection{Three-party Generalised Svetlichny Correlators}\label{sec24a}

Since $\mathcal{S}_{\mathcal{F}}$ is a convex polytope, it will be defined as the intersection of half-spaces defined by a set of linear inequalities in analogy with the facet Bell inequalities. Svetlichny actually originally described his set of linear inequalities of correlators. We shall now take this original approach and describe Svetlichny correlations in terms of correlators where $\mathcal{S}$ is the space of Svetlichny correlators for three parties. The following result captures this space $\mathcal{S}$ in terms of the description of functions that we have used in the last two sections. 

\begin{prop}\label{prop241}
The space $\mathcal{S}$ of Svetlichny correlators for three parties is the convex hull of vertices $p(k|\textbf{s})=\delta^{k}_{f(\textbf{s})}$ of $\mathcal{P}$ corresponding to bipartite linear functions $f(\textbf{s})$.
\end{prop}

\textit{Proof}: If we take a probability distribution of the form in \eqref{svet1}, then it can itself be written as a convex combination of deterministic probabilities of the form
\begin{equation}\label{svet3}
p(\textbf{m}|\textbf{s})=\delta^{\{m_{j},m_{k}\}}_{g^{1}(s_{j},s_{k})}\delta^{m_{l}}_{g^{2}(s_{l})},
\end{equation}
for the maps $g^{1}:\mathbb{Z}_{c_{j}}\times\mathbb{Z}_{c_{k}}\rightarrow\mathbb{Z}_{d}\times\mathbb{Z}_{d}$ and $g^{2}:\mathbb{Z}_{c_{l}}\rightarrow\mathbb{Z}_{d}$ with $j\neq k\neq l\in\{1,2,3\}$. The probability in \eqref{svet3} is defined for all possible maps $g^{1}$ and $g^{2}$, therefore we can rewrite \eqref{svet1} as a convex combination of these deterministic probabllities and permutations of $\{1,2,3\}$ to give
\begin{equation}\label{svet4}
p(\textbf{m}|\textbf{s})=\sum_{\{j,k,l\}\in\sigma\{1,2,3\}}\sum_{g^{1},g^{2}}p_{g^{1},g^{2}}\delta^{\{m_{j},m_{k}\}}_{g^{1}(s_{j},s_{k})}\delta^{m_{l}}_{g^{2}(s_{l})},
\end{equation}
where $p_{g^{1},g^{2}}\geq 0$ is defined over all maps such that $\sum_{g^{1},g^{2}}p_{g^{1},g^{2}}=1$. Therefore $\mathcal{S}_{\mathcal{F}}$ is the convex hull of extreme points defined by all possible maps of the form $g^{1}$ and $g^{2}$ for all different labellings of parties. 

For correlators, we take the sum modulo $d$ of all outcomes. Taking the sum $[m_{j}+m_{k}]_{d}$ results in all maps of the form $g^{1}$ now becoming all maps of the form $f^{1}:\mathbb{Z}_{c_{j}}\times\mathbb{Z}_{c_{k}}\rightarrow\mathbb{Z}_{d}$. Finally, the sum of all outcomes is now $[m_{j}+m_{k}+m_{l}]_{d}=f(\textbf{s})=[f^{1}(\textbf{s})+f^{2}(\textbf{s})]_{d}$ where $f^{2}=g^{2}$. These functions $f(\textbf{s})$ are by definition bipartite linear functions and so $\mathcal{S}$ is the convex hull of correlators resulting from bipartite linear functions. $\square$
\newline

A facet Svetlichny inequality is a linear inequality that defines a facet of $\mathcal{S}$ in analogy with the facet Bell inequalities. One of the original facet Svetlichny inequalities for the setting with three parties, $c_{j}=d=2$ for all $j$ can be written in terms of correlators as \cite{svet}
\begin{eqnarray}\label{svetineq}
p(1|000)+p(1|001)+p(1|010)-p(1|011)& \nonumber \\
+p(1|100)-p(1|101)-p(1|110)-p(1|111)&\leq 2.
\end{eqnarray}
Interestingly, despite the fact that we allow any possible correlation to be shared between two of the three parties, correlators in $\mathcal{Q}$ still violate \eqref{svetineq} with the quantum (Tsirelson-Svetlichny) upper bound $1+\sqrt{2}$ \cite{svet}. Whilst quantum correlator vectors may be outside the space $\mathcal{S}$, this Svetlichny polytope is not strictly smaller than the space of quantum correlators, i.e. some vertices of $\mathcal{S}$ are not achievable with quantum correlators.

\subsection{Multipartite Svetlichny Correlators}\label{sec24b}

The above discussion has been restricted to Svetlichny's original work for three parties. It is natural to ask how this approach generalises to more than three parties. One could suggest a model where we allow only at most two out of $n$ parties to share whatever correlation they wish and then share local hidden variables with the other $(n-2)$ parties. We will go further, and in line with other approaches (e.g. \cite{bancal,bancal2}), partition $n$ parties into two sub-sets and parties in each of the two sub-sets is allowed to share whatever correlations they wish (signalling or otherwise). Then each partition only shares a local hidden variable $\lambda\in\Lambda$ (with probability distribution $p(\lambda)d\lambda$) with the other partition to  obtain correlations of the form:
\begin{equation}\label{gensvet}
p(\textbf{m}|\textbf{s})=\int_{\Lambda}p(\lambda)d\lambda p(\mathbf{m}^{\{j|j\in\mathcal{J}\}}|\mathbf{s}^{\{j|j\in\mathcal{J}\}},\lambda)p(\mathbf{m}^{\{j|j\in\mathcal{J}^{c}\}}|\mathbf{s}^{\{j|j\in\mathcal{J}^{c}\}},\lambda),
\end{equation}
where $n$ parties are partitioned into sub-sets $\mathcal{J}$ and $\mathcal{J}^{c}$. 

As with three parties, we allow convex combinations of distributions in \eqref{gensvet} for all $(2^{n-1}-1)$ different partitions into strict sub-sets $\mathcal{J}$ and $\mathcal{J}^{c}$. Correlators resulting from this generalised Svetlichny model can again be expressed as a convex polytope as a generalisation of Proposition \ref{prop241}; the following result now captures this generalisation.

\begin{thm}\label{thm241}
The space $\mathcal{S}$ of generalised Svetlichny correlators for $n$ parties is the convex hull of vertices $p(k|\textbf{s})=\delta^{k}_{f(\textbf{s})}$ of $\mathcal{P}$ corresponding to bipartite linear functions $f(\textbf{s})$.
\end{thm}

\textit{Proof}: The correlations in \eqref{gensvet}, as with the three-party case, can be written as a convex combination of deterministic probabilities resulting from deterministic maps labelled $g^{1}$ and $g^{2}$:
\begin{equation}
p(\textbf{m}|\textbf{s})=\sum_{g^{1},g^{2}}p_{g^{1},g^{2}}\delta_{g^{1}(\mathbf{s}^{\{j|j\in\mathcal{J}\}})}^{\mathbf{m}^{\{j|j\in\mathcal{J}\}}}\delta_{g^{2}(\mathbf{s}^{\{j|j\in\mathcal{J}^{c}\}})}^{\mathbf{m}^{\{j|j\in\mathcal{J}^{c}\}}},
\end{equation}
where $g^{1}:\bigoplus_{j\in\mathcal{J}}\mathbb{Z}_{c_{j}}\rightarrow\mathbb{Z}_{d}^{|\mathcal{J}|}$ and $g^{2}:\bigoplus_{j\in\mathcal{J}^{c}}\mathbb{Z}_{c_{j}}\rightarrow\mathbb{Z}_{d}^{|\mathcal{J}^{c}|}$ with $p_{g^{1},g^{2}}\geq 0$ and $\sum_{g^{1},g^{2}}p_{g^{1},g^{2}}=1$. Now if we take the sum modulo $d$ of all outcomes then we obtain the following correlators:
\begin{equation}\label{svetcorr}
p(k|\textbf{s})=\sum_{f^{1},f^{2}}p_{f^{1},f^{2}}\delta^{k}_{[f^{1}(\textbf{s})+f^{2}(\textbf{s})]_{d}},
\end{equation}
with all possible maps of the form $f^{1}:\bigoplus_{j\in\mathcal{J}}\mathbb{Z}_{c_{j}}\rightarrow\mathbb{Z}_{d}$ and $f^{2}:\bigoplus_{j\in\mathcal{J}^{c}}\mathbb{Z}_{c_{j}}\rightarrow\mathbb{Z}_{d}$ and the distribution $p_{f^{1},f^{2}}\geq 0$ such that $\sum_{f^{1},f^{2}}p_{f^{1},f^{2}}=1$.

If we allow all possible correlators of the form \eqref{svetcorr} for all possible partitions into $\mathcal{J}$ and $\mathcal{J}^{c}$ then $\mathcal{S}$ is the convex hull of all deterministic correlators corresponding to functions $f(\textbf{s})=[f^{1}(\textbf{s})+f^{2}(\textbf{s})]_{d}$. These are all of the bipartite linear functions by definition. $\square$
\newline

The structure of bipartite linear functions gets translated from the three-party case to the $n$-party case. Despite the fact that we allowed signalling correlations within partitions of the $n$ parties, we can impose the non-signalling conditions on all parties once again. This means that even within a sub-set of parties, the correlations they share must satisfy special relativity. Interestingly, even if we apply this restriction, the space of Svetlichny correlators for many parties is still $\mathcal{S}$ as defined by Theorem \ref{thm241}. This is simply because all deterministic correlators (or vertices of $\mathcal{P}$) can be achieved with non-signalling probability distributions $\mathcal{NS}$. All the deterministic correlators associated with bipartite linear functions can be achieved with probability distributions in $\mathcal{NS}$. 

If one assumes that all correlations satisfy special relativity, then non-signalling correlations not achievable with Svetlichny-type correlations are said to be ``truly $n$-partite non-local'' \cite{bancal,barrett3}. They are ``non-local'' in the sense that across all partitions of $n$ parties, the correlations of the parties are not described by the parties sharing a local hidden variable. Therefore, the vertices of $\mathcal{P}$ that are not associated with bipartite linear functions can only result from truly $n$-partite non-local correlations. Of the non-signalling correlations in $\mathcal{NS}$, then for each of these vertices of $\mathcal{P}$ there is one truly $n$-partite non-local distribution, or vertex of $\mathcal{NS}$ as described by Theorem \ref{thm231}.

Instead of allowing all possible correlations within a sub-set of all parties or just allowing non-signalling correlations, one could allow correlations ``in-between'' that allow some, but not all forms of communication. Indeed, these issues have been investigated by Barrett and Pironio \cite{barrett5}. If one is only concerned with correlators, then whatever form of restricted, or unrestricted, communication within a partition of all parties, the space of Svetlichny-type correlators is $\mathcal{S}$ as described by Theorem \ref{thm241}. The space of correlators is conserved and we can always discuss the possibility of distinguishing between a model that permits, in part, an LHV description and something inconsistent with this model. 

\section{Chapter Summary}\label{sec25}

\noindent
In this chapter we have motivated and presented the study of Bell correlators in a natural generalisation of the Bell-CHSH test. We have also discussed how finding Bell inequalities that define the space of LHV correlations/correlators is in general a hard problem. Motivated by this, studying correlators instead of a full probability distribution, we reduce the size of the problem, if not reducing the general hardness.

The language of stochastic maps and functions has been key to describing the correlators resulting from particular theories (both physical and non-physical). This description of correlators in terms of \emph{computational expressiveness} is key to the central results of not only this chapter, but this entire thesis. To summarise, each potential theory has its own computational expressiveness and characterising this gains an insight into ``which computations the theory is capable of performing''. These ideas will be generalised in subsequent chapters to take into account data processing in Bell tests but the computational expressiveness insight will be key. Importantly, this computational point-of-view on correlators has allowed us to characterise the well-studied structures of LHV correlators in a new language.

This interpretation of correlators in terms of computation has also produced new results. We showed that vertices of the polytope of all correlators can correspond uniquely to vertices of the non-signalling polytope. As well as this, we have described the space of Svetlichny correlators in terms of computational expressiveness. Again, this description of Svetlichny correlations gives us a new insight into well-studied areas of research. 

\chapter{Constructing Bell Inequalities and Quantum Violations}\label{chap3}

\noindent
In the previous chapter, we focussed mostly on the description of the local hidden variable (LHV) polytope in terms of its vertices. Now we shift to a facet representation of the LHV polytope in terms of the facet Bell inequalities: linear inequalities defining the facets of this polytope \cite{fine,pitowski}. If a correlator is outside of the polytope it must necessarily violate at least one of these inequalities. However, recall that finding them is a hard problem. 

A Bell inequality is a linear inequality of the following form
\begin{equation}\label{bell1}
\sum_{\textbf{s}}\sum_{k=1}^{(d-1)}\beta_{k,\textbf{s}}p(k|\textbf{s})\leq \gamma_{\mathcal{L}},
\end{equation}
for some real coefficients $\beta_{k,\textbf{s}}$ where $\gamma_{\mathcal{L}}$ is the tight upper bound for all LHV correlators in $\mathcal{L}$ \footnote{In the literature, tight Bell inequalities are synonymous with facet Bell inequalities. Our use of the word tight reflects that the Bell inequality intersects the LHV polytope at (at least) one of its extreme points.}. We introduce the vernacular that a ``Bell expression'' is the left-hand-side of \eqref{bell1}. We make the distinction between Bell expression and Bell inequality as we can substitute correlators not in $\mathcal{L}$ into a Bell expression and they could violate a Bell inequality. 

We optimize over values $\beta_{k,\textbf{s}}$ and $\gamma_{\mathcal{L}}$ in \eqref{bell1} to find the facet Bell inequalities. But this optimization, in the worst case, is a hard computational task. In this chapter we look for these facet Bell inequalities but only manage to find them for a select number of scenarios on a desktop PC using Polymake \cite{polymake}. We give some indications of the possible connections between the violations of facet inequalities and the possibility of performing a non-$n$-partite linear function. However, this connection is not completely clear as the structure of $\mathcal{L}$ is in general, rather complicated. On the other hand, we review the results of Werner, Wolf, \.{Z}ukowski and Brukner \cite{ww,zb} in the $n$ party, $2$ input, $2$ output scenario and relate the structure of $\mathcal{L}$ in this scenario to a particular class of Boolean functions. 

In spite of the difficulty in understanding the structure of $\mathcal{L}$ and even finding the facet Bell inequalities, we find a general class of Bell inequalities that have a natural computational perspective. We call these inequalities non-trivial Bell inequalities. They are non-trivial in the sense that they provide a separation between all possible correlators in $\mathcal{L}$ and all possible correlators in $\mathcal{P}$. We go on to relate these inequalities to an information processing paradigm called a ``non-local game'' \cite{nonlocal}. We then use the construction of a non-local game to derive more of these non-trivial Bell inequalities.

Finally in this chapter, we make interesting connections between the discussion of Bell inequalities and Measurement-based Quantum Computing (MBQC) \cite{mbqc1,mbqc2,mbqc3,mbqc4}. In particular, we show that a sub-class of computations in Briegel and Raussendorf's construction of MBQC \cite{mbqc1,mbqc2} can be cast as non-local games. Through the language of non-local games, we relate these quantum computations to non-trivial Bell inequalities. All of these connections truly highlight the rich interplay between the foundations of quantum mechanics and its applications. 

The original material in sections \ref{sec31} and \ref{sec32} along with subsections  \ref{sec33a} and \ref{sec33b} were completed in collaboration with Joel Wallman and Dan Browne and published in part as \cite{hoban3}. The subsections of \ref{sec31e} and \ref{sec32c} consist of rederivations of results in \cite{ww} with a focus on the computational description of correlators. The original work in subsection \ref{sec33c} and section \ref{sec34} were done in collaboration with Earl Campbell, Klearchos Loukopoulos and Dan Browne and published as \cite{hoban1}. 

\subsection{Notation}\label{sec30a}

From now on, we simplify the scenarios of Bell tests that we consider by having the number of inputs at each site being the same, i.e. $c_{j}=c_{j'}$ for all $j\neq j'$. We introduce the notation $(n,c,d)$ to describe Bell tests with $n$ parties, $c$ inputs and $d$ outputs at each site. We also carry over the notation from chapter \ref{chap2} of $\mathcal{L}$, $\mathcal{S}$, $\mathcal{Q}$ and $\mathcal{P}$ being the LHV polytope, the Svetlichny polytope, the space of quantum and all possible correlators respectively for each scenario $(n,c,d)$.

The majority of the remainder of this thesis will be devoted to the study of the $(n,2,2)$ scenario. We privilege this scenario by assigning it a particular notation not shared by any others. Since the number of the inputs at each site is the same, inputs are always $\textbf{s}\in\bigoplus_{j=1}^{n}\mathbb{Z}_{c_{j}}=\mathbb{Z}_{c}^{n}$. As a result of this simplification, we will no longer use the notation $\bigoplus$ to describe the Cartesian product of groups $\mathbb{Z}_{c_{j}}$. We will use $\bigoplus$ to denote summation modulo $2$, i.e. $\bigoplus=\left[\sum...\right]_{2}$. This notation is used only in the $(n,2,2)$ scenario along with the notation $\oplus$ to describe addition modulo $2$, i.e. $\oplus=\left[...+...\right]_{2}$. Modulo $2$ multiplication between elements in $\mathbb{Z}_{2}$ is exactly multiplication of these elements for standard arithmetic. Therefore, for the $(n,2,2)$ scenario and only this scenario we re-write expressions in modulo $2$ arithmetic in terms of this notation. For example, the expression $\left[x_{1}x_{2}+x_{3}+1\right]_{2}$ becomes $x_{1}x_{2}\oplus x_{3}\oplus 1$, and, $\left[\left(\sum_{j=1}^{4}x_{j}\right)+x_{5}+1\right]_{2}$ becomes $\left(\bigoplus_{j=1}^{4}x_{j}\right)\oplus x_{5}\oplus 1$.

For scenarios other than $(n,2,2)$, we retain the notation from the previous chapter. That is, all arithmetic in $\left[...\right]_{x}$ is modulo $x$ arithmetic. Even if either $c$ or $d$ is equal to $2$ (but not both), we will use the notation $\left[...\right]_{2}$ for modulo $2$ arithmetic.

\section{Facet Bell Inequalities}\label{sec31}

\noindent
In this section, we will discuss the facet Bell inequalities for particular $(n,c,d)$ scenarios. We used the Polymake package of algorithms to find the facet Bell inequalities for a small number of cases \cite{polymake}. These are the $(n,c,d)$ scenarios where finding the inequalities was computationally tractable on a desktop PC\footnote{iMac with 2.4 GHz Intel Core 2 Duo (TM) Processor and 2 GB 800 MHz DDR2 SDRAM.}. We will show that these inequalities can be grouped together into symmetries, or in group theoretical terms, orbits; these orbits are generated by operations that preserve the region $\mathcal{L}$ \cite{pitowski2}. For the number of $(n,c,d)$ scenarios studied, we will describe elements in these orbits. Then we discuss the facet Bell inequalities for the $(n,2,2)$ scenario; there is a closed-form expression for these inequalities \cite{ww,zb}. 

In Table \ref{tab:tab1} we have listed the number of facet Bell inequalities for a few scenarios that could be computed using Polymake. Included in the number of facet Bell inequalities are the $c^{n}$ normalization and $(d-1)c^{n}$ positivity inequalities that define $\mathcal{P}$. Despite these $dc^{n}$ inequalities, there are still a significant number of inequalities remaining. On the other hand, Pitowsky has shown that correlation polytopes have certain symmetries \cite{pitowski2}. These symmetries are generated by operations on the inputs and outputs as well as permutations of parties. The group of these symmetry operations generates \emph{orbits} of facet Bell inequalities\footnote{We are using the terminology used by Werner and Wolf \cite{ww}.}. Every facet Bell inequality in each orbit can be mapped to every other inequality in that orbit via these symmetry operations. Therefore, we do not need to consider every single facet Bell inequality for each $(n,c,d)$ scenario but only one inequality in each orbit. In the following subsection we consider these symmetry operations.

\begin{table}
\begin{center}
 \begin{tabular}{| c | c | c | c | c |}
 \hline
 n & c & d & \# Vertices &\# Facet Bell inequalities \\ \hline
 2 & 2 & 2 & 8 & 16 \\ 
 2 & 2 & 3 & 27 & 66 \\
 2 & 2 & 4 & 64 & 216 \\
 2 & 2 & 5 & 125 & 1020 \\
 3 & 2 & 2 & 16 & 256 \\
 3 & 2 & 3 & 81 & 125,412 \\
 2 & 3 & 2 & 32 & 90 \\
 2 & 4 & 2 & 128 & 27,968 \\
 \hline
 \end{tabular}
\end{center}
\caption{A table of number of facet Bell inequalities for each scenario $(n,c,d)$ and the number of vertices for the LHV polytope.}
\label{tab:tab1}
\end{table}

\subsection{Symmetries of the LHV Polytope}\label{sec31a}

Pitowsky has shown that given a facet Bell inequality for an LHV correlation polytope, we may find more inequalities by some simple operations on data $\textbf{m}$ and $\textbf{s}$ \cite{pitowski2}. These operations $G$ map from the set $\mathcal{E}$ of extreme points of $\mathcal{L}$ to themselves, i.e. $G:\mathcal{E}\rightarrow\mathcal{E}$. By convexity, we only need to consider the extreme points. The symmetry operations $G$ that produce these maps are the following:
\begin{enumerate}
\item permutations of parties - $\{s_{i},s_{j},...,s_{n}\}\rightarrow\{s_{i'},s_{j'},...,s_{n'}\}$ where $k'=\sigma(k)$ is an element of the permutation group $S^{n}$ of order $n$;
\item relabeling of measurement scenarios - $s_{j}\rightarrow s_{j}+a_j$ for some $a_j\in\mathbb{Z}_{c}$;
\item relabeling of measurement outcomes - $m_{j}\rightarrow m_{j}+b(s_{j},j)$ where $b(s_{j},j)\in\mathbb{Z}_{d}$.
\end{enumerate}
 The operations $G$ and their products $GG'$ (for either $G\neq G'$ or $G=G'$) form a group $\mathbb{G}$ such that $G\in\mathbb{G}$. There are $n!$ permutations of $n$ parties and $c^{n}$ ways of relabeling measurement scenarios. Since for each input $s_{j}$ we add a value $b(s_{j})$, for each input $\textbf{s}$, $b(\textbf{s})=\sum_{j=1}^{n}b(s_{j})$ is added to $\sum_{j=1}^{n}m_{j}$. There will be at most $d^{cn}$ values of $b(\textbf{s})$. In total, there are at most $n!c^{n}d^{cn}$ elements of $\mathbb{G}$ in order for there to be closure\footnote{In principle, the number of operations could be smaller as the values of $b(\textbf{s})$ may be overcomplete for all possible transformations. For example, in \cite{ww} the cardinality of $\mathbb{G}$ is $n!2^{2n+1}$.}.

The $n$-partite linear functions are closed under all of these operations. Using the facet-defining condition, the vertices of $\mathcal{L}$ that saturate a facet Bell inequality must be equivalent to another set of vertices in $\mathcal{L}$; this new set also saturates a facet Bell inequality. In group theoretic terms, if we have one facet Bell inequality and perform all possible sequences of operations $G$, then the set of facet Bell inequalities produced by these operations forms an \emph{orbit} (see the use of terminology in \cite{ww}). In Table \ref{tab:tab2} we have listed the number of orbits for each of the scenarios in Table \ref{tab:tab1}. These orbits were numerically found using a search algorithm on all of the facet Bell inequalities. For each instance of $(n,c,d)$, it was found that one of the orbits consists of the normalisation and positivity inequalities; we call this orbit the ``trivial orbit''. Orbits which do not include the normalisation and positivity inequalities are called ``non-trivial orbits''.

\begin{table}
\begin{center}
 \begin{tabular}{| c | c | c | c | c |}
 \hline
 n & c & d & \# Facet Bell inequalities &\# Orbits \\ \hline
 2 & 2 & 2 & 16 & 2 \\ 
 2 & 2 & 3 & 66 & 2 \\
 2 & 2 & 4 & 216 & 4 \\
 2 & 2 & 5 & 1020 & 5 \\
 3 & 2 & 2 & 256 & 5 \\
 3 & 2 & 3 & 125,412 & 63 \\
 2 & 3 & 2 & 90 & 2 \\
2 & 4 & 2 & 27,968 & 15 \\
 \hline
 \end{tabular}
\end{center}
\caption{The number of orbits for each scenario $(n,c,d)$ under the symmetry operations described in the text. One of the orbits for each scenario is the orbit of normalization and positivity conditions.}
\label{tab:tab2}
\end{table}

For each of the $(n,c,d)$ scenarios, we only need to consider one inequality from each orbit. For the $(2,2,2)$, $(2,2,3)$ and $(2,3,2)$ scenarios, there is only one non-trivial orbit. In each of these scenarios, we then only need to consider one inequality. If one of these inequalities in each orbit is violated by a quantum correlator, then the above symmetry operations can be applied to that quantum correlator so that it will violate every other inequality in said orbit. The possibility of violation of facet Bell inequalities with quantum correlators is, as a result, rendered easier to study.

For the $(2,2,2)$ scenario, as Fine has also shown in \cite{fine}, the only facet Bell inequality we need to consider is the CHSH inequality \cite{chsh}. In the following subsection we consider other facet Bell inequalities for $n=2$. We show that the CHSH inequality and a generalisation in $d$ (for $c=2$) of this inequality (the CGLMP inequality \cite{cglmp}) between them generate a lot of the structure of $\mathcal{L}$. In later subsections \ref{sec31d} and \ref{sec31e} we will discuss the tripartite and multipartite scenario (i.e. for $n>2$). First we briefly introduce some new notation.

\subsection{Notation for Bell inequalities}\label{sec31b}

We now introduce a piece of notation to describe all Bell inequalities. If we write vectors $\vec{k}$ of correlators that have elements $p(k|\textbf{s})$, we can express an inequality as an inner product. The real pre-factors $\beta_{k,\textbf{s}}$ of \eqref{bell1} are elements of an $(d-1)c^{n}$-length row vector $\vec{b}\in\mathbb{R}^{(d-1)c^{n}}$. Therefore, every inequality results from the Euclidean inner product $\vec{b}\cdot\vec{p}\leq\gamma_{\mathcal{L}}$ of these two vectors. 

We adopt a convention to order the elements $\beta_{k,\textbf{s}}$ of $\vec{b}$ from left-to-right starting with $\beta_{1,\textbf{0}}$ and ending with $\beta_{(d-1),\textbf{c}}$ with $\textbf{c}=\{(c-1),(c-1),...,(c-1)\}$, the digit-string of all inputs being $(c-1)$. To be explicit, each digit-string $\textbf{s}\in\mathbb{Z}_{c}^{n}$ can be written as an integer in $\mathbb{Z}$, the set of positive integers. Digit-strings $\textbf{s}\in\mathbb{Z}_{c}^{n}$ can be ordered in terms of these integers in $\mathbb{Z}$. For example for $c=2$, the digit-string $\textbf{s}=\{1,0,0\}$ corresponds to the integer $4$ and for $c=3$ the same digit-string is equal to $9$. We order elements $\beta_{1,\textbf{0}}$ from left-to-right for increasing values of $k\in\mathbb{Z}_{d}$ for each ordered value of $\textbf{s}$.

To give a concrete example, the CHSH inequality \cite{chsh}
\begin{equation}
p(1|00)+p(1|01)+p(1|10)-p(1|11)\leq 2,
\end{equation}
corresponds to the vector $\vec{b}=(\beta_{1,\{0,0\}},\beta_{1,\{0,1\}},\beta_{1,\{1,0\}},\beta_{1,\{1,1\}})=(1,1,1,-1)$. We will employ this notation for specific values $n$, $c$ and $d$. For brevity, in more general expressions we may choose to write the inequality in terms of the sum in \eqref{bell1}. In the next subsection we will write both in terms of the sum in \eqref{bell1} and the vector notation introduced above. 

\subsection{Bipartite facet Bell inequalities}\label{sec31c}

In this subsection we will restrict ourselves to the $n=2$ scenario for particular values of $c$ and $d$. The CGLMP inequality \cite{cglmp} is a facet Bell inequality for all $d$ in $(2,2,d)$ scenarios, as shown by Masanes \cite{masanes}. For all $d$, this inequality can be written as
\begin{equation}
\mathcal{C}_{\textrm{CGLMP}}=d\times p(1|0,0)- \sum_{\textbf{s}}(-1)^{s_{1}+s_{2}}p(1|\textbf{s})+
\sum_{\textbf{s}}(-1)^{s_{1}+s_{2}}\sum_{k=2}^{d-1}(d-k-1)p(k|\textbf{s})\leq d.
\end{equation}
The CHSH inequality is exactly this inequality when $d=2$. For $d=2$, $3$, the only non-trivial orbit is generated by the CGLMP inequality. Whilst for $d=4$, $5$ the CGLMP inequality generates one of $(d-1)$ non-trivial orbits. For all possible correlators in $\mathcal{P}$, the maximal value of the left-hand-side of the CGLMP inequality is $2d-1$, thus violating it. In fact, for all $d$, this maximal violation of the CGLMP is obtained by a vertex of $\mathcal{P}$ corresponding to the function $f(\textbf{s})=[s_{1}s_{2}+1]_{d}$, i.e. the correlator $p(k|\textbf{s})=\delta^{k}_{[s_{1}s_{2}+1]_{d}}$. 

In the $(2,2,2)$ scenario there are $2^{4}-2^{3}=8$ non-$n$-partite linear functions and also $8$ inequalities in the non-trivial orbit of the CHSH inequality. This is no coincidence as every Bell inequality in this orbit in maximally violated by a vertex of $\mathcal{P}$ corresponding to a non-$n$-partite linear function. This also occurs for the $(2,2,3)$ scenario where there are $3^{4}-3^{3}=54$ non-$n$-partite linear functions and $54$ inequalities in the orbit of the CGLMP inequality. It can also be checked that every inequality in this orbit is violated by a different non-$n$-partite linear function.

For $(2,2,4)$, one of the orbits is generated by a generalisation of the CHSH inequality
\begin{equation}\label{genchsh4}
\mathcal{C}^{1}_{d=4}=\sum_{\textbf{s}}(-1)^{s_{1}s_{2}}\left[p(1|\textbf{s})+p(3|\textbf{s})\right]\leq 2.
\end{equation}
This expression is essentially the CHSH inequality if each party groups their outcomes $m_{j}$ into modulo $2$ terms. Since $1$ mod $2$ is equal to $3$ mod $2$, each party just maps from modulo $4$ arithmetic to modulo $2$. For all possible correlators in $\mathcal{P}$, the Bell expression in inequality \eqref{genchsh4} achieves the value of $3$. This value is achieved for two vertices of $\mathcal{P}$ corresponding to functions $f(\textbf{s})=[s_{1}s_{2}+1]_{2}$ or $f(\textbf{s})=[s_{1}s_{2}+1]_{4}$. Therefore the one-to-one relationship between inequality and maximal violation from a vertex of $\mathcal{P}$ breaks down for $d=4$ (and also $d=5$). This is confirmed by the number of facet Bell inequalities in non-trivial orbits for $(2,2,4)$ being $216-26=200$ whereas the number of non-$n$-partite linear functions is $4^{4}-4^{3}=192$.

The third and final non-trivial orbit for $(2,2,4)$ is generated by the following inequality (expressed in the notation described earlier):
\begin{equation}
\mathcal{C}^{2}_{d=4}=\left(1,2,1,1,2,1,1,2,1,-1,-2,-1\right)\cdot\vec{k}\leq 4.
\end{equation}
It is worth noting that this can be constructed by adding $\sum_{\textbf{s}}2(-1)^{s_{1}s_{2}}p(2|\textbf{s})$ to the left-hand-side of the previous inequality \eqref{genchsh4}. The maximal value of $6$ of the left-hand-side (i.e. Bell expression) results from the vertex of $\mathcal{P}$ corresponding to the function $f(\textbf{s})=[2s_{1}s_{2}+2]_{4}$.

For $(2,2,5)$, there are $4$ non-trivial orbits. One of these is generated by the CGLMP inequality and the other three are given by
\begin{align}
\mathcal{I}_{1} &= \frac{1}{2}\left(6,2,3,4,4,-2,2,1,4,-2,2,1,-4,2,-2,-1\right)\cdot\vec{k}\leq 5,	\nonumber\\
\mathcal{I}_{2} &= \left(3,1,-1,-3,2,-1,-4,-2,2,-1,-4,-2,-2,1,4,2\right)\cdot\vec{k}\leq 5,	\nonumber\\
\mathcal{I}_{3} &= \left(2,-1,1,-2,3,1,-1,2,3,1,-1,2,-3,-1,1,-2\right)\cdot\vec{k}\leq 5.
\end{align}
The inequality for the Bell expression $\mathcal{I}_1$ and the CGLMP inequality are maximally violated by the vertex corresponding to $f(\textbf{s})=[s_{1}s_{2}+1]_{5}$. The Bell expressions $\mathcal{I}_2$ and $\mathcal{I}_3$ are maximally violated by the vertex corresponding to $f(\textbf{s})=[2s_{1}s_{2}+1]_{5}$. As we can seen there is a corresponding function for each of these inequalities that leads to a maximal violation. 

We now consider scenarios with $c>2$ but with $d=2$. As can be seen from Table \ref{tab:tab2} for the $(2,3,2)$ scenario there is only one non-trivial orbit. The Bell inequality generating this orbit is another generalisation of the CHSH inequality:
\begin{equation}
\mathcal{C}_{c=3}=\sum_{\textbf{s}}(-1)^{s_{1}s_{2}}\prod_{j=1}^{2}(\delta^{s_{j}}_{0}+\delta^{s_{j}}_{1})p(1|\textbf{s})\leq 2.
\end{equation}

\begin{table}
\begin{center}
 \begin{tabular}{| c | c c c c c c c c c c c c c c c c |}
 \hline
 & & & & & & & & $\vec{b}$ & & & & & & & & \\ \hline
 $\mathcal{B}_{1}$& 2&2&1&1&2&-1&-1&-2&1&-1&-2&2&1&-2&2&1\\
 $\mathcal{B}_{2}$& 2&2&1&1&2&-1&-1&-2&1&-2&2&1&1&-1&-2&2\\
 $\mathcal{B}_{3}$& 2&2&1&1&2&-1&-2&-1&1&-2&1&2&1&-1&2&-2\\
 $\mathcal{B}_{4}$& 2&2&1&1&1&-1&2&-2&1&-2&1&2&2&-1&-2&-1\\
 $\mathcal{B}_{5}$& 2&2&1&1&1&-2&2&1&1&-1&-2&2&2&-1&-1&-2\\
 $\mathcal{B}_{6}$& 2&1&1&0&1&-1&-1&1&1&-1&-1&-1&0&1&-1&0\\
 $\mathcal{B}_{7}$& 2&1&1&0&1&-1&-1&1&0&1&-1&0&1&-1&-1&-1\\
 $\mathcal{B}_{8}$& 2&1&1&0&0&1&-1&0&1&-1&-1&1&1&-1&-1&-1\\
 $\mathcal{B}_{9}$& 2&1&0&1&1&-1&1&-1&0&1&0&-1&1&-1&-1&-1\\
 $\mathcal{B}_{10}$& 2&1&0&1&0&1&0&-1&1&-1&1&-1&1&-1&-1&-1\\
 $\mathcal{B}_{11}$& 2&0&1&1&0&0&1&-1&1&1&-1&-1&1&-1&-1&-1\\
$\mathcal{C}^{1}_{c=4}$& 1&1&0&0&1&-1&0&0&0&0&0&0&0&0&0&0\\
$\mathcal{C}^{2}_{c=4}$& 1&1&0&0&0&0&0&0&1&-1&0&0&0&0&0&0\\
$\mathcal{C}^{3}_{c=4}$& 1&0&1&0&0&0&0&0&1&0&-1&0&0&0&0&0\\
 \hline
 \end{tabular}
\end{center}
\caption{The facet Bell inequality expressions that each belong to a particular non-trivial orbit for $(2,4,2)$. Each row corresponds to a particular inequality belonging to a different symmetry class. Each column of $\vec{b}$ is an element of this vector that forms an inner product with $\vec{p}$. The LHV upper bound for inequalities $\mathcal{B}_{1}$ to $\mathcal{B}_{5}$ is $8$ and $4$ for $\mathcal{B}_{6}$ to $\mathcal{B}_{11}$.}
\label{tab:tab4}
\end{table}

For the $(2,4,2)$ scenario, three of these non-trivial orbits are forms of the CHSH inequality embedded in the larger number of inputs. For completeness, we have listed all $14$ Bell inequalities in Table \ref{tab:tab4}. We now explicitly write out one of these inequalities:
\begin{equation}
\mathcal{C}^{1}_{c=4}=\sum_{\textbf{s}}(-1)^{s_{1}s_{2}}\prod_{j=1}^{2}(\delta^{s_{j}}_{0}+\delta^{s_{j}}_{1})p(1|\textbf{s})\leq 2.
\end{equation}
which is almost exactly the same as $\mathcal{C}^{1}_{c=3}$. The other two inequalities, $\mathcal{C}^{2}_{c=3}$ and $\mathcal{C}^{3}_{c=3}$ are similar to this inequality except with altered delta functions for $\mathcal{C}^{2}_{c=3}$ via the substitutions:
\begin{equation}
\prod_{j=1}^{2}(\delta^{s_{j}}_{0}+\delta^{s_{j}}_{1})\rightarrow(\delta^{s_{1}}_{0}+\delta^{s_{1}}_{2})(\delta^{s_{2}}_{0}+\delta^{s_{2}}_{1}),
\end{equation}
and for $\mathcal{C}^{3}_{c=3}$:
\begin{equation}
\prod_{j=1}^{2}(\delta^{s_{j}}_{0}+\delta^{s_{j}}_{1})\rightarrow(\delta^{s_{1}}_{0}+\delta^{s_{1}}_{2})(\delta^{s_{2}}_{0}+\delta^{s_{2}}_{2}).
\end{equation}
We can see that the CHSH inequality generates a lot of the structure of the LHV polytope in the bipartite scenario. In general though, we have given some insight into the richness of structure of $\mathcal{L}$. This might give some indication why finding the facet Bell inequalities is a complicated task. All of this discussion is even before we consider more than $2$ parties. In the following subsection we discuss the $n=3$ case. Despite not having as many results in this scenario due to the scaling of the size of $\mathbb{R}^{(d-1)c^{n}}$ in $n$, we show some of the structure of $\mathcal{L}$ can be obtained from the $n=2$ scenario.

\subsection{Tripartite facet Bell inequalities}\label{sec31d}

We have given an indication that facet Bell inequalities for $n=c=2$ have a computational interpretation. Every facet Bell inequality we have found is maximally violated uniquely by a vertex of $\mathcal{P}$ when $d=2$, $3$, and $5$. In this sense the violation of a facet Bell inequality can quantify how computationally powerful a theory is. For situations with $n>2$, this becomes more complicated even for $n=3$ and $c=d=2$. The Mermin inequality \cite{mermin} which we introduced in the first chapter (see section \ref{sec12}) can be expressed as
\begin{equation}\label{merm}
p(1|000)+p(1|011)+p(1|101)-p(1|110)\leq 2,
\end{equation}
and forms a non-trivial orbit \cite{ww}. This inequality is maximally violated by more than one vertex of $\mathcal{P}$. If expressed in terms of expectation values of measurements, it can be generated from the CHSH inequality by a form of substitution \cite{ww}. WW showed that all inequalities for $(n,2,2)$ can be generated by this substitution \cite{ww}. We now discuss a possible method of doing this for $(3,2,3)$.

Analogously to the Mermin inequality \eqref{merm}, we define a CGLMP inequality for three parties using the two party inequality. We have three parties but now we only consider non-zero terms in a Bell inequality when the third party's input is $s_{3}=0$. For LHV correlators $p(k|s_{1},s_{2},0)$ the $n$-partite linear functions that can be achieved are $f(\textbf{s})=[\alpha_{1}s_{1}+\alpha_{2}s_{2}+\alpha_{3}]_{3}$ with $\alpha_{1},\alpha_{2},\alpha_{3}\in\mathbb{Z}_{d}$: the $n$-partite linear functions on two variables $s_{1}$ and $s_{2}$. Since the CGLMP inequality is facet-defining for the region of LHV correlators for two parties, or variables $s_{1}$ and $s_{2}$, it is facet-defining for this space of the $n=3$ correlators for $s_{3}=0$. Then we can write the tripartite CGLMP inequality as
\begin{eqnarray}
\mathcal{C}^{'}_{\textrm{CGLMP}}=d\times p(1|0,0,0)- 
\sum_{\textbf{s}}(-1)^{s_{1}+s_{2}}p(1|s_{1},s_{2},0)& \nonumber \\
+\sum_{\textbf{s}}(-1)^{s_{1}+s_{2}}\sum_{k=2}^{d-1}(d-k-1)p(k|s_{1},s_{2},0)&\leq d.
\end{eqnarray}
For the case of $(3,2,3)$, this tripartite CGLMP inequality is facet-defining and forms an orbit of $324$ inequalities. There are $61$ other non-trivial orbits for $(3,2,3)$. Inequalities from each of these orbits can be found in the supplementary material in \cite{hoban3}. Interestingly though, the Mermin inequality \eqref{merm} above which can be rewritten as:
\begin{equation}\label{mermin}
p(1|000)+p(1|011)+p(1|101)-p(1|110)=\sum_{\textbf{s}}\delta^{s_{3}}_{s_{1}\oplus s_{2}}(-1)^{s_{1}s_{2}}p(1|\textbf{s})\leq 2,
\end{equation}
does not generalize directly to the $(3,2,3)$ scenario. If we were to naively write the generalisation as
\begin{eqnarray}
\mathcal{C}^{''}_{\textrm{CGLMP}}=d\times p(1|0,0,0)- 
\sum_{\textbf{s}}\delta^{s_{3}}_{[s_{1}+ s_{2}]_{2}}(-1)^{s_{1}+s_{3}}p(1|\textbf{s})& \nonumber \\
+\sum_{\textbf{s}}\delta^{s_{3}}_{[s_{1}+ s_{2}]_{2}}(-1)^{s_{1}+s_{3}}\sum_{k=2}^{d-1}(d-k-1)p(k|\textbf{s})&\leq d,
\end{eqnarray}
then the right-hand-side is not $d=3$ in the $(3,2,3)$ scenario but $2d-1=5$, the algebraic upper bound for all possible correlators and not just LHV correlators. This upper bound of $5$ is achieved by vertices of $\mathcal{P}$ corresponding to the function $f(\textbf{s})=[s_{1}s_{2}+1]_{3}$. However, if parties produce the $n$-partite linear function $f(\textbf{s})=[2s_{1}+2s_{2}+s_{3}+1]_{3}$ and the only non-zero terms in the above inequality occur when $[s_{1}+s_{2}]_{2}=s_{3}$, then $f(\textbf{s})=[2s_{1}+2s_{2}+[s_{1}+ s_{2}]_{2}+1]_{3}=[s_{1}s_{2}+1]_{3}$.

Despite the fact that some of the facet Bell inequalities can be obtained from bipartite inequalities, understanding the full structure of $\mathcal{L}$ is a difficult task in general. For example, the straightforward substitution of the CGLMP inequality into expressions for $(3,2,3)$ still leaves a large number of orbits without characterisation. On the other hand, $\mathcal{L}$ in the $(n,2,2)$ scenario is well-understood as a hyperoctahedron \cite{ww,zb}. The facet Bell inequalities can be described in terms of Boolean functions where each facet inequality results from each particular Boolean function. In the following subsection we review the insight obtained by Werner and Wolf \cite{ww} as well as \.{Z}ukowski and Brukner \cite{zb}.

\subsection{Multipartite facet inequalities for $(n,2,2)$}\label{sec31e}

So far we have found facet Bell inequalities numerically. The size and hardness of the problem means that as $n$ gets larger, finding the facet inequalities quickly becomes intractable on a desktop PC. Convex polytopes are generalisations of the polyhedra and the geometry of these objects has been studied for thousands of years \cite{grunbaum}. A natural question to ask is whether there are analytical tools in convex geometry that can help us define $\mathcal{L}$ in terms of linear inequalities? This is not immediately obvious in the case of general $(n,c,d)$ but the case of $(n,2,2)$ has been amenable to this approach. Werner and Wolf (WW) independently with \.{Z}ukowski and Brukner (\.{Z}B) have shown that in this specific case, $\mathcal{L}$ is a hyperoctahedron \cite{ww,zb}.

Out of preference, we follow the WW construction of facet Bell inequalities \cite{ww}. Augmenting this approach we will use a central result from the previous chapter that $\mathcal{L}$ is the convex hull of $n$-partite linear functions. For the $(n,2,2)$ scenario, these $n$-partite linear functions are the \emph{linear Boolean functions}. The linear Boolean functions are a class of functions that have existed in the study of computer science and propositional logic well before our usage here. For example, linear Boolean functions are generated in error correction such as with the Hamming code \cite{hamming}. The following result demonstrates yet another application of the study of linear Boolean functions.

\begin{cor}\label{cor311}
The space $\mathcal{L}$ of LHV correlators in the $(n,2,2)$ scenario is the convex hull of linear Boolean functions.
\end{cor}

\textit{Proof}: Since this corollary is a special case of Theorem \ref{thm221} we just need to show that for the $(n,2,2)$ scenario, all the $n$-partite linear functions are the linear Boolean functions. Linear Boolean functions $f(\textbf{s})$ for an $n$-length bit-string $\textbf{s}$ can be written in terms of the Algebraic Normal Form (ANF) as:
\begin{equation}\label{linear}
f(\textbf{s})=\left(\bigoplus_{j=1}^{n}a_{j}s_{j}\right)\oplus b,
\end{equation}
where $a_{j}$, $b\in\{0,1\}$. Whereas, an $n$-partite linear function $g(\textbf{s})$ in this scenario can be written as
\begin{equation}
g(\textbf{s})=\bigoplus_{j=1}^{n}g_{j}(s_{j}),
\end{equation}
for single-site map $g_{j}:\mathbb{Z}_{2}\rightarrow\mathbb{Z}_{2}$. Crucially, as a special case, all single-site Boolean functions of this form can be expressed as $g_{j}(s_{j})=b_{j}\oplus a_{j}\delta^{s_{j}}_{1}= a_{j}s_{j}\oplus b_{j}$ since $\delta^{s_{j}}_{1}=s_{j}$ for $a_{j}$, $b_{j}\in\{0,1\}$. Therefore, take the sum modulo $2$ of all of these maps and setting $b=\bigoplus_{j=1}^{n}b_{j}$ returns the expression in \eqref{linear}. $\square$
\newline

The above corollary is a rederivation of the LHV convex polytope that was derived by WW and \.{Z}B \cite{ww,zb}. However, this rederivation is in terms of a language of \emph{computational expressiveness} whereas the original derivation is in the language of expectation values of measurements with outcomes $\pm 1$. The ``linearity'' (in the Boolean function sense of the word) is not explicit but buried in the mathematical derivation of $\mathcal{L}$. The language of computational expressiveness sheds a new light on an old result and this new perspective will become central to a lot of discussion in this chapter; the next chapter will also have Corollary \ref{cor311} at its heart.

As mentioned above, both constructions  due to WW and \.{Z}B use the expectation values of $\mathbb{E}(\textbf{s})=p(0|\textbf{s})-p(1|\textbf{s})$ rather than the correlators themselves. However, due to the ``law of the excluded middle'' giving $\mathbb{E}(\textbf{s})=1-2p(1|\textbf{s})$, expectation values and correlators are in one-to-one correspondence. For brevity of reproduction of results, we will work in terms of $\mathbb{E}(\textbf{s})$ and then map back to correlators $p(1|\textbf{s})$ at a final stage. 

Taking on this notation, we construct all Bell inequalities in the $(n,2,2)$ scenario in the following way \cite{ww}:
\begin{equation}\label{ww}
\left|\sum_{\textbf{s}}\beta_{\textbf{s}}\mathbb{E}(\textbf{s})\right|\leq 1,
\end{equation}
such that the real coefficients $\beta_{\textbf{s}}$ always give $1$ for LHV correlators. We are just choosing a normalisation convention without loss of generality. By convexity we only need to consider the extreme points of $\mathcal{L}$ which correspond to the linear Boolean functions. We rewrite these extreme points $\mathbb{E}(\textbf{s})_{E}$ in terms of the expectation values, i.e. $\mathbb{E}(\textbf{s})_{E}=\sum_{k}(-1)^{k}\delta^{k}_{g(\textbf{s})}=(-1)^{g(\textbf{s})}$ where $g(\textbf{s})$ is a linear Boolean function. We are also only interested in extreme points that maximally saturate the upper bound as these extreme points will define a facet. Putting all of this information together, we can rewrite \eqref{ww} as:
\begin{equation}\label{ww1}
\sum_{\textbf{s}}\beta_{\textbf{s}}(-1)^{g(\textbf{s})}=(-1)^{\gamma_{g(\textbf{s})}},
\end{equation}
where $\gamma_{g(\textbf{s})}\in\{0,1\}$ depends on the linear Boolean function. The linear Boolean functions can be written as $g(\textbf{s})=(\bigoplus_{j=1}^{n}a_{j}s_{j})\oplus b$ but the overall sign $(-1)^{b}$ leaves \eqref{ww} unaffected. Therefore we only need to consider linear Boolean functions with $b=0$, thus leaving $2^{n}$ such functions. In order to show that the inequalities in \eqref{ww1} are facet-defining, then we must form affinely independent $2^{n}$-length vectors with elements $(-1)^{g(\textbf{s})}$ for each $\textbf{s}$. To demonstrate affine independence we utilise the following lemma.

\begin{lem}\label{lem311}
For $2^{n}$-length vectors $\vec{g}\in\mathbb{R}^{2^{n}}$ with elements $g(\textbf{s})$ being the non-constant linear Boolean functions, a set of $(2^{n}-1)$ vectors $\vec{g}$ are linearly independent as long as no two vectors, $\vec{g}^{1}$ and $\vec{g}^{2}$, corresponding to two linear Boolean functions $g^{1}(\textbf{s})$ and $g^{2}(\textbf{s})$ respectively, have all elements $g^{1}(\textbf{s})=g^{2}(\textbf{s})\oplus 1$.
\end{lem}

\textit{Proof}: We demonstrate linear independence by mapping linear Boolean functions from $\mathbb{Z}_{2}$ to $\mathbb{R}$. Every linear Boolean function can always be expressed as $g(\textbf{s})=(\bigoplus_{j=1}^{n}a_{j}s_{j})\oplus b$ and is non-constant as long as at least one value of $a_{j}$ is non-zero. For two functions $g^{1}(\textbf{s})=g^{2}(\textbf{s})\oplus 1$, $b$ is $1$ for either of the functions and $0$ for the other. Mapping from $\mathbb{Z}_{2}$ to $\mathbb{R}$, we can write $g^{2}(\textbf{s})\oplus 1$ as
\begin{equation}
g^{2}(\textbf{s})\oplus 1=1-g^{2}(\textbf{s}).
\end{equation}
Therefore if we prove that the $(2^{n}-1)$ non-constant linear Boolean functions $g(\textbf{s})=(\bigoplus_{j=1}^{n}a_{j}s_{j})$ produce $(2^{n}-1)$ linearly independent vectors $\vec{g}$, then this holds if functions are $g(\textbf{s})\oplus 1$. This is true if the set of linear Boolean functions does not include two functions $g^{1}(\textbf{s})$ and $g^{2}(\textbf{s})$ where $g^{2}(\textbf{s})=g^{1}(\textbf{s})\oplus 1$ for all $\textbf{s}$.

 First, all variables $s_{j}$ will produce vectors $\vec{g}$ that are linearly independent from all $\vec{g}$ resulting from $s_{k}$ by construction where $k\neq j$. As a shorthand, we say that a function is linearly independent from other functions if the associated vectors $\vec{g}$ are linearly independent. We show that linear Boolean functions dependent on more than one variable $s_{j}$ are linearly independent. We start with the linear function $s_{1}\oplus s_{2}$ which can be rewritten as
\begin{equation}
s_{1}\oplus s_{2}=s_{1}+s_{2}-2s_{1}s_{2}.
\end{equation}
This expression is linearly independent from functions $s_{1}$ and $s_{2}$ due to the $s_{1}s_{2}$ term being multiplicative. This function will also be linearly independent from all functions $s_{j}\oplus s_{k}\neq s_{1}\oplus s_{2}$ due to $s_{j}s_{k}$ being linearly independent from $s_{1}s_{2}$. Having shown that all linear Boolean functions dependent on $2$ variables and $1$ variable are all linearly independent from each other, we proceed inductively. For functions dependent on $3$ variables $s_{j}$, eg. $g(\textbf{s})=s_{1}\oplus s_{2}\oplus s_{3}$, we can again map this function into standard arithmetic as
\begin{equation}
s_{1}\oplus s_{2}\oplus s_{3}=s_{1}+s_{2}+s_{3}-2(s_{1}s_{2}+s_{1}s_{3}+s_{2}s_{3})+4s_{1}s_{2}s_{3}.
\end{equation}
Again $s_{1}s_{2}s_{3}$ is linearly independent from all terms $s_{j}s_{k}$ for $j\neq k$ and single variable terms $s_{j}$, as well as all linear functions $s_{j}\oplus s_{k}\oplus s_{l}\neq s_{1}\oplus s_{2}\oplus s_{3}$. Proceeding inductively for each function $g(\textbf{s})=(\bigoplus_{j=1}^{n}a_{j}s_{j})$ with $q$ non-zero values of $a_{j}$, writing $g(\textbf{s})$ in standard arithmetic we have the product of these $q$ elements of $\textbf{s}$. This product of $q$ elements of $\textbf{s}$ is linearly independent from all other products of $q$ and $q'<q$ elements of $\textbf{s}$. The linear Boolean function $g(\textbf{s})=(\bigoplus_{j=1}^{n}s_{j})$ can thus be written as
\begin{equation}
\bigoplus_{j=1}^{n}s_{j}=\frac{1}{2}\left[1-\prod_{j}^{n}(1-2s_{j})\right].
\end{equation}
This function is finally linearly independent from all other linear Boolean functions due to the term $\prod_{j=1}^{n}s_{j}$. Therefore all of the non-constant linear Boolean functions $g(\textbf{s})=\bigoplus_{j=1}^{n}a_{j}s_{j}$ produce vectors $\vec{g}$ that are linearly independent. $\square$
\newline

As a result of this lemma, the extreme points $(-1)^{g(\textbf{s})}$ for the linear Boolean functions $g(\textbf{s})=\bigoplus_{j=1}^{n}a_{j}s_{j}$ are affinely independent. The dimension of $\mathcal{P}$ for $(n,2,2)$ is $2^{n}$, and so \eqref{ww1} is facet-defining if this expression is satisfied for all of these linear Boolean functions. 

The key observation made by WW is that \eqref{ww1} is a discrete Fourier Transform and its inverse is
\begin{equation}\label{beta}
\beta_{\textbf{s}}=\frac{1}{2^{n}}\sum_{g(\textbf{s})}(-1)^{\gamma_{g(\textbf{s})}}(-1)^{g(\textbf{s})}
\end{equation}
which is now a sum over all linear Boolean functions $g(\textbf{s})=(\bigoplus_{j=1}^{n}a_{j}s_{j})$. Therefore, for each facet Bell inequality we now have some choice of the variables $\gamma_{g(\textbf{s})}\in\{0,1\}$ for all functions $g(\textbf{s})$. There are then $2^{2^{n}}$ possible choices of these $2^{n}$ values of $\gamma_{g(\textbf{s})}$. We now express \eqref{ww1} in terms of correlators $p(1|\textbf{s})$ instead of expectation values $\mathbb{E}(\textbf{s})$,
\begin{equation}\label{wwfacet}
-\sum_{\textbf{s}}\beta_{\textbf{s}}p(1|\textbf{s})\leq \frac{1-\sum_{\textbf{s}}\beta_{\textbf{s}}}{2}\in\{0,1\}.
\end{equation}
The sum of coefficients $\sum_{\textbf{s}}\beta_{\textbf{s}}$ is equal to $\pm1$ as it is equal to $(-1)^{\gamma_{g(\textbf{s})}}$ when $g(\textbf{s})=0$ for all $\textbf{s}$. There are therefore $2^{2^{n}}$ facet Bell inequalities in the $(n,2,2)$ scenario of the form in \eqref{wwfacet}.

In the $(n,2,2)$ scenario, we show that if we deal with expectation values we can derive all of the facet inequalities. As mentioned above, all of these inequalities can be obtained through substitution of the CHSH inequality in terms of expectation values \cite{ww}. The CHSH inequalities are expressed as a polynomial in measurement operators on two sites, called ``Bell polynomials''. Every other inequality for $n>2$ are multiples of these polynomials with measurement operators on other sites. This substitution of the CHSH inequality is clear in the expectation value scenario but not so clear in the correlator description. Despite this drawback, the insight we gain from Lemma \ref{lem311} allows us to demonstrate that a particular inequality for each $n$ is facet-defining as we now show.

We have utilised a form of substitution in constructing tripartite CGLMP inequalities by having non-zero terms in the inequality when the input satisfies a particular constraint, e.g. $s_{1}=s_{2}$. But not all inequalities in the $(n,2,2)$ scenario can be constructed from the CHSH inequality by this simple method. For example, the following facet Bell inequality in the $(3,2,2)$ scenario as found by WW \cite{ww},
\begin{eqnarray}\label{newineq}
\frac{1}{4}\left[p(1|000)+p(1|001)+p(1|010)+p(1|011)\right]& &\nonumber\\
+\frac{1}{4}\left[p(1|100)+p(1|101)+p(1|110)-3p(1|111)\right]&\leq&1
\end{eqnarray}
has non-zero coefficients for all inputs $\textbf{s}$. However, we can generalise this inequality to $n$ parties utilising the result from Lemma 1 (in a slightly modified form). A generalisation of this inequality is
\begin{equation}\label{gennewineq}
\frac{1}{2^{n-1}}\left(-2^{n-1}p(1|\textbf{1})+\sum_{\textbf{s}}p(1|\textbf{s})\right)\leq 1.
\end{equation}
It is worth noting that this inequality not only reduces to \eqref{newineq} for $n=3$, but also the CHSH inequality for $n=2$. 

We observe that the upper bound on the right-hand-side of \eqref{gennewineq} is saturated for all $(2^{n}-1)$ linear Boolean functions $g(\textbf{s})=\left(\bigoplus_{j=1}^{n}a_{j}s_{j}\right)\oplus b\neq 0$ where $g(\textbf{1})=0$. The upper bound is also saturated when $g(\textbf{s})=1$ for all $\textbf{s}$. These $2^{n}$ linear Boolean functions are also affinely independent by the argument of Lemma \ref{lem311}. For a particular linear Boolean function $g^{1}(\textbf{s})$, only one out of the two functions $g^{1}(\textbf{s})$ and $g^{1}(\textbf{s})\oplus 1$ satisfy the condition that $g(\textbf{1})=0$. Therefore the set of $(2^{n}-1)$ linear Boolean functions $g(\textbf{s})=\left(\bigoplus_{j=1}^{n}a_{j}s_{j}\right)$ with some of these functions having $1$ added mod $2$, will be the set $g(\textbf{s})=\left(\bigoplus_{j=1}^{n}a_{j}s_{j}\right)\oplus b\neq 0$ where $g(\textbf{1})=0$. By Lemma \ref{lem311}, the former set forms a linearly independent set of functions, and we can just add the constant function $g(\textbf{s})=1$ for all $\textbf{s}$ to make an affinely independent set.

We have used an insight from the computational perspective of LHV correlators in the $(n,2,2)$ scenario to define a facet Bell inequality for all $n$. Interestingly, for all correlators in $\mathcal{P}$, the inequality in \eqref{gennewineq} is \emph{only} maximally violated by the correlator $p(1|\textbf{s})=\delta^{\prod_{j=1}^{n}s_{j}\oplus 1}_{1}=\prod_{j=1}^{n}s_{j}\oplus 1$ corresponding to the function $f(\textbf{s})=\prod_{j=1}^{n}s_{j}$ for all $n$. This is contrary to the Mermin inequality which is maximally violated by more than one correlator in $\mathcal{P}$. 

So far in the discussion in this chapter, we have described facet Bell inequalities. They define the space of $\mathcal{L}$. They also guarantee that if a correlator is outside of $\mathcal{L}$, it must violate one of these facet Bell inequalities. We have shown throughout that this violation can be achieved (uniquely or otherwise) maximally by particular vertices of $\mathcal{P}$. Heuristically then, a violation of a Bell inequality can be associated with a \emph{computational advantage}. The advantage being that non-LHV correlators can be associated with computations of non-$n$-partite linear functions. This insight will be utilised in section \ref{sec33} where Bell inequalities may not be facet-defining, which can highlight the computational advantage of non-LHV theories.

Of the possible theories that can be associated with non-LHV correlators, quantum theory is currently the only working theory. Whether the predictions of quantum theory in the form of a violation of a Bell inequality can be verified in a laboratory will be discussed in chapter \ref{chap4}. In the next section, we will discuss quantum correlators, or the space $\mathcal{Q}$. We will explore methods used to find the maximal violations of Bell inequalities possible with quantum theory. This will give some indication of the extreme points of the space $\mathcal{Q}$. 

\section{Quantum Violations of Bell Inequalities}\label{sec32}

\noindent
We have described the structure of $\mathcal{L}$ in terms of the facet Bell inequalities. We now give some indication of the structure of $\mathcal{Q}$. By giving an indication, we mean that we find the maximal violation of the facet Bell inequalities. It is still an open question of defining the extreme points of $\mathcal{Q}$ in general. In the specific $(n,2,2)$ scenario, WW have described the extreme points of the quantum region \cite{ww}, but otherwise, we can only numerically find particular extreme points.

In this section, numerical methods \cite{navascues,kaszlikowski} used to find the maximum quantum values of a Bell expression are reviewed. Using these methods we present numerical values for the bipartite facet Bell expressions we found in subsection \ref{sec31c}. In particular, we find the maximum quantum values for an expression in each orbit. Therefore, finding this value for an expression in an orbit also finds the quantum value for all expressions in that orbit. This is because the set of quantum correlators are also unaffected by the local operations on values $\textbf{m}$ and $\textbf{s}$ and permutations of parties. 

We also comment on the relationship between entanglement and violation of bipartite facet Bell inequalities. We show that the maximal quantum violation may not be achieved by a maximally entangled quantum state. Although a violation is a ``witness'' of entanglement (see section \ref{sec14}), more entanglement may not mean more non-classicality.

We present the result of WW that all extreme points of $\mathcal{Q}$ have a closed form \cite{ww}. The maximum quantum value of all Bell expressions is an optimization over these points. What is more, these maximal expressions can be obtained from projective measurements on the $n$-party Greenberger-Horne-Zeilinger (GHZ) state \cite{ghz,ww}. The GHZ state can be considered as a natural, if ambiguous \cite{plenio}, multipartite generalization of the maximally entangled state.

\subsection{Numerical Methods for finding Violations of Bell Inequalities}\label{sec32a}

In the literature, there are two main methods of finding violations of Bell inequalities. The first approach which we call the ``multiport beam-splitter'' or MBS approach \cite{kaszlikowski,durt}. This method fixes the quantum state shared by both parties as the maximally entangled state $|\Psi\rangle=\frac{1}{\sqrt{d}}\sum_{j=0}^{(d-1)}|jj\rangle$. We then optimize over projective measurements made by each party to find a lower bound of the maximum quantum violation of a Bell inequality, if a violation occurs. 

A second, more general approach for finding a quantum violation of a Bell inequality involves semi-definite programming (SDP) \cite{boyd}. Therefore we call this approach the ``SDP approach'' as developed by Navascu\'{e}s, Pironio and Ac\'{i}n \cite{navascues2,navascues}. This approach involves constructing a positive semi-definite Gram matrix of (sequences of) correlations. The Bell expression is then a linear function on elements of this matrix and we maximize this linear, or ``objective'' function. This second approach produces an upper bound on the violation of a Bell inequality. However, if the Gram matrix satisfies a certain property (called a rank loop) then the maximized objective is equal to the maximal violation of a Bell inequality \cite{navascues2}. On the other hand, if we do not satisfy this property if the lower bound produced by the MBS approach is equal to the upper bound of the SDP approach then we have found the maximum quantum violation.

Both of these approaches have been developed in the bipartite scenario but can be extended to the multipartite scenario \cite{navascues2,zukowski}. Naturally though, with an increasing number of parties, the optimization for both approaches becomes harder for a desktop PC. In this subsection, we only use these two methods for finding bipartite quantum violations, so we will only describe them in these two scenarios. We now proceed to describe each approach in more detail.

The MBS approach is described as follows \cite{kaszlikowski,durt}. The quantum state shared by two parties is first fixed as the $d^{2}$-dimensional maximally entangled state $|\Psi\rangle=\frac{1}{\sqrt{d}}\sum_{j=0}^{d-1}|jj\rangle$ and both parties attain measurement outcomes associated with projectors $|\mu_{j}\rangle_{s_{j}}\langle \mu_{j}|_{s_{j}}=V_{s_{j}}|k\rangle\langle k|V_{s_{j}}^{\dagger}$, where $\{|k\rangle|k\in\mathbb{Z}_{d}\}$ is the standard basis of $\mathcal{H}^D$. The $V_{s_{j}}$ is a unitary matrix and can be written as $V_{s_{j}}=FD_{s_{j}}$ where $F$ is the $d$-by-$d$ Quantum Fourier Transform matrix with elements for the $j$th row and $k$th column $F_{j,k}=\frac{1}{\sqrt{d}}e^{\frac{2\pi i}{d}(j-1)(k-1)}$. The $d$-by-$d$ matrix $D_{s_{j}}$ is a diagonal matrix $D_{s_{j}}=\textrm{diag}(e^{i\phi_{1}(s_{j})},e^{i\phi_{2}(s_{j})},...,e^{i\phi_{d}(s_{j})})$ with $\phi_{j}(s_{j})$ as real phases. Therefore we optimise over these phases $\phi_{j}(s_{j})$ to numerically maximize the quantum violation for the maximally entangled state.

This first approach can be modified further by altering the quantum state after optimization of the phases $\phi_{j}(s_{j})$, as indicated by Acin et al \cite{acin2}. We first obtain the optimal angles $\phi_{j}(s_{j})$ found for the maximally entangled quantum state. We then substitute these optimal angles into the projectors $V_{s_{j}}|k\rangle\langle k|V_{s_{j}}^{\dagger}$. Then we construct the Bell expression in terms of these optimal projectors giving
\begin{eqnarray}
\sum_{\textbf{s}}\beta_{\textbf{s}}p(k|\textbf{s})&=&\langle\psi|\left(\sum_{\textbf{s},\textbf{m}}\beta_{\textbf{s}}\delta^{k}_{\left[m_{1}+ m_{2}\right]_{d}}|\mu_{1}\rangle_{s_{1}}\langle \mu_{1}|_{s_{1}}\otimes|\mu_{2}\rangle_{s_{2}}\langle \mu_{2}|_{s_{2}}\right)|\psi\rangle \nonumber \\
&=&\langle\psi|\mathcal{W}|\psi\rangle
\end{eqnarray}
where $|\psi\rangle$ is not necessarily the maximally entangled state $|\Psi\rangle$. Finding the largest possible quantum value of the Bell expression is then a case of finding the largest eigenvalue of $\mathcal{W}$. Acin et al used this method to find a larger quantum violation of the CGLMP inequality for $(2,2,3)$ with a non-maximally entangled state \cite{acin2}. We will discuss the connection between entanglement and Bell inequality violation in subsection \ref{sec32b}.

We now briefly present the SDP approach. Central to the SDP approach is the construction of a positive semi-definite Gram matrix $\Gamma$. The elements $\Gamma_{jk}$ of this matrix are $\Gamma_{jk}=\langle\psi|O_{j}^{\dagger}O_{k}|\psi\rangle$ where $O_{j}$ is a linear combination of products of projectors $E_{m_{j},s_{j}}$ that depend on $m_{j}$ and $s_{j}$ at each $j$th site. These projectors correspond directly to probabilities of getting $m_{j}$ given $s_{j}$, i.e. $p(m_{j}|s_{j})=\langle E_{m_{j},s_{j}}\rangle$. The projectors act on an arbitrary dimension Hilbert space which is shared by all parties. They also satisfy $\langle\psi|E_{m_{j},s_{j}}|\psi\rangle\geq 0$ for all states $|\psi\rangle$, $E_{m_{j},s_{j}}=E_{m_{j},s_{j}}^{\dagger}$ (Hermiticity), $E_{m_{j},s_{j}}E_{m'_{j},s_{j}}=\mathbb{I}\delta^{m_{j}}_{m'_{j}}$ (orthogonality) and $\sum_{m_{j}}E_{m_{j},s_{j}}=\mathbb{I}$.

We associate the degree of this product (i.e. the number of terms in the product of projectors) with a set of quantum operators, i.e. for degree of products being $\nu$ we have the set $\mathbb{Q}_{\nu}$. For example, the set $\mathbb{Q}_{1}$ can be associated with the identity matrix $\mathbb{I}$ and $O_{j}$ consisting solely of linear combinations of single projectors $E_{m_{j},s_{j}}$. $\mathbb{Q}_{2}$ is the set of $2$-term products $E_{m_{j},s_{j}}E_{m'_{j},s'_{j}}$ for $s'_{j}\neq s_{j}$ and $E_{m_{j},s_{j}}E_{m'_{j},s_{j'}}$ for  $j \neq j'$. Another set of interest is $\mathbb{Q}'_{2}$, an intermediate set\footnote{In \cite{navascues}, this set is written as $\mathbb{Q}_{1+AB}$ where $A$ and $B$ represent two parties, and the set includes pairwise products of the projectors for each party.} between $\mathbb{Q}_{1}$ and $\mathbb{Q}_{2}$, where we have all the operators which are the pairwise product of projectors between parties $j$ and $j'$ where $j'\neq j$.

The set $\mathbb{Q}_{\infty}$ of all products of projectors is then the set of all values of $\langle\psi|O_{j}^{\dagger}O_{k}|\psi\rangle$ possible with quantum mechanics. However, it is possible that $\Gamma_{\nu}$, the Gram matrix of operators associated with $\mathbb{Q}_{\nu}$ may already contain all values $\langle\psi|O_{j}^{\dagger}O_{k}|\psi\rangle$ in $\mathbb{Q}_{\infty}$. If this occurs then the rank of $\Gamma_{\nu}$ is equal to the rank of $\Gamma_{\nu-1}$, resulting in a ``rank loop''. For more detail see \cite{navascues}.

To find the quantum upper bound for a Bell expression we perform the following semi-definite program:
\begin{eqnarray}
\textrm{maximize}&\text{   }\textrm{tr}(B^{\textrm{T}}\Gamma)\nonumber \\
\textrm{subject to}&\text{   }\Gamma\geq 0 \nonumber \\
&\text{   }\textrm{tr}(F^{\textrm{T}}_{j}\Gamma)=0\text{,   }j\in\{0,1,...,x\},
\end{eqnarray}
where $B$ is a matrix of the coefficients of the Bell expression for each probability $p(\textbf{m}|\textbf{s})=\langle \prod_{j=1}^{n}E_{m_{j},s_{j}}\rangle$. The matrices $F_{j}$ are $x$ linear constraints on elements of $\Gamma$. Of course, $\Gamma_{\infty}$ will be infinitely large, so if we restrict at first to $\Gamma_{1}$, we obtain an upper bound on $\textrm{tr}(B^{\textrm{T}}\Gamma)$ for all quantum probabilities. It is an upper bound as there are fewer constraints on the elements of $\Gamma$, and so $\Gamma$ might not be compatible with quantum physics. The bound can then be subsequently lowered if we consider matrices $\Gamma'_{2}$ (corresponding to the set $\mathbb{Q}'_{2}$) which will impose more constraints on the products of projectors compatible with quantum physics. 

The value of $\textrm{tr}(B^{\textrm{T}}\Gamma)$ will be the true quantum value (up to numerical error) if we have a rank loop as described above. Semi-definite programming forms part of the subject of convex optimization \cite{boyd}. There are algorithms for dealing with semi-definite programming such as those in the packages of YALMIP \cite{yalmip} and SeDuMi \cite{sedumi}. We utilise these numerical methods to find quantum bounds of Bell expressions, and also to look for a rank loop. However, we do not need to look for a rank loop if the value of $\textrm{tr}(B^{\textrm{T}}\Gamma)$ is equal to the lower bound of the MBS method within numerical error. These two methods then give us an indication of the extreme points of $\mathcal{Q}$.

In the construction of the SDP approach we did not explicitly say that $n=2$. Indeed this method can be utilised in the multipartite case but in order to have the correlations of $n$ parties in $\Gamma$, one needs to go to at least $\mathbb{Q}_{\lceil \frac{n}{2}\rceil}$. The MBS approach can also be generalised to the multipartite scenario but again the problem becomes more complicated. In the following subsection we will utilise both the MBS and SDP approaches to find the maximal quantum violations of all bipartite facet Bell inequalities. Therefore, consideration of multipartite generalisations will not be relevant for our discussion.

\subsection{Bipartite Quantum Violations and Entanglement}\label{sec32b}

We now describe the maximal quantum violations of facet Bell inequalities for $n=2$. We used both methods described in the previous subsection first finding a lower bound using the MBS approach and then the SDP approach to confirm that this is the maximal value. We list all of the maximal quantum violations for $n=2$ facet Bell inequalities numerically in Table \ref{tab:tab3}. The numerical error in these values is of the order of $\pm10^{-9}$ and maximal violations resulting from both the MBS and SDP approaches agree within this error. Also in Table \ref{tab:tab3} we have indicated which maximal violations result from the maximally entangled state $|\Psi\rangle=\frac{1}{\sqrt{d}}\sum_{j=0}^{(d-1)}|jj\rangle$. For $d\neq 2$, there are instances where maximal violation is not a result of maximal entanglement. 

\begin{table}
\begin{center}
 \begin{tabular}{| c | c | c | c | c | c | c | c |}
 \hline
 n & c & d & Orbit & LHV bound & Quantum bound & Entanglement\\ \hline
 2 & 2 & 2 & $\mathcal{C}_{d=2}$ & $2$ & $2.4142^{\dagger}$ & 1.000\\ \hline
 2 & 2 & 3 & $\mathcal{C}_{CGLMP}$ &  $3$ & $3.9149$  &1.555\\ \hline
 2 & 2 & 4 & $\mathcal{C}_{CGLMP}$ &  $4$ & $5.4594$ & 1.938\\
 2 & 2 & 4 & $\mathcal{C}^{1}_{d=4}$ &  $2$ & $2.4142$ & 1.000\\
 2 & 2 & 4 & $\mathcal{C}^{2}_{d=4}$ &  $4$ & $4.8284^{\dagger}$ & 2.000\\ \hline
 2 & 2 & 5 & $\mathcal{I}_{1}$ &  $5 $ & $6.3145$ &$2.310$\\
 2 & 2 & 5 & $\mathcal{I}_{2}$ &  $5 $ & $7.6290 $ &$2.310$\\
 2 & 2 & 5 & $\mathcal{I}_{3}$ &  $5 $ & $7.0314$ &$2.230$\\
 2 & 2 & 5 & $\mathcal{C}_{CGLMP}$ &  $5 $ & $7.0314$ &$2.230$\\ \hline
 2 & 3 & 2 & $\mathcal{C}_{c=3}$ &  $2 $ & $2.4142^{\dagger}$ &1.000 \\ \hline
 2 & 4 & 2 & $\mathcal{B}_{1}$ to $\mathcal{B}_{5}$&  $8 $&$9.7570^{\dagger}$ &1.000\\
 2 & 4 & 2 & $\mathcal{B}_{6}$ to $\mathcal{B}_{11}$&  $4 $&$5.0825^{\dagger}$ &1.000\\
 2 & 4 & 2 & $\mathcal{C}^{1}_{c=4}$ to $\mathcal{C}^{3}_{c=4}$&  $2 $&$2.4142^{\dagger}$ &1.000\\
 \hline
 \end{tabular}
\end{center}
\caption{We list the bipartite maximal quantum violations for particular facet Bell inequalities for $c$ and $d$. We have grouped the orbits of inequalities $\mathcal{B}_{1}$ to $\mathcal{B}_{5}$ (increasing numerically in the label of the inequality as they have the same LHV and quantum upper bounds. The same grouping also applies for inequalities $\mathcal{B}_{6}$ to $\mathcal{B}_{11}$. Those violations that are achieved with the bipartite maximally entangled state of $d^2$ dimension are labelled with a $\dagger$. We also present the numerical calculation of entropy of entanglement for the pure state associated with each maximal violation. Recall that the maximally entangled state will have entanglement $\log_{2}(d)$.}
\label{tab:tab3}
\end{table}

While the construction of Bell inequalities was initially partly motivated by the issue of entanglement, the connection between entanglement and violation is not completely clear. A violation of a Bell inequality indicates that measurements are made on an entangled state, but entanglement does not necessarily result in a violation of a particular inequality \cite{werner}. For the CHSH inequality, the maximal violation allowed by quantum mechanics is produced by the maximally entangled state \cite{tsirelson}. As we can see from Table \ref{tab:tab3}, this is not true in general. Also in Table \ref{tab:tab3}, we have calculated the entanglement of the pure state that maximally violates each inequality. The entanglement of bipartite pure states $|\psi\rangle\langle\psi|\in\mathcal{H}^{d}\otimes\mathcal{H}^{d}$ in $d^2$-dimensional Hilbert space is calculated from the entropy of entanglement $E(|\psi\rangle\langle\psi|)$ \cite{plenio}. Interestingly from Table \ref{tab:tab3}, the entanglement of the state that maximally violates the CGLMP inequality decreases with $d$.

It has been established previously that a violation of a Bell inequality and entanglement are two different, but related issues \cite{vidick,liang}. For example, statistics that violate a Bell inequality can be seen as a ``resource'' for demonstrating non-classicality, and entanglement can also be seen as a resource (see section \ref{sec13}). It has been shown that these two resources are different if one wants to use one resource to simulate the statistics of the other \cite{brunner}.

In subsection \ref{sec32a}, we mentioned that if one attains a rank loop between a Gram matrix $\Gamma_{\nu}$ and another Gram matrix $\Gamma_{\nu-1}$, then the quantum value of Bell expression has reached its maximum value for $\Gamma_{\nu}$. For all of the examples in Table \ref{tab:tab3}, there was a rank loop found between $\Gamma'_{2}$ and $\Gamma_{1}$. This observation is confirmed for the CGLMP inequalities by results obtained by Navascu\'{e}s, Pironio and Ac\'{i}n \cite{navascues}. This leads us to conjecture that the maximal quantum value resulting from $\mathcal{Q}$ for all bipartite Bell expressions for correlators is obtained from correlations in the set $\mathbb{Q}'_{2}$.

In this subsection we have indicated that all of the bipartite facet Bell inequalities found in this chapter are violated by quantum correlators. However, the maximum possible violation is not achieved by the relevant maximally entangled state. This implies it might not be favourable to use a maximally entangled state for the largest violation. This behaviour has also been observed when considering Bell inequalities expressed in terms of elements of the full probability distribution $p(\textbf{m}|\textbf{s})$ \cite{vidick,liang}. In the next subsection, we describe the quantum region $\mathcal{Q}$ for the $(n,2,2)$ scenario. The connection between maximal violation and quantum state is also far clearer for all $n$; it results from the GHZ state \cite{ghz}. The GHZ state for $n=2$ case is the maximally entangled state for $d=2$.

\subsection{Quantum Upper Bounds of $(n,2,2)$ Bell Inequalities}\label{sec32c}

We now consider the maximal quantum violation of any Bell inequality in the $(n,2,2)$ scenario. We state the following result (as obtained by WW \cite{ww}) in terms of the maximal quantum value of a Bell expression.

\begin{thm}\label{thm321} 
The maximal quantum value of a Bell expression for the $(n,2,2)$ scenario is
\begin{equation}\label{quantumbound}
\sum_{\textbf{s}}\beta_{\textbf{s}}p(1|\textbf{s})=\underset{\{\theta_{j}\}}{\textrm{sup}}\left[\left(\sum_{\textbf{s}}\frac{\beta_{\textbf{s}}}{2}\right)+\left|\sum_{\textbf{s}}\frac{\beta_{\textbf{s}}}{2}e^{i(\sum_{j=1}^{n}s_{j}\theta_{j})}\right|\right]
\end{equation}
where $\theta_{j}$ are $n$ angles, or real parameters. These maximal quantum values result from von Neumann measurements on the GHZ state:
\begin{equation}
|\textrm{GHZ}\rangle=\frac{1}{\sqrt{2}}\left(|0\rangle^{\otimes n}+|1\rangle^{\otimes n}\right).
\end{equation}
\end{thm}

\textit{Proof}: We map from the correlators $p(1|\textbf{s})$ to the expectation values $\mathbb{E}(\textbf{s})$ for measurements, or observables having outcomes $\pm 1$. For quantum correlators, the measurements are Hermitian operators $\hat{M}_{s_{j}}=Q^{0}_{s_{j}}-Q^{1}_{s_{j}}$ where $Q^{m_{j}}_{s_{j}}$ are the projectors corresponding to outcome $m_{j}$. Therefore, $-\mathbb{I}\leq \hat{M}_{s_{j}}\leq\mathbb{I}$ and $\hat{M}_{s_{j}}^{2}=\mathbb{I}$ where $\mathbb{I}$ is the identity matrix. The expectation value is then for all pure states $|\psi\rangle$, $\mathbb{E}(\textbf{s})=\langle \psi|\bigotimes_{j=1}^{n}\hat{M}_{s_{j}}|\psi\rangle$, which can be substituted into a Bell expression to achieve the maximal quantum value
\begin{equation}
\sum_{\textbf{s}}\beta_{\textbf{s}}p(1|\textbf{s})=\underset{\{|\psi\rangle,\hat{M}_{s_{j}}\}}{\textrm{sup}}\frac{1}{2}\left[\sum_{\textbf{s}}\beta_{\textbf{s}}-\sum_{\textbf{s}}\beta_{\textbf{s}}\langle \psi|\bigotimes_{j=1}^{n}\hat{M}_{s_{j}}|\psi\rangle\right].
\end{equation}
It remains then to minimize the expression $\sum_{\textbf{s}}\beta_{\textbf{s}}\langle \psi|\bigotimes_{j=1}^{n}\hat{M}_{s_{j}}|\psi\rangle$ over all states and choice of measurements. This equates to finding the minimum eigenvalue of the operator $\sum_{\textbf{s}}\beta_{\textbf{s}}\bigotimes_{j=1}^{n}\hat{M}_{s_{j}}$, or the operator norm $\|...\|$ of $-\sum_{\textbf{s}}\beta_{\textbf{s}}\bigotimes_{j=1}^{n}\hat{M}_{s_{j}}$. To find this operator norm, we need to diagonalize the operator and we can do this in the following way since the identity commutes with all operators:
\begin{eqnarray}\label{quantumupper}
\underset{\{|\psi\rangle,\hat{M}_{s_{j}}\}}{\textrm{sup}}\left[-\sum_{\textbf{s}}\beta_{\textbf{s}}\langle \psi|\bigotimes_{j=1}^{n}\hat{M}_{s_{j}}|\psi\rangle\right]&=&\|-\bigotimes_{j=1}^{n}\hat{M}_{0}\|\|\sum_{\textbf{s}}\beta_{\textbf{s}}\bigotimes_{j=1}^{n}(\hat{M}_{0}\hat{M}_{1})^{s_{j}}\|\nonumber\\
&=&\|\sum_{\textbf{s}}\beta_{\textbf{s}}\bigotimes_{j=1}^{n}(U_{s_{j}})^{s_{j}}\|\nonumber\\
&=&\underset{\{\theta_{j}\}}{\textrm{sup}}\left|\sum_{\textbf{s}}\beta_{\textbf{s}}e^{i(\sum_{j=1}^{n}s_{j}\theta_{j})}\right|
\end{eqnarray}
We obtain this sum of complex terms as $\hat{M}_{0}\hat{M}_{1}=U_{j}$ is a unitary matrix as $(\hat{M}_{0}\hat{M}_{1})^{\dagger}=\hat{M}_{1}\hat{M}_{0}$ and $(\hat{M}_{0}\hat{M}_{1})^{\dagger}\hat{M}_{0}\hat{M}_{1}=\mathbb{I}$. The last line is then the norm of a linear combination of unitary matrices.

We now show that the value of \eqref{quantumupper} is attained by observables $\hat{M}_{s_{j}}$ on the GHZ state $|\textrm{GHZ}\rangle=\frac{1}{\sqrt{2}}\left(|0\rangle^{\otimes n}+|1\rangle^{\otimes n}\right)$. We prove this by construction where each party's measurement is
\begin{equation}\label{measure}
\hat{M}_{s_{j}}=e^{i(\phi+s_{j}\theta_{j})}|1\rangle\langle 0|+e^{-i(\phi+s_{j}\theta_{j})}|0\rangle\langle 1|.
\end{equation}
With these measurements, we obtain the following expectation values
\begin{equation}
-\sum_{\textbf{s}}\beta_{\textbf{s}}\langle GHZ|\bigotimes_{j=1}^{n}\hat{M}_{s_{j}}|GHZ\rangle=-\sum_{\textbf{s}}\beta_{\textbf{s}}\cos{\left(n\phi+\sum_{j=1}^{n}s_{j}\theta_{j}\right)}.
\end{equation}
We can write the expression over which we take the supremum in \eqref{quantumupper} as
\begin{equation}
\left|\sum_{\textbf{s}}\beta_{\textbf{s}}e^{i(\sum_{j=1}^{n}s_{j}\theta_{j})}\right|=\sum_{\textbf{s}}\beta_{\textbf{s}}\textrm{Re}\left(e^{i(\psi+\sum_{j=1}^{n}s_{j}\theta_{j})}\right)=\sum_{\textbf{s}}\beta_{\textbf{s}}\cos{(\psi+\sum_{j=1}^{n}s_{j}\theta_{j})}.
\end{equation}
We choose $\phi=\frac{\psi+\pi}{n}$ and so the optimal values of $\theta_{j}$ in \eqref{quantumupper} can be substituted into the measurement in \eqref{measure}. Therefore, these measurements on a GHZ state attain the maximum quantum upper bound of a Bell expression. $\square$
\newline

A corollary of this theorem is that since the quantum correlators 
\begin{equation}
p(1|\textbf{s})=1-2\cos{(\psi+\sum_{j=1}^{n}s_{j}\theta_{j})}
\end{equation}
can be optimized to maximally violate a Bell inequality, these correlators are extreme points of $\mathcal{Q}$ for all $\psi$, $\theta_{j}$. The space $\mathcal{Q}$ must contain every one of these extreme points, and so is the convex hull of these correlators \cite{ww}.

We now illustrate how the above theorem can be used to find the maximal quantum violation for the CHSH and Mermin inequality respectively. The phase values $\{\psi,\theta_{j}|j\in\{1,...,n\}\}$ for the CHSH inequality are $\psi=-\frac{\pi}{4}$ and $\theta_{1}=\theta_{2}=\frac{\pi}{2}$. Substituting this into \eqref{quantumupper}, we obtain Tsirelson's bound $1+\sqrt{2}$ \cite{tsirelson}. For the Mermin inequality \eqref{mermin}, $\psi=0$ and $\theta_{1}=\theta_{2}=-\theta_{3}=-\frac{\pi}{2}$ we have the maximal quantum (and algebraic) upper bound of $3$. The quantum violations for facet Bell inequalities in the $(3,2,2)$ and $(4,2,2)$ cases are listed in \cite{ww}.

In this section, we focussed on the quantum violation of facet Bell inequalities in various $(n,c,d)$ scenarios. However, all of the methods described so far apply to any Bell inequality, facet-defining or otherwise. We have used the facet Bell inequalities to show that in all of the scenarios investigated, $\mathcal{Q}$ is strictly larger than $\mathcal{L}$. The facet Bell inequalities are associated with their own difficulty; we have only shown that $\mathcal{Q}$ is larger than $\mathcal{L}$ for a small number of scenarios where we could actually find the facet Bell inequalities. On the other hand, if we suspend the necessity for the facet-defining condition and demonstrate a violation of an arbitrary Bell inequality, then $\mathcal{Q}$ is still strictly larger than $\mathcal{L}$. In the next section, we will consider Bell inequalities that are not facet-defining and show that they are of importance for considering quantum correlations. These inequalities are also of relevance when considering information processing tasks.

\section{Non-trivial Bell Inequalities}\label{sec33}

\noindent
Bell inequalities were first constructed in order to show that the statistics resulting from LHV theories \cite{bellineq,chsh} are constrained; this constraint does then not apply to quantum theory. The facet Bell inequalities go further and not only constrain LHV statistics but also \emph{define} the space of LHV correlators. We have indicated that to find these region-defining inequalities is a difficult task. However, if we just want to find Bell inequalities that distinguish between LHV and non-LHV correlators, satisfying the facet-defining condition is not necessary. We say that Bell inequalities are ``non-trivial'' if there are correlators in $\mathcal{P}$ that violate it, i.e.
\begin{equation}\label{bell2}
\sum_{\textbf{s}}\sum_{k=1}^{(d-1)}\beta_{k,\textbf{s}}p(k|\textbf{s})\leq \gamma_{\mathcal{L}} <\gamma_{\mathcal{P}},
\end{equation}
with $\beta_{k,\textbf{s}}$ as real pre-factors and $\gamma_{\mathcal{L}}$ as the upper bound resulting from all correlators in $\mathcal{L}$; $\gamma_{\mathcal{P}}$ is the upper bound of the inequality for all possible correlators in $\mathcal{P}$. As indicated above, for a non-trivial Bell inequality, there is the strict separation $\gamma_{\mathcal{L}}<\gamma_{\mathcal{P}}$.

We describe an explicit set of Bell inequalities that are non-trivial. We then employ a connection between these inequalities and an information processing task called a ``non-local game'' \cite{nonlocal} to derive an infinite number of non-trivial Bell inequalities. We begin our discussion in the simplest scenario by discussing the CHSH inequality and utilise its ``computational nature'' \cite{vandam}. We show that the intuition of the CHSH inequality as measuring the ability to perform a non-linear Boolean function with classical correlations can be applied to all scenarios. Again, central to our discussion is the computational perspective of LHV correlators. We utilise the limited computational expressiveness of LHV theories to derive consequences of this limitation. 

\subsection{Non-trivial Inequalities as Generalisations of the CHSH Inequality}\label{sec33a}

When the CHSH inequality \cite{chsh} was originally derived, the characterisation of correlations in terms of convex polytopes had not yet been considered. It may be considered a happy coincidence that this inequality is facet-defining for the LHV polytope. Despite being placed in the context of convex polytopes, the CHSH inequality has been redefined in the context of non-local games \cite{nonlocal} as we shall discuss in the next subsection. Such a versatile inequality also has a computational perspective that helps understand why it puts a restriction on LHV correlators \cite{vandam}. We will exploit this perspective to derive a generalisation of the CHSH inequality for all $(n,c,d)$ scenarios.

In order to describe this computational perspective we again write out the CHSH inequality
\begin{equation}
p(1|00)+p(1|01)+p(1|10)-p(1|11)\leq 2,
\end{equation}
and make the substitution $p(1|11)=1-p(0|11)$, to obtain
\begin{eqnarray}
p(1|00)+p(1|01)+p(1|10)+p(0|11) &=& \nonumber \\
\sum_{\textbf{s}}\sum_{k=0}^{1}\delta^{k}_{s_{1}s_{2}\oplus 1}p(k|\textbf{s})&\leq &3.
\end{eqnarray}
LHV correlators $p(k|\textbf{s})$ are contained in the convex hull of linear Boolean functions $g(\textbf{s})$ on $\textbf{s}$. So, $p(k|\textbf{s})=\sum_{g(\textbf{s})}p_{g(\textbf{s})}\delta^{k}_{g(\textbf{s})}$ with $p_{g(\textbf{s})}\geq 0$ and $\sum_{g(\textbf{s})}p_{g(\textbf{s})}=1$. Then, by convexity, the following expression must be satisfied for all linear Boolean functions $g(\textbf{s})$ in the $(2,2,2)$ scenario:
\begin{eqnarray}
\sum_{\textbf{s}}\sum_{k=0}^{1}\delta^{k}_{s_{1}s_{2}\oplus 1}\delta^{k}_{g(\textbf{s})}&=&\nonumber \\
\sum_{\textbf{s}}\delta^{g(\textbf{s})}_{s_{1}s_{2}\oplus 1}&\leq &3.
\end{eqnarray}
By listing all possible functions $g(\textbf{s})$ and seeing when they overlap with $s_{1}s_{2}\oplus 1$, we see that the maximum overlap is $3$. We can then rewrite the original CHSH inequality in terms of correlators $p(1|\textbf{s})$ and this derivation of the LHV upper bound $\gamma_{\mathcal{L}}$
\begin{equation}
p(1|00)+p(1|01)+p(1|10)-p(1|11)\leq\underset{g(\textbf{s})}{\textrm{max}}\left[\sum_{\textbf{s}}\delta^{s_{1}s_{2}\oplus 1}_{g(\textbf{s})}\right]-1.
\end{equation}
Essentially, this inequality ``measures'' the inability for LHV correlators to achieve the non-$n$-partite linear function $s_{1}s_{2}\oplus 1$ deterministically \cite{vandam}. If LHV theories could achieve this function deterministically then $\gamma_{\mathcal{L}}=3$ as $\sum_{\textbf{s}}\delta^{g(\textbf{s})}_{f(\textbf{s})}=4$. This is, however, not possible and this is the upper bound $\gamma_{\mathcal{P}}$ for all correlators in $\mathcal{P}$ so $\gamma_{\mathcal{P}}>\gamma_{\mathcal{L}}$.

The CHSH inequality is not the only example of a well-studied Bell inequality that can be written in terms of the overlap between a non-$n$-partite linear and $n$-partite linear function. The Svetlichny inequality \cite{svet} as mentioned in chapter \ref{chap2}, section \ref{sec24},
\begin{eqnarray}\label{nontrivsvet}
p(1|000)+p(1|001)+p(1|010)-p(1|011)& \nonumber \\
+p(1|100)-p(1|101)-p(1|110)-p(1|111)&\leq 2,
\end{eqnarray}
can be rewritten as
\begin{equation}
\sum_{\textbf{s}}\sum_{k=0}^{1}\delta^{k}_{s_{1}s_{2}\oplus s_{1}s_{3}\oplus s_{2}s_{3}\oplus 1}p(k|\textbf{s})\leq 6,
\end{equation}
after making the substitution of $p(1|\textbf{s})=1-p(0|\textbf{s})$ when the prefactors in \eqref{nontrivsvet} are $-1$; this is the case when $s_{1}s_{2}\oplus s_{1}s_{3}\oplus s_{2}s_{3}\oplus 1=0$. Again, by convexity the upper bound of this inequality just results in the maximum overlap $\sum_{\textbf{s}}\delta^{g(\textbf{s})}_{s_{1}s_{2}\oplus s_{1}s_{3}\oplus s_{2}s_{3}\oplus 1}$ for all linear Boolean functions $g(\textbf{s})$ for $(3,2,2)$. The function $f(\textbf{s})=s_{1}s_{2}\oplus s_{1}s_{3}\oplus s_{2}s_{3}\oplus 1$ is a non-linear Boolean function and so the overlap $\sum_{\textbf{s}}\delta^{g(\textbf{s})}_{f(\textbf{s})}$ by definition will always be lower than $2^{n}=8$. Again, we can rewrite the above Svetlichny inequality as
\begin{eqnarray}
& &p(1|000)+p(1|001)+p(1|010)-p(1|011)+p(1|100)-p(1|101)\nonumber \\
& &-p(1|110)-p(1|111)\nonumber \\
&\leq& \underset{g(\textbf{s})}{\textrm{max}}\left[\sum_{\textbf{s}}\delta^{g(\textbf{s})}_{s_{1}s_{2}\oplus s_{1}s_{3}\oplus s_{2}s_{3}\oplus 1}\right]-4=2.
\end{eqnarray}
For all possible correlators in $\mathcal{P}$, the upper bound is then $4$ thus it is a non-trivial Bell inequality, as expected. However, it is not a facet Bell inequality for the region $\mathcal{L}$, but facet-defining for the Svetlichny region, $\mathcal{S}$. The region $\mathcal{S}$ is a sub-region of $\mathcal{P}$ but larger than $\mathcal{L}$, therefore bounds the region $\mathcal{L}$. Non-trivial Bell inequalities can then provide a useful tool to bound $\mathcal{L}$ away from the whole space $\mathcal{P}$.

The CHSH and Svetlichny inequalities above notably utilise the fact that linear Boolean functions cannot be equal to non-linear Boolean functions for all inputs $\textbf{s}$. Given that LHV correlators are associated with the former and not the latter, we can write down inequalities of the following form for all scenarios $(n,c,d)$:
\begin{equation}\label{nontrivial}
\sum_{\textbf{s}}\sum_{k=0}^{(d-1)}\delta^{k}_{f(\textbf{s})}p(k|\textbf{s})\leq\underset{g(\textbf{s})}{\textrm{max}} \sum_{\textbf{s}}\delta^{f(\textbf{s})}_{g(\textbf{s})},
\end{equation}
for all non-$n$-partite linear functions $f(\textbf{s})$ and $n$-partite linear functions $g(\textbf{s})$. The above inequality in \eqref{nontrivial} is defined for all correlators $p(k|\textbf{s})$ and not the normalised set of correlators for $k\in\{1,2,...,(d-1)\}$. Therefore in order to describe this inequality in terms of normalised correlators, i.e. vectors in $\mathcal{P}$, we impose the normalisation condition that $1-\sum_{k=1}^{(d-1)}p(k|\textbf{s})=p(0|\textbf{s})$. The expression on the left-hand-side of \eqref{nontrivial} becomes
\begin{equation}
\sum_{\textbf{s}}\left[\delta^{0}_{f(\textbf{s})}\left(1-\sum_{k=1}^{(d-1)}p(k|\textbf{s})\right)+\sum_{k=1}^{(d-1)}\delta^{k}_{f(\textbf{s})}p(k|\textbf{s})\right]\leq\underset{g(\textbf{s})}{\textrm{max}} \sum_{\textbf{s}}\delta^{f(\textbf{s})}_{g(\textbf{s})},
\end{equation}
which can be rewritten in a form similar to the CHSH inequality,
\begin{equation}\label{nontrivial2}
\sum_{\textbf{s}}\left[\sum_{k=1}^{(d-1)}\left(\delta^{k}_{f(\textbf{s})}-\delta^{0}_{f(\textbf{s})}\right)p(k|\textbf{s})\right]\leq\underset{g(\textbf{s})}{\textrm{max}} \sum_{\textbf{s}}\left(\delta^{f(\textbf{s})}_{g(\textbf{s})}-\delta^{0}_{f(\textbf{s})}\right).
\end{equation}
The upper bound $\gamma_{\mathcal{L}}$ then is strictly smaller than $\gamma_{\mathcal{P}}=c^{n}-\sum_{\textbf{s}}\delta^{0}_{f(\textbf{s})}$, so the inequality is non-trivial. As we have already demonstrated, the CHSH inequality and Svetlichny inequality are examples of these non-trivial inequalities. As with the Svetlichny inequality, they are not necessarily facet inequalities for $\mathcal{P}$, but necessarily bound the region $\mathcal{L}$. A non-trivial Bell inequality must also intersect $\mathcal{L}$ at, at least, one vertex otherwise the right-hand-side of the inequality in \eqref{nontrivial2} is not tight. 

Not only are these inequalities interesting because of their ability to bound $\mathcal{L}$, but they have a role in information processing tasks. The particular task of relevance is a non-local game \cite{nonlocal}. One can be successful at such a game if they violate a Bell inequality, hence the use of ``non-local'', as in non-LHV resources. One wants to achieve some task (expressed as a game) with as great a probability as possible. Games in general are of interest in computer science and in fields of applied mathematics such as economics \cite{game}. In some non-local games such as the ``XOR games'' \cite{nonlocal}, the Bell inequality can quantify the probability of achieving a task and so have a natural role in these games. We use the language and structure of non-local game to describe an \emph{infinite} number of non-trivial Bell inequalities for \emph{each} scenario $(n,c,d)$.

\subsection{Non-local Games}\label{sec33b}

We have discussed the operational perspective of Bell tests where we have many parties each with inputs and outputs. Many information processing tasks can be abstracted to a process with an input, and a transformation of the input to produce an output. We now focus on one particular task that has a natural connection to Bell tests, the non-local game (NLG) \cite{nonlocal}. In this section we discuss the set-up of an NLG and how it is relevant to the discussion of constructing non-trivial Bell inequalities. In the next section, NLG will again be discussed and made relevant to the subject of MBQC. Therefore, these games are of relevance to a great deal of discussion to both Bell tests and this thesis in particular. They also give an interesting computational perspective on Bell tests that has been of interest to the quantum information science community. 

We now describe a particular NLG with $n$ parties, or ``players'' as they are often called. These $n$ parties do not communicate with each other, and so for all intents and purposes, are space-like separated as in Bell tests. As well as these $n$ parties, there is another party that is not a player, but a ``referee''. A referee can be seen as the experimenter in a Bell test who calculates the correlators $p(k|\textbf{s})$. However, one distinct aspect in NLG from Bell tests is in the role of the referee as the person who distributes inputs to the $n$ parties as well as retrieving their outputs. In the format of Bell tests that we have discussed so far, the inputs at each site are generated randomly by the parties themselves, in NLG this is not the case. To summarise, $n$ parties each receive an input from the referee and then generate an output which they send to the referee. The referee finally calculates some function on the outputs; the objective of these parties is to maximize the mean probability (for all inputs) of this function being equal to some desired value. We now specify the particular NLG that is of relevance to our discussion:
\begin{enumerate}
\item A referee sends the input digit-string $\textbf{s}=\mathbb{Z}_{c}^{n}$ to the $n$ non-communicating parties. The inputs $\textbf{s}$ are sent with probability distribution $\pi(\textbf{s})$ such that $\sum_{\textbf{s}}\pi(\textbf{s})=1$ and all $\pi(\textbf{s})\geq 0$;\\

\item All $n$ parties generate an output digit-string $\textbf{m}=\mathbb{Z}_{d}^{n}$ which is sent to the referee;\\

\item The referee calculates the sum modulo $d$ of all outcomes $[\sum_{j=1}^{n}m_{j}]_{d}=k$;\\

\item The goal of the game is for the players to maximise the average success probability of $\left[\sum_{j=1}^{n}m_{j}\right]_{d}=f(\textbf{s})$ for some function $f:\mathbb{Z}^{n}_{c}\rightarrow\mathbb{Z}_{d}$.\\
\end{enumerate}
Examples of these games include the well-studied multi-party ``XOR games'' where $c=d=2$ \cite{nonlocal}. The average success probability $\bar{p}_{f(\textbf{s})}$ of achieving $k=\left[\sum_{j=1}^{n}m_{j}\right]_{d}=f(\textbf{s})$ can be written in terms of the correlators $p(k|\textbf{s})$:
\begin{equation}\label{nlgprob}
\bar{p}_{f(\textbf{s})}=\sum_{\textbf{s}}\pi(\textbf{s})\sum_{k=0}^{(d-1)}\delta^{k}_{f(\textbf{s})}p(k|\textbf{s}).
\end{equation}
In order to maximize this average success probability, we want to find the optimal correlators $p(k|\textbf{s})$. We can then distinguish between the maximum average success probability $\bar{p}^{\mathcal{L}}_{f(\textbf{s})}$ and $\bar{p}^{\mathcal{Q}}_{f(\textbf{s})}$ resulting from quantum and classical (or LHV) correlators respectively. If $\bar{p}^{\mathcal{L}}_{f(\textbf{s})}<\bar{p}^{\mathcal{Q}}_{f(\textbf{s})}$, it is optimal to use quantum resources instead of classical resources. Also, \eqref{nlgprob} produces a Bell inequality if the correlators result from LHV theories which is upper bounded by $\bar{p}^{\mathcal{L}}_{f(\textbf{s})}$; if $\bar{p}^{\mathcal{L}}_{f(\textbf{s})}<\bar{p}^{\mathcal{Q}}_{f(\textbf{s})}$, we have a violation of this Bell inequality. One way that we can possibly have a separation $\bar{p}^{\mathcal{L}}_{f(\textbf{s})}<\bar{p}^{\mathcal{Q}}_{f(\textbf{s})}$, is if the function $f(\textbf{s})$ is a non-$n$-partite linear function. If $f(\textbf{s})$ is an $n$-partite linear function, then $\bar{p}^{\mathcal{L}}_{f(\textbf{s})}=1$.

To make the connection to Bell inequalities explicit, if $\pi(\textbf{s})=\pi(\textbf{s}')=\frac{1}{c^{n}}$ for all $\textbf{s}\neq\textbf{s}'$, then we obtain a modification of the non-trivial Bell inequalities \eqref{nontrivial} discussed in the previous subsection. We can rewrite the inequality in \eqref{nontrivial2} in terms of $\bar{p}^{\mathcal{L}}_{f(\textbf{s})}$ for $f(\textbf{s})$ being a non-$n$-partite linear function:
\begin{equation}\label{nontrivial3}
\frac{1}{c^{n}}\sum_{\textbf{s}}\left[\sum_{k=1}^{(d-1)}\left(\delta^{k}_{f(\textbf{s})}-\delta^{0}_{f(\textbf{s})}\right)p(k|\textbf{s})\right]\leq\bar{p}^{\mathcal{L}}_{f(\textbf{s})}-\frac{1}{c^{n}}\sum_{\textbf{s}}\delta^{0}_{f(\textbf{s})},
\end{equation}
where
\begin{equation}
\bar{p}^{\mathcal{L}}_{f(\textbf{s})}=\underset{g(\textbf{s})}{\textrm{max}} \frac{1}{c^{n}}\sum_{\textbf{s}}\delta^{f(\textbf{s})}_{g(\textbf{s})}<1.
\end{equation}
For \eqref{nlgprob}, if the correlators are all possible correlators in $\mathcal{P}$, then the maximum average success probability is $\bar{p}^{\mathcal{P}}_{f(\textbf{s})}=\sum_{\textbf{s}}\pi(\textbf{s})=1$. Therefore, the fact that $\bar{p}^{\mathcal{L}}_{f(\textbf{s})}<\bar{p}^{\mathcal{P}}_{f(\textbf{s})}$ indicates that the inequality \eqref{nontrivial3} is non-trivial. 

In order to establish a non-trivial Bell inequality, we need to find a probability distribution $\pi(\textbf{s})$ such that $\bar{p}^{\mathcal{L}}_{f(\textbf{s})}<\bar{p}^{\mathcal{P}}_{f(\textbf{s})}$. In the following result, we describe an infinite number of simple probability distributions such that we can generate a non-trivial Bell inequality.

\begin{prop}\label{prop331}
All inequalities of the form
\begin{equation}
\sum_{\textbf{s}}\pi(\textbf{s})\left[\sum_{k=1}^{(d-1)}\left(\delta^{k}_{f(\textbf{s})}-\delta^{0}_{f(\textbf{s})}\right)p(k|\textbf{s})\right]\leq\bar{p}^{\mathcal{L}}_{f(\textbf{s})}-\sum_{\textbf{s}}\pi(\textbf{s})\delta^{0}_{f(\textbf{s})},
\end{equation}
are non-trivial Bell inequalities for all non-zero probabilities $\pi(\textbf{s})$ if $f(\textbf{s})$ is a non-$n$-partite linear function.
\end{prop}

\textit{Proof}: In order to prove this we just need to show that $\bar{p}^{\mathcal{L}}_{f(\textbf{s})}<\bar{p}^{\mathcal{P}}_{f(\textbf{s})}=1$ for the probability distribution $\pi(\textbf{s})$ being non-zero for all values of $\textbf{s}$. That is,
\begin{equation}\label{nontriv}
\underset{g(\textbf{s})}{\textrm{max}} \sum_{\textbf{s}}\pi(\textbf{s})\delta^{f(\textbf{s})}_{g(\textbf{s})}<1,
\end{equation}
which is true as $0<\pi(\textbf{s})<1$ for each $\textbf{s}$ by definition and $\sum_{\textbf{s}}\delta^{f(\textbf{s})}_{g(\textbf{s})}<1$. $\square$
\newline

For a distribution $\pi(\textbf{s})$ satisfying $0<\pi(\textbf{s})<1$ for each $\textbf{s}$ and $\sum_{\textbf{s}}\pi(\textbf{s})=1$, we can construct $\left(d^{c^{n}}-d^{n(c-1)+1}\right)$ non-trivial Bell inequalities: the number of non-$n$-partite linear functions $f(\textbf{s})$. We are able to construct an infinite number of non-trivial Bell inequalities parametrized by $\pi(\textbf{s})$ utilizing a computational perspective on Bell tests. Crucially though, the non-trivial Bell inequalities in \eqref{nontriv} are not dependent on an NLG construction, they exist outside of NLG. More specifically, the probability distribution $\pi(\textbf{s})$ is just a positive, non-zero weighting on correlators for a particular $\textbf{s}$. In the context of NLG, the inequalities in \eqref{nontriv} are related to the success probability of the game, but outside of this context we still have an infinite number of non-trivial Bell inequalities. 

So far our discussion has applied to all possible scenarios $(n,c,d)$ and we have stated general results for all these scenarios. In the discussion in the next subsection we will focus on a particular example of non-trivial Bell inequality in $(n,2,2)$ for all $n$ of the form in \eqref{nontrivial2}. This example is a generalisation of the CHSH inequality to $n$ parties. In contrast to the CHSH inequality, we will show that the upper bound of the quantum correlators is no better than the LHV upper bound.

\subsection{$(n,2,2)$ scenario and the $n$-partite NAND function}\label{sec33c}

It is natural at this point to question the motivation for finding non-trivial Bell inequalities when facet Bell inequalities are more useful for determining the consequences of LHV correlators. As well as the motivations from computational complexity that finding facet Bell inequalities is hard, the utility of bounding $\mathcal{L}$ and connection to information processing tasks, we have further motivation in the $(n,2,2)$ scenario. We show that in this scenario, facet Bell inequalities can be related to the non-trivial Bell inequalities described earlier. 

As well as this general discussion, we will give an example of a non-trivial Bell inequality of the form \eqref{nontrivial2} for the $(n,2,2)$ scenario. This non-trivial Bell inequality for all $n$ is a direct generalisation of the CHSH inequality. Essentially it is associated with a non-$n$-partite linear function, itself a generalisation of the function $s_{1}s_{2}\oplus 1$ for $n$ parties. We will use this function also to say something about MBQC in the section \ref{sec34}. We will show that for more than $2$ parties this non-trivial inequality cannot be violated by quantum correlators.

Firstly, we rewrite the non-trivial Bell inequalities in \eqref{nontrivial3} in the specific $(n,2,2)$ scenario as
\begin{equation}\label{nontrivial4}
\sum_{\textbf{s}}\pi(\textbf{s})(-1)^{f(\textbf{s})+1}p(1|\textbf{s})\leq\underset{g(\textbf{s})}{\textrm{max}} \sum_{\textbf{s}}\pi(\textbf{s})\left(\delta^{f(\textbf{s})}_{g(\textbf{s})}-\delta^{0}_{f(\textbf{s})}\right),
\end{equation}
but now we allow probabilities $0\leq \pi(\textbf{s})\leq 1$. If we allow probabilities $\pi(\textbf{s})\in\{0,1\}$ then we might not have non-trivial Bell inequalities as we can choose probabilities such that $\pi(\textbf{s})=0$ when $f(\textbf{s})\neq g(\textbf{s})$ and non-zero otherwise. In this instance, $\sum_{\textbf{s}}\pi(\textbf{s})\delta^{f(\textbf{s})}_{g(\textbf{s})}=1$, hence $\gamma_{\mathcal{L}}=\gamma_{\mathcal{P}}$ and we do not have a non-trivial Bell inequality. However, for a choice of function $f(\textbf{s})$ and probability distribution $\pi(\textbf{s})$ we can construct a facet Bell inequality. Since all prefactors $\beta_{\textbf{s}}$ of a facet Bell inequality are real, then they can be rewritten as $\beta_{\textbf{s}}=|\beta_{\textbf{s}}|\textrm{sign}(\beta_{\textbf{s}})$. We now fix values as $\pi(\textbf{s})=\frac{|\beta_{\textbf{s}}|}{\sum_{\textbf{s}}|\beta_{\textbf{s}}|}$ and $\textrm{sign}(\beta_{\textbf{s}})=(-1)^{f(\textbf{s})+1}$, then we multiply both sides of \eqref{nontrivial4} with $\sum_{\textbf{s}}|\beta_{\textbf{s}}|$ to obtain the inequality of the form in \eqref{bell1}. If $f(\textbf{s})$ is an $n$-partite linear function in \eqref{nontrivial4}, then we cannot define a non-trivial Bell inequality and so cannot define a non-trivial, facet Bell inequality. Finding the facet Bell inequalities for $(n,2,2)$ is then a case of finding a probability distribution $\pi(\textbf{s})$ where we satisfy the facet-defining condition.

For example, the inequality \eqref{newineq}:
\begin{eqnarray}\label{wwnew}
\frac{1}{4}\left[p(1|000)+p(1|001)+p(1|010)+p(1|011)\right]& &\nonumber\\
+\frac{1}{4}\left[p(1|100)+p(1|101)+p(1|110)-3p(1|111)\right]&\leq&1,
\end{eqnarray}
can be rewritten in the form of \eqref{nontrivial4} with $\pi(\textbf{s})=\frac{1}{10}$ for all $\textbf{s}\neq\{1,1,1\}$ and $\pi(\textbf{s})=\frac{3}{10}$ for $\textbf{s}=\{1,1,1\}$ and $f(\textbf{s})=s_{1}s_{2}s_{3}\oplus 1$. With these substitutions \eqref{wwnew} can be retrieved from \eqref{nontrivial4} but now both sides of the inequality in \eqref{wwnew} are multiplied by $\frac{1}{10}$. Now, $\underset{g(\textbf{s})}{\textrm{max}} \sum_{\textbf{s}}\pi(\textbf{s})\left(\delta^{f(\textbf{s})}_{g(\textbf{s})}-\delta^{0}_{f(\textbf{s})}\right)=\frac{4}{10}$ so that $\bar{p}_{f(\textbf{s})}^{\mathcal{L}}=\frac{7}{10}$. The latter result can be seen from the fact that the $n$-partite linear function $g(\textbf{s})=1$ overlaps with $f(\textbf{s})$ for $7$ values of $\textbf{s}$; these are the inputs $\textbf{s}\neq\textbf{1}$. We have briefly shown that at least one of the non-trivial Bell inequalities as described in Proposition \ref{prop332} is also a facet Bell inequality.

We have shown that for a function $f(\textbf{s})$ we can find a probability distribution $\pi(\textbf{s})$ where the resulting non-trivial Bell inequality in \eqref{nontrivial4} is a facet Bell inequality. We now take a different approach and fix the probability distribution to be $\pi(\textbf{s})=\frac{1}{2^{n}}$ for all inputs $\textbf{s}$. This can be seen as the probability distribution of the Bell test being an NLG with inputs chosen randomly. 

Given this probability distribution, we consider a function $f(\textbf{s})$ for all $n$ in $(n,2,2)$. This function is a natural generalisation of the function $f(\textbf{s})=s_{1}s_{2}\oplus 1$ corresponding to the function defining the CHSH inequality. This function will be discussed later with reference to quantum computing and so we define it now.

\begin{defn}\label{defn331}
The \textbf{n-partite NAND function} is $f_{2}(\textbf{s})=\prod_{j=1}^{n}s_{j}\oplus1$ acting on bit-string $\textbf{s}\in\mathbb{Z}^{n}_{2}$.
\end{defn}

A NAND function is defined on two bits $s_{1}$ and $s_{2}$ as $f(\textbf{s})=s_{1}s_{2}\oplus 1$. This is exactly the function that we used when describing the CHSH inequality earlier in this chapter. The NAND function is the negation (or NOT) of the AND function $f(\textbf{s})=s_{1}s_{2}$ \cite{pap}. The $n$-partite NAND function consists of the entire NOT of a number of AND functions between variables in $\textbf{s}$. What is clear is that it is a non-linear Boolean function due to the multiplication between elements of $\textbf{s}$. It is also the function describing the facet Bell inequality \eqref{wwnew} above. 

For the $n$-partite NAND function and uniform probability distribution we obtain a non-trivial Bell inequality of the form
\begin{equation}\label{npartite}
\frac{1}{2^{n}}\sum_{\textbf{s}}(-1)^{f_{2}(\textbf{s})+1}p(1|\textbf{s})\leq\frac{2^{n}-2}{2^{n}}.
\end{equation}
The upper bound on the right-hand-side is due to the fact that $\sum_{\textbf{s}}\delta^{f_{2}(\textbf{s})}_{g(\textbf{s})}=(2^{n}-1)$ if $g(\textbf{s})=1$, and all linear Boolean functions $g(\textbf{s})$ are never always equal to $f_{2}(\textbf{s})$. We have shown that this inequality for $n=3$ is related to the facet Bell inequality \eqref{wwnew}, and this relation extends to all $n$ in \eqref{gennewineq}. We now show that this natural generalisation of the CHSH inequality has no quantum violation whatsoever for $n\geq 3$.

\begin{prop}\label{prop332}
The non-trivial Bell inequality \eqref{npartite} for the $n$-partite NAND function for uniform probability distribution $\pi(\textbf{s})=2^{-n}$ for all $\textbf{s}$ is not violated by quantum correlators for $n\geq 3$.
\end{prop}

\textit{Proof}: The quantum upper bound for the inequality \eqref{npartite} can be calculated from \eqref{quantumupper} to obtain:
\begin{eqnarray}
2\sum_{\textbf{s}}(-1)^{f_{2}(\textbf{s})+1}p(1|\textbf{s})&=&\underset{\{\theta_{j}\}}{\textrm{sup}}\left[\left(\sum_{\textbf{s}}(-1)^{f_{2}(\textbf{s})+1}\right)+\left|\sum_{\textbf{s}}(-1)^{f_{2}(\textbf{s})+1}e^{i(\sum_{j=1}^{n}s_{j}\theta_{j})}\right|\right]\nonumber \\
&=&2^{n}-2+\underset{\{\theta_{j}\}}{\textrm{sup}}\left|\sum_{\textbf{s}}(-1)^{f_{2}(\textbf{s})+1}e^{i(\sum_{j=1}^{n}s_{j}\theta_{j})}\right|.
\end{eqnarray}
If there is a quantum violation of the inequality in \eqref{npartite} then the following relationship must be satisfied:
\begin{equation}\label{ineq1}
\underset{\{\theta_{j}\}}{\textrm{sup}}\left|\sum_{\textbf{s}}(-1)^{f_{2}(\textbf{s})+1}e^{i(\sum_{j=1}^{n}s_{j}\theta_{j})}\right|>2^{n}-2
\end{equation}
We may, without loss of generality, restrict $\theta_j$ to the range $\theta_{j}\in(-\pi,\pi)$. We simplify inequality \eqref{ineq1}, using the fact that $(-1)^{f_{2}(\textbf{s})+1}=1$ for all bit strings $\textbf{s}$ except when $\textbf{s}=\textbf{1}$, to write
\begin{eqnarray}\label{ineq2}
\underset{\{\theta_{j}\}}{\textrm{sup}}\left|\sum_{\textbf{s}}\prod _{k=1}^{n} e^{i s_{j}\theta_{j} }-2e^{i\sum_{k}^{n}{\theta_{k}}}\right|&=&\nonumber \\
\underset{\{\theta_{j}\}}{\textrm{sup}}\left|2^{n}\prod_{j=1}^{n}\cos\left(\frac{\theta_{j}}{2}\right)-2e^{i\sum_{k}^{n}{\theta_{k}}}\right|&>&2^{n}-2.
\end{eqnarray}
We now adopt a geometric argument. The goal is to maximize the modulus of a sum of two complex numbers. These numbers may be represented, on the plane, as two sides of a triangle. The first side has length $2^{n}\prod_{j=1}^{n}\cos(\frac{\theta_{j}}{2})$, the second is of length 2 and the angle between these sides is $\sum_{k}^{n}\frac{\theta_{k}}{2}$.  We complete the proof by showing that when $n>7$, the length of the third side of the triangle can never exceed $2^n-2$, and hence \eqref{ineq2} is never satisfied.

We proceed by assuming the opposite of what we want to prove and demonstrating a contradiction. Via the triangle inequality, for this inequality to be satisfied, the length of the base of the triangle must be greater than $2^n-4$, and thus
\begin{equation}\label{inequality}
	\prod_{j=1}^{n}\cos\left(\frac{\theta_{j}}{2}\right)>1-2^{2-n}.
\end{equation}
Since $\theta_{j}\in(-\pi,\pi)$ all terms in the product are non-negative, hence we can impose the weaker condition for \eqref{inequality} that $\forall\theta_{j}$, $\cos\left(\frac{\theta_{j}}{2}\right)>1-2^{2-n}$. This implies that $|\sum_{j=1}^{n}\frac{\theta_{j}}{2}|<n \arccos(1-2^{2-n})$.

Proceeding geometrically, we now use the cosine rule to express the third side of the triangle (representing the modulus in \eqref{ineq1}), and this expression must satisfy
\begin{equation}\label{cosiner}
4+2^{2n}\prod_{j=1}^{n}\cos^{2}\left(\frac{\theta_{j}}{2}\right)-2^{n+2}\cos\left(\sum_{k=1}^{n}\frac{\theta_{k}}{2}\right)\prod_{l=1}^{n}\cos\left(\frac{\theta_{l}}{2}\right)>4+2^2n-2^{n+2},
\end{equation}
to obtain a quantum violation. Since $\theta_{j}\in(-\pi,\pi)$, $\prod_{l=1}^{n}\cos(\frac{\theta_{l}}{2})$ is non-negative, and hence a violation can only be achieved if $\cos(\sum_{k=1}^{n}\frac{\theta_{k}}{2})$ is negative which implies  $|\sum_{k=1}^{n}\frac{\theta_{k}}{2}|>\frac{\pi}{2}$. Using this, we achieve
\begin{equation}
	n \arccos(1-2^{2-n})>\pi/2
\end{equation}
or equivalently
\begin{equation}\label{ineqc}
\cos\left(\frac{\pi}{2n}\right)>(1-2^{2-n}).
\end{equation}
This inequality is only satisfied for integers $n\le 7$, hence, due to the contradiction with our initial assumption, for $n>7$ the quantum and classical bounds of the non-trivial Bell inequality \eqref{npartite}. Direct numerical verification of the bounds, via equation \eqref{quantumupper} for $n<7$ indicates that the bounds coincide for all integer values $3\le n\le 7$, thus completing the proof. $\square$
\newline

This proof demonstrates that, $\mathcal{Q}$ is smaller than $\mathcal{P}$ for this scenario. It also demonstrates that quantum correlators are not always useful in every non-local game. If we modify the probability distribution $\pi(\textbf{s})$ by weighting the input $\textbf{s}=\textbf{1}$ more than other inputs, we can regain a quantum advantage as with the inequality in \eqref{gennewineq}.

This example of a non-trivial Bell inequality not being violated by quantum mechanics is not isolated. For example, the following non-trivial Bell inequality for $(2,3,3)$ corresponding to the function $f(\textbf{s})=\left[s_{1}^{2}s_{2}^{2}+1\right]_{3}$ with uniform probability distribution $\pi(\textbf{s})=\frac{1}{9}$:
\begin{eqnarray}\label{nontrivialthree}
\frac{1}{9}\left(p(1|00)+p(1|01)+p(1|02)+p(110)+p(2|11)\right)&\nonumber \\
+\frac{1}{9}\left(p(2|12)+p(1|20)+p(2|21)+p(2|22)\right)\leq& \frac{8}{9}
\end{eqnarray}
is also not violated by quantum correlators $\mathcal{Q}$. This upper bound of $\frac{8}{9}$ was found using both the MBS and SDP approach. Just like the $n$-partite NAND function, the function $f(\textbf{s})=\left[s_{1}^{2}s_{2}^{2}+1\right]_{3}$ differs from all possible $n$-partite linear functions for only one value of $\textbf{s}$.

We have shown that quantum resources are not always better than classical resources when trying to maximize the mean probability of winning an NLG. This is not new as Linden et al \cite{linden} devised a model of ``non-local computation'' where quantum resources do no better than classical, or LHV resources. The resulting Bell inequality defining over probabilities $p(\textbf{m}|\textbf{s})$ from this model is also not facet-defining. Perhaps more interesting, Almeida et al found an NLG where quantum resources do no better than classical resources \cite{almeida}; and for $n\geq 3$, this game defines a facet Bell inequality of $\mathcal{L}_{\textbf{F}}$ for probabilities $p(\textbf{m}|\textbf{s})$. Investigating the limitations of quantum correlations therefore seems to be just as interesting as finding its advantages. 

We will use this function in the next subsection to say something about MBQC \cite{mbqc1,mbqc2}. In particular, we look at a restricted class of computations in MBQC and map this class into the framework of Bell tests. We employ the Bell test as an NLG but in the language of games, there is a ``promise'' on the inputs \cite{nonlocal}. That is, the inputs are the result of some pre-processing on a bit-string \cite{anders}. This pre-processing has a well-defined role in MBQC and we use our NLG to show that our restricted class of MBQC is not equivalent to a universal Quantum Computer. The key to all of these insights is the computational perspective of the space $\mathcal{L}$.

\section{Non-adaptive Measurement-based Quantum Computing}\label{sec34}

\noindent
MBQC as formulated by Raussendorf and Briegel \cite{mbqc1} has been one of the great breakthroughs in quantum computing. Whereas the original circuit model of quantum computing requires the ability to perform unitary operators over the length of the computation \cite{nielsen}, MBQC reduced this to state preparation and sequential single-site (single-qubit) measurements \cite{mbqc1}. The state that is prepared is a multipartite entangled state, e.g. the ``cluster state'' \cite{mbqc2}. We immediately see that MBQC is more in the vein of a Bell test, which (for quantum correlators) consists of the preparation of a potentially entangled state and then single-site measurements on each part of this state. In this section, we show that the connection is more concrete than just this superficially shared language. 

Briegel and Raussendorf showed that adaptivity is a key component of their formulation of MBQC \cite{mbqc1,mbqc2}, in order that all possible quantum circuits are implemented deterministically. A natural question is what happens when we remove adaptivity? If we do not have adaptivity, then all measurements can take place simultaneously. This also simplifies the technological implementation of an MBQC, where a state only needs to be prepared and then measured instantly. Adaptivity means that a state needs to be stored for a non-negligible amount of time between measurement rounds. 

\begin{figure}
	\centering
		\includegraphics[width=0.75\textwidth]{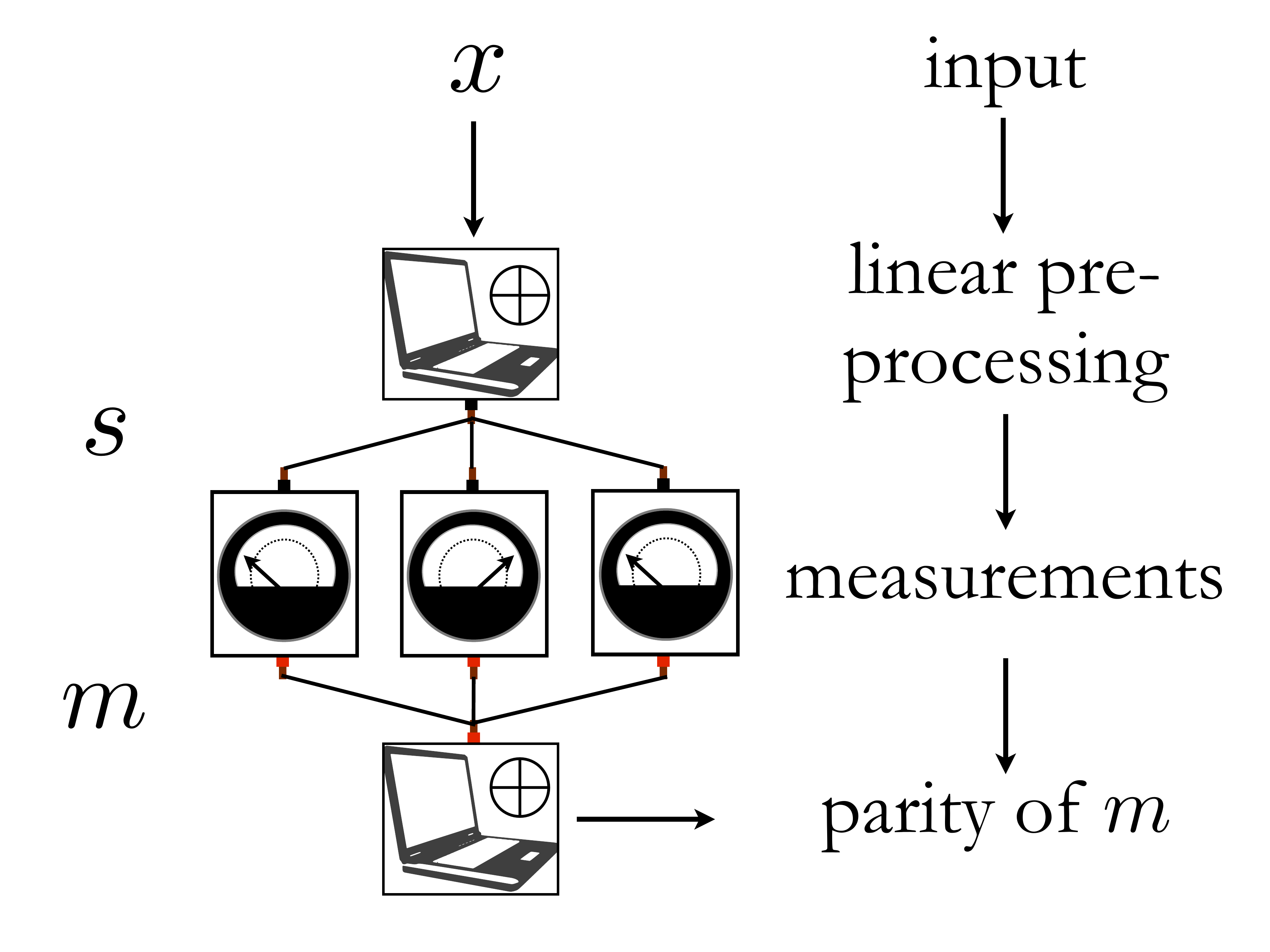}
			\caption{Non-adaptive MBQC consists of pre-processing on data, this data is then sent to the parties who make a single round of measurements. The classical control computer as well as performing pre-processing, processes the measurement outcomes. (Copyright: Institute of Physics, 2011).}
			\label{fig:belltest}
\end{figure}

We now define the class of computations in MBQC without adaptivity or as we will call it, \textbf{nMBQC}. We do not place restrictions on the measurements, state prepared or number of sites. The element which remains the same in Raussendorf and Briegel's formulation of MBQC is the control computer \cite{mbqc2,anders}. The control computer can only implement XOR gates, or addition modulo $2$, on bits \cite{anders} (see section \ref{sec13} and \ref{sec14}). However, the necessity for particular measurements and states is not as well-defined; the cluster state is an example of a useful resource \cite{mbqc2}, but there are other examples \cite{hein,vandennest}. We now define nMBQC as an abstract model. See Figure \ref{fig:belltest} for an accompanying schematic of nMBQC.

\begin{defn}\label{defn341}
The model of \textbf{Non-adaptive Measurement-Based Quantum Computing}, or nMBQC, involves the preparation of an $n$-partite quantum state $|\psi\rangle$ and a classical control computer $\mathcal{C}$. The computer $\mathcal{C}$ receives a bit-string $\textbf{x}$ of length $|\textbf{x}|$ with uniform probability $\frac{1}{2^{|\textbf{x}|}}$. The control computer performs arbitrary XOR gates on a bit-string $\textbf{x}$ and communicates the choice of measurement $s_{j}(\textbf{x})$ to each $j$th site. There is a single round of measurements on all $n$ non-communicating sites. The control computer receives the measurement outcomes from each site as bits $m_{j}$ and computes the parity of $\textbf{m}$: $\bigoplus_{j=1}^{n}m_{j}$.
\end{defn}

The goal of this model is then to \emph{deterministically} perform some Boolean function $f(\textbf{x})$ \emph{efficiently} on the original bit-string $\textbf{x}$ for all $\textbf{x}$. By efficient, we use the computational complexity convention that the amount of resources, in this case $n$, is polynomial in the size $|\textbf{x}|$ of the input $\textbf{x}$. We can ask what the \emph{worst-case} number of resources, or measurements sites to perform this function so that the function can be performed for all instances of $\textbf{x}$. In our definition above we add a uniform probability distribution on all inputs. This uniform probability distribution becomes relevant if we cannot perform a function deterministically, but want to maximize the probability of performing a function. The uniformity condition on all instances of $\textbf{x}$ means there is no bias on any particular bit-string $\textbf{x}$, since it may be easier to compute a function $f(\textbf{x})$ for particular instances of $\textbf{x}$. Some results in the following discussion (such as \ref{thm341}) do not require us to consider a distribution at all but we introduce it to cement a connection to Bell inequalities later on. 

We have so far not mentioned any constraint on how our resource is constructed. Perhaps our resource can only be produced using exponential \emph{quantum computations}. We place no constraint and just assume that the resource quantum state is ``presented to us'' and we make measurements on it. In fact, the optimal resource for nMBQC can be generated efficiently by a quantum computer. We will show that the optimal resource for all computations in nMBQC is the GHZ state, and this state can be generated efficiently \cite{hein}.
 
The hope is that even in this model of nMBQC we might be able to perform (at least) all efficient classical computations (i.e. in the complexity class P \cite{pap}) efficiently. In this section, we show that this is not possible and the model is quite limited. Even then, this model can simulate the statistics of Clifford circuits \cite{mbqc2,mbqc4}, which are not believed even to be universal for classical computing \cite{aaronson} even though they produce entanglement. Computations in nMBQC are also in a recently studied class of limited quantum computations called ``Instantaneous Quantum Polytime'' (IQP) \cite{shepherd}. It is possible that IQP is not capable of simulating a full quantum computer, but IQP circuits are also not believed to be simulatable efficiently with a classical computer \cite{bremner}. 

We present an analogous result to that for IQP, but for nMBQC where there are functions that can be performed with greater mean success probability with quantum than classical resources. This result is in terms of computational expressiveness, rather than computational complexity. It also results from the fact that the model of nMBQC can be expressed as an NLG \cite{hoban1}. Non-trivial Bell inequalities can be derived from these games, and a violation of these Bell inequalities implies a computational advantage with quantum resources in nMBQC.

\subsection{nMBQC, NLG and non-trivial Bell inequalities}\label{sec34a}

We now formalise nMBQC and consider the tools required for our analysis. Firstly, the ``goal'' of nMBQC is to perform functions $f(\textbf{x})$ both deterministically and efficiently in $|\textbf{x}|$ for all instances of $\textbf{x}$. This means that the mean success probability $\bar{p}(\bigoplus_{j=1}^{n}m_{j}=f(\textbf{x}))$ of performing a function is
\begin{equation}
\bar{p}(\bigoplus_{j=1}^{n}m_{j}=f(\textbf{x}))=\frac{1}{2^{|\textbf{x}|}}\sum_{\textbf{x}}p(\bigoplus_{j=1}^{n}m_{j}=f(\textbf{x})|\textbf{x})=1.
\end{equation}
If this value is less than unity, a function $f(\textbf{x})$ cannot be performed deterministically. We now relate this probability to the correlators $p(1|\textbf{s})$, i.e. the statistics of obtaining $\bigoplus_{j=1}^{n}m_{j}=1$ given inputs $\textbf{s}$. In nMBQC, the inputs $s_{j}$ are linear Boolean functions in $\textbf{x}$, i.e. XOR gates performed on elements of $\textbf{x}$. Also, without loss of generality, we consider inputs $s_{j}$ being of the form $\bigoplus_{j=1}^{|\textbf{x}|}a_{j}x_{j}\oplus b$ for $a_{j}\in\{0,1\}$ and $b=0$. If $b=1$, each site can remove this constant from their input. We can then relate the bit-string $\textbf{s}$ to $\textbf{x}$ by an $|\textbf{x}|$-by-$n$ matrix $\textbf{P}$ representing the linear transformations on $\textbf{x}$ in mod $2$. That is, every string $\textbf{s}$ can be expressed as
\begin{equation}
\textbf{s}=\left(\textbf{P}\textbf{x}\right)_{\oplus}
\end{equation}
where $(...)_{\oplus}$ represents matrix multiplication modulo $2$. The strings $\textbf{s}$ and $\textbf{x}$ are then $n$-length and $|\textbf{x}|$-length column vectors respectively. If we use the example from the introduction with $3$ parties and have the input on the third party being $s_{3}=s_{1}\oplus s_{2}$, with the choice of measurement on site $1$ and $2$ being $s_{1}$ and $s_{2}$ respectively. If we fix $s_{1}=x_{1}$ and $s_{2}=x_{2}$ then $\textbf{P}=\bigl[ \begin{smallmatrix} 1&0\\ 0&1\\1&1 \end{smallmatrix} \bigr]$. Therefore, for each computation in nMBQC we fix the matrix $\textbf{P}$ that designates the computation performed by the control computer $\mathcal{C}$. The matrix $\textbf{P}$ does not contain a row consisting of all-zeroes so there $s_{j}$ is always dependent on elements of $\textbf{x}$.

The figure of merit $\bar{p}(\bigoplus_{j=1}^{n}m_{j}=f(\textbf{x}))$ can now be expressed in terms of correlators $p(1|\textbf{s})$ to obtain the following:
\begin{equation}
\bar{p}(\bigoplus_{j=1}^{n}m_{j}=f(\textbf{x}))=\frac{1}{2^{|\textbf{x}|}}\sum_{\textbf{x}}\sum_{\textbf{s}}\delta^{\textbf{s}}_{(\textbf{P}\textbf{x})_{\oplus}}\left(\delta^{f(\textbf{x})\oplus 1}_{1}+(-1)^{f(\textbf{x})\oplus 1}p(1|\textbf{s})\right).
\end{equation}
We can immediately see that the right-hand-side is of the form of a non-trivial Bell inequality for $(n,2,2)$. The probability distribution  $\pi(\textbf{s})$ is $\sum_{\textbf{x}}\frac{1}{2^{|\textbf{x}|}}\delta^{\textbf{s}}_{(\textbf{P}\textbf{x})_{\oplus}}$ but with the sum now over $\textbf{x}$ instead of $\textbf{s}$; $\textbf{x}$ is however uniquely related to $\textbf{s}$. The function $f(\textbf{x})$ is then precisely the function in a non-trivial Bell inequality. This non-trivial Bell inequality is then
\begin{eqnarray}\label{mbqc1}
\frac{1}{2^{|\textbf{x}|}}\sum_{\textbf{x}}\sum_{\textbf{s}}\delta^{\textbf{s}}_{(\textbf{P}\textbf{x})_{\oplus}}\left((-1)^{f(\textbf{x})\oplus 1}p(1|\textbf{s})\right)&\leq&\frac{1}{2^{|\textbf{x}|}}\underset{g(\textbf{s})}{\textrm{max}} \sum_{\textbf{x}}\sum_{\textbf{s}}\delta^{\textbf{s}}_{(\textbf{P}\textbf{x})_{\oplus}}\left(\delta^{f(\textbf{x})}_{g(\textbf{s})}-\delta^{0}_{f(\textbf{x})}\right)\nonumber \\
&\leq&\frac{1}{2^{|\textbf{x}|}}\underset{g(\textbf{x})}{\textrm{max}} \sum_{\textbf{x}}\left(\delta^{f(\textbf{x})}_{g(\textbf{x})}-\delta^{0}_{f(\textbf{x})}\right)
\end{eqnarray}
where $g(\textbf{x})$ are all possible linear Boolean functions on $\textbf{x}$. Since linear Boolean functions are $g(\textbf{s})=\bigoplus_{j=1}^{n}a_{j}s_{j}\oplus b$ for $a_{j}$, $b\in\{0,1\}$ and $s_{j}=[\left(\textbf{P}\textbf{x}\right)_{\oplus}]_{j}$ is a linear Boolean function on $\textbf{x}$. Therefore $g(\textbf{s})$ becomes an arbitrary linear Boolean function on $\textbf{x}$ being $g(\textbf{x})$. 

Crucially the right-hand-side of \eqref{mbqc1} is independent of the number of sites $n$ and is always strictly less than $1-\frac{1}{2^{|\textbf{x}|}}\sum_{\textbf{x}}\delta^{0}_{f(\textbf{x})}$ for $f(\textbf{x})$ being a non-linear Boolean function. This latter fact also means that $\bar{p}(\bigoplus_{j=1}^{n}m_{j}=f(\textbf{x}))<1 $ for classical resources. If $f(\textbf{x})$ is linear then $\bar{p}(\bigoplus_{j=1}^{n}m_{j}=f(\textbf{x}))=1$ for classical resources; this is because the only deterministic correlators $p(1|\textbf{s})=\delta^{f(\textbf{x})}_{1}$ possible in LHV theories are for the linear Boolean functions $f(\textbf{x})$. Raussendorf has also shown that classical, or more generally, ``noncontextual'' resources can only perform linear Boolean functions deterministically in MBQC with a single round of measurements \cite{raussendorf}.

We now focus on using quantum resources in nMBQC. If quantum resources are more useful in nMBQC than classical resources then the inequality in \eqref{mbqc1} is violated by quantum correlators. Also if $\bar{p}(\bigoplus_{j=1}^{n}m_{j}=f(\textbf{x}))=1$, then from \eqref{quantumbound}, 
\begin{eqnarray}
\underset{\{\theta_{j}\}}{\textrm{sup}}\frac{1}{2^{|\textbf{x}|+1}}\left[\left(\sum_{\textbf{x}}\sum_{\textbf{s}}\delta^{\textbf{s}}_{(\textbf{P}\textbf{x})_{\oplus}}(-1)^{f(\textbf{x})\oplus 1}\right)+\left|\sum_{\textbf{x}}\sum_{\textbf{s}}\delta^{\textbf{s}}_{(\textbf{P}\textbf{x})_{\oplus}}(-1)^{f(\textbf{x})\oplus 1}e^{i(\sum_{j=1}^{n}s_{j}\theta_{j})}\right|\right]\nonumber\\
=1-\frac{1}{2^{|\textbf{x}|}}\sum_{\textbf{x}}\sum_{\textbf{s}}\delta^{\textbf{s}}_{(\textbf{P}\textbf{x})_{\oplus}}\delta^{0}_{f(\textbf{x})}\nonumber \\
\end{eqnarray}
We make the substitution of $(-1)^{f(\textbf{x})\oplus 1}=1-2\delta^{0}_{f(\textbf{x})}$ to obtain
\begin{equation}
\underset{\{\theta_{j}\}}{\textrm{sup}}\frac{1}{2^{|\textbf{x}|}}\left|\sum_{\textbf{x}}\sum_{\textbf{s}}\delta^{\textbf{s}}_{(\textbf{P}\textbf{x})_{\oplus}}(-1)^{f(\textbf{x})\oplus 1}e^{i(\sum_{j=1}^{n}s_{j}\theta_{j})}\right|\nonumber=1.
\end{equation}
This condition reduces to $(-1)^{f(\textbf{x})\oplus 1}=e^{i(\sum_{j=1}^{n}[(\textbf{P}\textbf{x})_{\oplus}]_{j}\theta_{j})}$ where $s_{j}=[(\textbf{P}\textbf{x})_{\oplus}]_{j}$. In the following result, we indicate that this condition can always be satisfied if $n$ is at most equal to $2^{|\textbf{x}|}-1$. Therefore, all Boolean functions can be performed deterministically with quantum resources in nMBQC. The issue of efficiency will be discussed further into this section.

\begin{thm}\label{thm341}
Every Boolean function $f(\textbf{x})$ can be performed deterministically in nMBQC for at most $n=2^{|\textbf{x}|}-1$ parties.
\end{thm}

\textit{Proof}: As mentioned, every function $f(\textbf{x})$ can be achieved deterministically if $(-1)^{f(\textbf{x})\oplus 1}$ is equal to $e^{i(\sum_{j=1}^{n}[(\textbf{P}\textbf{x})_{\oplus}]_{j}\theta_{j})}$. This will be satisfied if each expression $\sum_{j=1}^{n}[(\textbf{P}\textbf{x})_{\oplus}]_{j}\theta_{j}$ for $\textbf{x}\neq\textbf{0}$ is linearly independent from every other expression corresponding to each $\textbf{s}$. To show this we just need to establish that all vectors $\textbf{s}=(\textbf{P}\textbf{x})_{\oplus}$ are linearly independent over $\mathbb{R}$. In other words, if we construct the $2^{|\textbf{x}|}-1$-by-$n$ matrix $\textbf{S}$ where rows are the vectors $\textbf{s}^{\textrm{T}}$ for $\textbf{s}\neq\textbf{0}$, then $\textbf{S}$ must have rank $2^{|\textbf{x}|}-1$.

Every column of $\textbf{S}$ has elements $g(\textbf{x})$ where $g(\textbf{x})=\bigoplus_{j=1}^{n}a_{j}x_{j}$ is a linear Boolean function. We showed in Lemma \ref{lem311} that $2^{|\textbf{x}|}-1$ vectors which have the elements being a different linear Boolean function of this form are linearly independent over $\mathbb{R}$. Therefore if each column of $\textbf{S}$ corresponds to a different linear Boolean function $g(\textbf{x})=\bigoplus_{j=1}^{n}a_{j}x_{j}$ on $\textbf{x}$, then the rank of $\textbf{S}$ is $2^{|\textbf{x}|}-1$. $\square$
\newline

Therefore, with quantum resources a function $f(\textbf{x})$ can be computed. However, this result has only upper-bounded the resources $n$ required to compute functions and this upper bound is inefficient. We now show that this upper bound is tight for all possible functions by using the example of the $n$-partite NAND function. We then use the following result to show that adaptivity is a crucial ingredient in MBQC. 

\begin{thm}\label{thm342}
The $n$-partite NAND function can only be performed deterministically in nMBQC for $n=2^{|\textbf{x}|}-1$ parties.
\end{thm}

\textit{Proof}: For ease of calculation, we prove this theorem for the $n$-partite NAND function with a NOT on each element of $\textbf{x}$, i.e. $f(\textbf{x})=\prod_{j=1}^{|\textbf{x}|}(x_{j}\oplus 1)\oplus 1$. However, the two functions are equivalent in our model as the control computer $\mathcal{C}$ can perform a NOT operation on each element of $\textbf{x}$. We prove this theorem by assuming that this function can be performed deterministically with $2^{|\textbf{x}|}-2$ parties, and then obtain a contradiction. We first simplify the proof, instead of considering all possible $2^{|\textbf{x}|}-2$-by-$|\textbf{x}|$ $\textbf{P}$ matrices, we only need to consider one particular matrix $\textbf{Q}$. This matrix $\textbf{Q}$ is the matrix with all rows being bit-strings $\mathbb{Z}_{2}^{|\textbf{x}|}$ not equal to either $\textbf{0}$ or $\textbf{1}$. Any matrix $\textbf{P}$ not equal to $\textbf{Q}$ can be turned into $\textbf{Q}$ in the following way $(\Pi\textbf{P}\textbf{M})_{\oplus}$ where $\Pi$ is a $2^{|\textbf{x}|}-2$-by-$2^{|\textbf{x}|}-2$ permutation matrix and $\textbf{M}$ is any binary, invertible $|\textbf{x}|$-by-$|\textbf{x}|$ matrix. This is because $\textbf{P}\neq\textbf{Q}$ then $\textbf{P}$ contains a row equal to $\textbf{1}$ and does not contain a bit-string $\textbf{y}\in\mathbb{Z}_{2}^{|\textbf{x}|}$. Therefore we use $\textbf{M}$ to map $\textbf{y}$ to $\textbf{1}$ by right multiplication $(\textbf{M}\textbf{y})_{\oplus}=\textbf{1}$ and $(\textbf{P}\textbf{M})_{\oplus}$ contains all the same rows as $\textbf{Q}$ but not necessarily in the same ordering. To establish the same ordering we left-multiply $(\textbf{P}\textbf{M})_{\oplus}$ by the permutation matrix $\Pi$ so that $(\Pi\textbf{P}\textbf{M})_{\oplus}=\Pi(\textbf{P}\textbf{M})_{\oplus}$.

The permutation matrix $\Pi$ just is equivalent to permuting all $n$ parties and so will leave the probability of performing a function invariant. If the function $f(\textbf{x})=\prod_{j=1}^{|\textbf{x}|}(x_{j}\oplus 1)\oplus 1$ is performed deterministically with quantum resources then $e^{i(\sum_{j=1}^{n}[(\textbf{P}\textbf{x})_{\oplus}]_{j}\theta_{j})}=1$ for $\textbf{x}=\textbf{0}$ and
\begin{equation}\label{equiv}
e^{i(\sum_{j=1}^{n}[(\textbf{P}\textbf{x})_{\oplus}]_{j}\theta_{j})}=-1
\end{equation}
for all $\textbf{x}\neq\textbf{0}$. Since $(\textbf{P}\textbf{M}\textbf{0})_{\oplus}=(\textbf{P}\textbf{0})_{\oplus}$ then $e^{i(\sum_{j=1}^{n}[(\textbf{P}\textbf{M}\textbf{x})_{\oplus}]_{j}\theta_{j})}=1$ for $\textbf{x}=\textbf{0}$. The set of strings $\{\textbf{x}|\textbf{x}\neq\textbf{0}\}$ is equal to the set $\{(\textbf{M}\textbf{x})_{\oplus}|\textbf{x}\neq\textbf{0}\}$, then \eqref{equiv} is satisfied for both $\textbf{P}$ and $(\textbf{P}\textbf{M})_{\oplus}$. Therefore satisfying determinism for one matrix such as $\textbf{Q}$ is equivalent to satisfying determinism for all $\textbf{P}$ matrices.

We now show that deterministically performing $f(\textbf{x})=\prod_{j=1}^{|\textbf{x}|}(x_{j}\oplus 1)\oplus 1$ is impossible for the matrix $\textbf{Q}$. First we observe that if determinism is satisfied then from \eqref{equiv} we must satisfy
\begin{equation}
\sum_{j=1}^{n}[(\textbf{Q}\textbf{x})_{\oplus}]_{j}\theta_{j}=\pi(2t_{\textbf{x}}+1),
\end{equation}
for all $\textbf{x}\neq\textbf{0}$ where $t_{\textbf{x}}$ is an integer. We can construct a sum over all bit-strings $\textbf{x}$ (including $\textbf{0}$) which alternates in sign:
\begin{equation}\label{equiv2}
\sum_{j=1}^{2^{|\textbf{x}|}-2}\left[\sum_{\textbf{x}}(-1)^{W(\textbf{x})}[(\textbf{Q}\textbf{x})_{\oplus}]_{j}\theta_{j}\right]=\sum_{\textbf{x}\neq\textbf{0}}(-1)^{W(\textbf{x})}[\pi(2t_{\textbf{x}}+1)],
\end{equation}
where $W(\textbf{x})$ is the Hamming weight \cite{hamming} of $\textbf{x}$, i.e. the number of non-zero elements of $\textbf{x}$. We collect terms that have the same Hamming weight $W(\textbf{x})$ on the right-hand-side of \eqref{equiv2} and set $y=W(\textbf{x})$. Defining $t_{y}=\sum_{\textbf{x};W(\textbf{x})=y}t_{\textbf{x}}$ then the right-hand-side of \eqref{equiv2} is equal to:
\begin{equation}
\sum_{y=1}^{|\textbf{x}|}(-1)^{y}\pi\left(\frac{|\textbf{x}|!}{y!(|\textbf{x}|-y)!}+2t_{y}\right)=\pi(2t-1),
\end{equation}
where $t=\sum_{y=1}^{|\textbf{x}|}(-1)^{y}t_{y}$ is some integer.

We now show that the left-hand-side of \eqref{equiv2} is actually equal to zero, thus leading to the contradiction that $\pi(2t-1)=0$ indicating that \eqref{equiv} is not true for all $\textbf{x}\neq\textbf{0}$ and determinism is not achieved. We can express the $j$th element of the sum in \eqref{equiv} as
\begin{equation}\label{equiv3}
\sum_{\textbf{x}}(-1)^{W(\textbf{x})}[(\textbf{Q}\textbf{x})_{\oplus}]_{j}=\sum_{\textbf{x}}(-1)^{W(\textbf{x})}\left(\bigoplus_{k=1}^{|\textbf{x}|}Q_{j,k}x_{k}\right)
\end{equation}
where $Q_{j,k}$ the element of $\textbf{Q}$ corresponding to the $j$th row and $k$th column. Since every row of $\textbf{Q}$ has at least one element $Q_{j,k}=0$ for a particular value of $k$, then for two bit-strings $\textbf{x}'$ and $\textbf{x}''$ that only differ in their $k$th element, $\left(\bigoplus_{k=1}^{|\textbf{x}|}Q_{j,k}x_{k}'\right)=\left(\bigoplus_{k=1}^{|\textbf{x}|}Q_{j,k}x_{k}''\right)$. However, the two Hamming weights $W(\textbf{x}')$ and $W(\textbf{x}'')$ corresponding respectively to these bit-strings differ by $1$ resulting in
\begin{equation}
(-1)^{W(\textbf{x}')}\left(\bigoplus_{k=1}^{|\textbf{x}|}Q_{j,k}x'_{k}\right)+(-1)^{W(\textbf{x}'')}\left(\bigoplus_{k=1}^{|\textbf{x}|}Q_{j,k}x_{k}''\right)=0.
\end{equation}
Since all $2^{|\textbf{x}|}$ bit-strings $\textbf{x}$ can be paired into bit-strings that differ by one element, then \eqref{equiv3} must be equal to zero. Therefore, we have reached a contradiction and deterministic computation of $f(\textbf{x})=\prod_{j=1}^{|\textbf{x}|}(x_{j}\oplus 1)\oplus 1$ cannot be achieved with less than $2^{|\textbf{x}|}-1$ parties. $\square$
\newline

This result says that determinism in nMBQC comes potentially at the price of an exponential overhead in resources. Contrast the above result with the fact that the $n$-partite NAND function can be implemented deterministically and efficiently by a classical computer. In fact, we do not need the full computing power of P, but a smaller complexity class called NC$^{1}$ which is contained in P \cite{pap}. Since a quantum computer can implement all computations in P, we have the following corollary.

\begin{cor}\label{cor341}
It is impossible to efficiently achieve universal quantum computation
deterministically in nMQBC.
\end{cor}

It is interesting that Bell tests, and in particular, non-trivial Bell inequalities have something to say about quantum computers. We know that Bell tests have a role in quantum cryptography and communication complexity; they now have some role to play in quantum computation. This relationship between foundations and applications of quantum physics is not unidirectional. We now discuss in the following subsection, how these results for nMBQC say something about Bell tests and correlations. In particular, they convey generalisations of the GHZ paradox \cite{ghz} mentioned earlier and indicate that there exist generalisations of the PR box \cite{pr} that may not defined on all inputs $\textbf{s}$.

\subsection{Generalized GHZ Paradoxes and PR boxes}\label{sec34b}

The original GHZ paradox \cite{ghz} was constructed as a way to demonstrate the incompatability of quantum physics with a LHV theory, but without the use of a Bell inequality. In the original paradox as discussed in section \ref{sec12}, the following outcomes, translated into correlators:
\begin{eqnarray}\label{ghzpara}
p(1|000)=1\nonumber\\
p(1|011)=1\nonumber\\
p(1|101)=1
\end{eqnarray}
in an LHV theory deterministically predict that $p(1|110)=1$. These statistics belong to the statistics of an extreme point of $\mathcal{L}$. However, measurements on a GHZ state lead to a contradiction where the expressions in \eqref{ghzpara} are satisfied but $p(1|110)=0$. The LHV statistics in \eqref{ghzpara} result in a value of $2$ for the Mermin inequality \eqref{mermin}, but the quantum statistics result in a maximal algebraic violation of $3$. Therefore our result in Theorem \ref{thm341} can result in a GHZ paradox for $2^{|\textbf{x}|}-1$ parties. We can assign the following statistics in an LHV theory:
\begin{equation}\label{genghz}
p(1|\textbf{s};\textbf{s}=(\textbf{P}\textbf{x})_{\oplus})=1,
\end{equation}
for all $\textbf{x}\neq\textbf{1}$ and $\textbf{P}$ is the $2^{|\textbf{x}|}-1$-by-$|\textbf{x}|$ matrix with rows consisting of all $2^{|\textbf{x}|}-1$ bit-strings $\textbf{y}\in\mathbb{Z}_{2}^{|\textbf{x}|}$ not equal to $\textbf{0}$. If we put these statistics into the non-trivial Bell inequality in \eqref{mbqc1} corresponding to the $n$-partite NAND function, we obtain:
\begin{eqnarray}
\frac{1}{2^{|\textbf{x}|}}\sum_{\textbf{x}}\sum_{\textbf{s}}\delta^{\textbf{s}}_{(\textbf{P}\textbf{x})_{\oplus}}(-1)^{f_{2}(\textbf{x})+1}p(1|\textbf{s})&=&\frac{1}{2^{|\textbf{x}|}}\left(2^{|\textbf{x}|}-1-p(1|\textbf{s};\textbf{s}=(\textbf{P}\textbf{1})_{\oplus})\right)\nonumber \\
&\leq&\frac{2^{|\textbf{x}|}-2}{2^{|\textbf{x}|}}.
\end{eqnarray}
then for LHV theories we can only assign the probability $p(1|\textbf{s};\textbf{s}=(\textbf{P}\textbf{1})_{\oplus})=1$ deterministically. However, since with $2^{|\textbf{x}|}-1$ parties, we can perform the NAND function deterministically with quantum mechanics, we can satisfy both the probabilities in \eqref{genghz} and $p(1|\textbf{s};\textbf{s}=(\textbf{P}\textbf{1})_{\oplus})=0$, leading to a contradiction.

We did not need to make this argument utilising a Bell inequality as we could have just used the statistics of the LHV correlator producing the linear Boolean function $f(\textbf{x})=1$ deterministically. This deterministic correlator is the only correlator that satisfies all assignments in \eqref{genghz}. In this sense then, we have a GHZ paradox for all choices of $|\textbf{x}|$. 

Finally, when we introduce the pre-processing on inputs $\textbf{s}=(\textbf{P}\textbf{x})_{\oplus}$ and construct a non-trivial Bell inequality of the form in \eqref{mbqc1} then we do not consider all possible correlators $p(1|\textbf{s})$ but only those correlators where $\textbf{s}$ is defined by $\textbf{x}$ and $\textbf{P}$. As a result we only consider probabilities $p(\textbf{m}|\textbf{s})$ that also satisfy this relationship between $\textbf{s}$ and $\textbf{x}$. We can consider non-signalling probability distributions $p(\textbf{m}|\textbf{s})$ that are of the following form
\begin{equation}\label{genpr}
p(\textbf{m}|\textbf{s}; \textbf{s}=(\textbf{P}\textbf{x})_{\oplus}) = 
\begin{cases} \frac{1}{2^{n-1}} & \text{if $\bigoplus_{j=1}^{n}m_{j}=f(\textbf{x})$,}
\\
0 &\text{otherwise,}
\end{cases}
\end{equation}
for any non-linear Boolean function $f(\textbf{x})$. We are not concerned with inputs $\textbf{s}$ that do not satisfy $\textbf{s}=(\textbf{P}\textbf{x})_{\oplus}$, therefore, these distributions are not necessarily extreme points of $\mathcal{NS}$. The distributions may even be in the interior of $\mathcal{NS}$ but can be perceived as a generalisation of the PR box \cite{pr}, due to the fact that they maximally violate a Bell inequality for all correlators. 

For example, the Mermin inequality is maximally violated by correlators resulting from a GHZ state, but we can also achieve the same maximal violation with vertices of $\mathcal{NS}$. The correlations $p(\textbf{m}|\textbf{s})$ that result from the GHZ state do not form a vertex of $\mathcal{NS}$. For $\textbf{s}\notin\{\{000\},\{011\},\{101\},\{110\}\}$, the correlations $p(\textbf{m}|\textbf{s})$ resulting from the GHZ state do not resemble those of extreme points in $\mathcal{NS}$.

What Theorem \ref{thm342} implies, is that even though \eqref{genpr} is defined on a subset of inputs $\textbf{s}$, there exist non-signalling probability distributions for $n\leq 2^{|\textbf{x}|}-2$ that cannot be achieved by quantum mechanics. More specifically, if $f(\textbf{x})$ in \eqref{genpr} is the $n$-partite NAND function, since quantum physics cannot achieve this distribution for these values of $n$, they are as ``unphysical'' as the PR box. 

Theorem \ref{thm342} also implies that there are generalised PR boxes that can efficiently perform the $n$-partite NAND function in our nMBQC model. The fact that these unphysical resources can efficiently perform tasks unthinkable with physical resources has been analogously investigated in the field of communication complexity. An argument put forward first by Van Dam \cite{vandam} and then developed by Brassard et al \cite{cc}, is that if these unphysical, bipartite PR boxes exist then tasks in communication complexity are rendered ``trivial''. By trivial, we mean that only one bit of communication is required between two parties to achieve all Boolean functions. These ideas were also extended to the multipartite scenario \cite{reznik}. It could be argued that the result of Theorem \ref{thm342} complements the idea that quantum mechanics cannot simulate all non-signalling probability distributions because information processing would be rendered ``too easy''.

In this section we have discussed the interplay between the computational perspectives on Bell tests and computation itself. In particular, we looked at a restricted class of computations in MBQC, itself a promising avenue for quantum computing. We have used Bell tests to show that adaptivity is crucial in Briegel and Raussendorf's MBQC scheme \cite{mbqc1}. With adaptivity comes the possibility for parties to communicate to each other and the connection between computation and Bell tests can break down. In the next chapter, we hint at a method to re-establish this connection. 

\section{Chapter Summary}\label{sec35}

\noindent
When Bell first formulated his inequality he wanted to say something concrete about the interpretation of the wavefunction \cite{bell}. He established that if quantum mechanics is to be re-imagined as a local hidden variable theory, then a great deal of the theory's predictions would have to be ``thrown out''. Classical physics can be conceived as a local hidden theory, so there is an incompatibility between classical physics and quantum physics. This incompatibility is ``witnessed'' by a Bell inequality: a violation indicates incompatibility. It immediately tells us that quantum systems can do something that classical systems cannot.

It could be argued that it was inevitable that this tool for disambiguation between classical and non-classical would be used to show that quantum correlations can perform some tasks that classical correlations cannot. With the development of quantum information theory, Bell tests were approached with a new motivation: to find a quantum advantage for some quantum information processing tasks. For example, the application of Bell tests to cryptography \cite{acin3} and random number generation \cite{pironio} has been successful.

Quantum computation could produce an advantage over classical computers \cite{shor}. The proof that quantum computers are more powerful than classical computers would have an immense impact on the study of classical computational complexity as it would provide a separation in a conjectured hierarchy of computational models \cite{pap}. Since the Bell test produces a clear cut distinction between quantum and classical, it could be considered a useful tool for proving this separation in computational models. The difficulty lies in communication, a resource not allowed in Bell tests, but not prohibited in most models of computation.

Immediately one can suggest that we study models of computation that do not require or even limit communication. Communication complexity is a model of computation that limits communication \cite{vandam}, and non-local games do not allow communication between players but to the referee \cite{nonlocal}. Connections have been made to the latter with multi-prover interative proof systems, a model of computing based on the exchange of messages between parties in order to ascertain whether a potential solution to a problem is correct \cite{nonlocal}. Interactive proof systems have been shown to be extremely powerful, potentially far more powerful than computations in NP depending on the model \cite{jain}. If we want to say something about classical and quantum computers, then in these ``simpler'' models we will still want to place restrictions on communication. This motivates our study of MBQC circuits where the only communication allowed is between a classical computer and measurement sties, sites cannot communicate with each other and there is a single-round of measurements. 

In this chapter, we began by discussing the space of LHV correlators in terms of the facet Bell inequalities. Finding facet Bell inequalities is hard and in practice we could only find them for a limited number of $(n,c,d)$ scenarios. This motivated us to find a set of non-trivial Bell inequalities. These non-trivial inequalities were motivated by our computational insight into the space of LHV correlators, and were shown to be relevant for the study of non-local games (NLG). Finally, our restricted class of MBQC computations was shown to be cast as an NLG, and again made relevant to non-trivial Bell inequalities. Using the tools from the study of Bell inequalities, we showed that this restricted class of MBQC computations is not universal for quantum computing. However, in this model, due to the very nature of the Bell inequality, we showed that quantum resources can do something that classical resources cannot. 

We have shown that there are concrete connections between Bell tests and some models of computing. On the other hand, we have also shown that communication in the form of adaptivity is vital for MBQC. In the next chapter, we will indicate how to simulate communication in computations within the framework of a Bell test. Perhaps surprisingly, this communication simulation still allows the possibility for disambiguating quantum and classical resources.

\chapter{Data Post-selection in Bell Tests}\label{chap4}

\noindent
The Bell test has been around formally for decades. A natural question is `can we go beyond this formulation?' Of course, situations altering the number of parties, inputs and outputs have been studied. Despite these generalizations, the core of the gedankenexperiment still involves space-like separated parties making their measurements and then sending their data to be turned into statistics. However, in reality, data does not always emerge perfectly from experiments, and often it needs to be discarded. CHSH took this imperfection into account and added an extra assumption to the construction of Bell tests beyond Bell's formulation: the ``fair-sampling assumption'' \cite{chsh,clauser,berry}. This assumption essentially states that the experimental errors in performing a Bell test are independent of the choice of measurement at each site. In the history of experimental tests of Bell inequalities, this assumption has featured strongly, especially in optical tests \cite{freedman,shih,ou,rarity,tittel,weihs}.

Whilst the fair-sampling assumption may be rooted in common sense, we cannot assume, in general, that it is true. However, if we relax it then the discarding of data can be problematic. In particular, it can lead to the ``detection loophole'' \cite{pearle,garg} as it is now often referred. A ``loophole'' emerges when some imperfection in the experiment can allow LHV correlations to simulate quantum correlations. There a several sources of loopholes in experimental Bell tests, some more subtle that others. 

Two central constraints on the construction of Bell tests are measurement choice independence and space-like separation. If the latter is not respected in an experiment, then parties can communicate and from this communication, simulate whichever correlations they wish. Bell has emphasized himself how important that choice of measurement be completely random and independent of the parties' systems \cite{bell}. Barrett and Gisin have directly related the lack of measurement choice independence to simulating communication between parties. These central stipulations of the Bell test must be upheld if we want to restrict what is possible with LHV correlations.

Modern, photonic-based Bell tests allow for space-like separated measurements \cite{tittel,weihs}. The issue of freedom of measurement choice can tend towards philosophy, and the concept of ``free will''. These discussions are well beyond the scope of this thesis. It could be argued though that photon Bell tests can also address the need for random choice of measurements \cite{weihs}. As discussed in section \ref{sec14} of chapter \ref{chap1}, random numbers can be generated by quantum processes, potentially in a device independent manner \cite{colbeck,pironio}. Experimental groups have exploited this source of randomness to produce random measurements \cite{weihs}.

The issues raised by more systematic failures to implement Bell tests are problematic. The detection loophole is a more subtle source of problems. It can be seen to result from a form of ``post-selection''. Here we use the term post-selection as a means of accepting measurement data if it satisfies particular criteria\footnote{Post-selection in quantum information can often mean the acceptance of a quantum state after measurement, if a particular measurement outcome is achieved. Otherwise the quantum state is discarded.}. In the case of imperfect detection where our measurement devices (detectors) may or may not receive a measurement outcome (detection event), we can only calculate correlations for all parties if all parties have made a successful measurement. Therefore, we accept or post-select on measurement data if all sites successfully detected a measurement outcome. In the first section of this chapter, we will formalise these ideas in the $(n,2,2)$ scenario\footnote{These ideas can be extended to different scenarios, but for pedagogical clarity and the ease of producing new results we make this restriction.}.

This chapter concerns itself more generally with data post-selection in Bell tests. In particular, we introduce two forms of post-selection and associate a loophole with each form of post-selection. In section \ref{sec41} of this chapter, we discuss the form of post-selection in the presence of imperfect detection, whereas in section \ref{sec42} we consider post-selection in \emph{perfect} Bell tests. By the latter, we mean that we have perfect detection, space-like separation and freedom of measurement choice (the original gedankenexperiment) but introduce a form of post-selection on accepting measurement data. Whilst the post-selection in section \ref{sec41} is experimentally motivated, the post-selection in section \ref{sec42} is very much conceptually motivated. Despite their differing motivation there is an overlap in the language we use to describe the loopholes. This language is rooted in our \emph{computational} insight into LHV correlators. 

Interestingly, whilst post-selection on successful detection can lead to the detection loophole (as we shall show), the post-selection in section \ref{sec42} can be described as ``loophole-free''. As well as the latter constraining LHV correlators in the presence of post-selection, it can also \emph{enlarge} the space of quantum correlators. We also indicate that connections can be made between MBQC and our new form of data post-selection. Finally in section \ref{sec43}, we give some indications that generalising the results of section \ref{sec42} to different $(n,c,d)$ scenarios may become problematic, and no longer loophole-free. 

The original work in this chapter was developed in collaboration with Dan Browne. Section \ref{sec41} (except subsection \ref{sec41c}) is a rederivation of the work of Garg and Mermin in \cite{garg}, but now in our computational description of Bell tests. Subsection \ref{sec41c} consists of a new result generalising the work of Garg and Mermin to $n$ parties. In section \ref{sec43} all of the work was completed also in collaboration with Joel Wallman. Results in section \ref{sec42} have been published as \cite{hoban2} and some of the results in section \ref{sec43} have been published in \cite{hoban3}.

\subsection{Notation}\label{sec40a}

In this chapter, we will carry over the notation convention for modular arithmetic introduced in the last chapter. The first two sections of this chapter solely consider the $(n,2,2)$ scenario and so we use $\oplus$ and $\bigoplus$ to denote addition and summation modulo 2 for only the $(n,2,2)$ case. In section \ref{sec43} we consider the $(n,c,d)$ cases, and we enclose modulo $x$ arithmetic in brackets, i.e. $\left[...\right]_{x}$. For further clarification see section \ref{sec30a} of chapter \ref{chap3}.

\section{Post-selection and the Detection Loophole}\label{sec41}

\noindent
We know that the space of quantum correlators is larger than $\mathcal{L}$ by Bell's theorem. This is a mathematical statement and testing it in the laboratory has been a major endeavour and challenge in the past few decades \cite{chsh,freedman,shih,ou,rarity,tittel,weihs,rowe}. However, of these experiments, the majority have suffered from the detection loophole. Experiments such as \cite{rowe} that manage to overcome the detection loophole suffer from not having space-like separated measurements \cite{rowe}. There are currently no loophole-free Bell tests but there are promising routes for overcoming the detection loophole \cite{monroe, vertesi, sangouard}.

The issue of imperfect detection, culminating in the detection loophole is a subtle issue \cite{pearle, garg}. In a full treatment of a Bell test, a non-detection of an event is in itself an event. That is, if a measurement is the result of a detection and there are $d$ possible outcomes, a non-detection must be another outcome. We cannot rule out the possibility that an LHV theory can produce all $(d+1)$ outcomes. The fair-sampling assumption aims to exclude this possibility by saying that the non-detection event is independent of our choice of measurement \cite{chsh,clauser,berry}. This assumption cannot itself be tested. For example, we construct an explicit LHV model that violates the fair-sampling assumption but the statistics of detection are random at each site. We cannot extract the dependence on $\textbf{s}$ from the statistics alone. We do not therefore impose the fair-sampling assumption in our discussion.

Having imperfect detectors does not necessarily mean that LHV correlators can completely simulate quantum correlators. Recall that this simulation is how we describe a loophole, but we shall make this notion more rigorous in subsequent discussion. Work by Pearle \cite{pearle} which was then developed by Garg and Mermin \cite{garg} showed that if the detector efficiency (the ratio of successful detection to all incoming events) at each site is above some threshold, then a loophole can be ruled out. This detection efficiency threshold has been subsequently lowered by further research \cite{eberhard,vertesi}.

A final, somewhat more applied, motivation for considering the detection loophole comes from quantum key distribution \cite{ekert}. We discussed device-independent quantum key distribution \cite{acin3,pironio2} in section \ref{sec14} of the first chapter. Recall that the security of device-independent quantum key distribution can be ensured by the violation of a Bell inequality. The intuition is as follows: an adversary trying to learn the generated secret key (thus able to decode any secret message) can learn it if the key is described by an LHV. The secret information is contained in some ``local'' information at each site which can be ``extracted'' by said adversary. If the secret information is generated by some correlations incompatible with an LHV theory, then an adversary cannot localise it and obtain it. The detection loophole allows an adversary to learn a secret key that can be generated by LHV resources via the loophole \cite{acin3}.

We structure this section so that we introduce and describe the detection loophole. Our novel insight into this loophole is to use the language of computational expressiveness to describe what LHV correlators can do in the presence of imperfect detection. We show that the post-selection of accepting measurement data based on successful detection induces a relationship between each party's shared hidden variables and inputs $s_{j}$. We use this discussion to derive the GM threshold detector efficiency, but also to generalise their result to $n$ parties. We show that this threshold can be lowered by going from $2$ to $n$ parties. A previous reduction in the threshold detector efficiency for $(2,2,2)$ have resulted from considering the full probability distribution and not correlators \cite{eberhard}.

In this section and the next, we will restrict ourselves to the study of Bell tests in the $(n,2,2)$ scenario. Therefore we will use Corollary \ref{cor311} of Theorem \ref{thm221} where $\mathcal{L}$ is the convex hull of the linear Boolean functions. If a correlator cannot be written as a convex combination of linear Boolean functions for all possible decompositions it must lie outside of $\mathcal{L}$. 

\subsection{The Detection loophole}\label{sec41a}

The action of discarding data means that the person carrying out a Bell test is playing an active role\footnote{With perfect detection, the experimenter only calculated the sum modulo $d$ of outcomes. This can be seen as an active role, however, we take active to mean that they can do something non-trivial with the data.}. Because of this active role, throughout this chapter, we will refer to an ``experimenter'' who does something non-trivial with the experimental data. We will describe the role of the experimenter in different contexts in more detail throughout this chapter. That is, what the experimenter can and cannot do will be prescribed.

How do we incorporate the issue of a non-detection event into an $(n,2,2)$ Bell test? Since the number of outputs of a successful measurement is binary, then the total number of outcomes is ternary, i.e. $m_{j}\in\mathbb{Z}_{3}$. What is an appropriate joint outcome, the sum modulo $2$ of all outcomes, or the sum modulo $3$? If we take the sum modulo $2$ then a non-detection will necessarily get mapped to an event with a successful detection. Can we still talk in terms of Boolean functions if the number of outcomes at each site is ternary? Is a loophole is caused by de facto moving out of the scope of Boolean functions?

We can resolve this discussion by redescribing the scenario only in terms of bit-strings. Now instead of each $j$th site outputting a single digit $m_{j}$, they output two bits $\{t_{j},m_{j}\}\in\mathbb{Z}_{2}\times\mathbb{Z}_{2}$. Here $t_{j}$ is a bit that indicates whether an event is successfully detected (represented by $1$) or not detected (represented by $0$). If $t_{j}=1$ for all $j$, then the experimenter takes the sum modulo $2$ of all outcomes $m_{j}$, if $t_{j}=0$ for at least one site $j$, we throw away all data. The elements $t_{j}$ make up an $n$-length bit-string $\textbf{t}$ and we accept $\textbf{m}$ if $\textbf{t}=\textbf{1}$, the string of all-ones. This discarding of data is a form of post-selection; we call this method of post-selection when $\textbf{t}=\textbf{1}$ ``detection post-selection''. 

\begin{defn}\label{defn411}
When the experimenter accepts, or post-selects on data $\textbf{m}$ and $\textbf{s}$ when $\textbf{t}=\textbf{1}$, this is \textbf{detection post-selection}. This data after post-selection is then used to calculate $\bigoplus_{j=1}^{n}m_{j}$.
\end{defn}

This action of post-selection as we shall show can be a way of introducing loopholes. Before we define a loophole we need to introduce the mathematical construction we need to define them.

The convex polytope $\mathcal{P}$ is the space of correlators $p(1|\textbf{s})$ that are perfectly detected, i.e. $\textbf{t}=\textbf{1}$ for all runs of an experiment. For imperfect detection, we need a new, more general space of correlators that are calculated \emph{after} post-selecting on $\textbf{m}$ and $\textbf{s}$ when $\textbf{t}=\textbf{1}$. We call this more general space $\tilde{\mathcal{P}}$ and if every run of an experiment produces $\textbf{t}=\textbf{1}$, then $\tilde{\mathcal{P}}=\mathcal{P}$. However, more generally, correlators are now defined in the following way
\begin{equation}\label{postcorrelators}
\tilde{p}(1|\textbf{s})=p(\bigoplus_{j=1}^{n}m_{j}=1|\textbf{s},\textbf{t}=\textbf{1}).
\end{equation}
$\tilde{\mathcal{P}}$ is now the space of correlators of the form \eqref{postcorrelators}. However, the space $\tilde{\mathcal{P}}$ for the $(n,2,2)$ setting can be defined in an analogous way to $\mathcal{P}$. That is, $\tilde{\mathcal{P}}$ is the convex hull of all correlators $\tilde{p}(1|\textbf{s})=\delta^{1}_{f(\textbf{s})}$ for any Boolean function $f:\mathbb{Z}^{n}_{2}\rightarrow\mathbb{Z}_{2}$. We have put no restriction on the probability of detection $p(\textbf{t}=\textbf{1})$, only that $\textbf{t}=\textbf{1}$. 

In the case for perfect detection, the space of LHV correlators is $\mathcal{L}$ as defined by Corollary \ref{cor311}. We define $\tilde{\mathcal{L}}$ as the space of correlators $\tilde{p}(1|\textbf{s})$ resulting from LHV correlators, computed after detection post-selection. Is the space $\tilde{\mathcal{L}}$ always the convex hull of linear Boolean functions on $\textbf{s}$? For perfect detectors where $\textbf{t}=\textbf{1}$ is always satisfied, then $\tilde{\mathcal{L}}=\mathcal{L}$. Another way of asking this is to write the CHSH inequalities in terms correlators $\tilde{p}(1|\textbf{s})$,
\begin{equation}
\tilde{p}(1|00)+\tilde{p}(1|00)+\tilde{p}(1|00)-\tilde{p}(1|00)\leq 2.
\end{equation}
If this inequality can be violated by correlators in $\tilde{\mathcal{L}}$ then the space $\tilde{\mathcal{L}}$ is no longer the convex hull of linear Boolean functions. We then associate this violation by LHV correlators (in the presence of imperfect detection) with a loophole in a Bell test. We now define this loophole.

\begin{defn}\label{defn412}
A \textbf{loophole} is introduced by an experimenter into a Bell test if after detection post-selection, the space $\tilde{\mathcal{L}}$ is larger than the convex hull of linear Boolean functions.
\end{defn}

The intuition behind this being a loophole is that if we have a quantum correlator $\vec{q}$ (obtained with perfect detection) being outside of $\mathcal{L}$, then it will violate a facet Bell inequality. However, if the detectors which obtained this quantum correlator become imperfect, then after detection post-selection, the resulting quantum correlator in $\tilde{\mathcal{P}}$ will again\footnote{We assume that the detection device is independent of the quantum state or choice of measurement made.} be equal to $\vec{q}$. The loophole means that $\vec{q}$ could now be in the space $\tilde{\mathcal{L}}$. It is possible that $\vec{q}$ could be outside of $\tilde{\mathcal{L}}$, but the facet Bell inequalities for $\mathcal{L}$ are possibly no longer relevant for informing us either way. We now show that loopholes are achievable with detection post-selection. In the following result we show that it is possible that $\tilde{\mathcal{L}}$ can no longer be confined to the convex hull of linear Boolean functions.

\begin{prop}\label{prop411}
For all LHV theories, $\tilde{\mathcal{L}}$ is larger than the convex hull of linear Boolean functions on $\textbf{s}$.
\end{prop}

\textit{Proof}: We construct the following specific model with $n$ sites. The $(n-1)$ sites for $j\in\{2,3,...,n\}$ have perfect detectors whereas the first site has an imperfect detector. The first detector outputs the detection bit as a function of an LHV $\lambda$ and its input, $t_{1}=s_{1}\oplus b(\lambda)\oplus 1$ so that $b(\lambda)\in\{0,1\}$, whereas $t_{2}=1$. When we post-select so that $t_{1}=1$ then $s_{1}=b(\lambda)$. The variable $b(\lambda)$ is shared by all parties, and the second party's measurement outcome $m_{2}=b(\lambda)s_{2}$. If for all $(n-1)$ sites where $j\neq 2$, the parties' measurement outcomes upon successful detection are $m_{j}=0$, then when $\textbf{t}=\textbf{1}$, $\bigoplus_{j=1}^{n}m_{j}=s_{1}s_{2}$. The resulting correlator is then $\tilde{p}(1|\textbf{s})=\delta^{s_{1}s_{2}}_{1}=s_{1}s_{2}$, which is a vertex outside of the convex hull of linear Boolean functions. $\square$
\newline

This demonstrates how post-selection can be problematic in Bell tests. A drawback of the proof of the above result is the asymmetry in the detectors between the first detector and the rest. If we were to switch the detectors in the experiment and still got the same imperfect detection at site $1$ then the rate of detection must be independent of the detector. The measure of detection is the detection efficiency $\eta$ which is the quotient of number of successful detections to the number of events incoming to the detector. We can obtain the efficiency of a detector if two sites each make measurements, and then condition the statistics of the detector upon the other detecting an event so that 
\begin{equation}\label{eta}
\eta=\frac{p(t_{1}=1|t_{2}=1)}{p(t_{1}=0|t_{2}=1)+p(t_{1}=1|t_{2}=1)}=\frac{p(\textbf{t}=\{1,1\})}{p(\textbf{t}=\{0,1\})+p(\textbf{t}=\{1,1\})}.
\end{equation}
We assume that the detector efficiency $\eta$ is the same for all $n$ sites. Situations with non-uniform $\eta$ amongst parties have been investigated (e.g. \cite{vertesi}) but is beyond the scope of our discussion here.

In the early literature discussing the detection loophole (e.g. \cite{pearle, garg}), the probabilities in \eqref{eta} are calculated from the number $N_{\textbf{t}}$ of events where $\textbf{t}$ occured. It is assumed that the number of events where $\textbf{t}=\textbf{0}$ is unobservable as they are non-events. Probabilities then become normalised relative to this inability to detect when $\textbf{t}=\textbf{0}$ and
\begin{equation}
p(\textbf{t})=\frac{N_{\textbf{t}}}{\sum_{\textbf{t}\neq\textbf{0}}N_{\textbf{t}}}.
\end{equation}
Then for the above discussion about $\eta$ for $2$ parties, $N_{\textbf{t}}$ being the number of events where $\textbf{t}\in\{0,1\}^{2}$ occurs, the total number of events is $N=N_{\textbf{1}}+N_{\{0,1\}}+N_{\{1,0\}}$. The probabilities in \eqref{eta} then are obtained in the limit where $N\rightarrow \infty$ giving the efficiency 
\begin{equation}
\eta=\frac{N_{\textbf{1}}}{N_{\textbf{1}}+N_{\{0,1\}}}.
\end{equation}
If we want $\eta$ to be the same for all sites then $N_{\{0,1\}}=N_{\{1,0\}}$. We also now impose that the statistics $p(t_{j}=1)$ should be independent of $s_{j}$. This is not as strong as the fair-sampling assumption and we can experimentally test whether single-site detection statistics are independent of $s_{j}$ \cite{garg}. This reinforces the intuition that the properties of a detector such as $\eta$ should be independent of whatever measurement we make.

In line with previous research such as in \cite{garg}, we now weigh the correlation statistics with the statistics of detection. Therefore correlators now take the form
\begin{equation}
\bar{p}(1|\textbf{s})=p(\textbf{t}=\textbf{1})\tilde{p}(1|\textbf{s}).
\end{equation}
These correlators $\bar{p}(1|\textbf{s})$ are not necessarily normalised so $\bar{p}(0|\textbf{s})\neq 1-\bar{p}(1|\textbf{s})$ in general. On the other hand, $p(\textbf{t}=\textbf{1})\tilde{p}(0|\textbf{s})=p(\textbf{t}=\textbf{1})(1-\tilde{p}(1|\textbf{s}))$. Expectation values of outcomes in the space $\tilde{\mathcal{P}}$ can be defined in exact analogy with the expectation values $\mathbb{E}(\textbf{s})$ over correlators in $\mathcal{P}$, giving
\begin{equation}
\tilde{\mathbb{E}}(\textbf{s})=1-2\tilde{p}(1|\textbf{s}).
\end{equation}
Expectation values $\bar{\mathbb{E}}(\textbf{s})$ for the correlators $\bar{p}(0|\textbf{s})$ and $\bar{p}(1|\textbf{s})$ can then be related to $\tilde{\mathbb{E}}(\textbf{s})$ to obtain
\begin{equation}
\bar{\mathbb{E}}(\textbf{s})=\bar{p}(0|\textbf{s})-\bar{p}(1|\textbf{s})=p(\textbf{t}=\textbf{1})(1-2\tilde{p}(1|\textbf{s}))=p(\textbf{t}=\textbf{1})\tilde{\mathbb{E}}(\textbf{s}).
\end{equation}
This relationship between expectation values will be utilised in the following section. In fact, because the correlators $\bar{p}(1|\textbf{s})$ are not normalised, it will be more useful to work in terms of the expectation values $\bar{\mathbb{E}}(\textbf{s})$. This means we only need to consider one number instead of both $\bar{p}(0|\textbf{s})$ and $\bar{p}(1|\textbf{s})$. 

In the following two subsections we will work in the new space $\bar{\mathcal{L}}_{\mathbb{E}}$ of the expectation values $\bar{\mathbb{E}}(\textbf{s})$ for LHV theories. In line with previous discussion, this space is a $2^{n}$ dimensional real space of vectors having the elements $\bar{\mathbb{E}}(\textbf{s})$. These elements can now be negative but their magnitudes are bounded by unity. The space $\bar{\mathcal{L}}_{\mathbb{E}}$ is a sub-space of $\bar{\mathcal{P}}_{\mathbb{E}}$ which is now the space of all possible vectors of expectation values $\bar{\mathbb{E}}(\textbf{s})$.

\subsection{Rederivation of the GM detection efficiency}\label{sec41b}

\noindent
We now address the $(2,2,2)$ scenario and use it to give an upper bound on the detection efficiency $\eta$ required in order to demonstrate a violation of the CHSH inequality in the presence of imperfect detectors. This upper bound was derived by GM \cite{garg} and has since been improved upon by Eberhard \cite{eberhard} in the Clauser-Horne inequality setting \cite{ch}. As an aside, it has been suggested that if we consider different Bell test settings, we can lower the detection efficiency required to violate any Bell inequality \cite{vertesi}. 

We will use our computational interpretation of correlators to rederive the GM upper bound on the threshold detection efficiency $\eta$. In order to do this, we first describe the space $\bar{\mathcal{L}}_{\mathbb{E}}$ of expectation values $\bar{\mathbb{E}}(\textbf{s})$. The following result now captures this space in terms of a vertex description.

\begin{prop}\label{prop412}
The space $\bar{\mathcal{L}}_{\mathbb{E}}$ is the convex hull of all expectation values $\bar{\mathbb{E}}(\textbf{s})=(-1)^{g(\textbf{s})}$ for $g(\textbf{s})$ being a linear Boolean function on $\textbf{s}$.
\end{prop}

\textit{Proof}: First, just like measurement outcomes $m_{j}$ resulting from LHV theories, the detection values $t_{j}$ can be, in general, written as $t_{j}=x_{j}(\lambda)s_{j}\oplus y_{j}(\lambda)$ for bits $x_{j}(\lambda)\in\{0,1\}$ depending on the local hidden variable $\lambda$. Therefore, if $y_{j}(\lambda)=1$ and $x_{j}(\lambda)=0$, then $p(\textbf{t}=\textbf{1})=1$ for all $j,\lambda$, otherwise if $y_{j}(\lambda)=x_{j}(\lambda)=0$, then $p(\textbf{t}=\textbf{1})=0$ again for all $j,\lambda$. So then for all LHV maps where $x_{j}(\lambda)=0$ for all $j,\lambda$, the probability of detection for a single-site is $p(t_{j}=1)=p(y_{j}(\lambda)=1)$.

Since $s_{j}$ is randomly generated, for $x_{j}(\lambda)=1$ for all $j,\lambda$, then $p(\textbf{t}=\textbf{1})\neq 1$ for all $y_{j}(\lambda)$. Finally, the map $t_{j}=s_{j}$ or $t_{j}=s_{j}\oplus 1$ is forbidden as this means there is a direct dependence in the statistics of detection with the choice of input. We then instead have maps $t_{j}=s_{j}\oplus y_{j}(\lambda)$ where $y_{j}(\lambda)$ is shared by both parties and generated randomly so that $p(t_{j}|s_{j})=p(t_{j}|s'_{j})$ for $s_{j}\neq s'_{j}$. As a result of $t_{j}$ being random, $p(\textbf{t}=\textbf{1})$ is at most equal to $\frac{1}{2}$. 

If one party employs the strategy of $t_{j}=s_{j}\oplus y_{j}(\lambda)$ and the other site produces the deterministic map $t_{j'}=1$ then $p(\textbf{t}=\textbf{1})=\frac{1}{2}$. However, the detection efficiency $\eta$ is not the same for both sides. We can maintain the same probability $p(\textbf{t}=\textbf{1})=\frac{1}{2}$ while making the detection efficiency the same for both sides if both parties share a random bit $z(\lambda)\in\{0,1\}$. When $z(\lambda)=0$, $t_{1}=s_{1}\oplus y_{1}(\lambda)$ and $t_{2}=1$, and when $z(\lambda)=1$, $t_{1}=1$ and $t_{2}=s_{1}\oplus y_{1}(\lambda)$. As $z(\lambda)$ is randomly generated then $p(\textbf{t}=\textbf{1})=\frac{1}{2}(\frac{1}{2}+\frac{1}{2})=\frac{1}{2}$. If $z(\lambda)$ were not random then we bias one of the strategies and $N_{\{0,1\}}\neq N_{\{1,0\}}$ for $N\rightarrow \infty$, which is forbidden. 

In this strategy where parties share $z(\lambda)$, one of the parties learns the other party's input $s_{j}$ as it is equal to a variable $y_{j}(\lambda)\oplus 1$ when $t_{j}=1$. If one party learns the other party's variable then they can compute the non-linear Boolean functions $f(\textbf{s})=(s_{1}\oplus a)(s_{2}\oplus b)\oplus c$ for $a$, $b$, $c\in\{0,1\}$ deterministically. Therefore, the parties can achieve the post-selected expectation value:
\begin{equation}
\tilde{\mathbb{E}}(\textbf{s})=(-1)^{f(\textbf{s})},
\end{equation}
with $f(\textbf{s})$ being the above non-linear Boolean function. This gives a value of 
\begin{equation}\label{postexp}
\bar{\mathbb{E}}(\textbf{s})=p(\textbf{t}=\textbf{1})\tilde{\mathbb{E}}(\textbf{s})=\frac{1}{2}(-1)^{f(\textbf{s})}.
\end{equation}
We take the convex combination of LHV strategies producing all allowed deterministic maps $t_{j}=x_{j}(\lambda)s_{j}\oplus y_{j}(\lambda)$ and then the possible deterministic values of $\tilde{\mathbb{E}}(\textbf{s})$ for each strategy. This then produces the expectation values:
\begin{equation}
\bar{\mathbb{E}}(\textbf{s})=\sum_{g(\textbf{s})}p_{g(\textbf{s})}(-1)^{g(\textbf{s})}+\sum_{f(\textbf{s})}\frac{p_{f(\textbf{s})}}{2}(-1)^{f(\textbf{s})}.
\end{equation}
with $f(\textbf{s})$ and $g(\textbf{s})$ being all of the non-linear and linear Boolean functions respectively. We have taken the convex combination with positive coefficients $p_{g(\textbf{s})}$, $p_{f(\textbf{s})}$ such that $\sum_{g(\textbf{s})}p_{g(\textbf{s})}+\sum_{f(\textbf{s})}p_{f(\textbf{s})}=1$. Thus $\bar{\mathcal{L}}_{\mathbb{E}}$ is at least as large as the convex hull of $\bar{\mathbb{E}}(\textbf{s})=(-1)^{g(\textbf{s})}$. If the expectation values in \eqref{postexp} are outside of this space then they will violate one of the CHSH inequalities
\begin{equation}
\left|\sum_{\textbf{s}}(-1)^{f(\textbf{s})}\bar{\mathbb{E}}(\textbf{s})\right|\leq 2,
\end{equation}
where $f(\textbf{s})$ can one of the non-linear Boolean functions $f(\textbf{s})=(s_{1}\oplus a)(s_{2}\oplus b)\oplus c$ for $a$, $b$, $c\in\{0,1\}$. If we use the strategy of allowing the maps $t_{j}=s_{j}\oplus y_{j}(\lambda)$, then $p(\textbf{t}=\textbf{1})=\frac{1}{2}$ even though $\left|\sum_{\textbf{s}}(-1)^{f(\textbf{s})}\tilde{\mathbb{E}}(\textbf{s})\right|\leq4$. This lack of violation for the CHSH inequalities therefore concludes the proof. $\square$
\newline

This result will give an upper bound on the efficiency $\eta$ required of detectors in order to establish that certain values of $\bar{\mathbb{E}}(\textbf{s})$ are not in $\bar{\mathcal{L}}_{\mathbb{E}}$. The result indicates the structure of $\bar{\mathcal{L}}_{\mathbb{E}}$ is the same as $\mathcal{L}$, and the CHSH inequalities are exactly the same, i.e.
\begin{eqnarray}\label{postchsh}
\bar{\mathbb{E}}(00)+\bar{\mathbb{E}}(01)+\bar{\mathbb{E}}(10)-\bar{\mathbb{E}}(11)&=&\nonumber\\
p(\textbf{t}=\textbf{1})\left(\tilde{\mathbb{E}}(00)+\tilde{\mathbb{E}}(01)+\tilde{\mathbb{E}}(10)-\tilde{\mathbb{E}}(11)\right)&\leq&2.
\end{eqnarray}

If we assume that the values $\tilde{E}(\textbf{s})$ are obtained from measurements on quantum systems, then the maximum quantum value of $\tilde{E}(00)+\tilde{E}(01)+\tilde{E}(10)-\tilde{E}(11)$ is Tsirelson's bound, $2\sqrt{2}$. In order to demonstrate a violation of the inequality \eqref{postchsh}, we must then satisfy $p(\textbf{t}=\textbf{1})>\frac{1}{\sqrt{2}}$.

We now relate the value of $p(\textbf{t}=\textbf{1})$ to the detection efficiency $\eta$ with the following expression:
\begin{equation}
p(\textbf{t}=\textbf{1})=\lim_{N\to\infty}\frac{N_{\textbf{1}}}{N_{\textbf{1}}+N_{\{0,1\}}+N_{\{1,0\}}}=\lim_{N\to\infty}\frac{N_{\textbf{1}}}{N_{\textbf{1}}+2N_{\{0,1\}}}=\frac{\eta}{2-\eta},
\end{equation}
since $N_{\{0,1\}}=N_{\{1,0\}}$. A value of $p(\textbf{t}=\textbf{1})>\frac{1}{\sqrt{2}}$ thus gives $\eta>\frac{2}{\sqrt{2}+1}\approx 0.8284$. This is exactly the detection efficiency derived by GM \cite{garg}. 

In GM's result of $\eta\approx 0.8284$, they use a Bell inequality derived for spin-$0$ particles \cite{merminschwarz}. In this original work, it is perhaps not clear, in general, how a loophole is avoided or created. We have explicitly shown the mechanism of how loopholes are formed and this is due to the emergence of non-linear Boolean functions in the event of post-selection. The beauty of our approach, as we shall show in the next subsection is that it can be generalised to $n$ parties; something not immediately attainable in the GM approach\footnote{One would need to find the facet Bell inequalities for $3$ or more spin-$0$ particles.}. In the following subsection, we describe this generalisation to $(n,2,2)$ scenarios. 

\subsection{Generalisation of the GM bound to Many Parties}\label{sec41c}

In the previous subsection, we showed that as long as detection efficiency is above some threshold then quantum physics can violate a Bell inequality. The threshold we derived was already attained by GM. Our rederivation makes the mechanism of loopholes very clear and also establishes the framework for generalising to more than two parties. In this subsection we now present a new result. 

We have shown that the GM threshold for detection efficiency is reached when the quantum systems achieve Tsirelson's bound. If the quantum systems do not achieve this bound then the detection efficiency needed to rule out an LHV description needs to be higher. That is if $\tilde{E}(00)+\tilde{E}(01)+\tilde{E}(10)-\tilde{E}(11)=2+\epsilon$ where $0<\epsilon\leq 2(\sqrt{2}-1)$ results from quantum correlators then the detection efficiency must satisfy $\eta>\frac{4}{4+\epsilon}$. A natural extension of this result is to find Bell inequalities in other Bell tests where the detection efficiency required is lower. Then a bigger range of quantum values of a Bell expression can be tolerated and rule out an LHV description.

This has also been investigated in the full probability distribution Bell setting (e.g. the Clauser-Horne Bell setting\cite{ch}). For example, Eberhard showed that for the CH inequality the minimum detection efficiency is given by $\eta>\frac{2}{3}$ \cite{eberhard}. This value has been subsequently lowered if one increases the number of measurement settings that one can choose from \cite{vertesi}. However, we are focussing on the $n$-party setting with $2$ inputs and $2$ outputs; we will explore a generalisation of the derivation of the GM bound to the $(n,2,2)$ setting and show that the threshold for $\eta$ decreases from $\eta\approx 0.8284$.

The intuition then is to find inequalities where the maximal quantum violation is larger than for the $(2,2,2)$ case. For the $(n,2,2)$ case, WW have shown \cite{ww} that the quantum violation of the Mermin-Klyshko inequalities \cite{mermin,belinskii,gisin} (and inequalities in its orbit) is the largest violation for any $(n,2,2)$ inequality.  There is only one vertex of $\mathcal{P}$ that maximally violates this inequality (for $n$ being even), as shown by Marcovitch and Reznik \cite{reznik}. We will restrict ourselves to the cases $(n,2,2)$ for $n$ being even. We shall describe the odd $n$ case as an extension of the even case. 

The vertex of $\mathcal{P}$ that maximally violates the Mermin inequality for $n$ being even is $p(1|\textbf{s})=\delta^{1}_{f(\textbf{s})}$ where $f(\textbf{s})=\bigoplus_{j=1}^{n-1}s_{j}(\bigoplus_{k=j+1}^{n}s_{k})$ \cite{reznik}. When we refer to $f(\textbf{s})$ in this subsection we mean this function in particular. If we were to allow communication then this function could be performed deterministically. One method would be if each $j$th party received the inputs $s_{k}$ for all $k$th parties where $(j+1)\leq k\leq n$ and $1\leq j\leq (n-1)$. Each party did not even need to learn every other party's input. This protocol also works if we cyclically permute the parties as the function $f(\textbf{s})$ is invariant under all permutations of parties. We now show that this communication protocol can be ``simulated'' if we perform detection post-selection. 

We now describe how we can achieve the vertex $\tilde{p}(1|\textbf{s})=\delta^{1}_{f(\textbf{s})}$ of $\tilde{\mathcal{P}}$ corresponding to the function $f(\textbf{s})=\bigoplus_{j=1}^{n-1}s_{j}(\bigoplus_{k=j+1}^{n}s_{k})$ with LHV correlators. We do this by simulating the above communication protocol using detection post-selection. We call this post-selection protocol the ``Mermin-Klyshko post-selection'' (MKP) protocol: each $k$th party for $2\leq k\leq n$ produces the map $t_{k}=s_{k}\oplus y_{k}(\lambda)\oplus 1$ where all $n$ parties share the $(n-1)$ bit-values $y_{k}(\lambda)$. Party $1$ produces the map $t_{1}=1$. As before, the variables $y_{k}(\lambda)$ are randomly generated. Therefore after detection post-selection, all parties have mapped the inputs $s_{k}$ for $j\neq 1$ onto the shared variables $y_{k}(\lambda)$. Then each $j$th party for $1\leq j\leq (n-1)$ outputs the value $m_{j}=s_{j}(\bigoplus_{k=j+1}^{n}y_{k}(\lambda))=s_{j}(\bigoplus_{k=j+1}^{n}s_{k})$ and the $n$th party outputs $m_{n}=0$. As a result, we obtain the correlator $\tilde{p}(1|\textbf{s})=\delta^{1}_{f(\textbf{s})}$. It is worth noting that we need all $(n-1)$ maps $t_{k}=s_{k}\oplus y_{k}(\lambda)\oplus 1$ so that the first party can obtain all other inputs. 

As in the previous subsection, in order to consider the detection efficiency we need to consider the space $\bar{\mathcal{L}}_{\mathbb{E}}$. We need to consider the probabilities $p(\textbf{t}=\textbf{1})$ for the LHV maps $t_{j}$. The MKP protocol produces $p(\textbf{t}=\textbf{1})=\frac{1}{2^{(n-1)}}$. However, $p(t_{1}=1)=1$ in this protocol. To counter this the $n$ parties share the variable $z(\lambda)\in\{1,2,...,n\}$ which corresponds to each cyclic permutation of the $n$ parties. This variable is randomly generated and then the $n$ parties produce the MKP protocol but for a particular cyclic permutation. As a result, $p(\textbf{t}=\textbf{1})=\frac{1}{n}\frac{1}{2^{(n-1)}}n=\frac{1}{2^{(n-1)}}$. 

Therefore, LHV theories can produce a convex combination of expectation values $\bar{\mathbb{E}}(\textbf{s})=(-1)^{g(\textbf{s})}$ and $\bar{\mathbb{E}}(\textbf{s})=\frac{1}{2^{(n-1)}}(-1)^{f(\textbf{s})}$ where $f(\textbf{s})$. We can substitute these expectation values in the Mermin-Klyshko inequality for even $n$
\begin{equation}
\frac{1}{2^{\frac{n}{2}-1}}\sum_{\textbf{s}}(-1)^{f(\textbf{s})}\bar{\mathbb{E}}(\textbf{s})\leq 2,
\end{equation}
For the expectation value $\bar{\mathbb{E}}(\textbf{s})=\frac{1}{2^{(n-1)}}(-1)^{f(\textbf{s})}$, the Bell expression takes the value $2^{2-\frac{n}{2}}$. This inequality is therefore not violated. For odd $n$, the Mermin-Klyshko inequality can be rewritten as \cite{reznik}
\begin{equation}
\frac{1}{2^{\frac{n-1}{2}-1}}\sum_{\textbf{s}}\delta^{s_{n}}_{\bigoplus_{j=1}^{(n-1)}s_{j}}(-1)^{f(\textbf{s})}\bar{\mathbb{E}}(\textbf{s})\leq 2.
\end{equation}
We can use the same argument for even $n$ to show that this inequality is not violated for any vector of expectation values in $\bar{\mathcal{L}}_{\mathbb{E}}$. First, one can use the MKP protocol, as for even $n$, to give $\bar{\mathbb{E}}(\textbf{s})=\frac{1}{2^{(n-1)}}(-1)^{f(\textbf{s})}$. This gives a value of $2^{\frac{3-n}{2}}$ and so does not lead to a violation. On the other hand, due to the delta function $\delta^{s_{n}}_{\bigoplus_{j=1}^{(n-1)}s_{j}}$, the function $f(\textbf{s})$ is now independent of $s_{n}$ and becomes $f'(\textbf{s})=\bigoplus_{j=1}^{n-2}s_{j}(\bigoplus_{k=j+1}^{n-1}s_{k}\oplus 1)$. This function can be achieved by $(n-1)$ parties carrying out the MKP protocol, thus producing a value of $2^{\frac{5-n}{2}}$ for the Bell expression. In summary then, the Mermin-Klyshko inequality is not violated for all expectation values in $\bar{\mathcal{L}}_{\mathbb{E}}$.

It now remains to express $p(\textbf{t}=\textbf{1})$ in terms of detector efficiency $\eta$. Again we assume that $\eta$ is the same for all sites and so can be calculated from the number counts $N_{\textbf{t}}$ (for $\textbf{t}\neq\textbf{0}$). Therefore taking the limit of $N=\sum_{\textbf{t}\neq\textbf{0}}N_{\textbf{t}}\to\infty$, the efficiency is
\begin{equation}\label{eta3}
\eta=\frac{N_{\{1,\textbf{t}'\}}}{N_{\{1,\textbf{t}'\}}+N_{\{0,\textbf{t}'\}}},
\end{equation}
where $\textbf{t}'\neq\textbf{0}$ is any of the bit-strings for all of the 2-party sub-sets of all $3$ parties. The notation $\{0,\textbf{t}'\}$ ($\{1,\textbf{t}'\}$) then says that the other bit not in the sub-set $\textbf{t}'$ is $0$ ($1$). We can obtain values of $N_{\{0,\textbf{t}'\}}$ in terms of $\eta$ and $N_{\{1,\textbf{t}'\}}$ and substitute them into an expression for $p(\textbf{t}=\textbf{1})$ (using recursion) to obtain
\begin{equation}
p(\textbf{t}=\textbf{1})=\frac{\eta^{n}}{1-(1-\eta)^{n}}.
\end{equation}
If we substitute the maximal quantum violation of the Mermin-Klyshko inequality $2^{\frac{n+1}{2}}$ for the expectation values $\bar{\mathbb{E}}(\textbf{s})$ then we have the following expressions $p(\textbf{t}=\textbf{1})=2^{\frac{1-n}{2}}$. Therefore, for $n=3$, detection efficiency must satisfy $\eta>\frac{1}{2}(\sqrt{21}-3)\approx 0.7913$ in order to demonstrate a loophole-free violation of a Bell inequality. Whilst this is a decrease from the GM bound, this value of $\eta$ does not decrease dramatically; for example for $n=25$, $\eta\gtrsim 0.7170$ but for $n=75$, $\eta\gtrsim 0.7104$. The bound of $\eta>\frac{2}{3}$ found by Eberhard (and subsequently improved) is more effective for a loophole-free Bell test \cite{eberhard}.

While these generalisations of the GM bound on $\eta$ may not be impressive compared to the current literature, our discussion has been motivated by a qualitative description of loopholes. We have also connected the detection loophole to communication protocols (cf. \cite{barrett2}). Detection post-selection can simulate communication between parties by correlating input data to shared hidden variables. We used this simulation of a communication protocol to derive these generalisations of the GM bound. We have also used our computational description of all possible LHV maps to make this loophole-producing mechanism clear. 

\subsection{Summary of Loopholes}\label{sec41d}

\noindent
We have discussed how experimental imperfections in Bell tests can lead to loopholes. We have briefly covered how loss of measurement freedom and no space-like separation can lead to loopholes. In more detail, we have discussed how the subtleties of the detection loophole can be made clearer with the language of Boolean functions. Our language in terms of computational expressiveness allowed us to redrive the GM bound and generalised it to $n$ parties.

Beyond the loopholes we have discussed already, we will now briefly mention another: the memory loophole \cite{barrett6}. The memory loophole emerges if parties retain their choice of input and subsequent output in a ``memory'' that can be communicated between parties in-between tests. From this memory, parties can make ``educated guesses'' about which measurement outputs to give for a particular input. This problem occurs from a finite number $N$ of Bell tests from which we produce correlation statistics. However, the loophole does not become an issue as $N\rightarrow\infty$ \cite{barrett6}, heuristically, the region of the LHV polytope outside of the linear Boolean functions disappears exponentially in $N$. Since we have assumed that all statistics from experiments are obtained in this limit, the memory loophole is not a conceptual, problematic issue.

In the next section, we look again at post-selection but not from an experimental point-of-view. We will assume that Bell tests are perfectly implemented in the laboratory. The post-selection introduced establishes a relationship between measurement data in a non-trivial fashion. We have shown that with detection post-selection, relationships are induced between hidden variables and measurement settings, thus leading to loopholes. In this new setting we will define a loophole in analogy to the definition in this section. Given this definition, we show that this new form of post-selection is free of loopholes. This new method is a way of conceptually modifying Bell tests but not modifying the implications of LHV theories.

\section{Loophole-free Post-selection and Quantum Correlators}\label{sec42}

\noindent
In the previous section, post-selection was a necessity in order to calculate correlators. For non-detection events, measurement outcomes are not defined so the sum modulo $2$ of outcomes could not be calculated. We now explore the use of post-selection utilised by the experimenter out of choice rather than necessity. We assume that the Bell test has perfect detectors and the experimenter does not need to use detection post-selection. Therefore, data is perfectly obtained by the experimenter but they still choose to discard some of this data. We will construct a new model to reflect this choice and discuss the possibility of loopholes in this model.

For all of the discussion so far in this thesis, the variable $\textbf{s}$ for each correlator $p(k|\textbf{s})$ has two functions: 1) it labels the inputs to all sites corresponding to the choice of measurement settings; 2) $\textbf{s}$ acts as a conditioning variable for the probability measure on all maps $\mathbb{Z}_{2}^{n}\rightarrow\mathbb{Z}_{2}$. In this section we will distinguish between these two roles by using post-selection on measurement data. This is done by relating measurement data to data that is independent of measurement settings or outcomes. We motivate this discussion by returning to the Mermin inequality \cite{mermin}:
\begin{equation}
p(1|000)+p(1|011)+p(1|101)-p(1|110)\leq 2
\end{equation}
and recall that as in the GHZ paradox, we are interested in correlators when $s_{3}=s_{1}\oplus s_{2}$. In the language of computer science, this is called a \emph{promise} on the inputs that they satisfy a particular relation \cite{nonlocal}. This inequality is also superficially similar to the CHSH inequality but now with a third party whose inputs are related to the other two sites. 

To make the connection to the CHSH inequality clearer, we notice that the linear Boolean functions that LHV theories can achieve if $s_{3}=s_{1}\oplus s_{2}$ are written as $m_{1}\oplus m_{2}\oplus m_{3}=\alpha s_{1}\oplus\beta s_{2}\oplus \gamma s_{3}\oplus \delta=\alpha' s_{1}\oplus\beta' s_{2}\oplus \delta$ for $\alpha$, $\beta$, $\alpha'$, $\beta'$, $\gamma$, $\delta\in\{0,1\}$. These functions are exactly the linear Boolean functions for the $(2,2,2)$ CHSH setting. Therefore,  the linear Boolean functions that satisfy the CHSH inequality also satisfy the Mermin inequality. The Mermin inequality can be seen as a manifestation of the CHSH inequality. 

If we reconsider experimental implementations of Bell tests, then how do parties obtain the input $s_{3}=s_{1}\oplus s_{2}$ if they are space-like separated from the other two parties? A possible solution is through data post-selection; the third party makes a completely random choice of $s_{3}$. After receiving all data $\textbf{m}$ and $\textbf{s}$ from all parties the experimenter only accepts data from all parties and calculate $p(1|\textbf{s})$ if $s_{3}=s_{1}\oplus s_{2}$; otherwise data is discarded. Since $s_{1}$ and $s_{2}$ are also randomly generated, the rate at which the experimenter discards the data will tend to $\frac{1}{2}$ for $N\rightarrow\infty$ runs of the experiment.

We will proceed to generalise this method of post-selection utilised in the GHZ paradox. Central to this approach will be the linear Boolean functions. In the example of the GHZ paradox, the experimenter post-selects on one input being a linear Boolean function. This keeps the computational power of the LHV correlators confined to these linear functions. We showed in the previous section that loopholes can lead to LHV correlators performing non-linear Boolean functions. In analogy with the detection loophole, the post-selection in the GHZ paradox can be seen to avoid a loophole. This is the central insight in this section and we will now develop these ideas rigorously.

\subsection{Post-selection, Linearity and Loopholes}\label{sec42a}

We now introduce some more general structure beyond the GHZ paradox. The experimenter now has some bit-string $\textbf{x}$ of length $|\textbf{x}|\leq n$. Referring back to the two roles of $\textbf{s}$ described above, $\textbf{x}$ now plays the role of conditioning variable (role 2). That is, instead of the stochastic maps $p(1|\textbf{s})$ being conditioned upon $\textbf{s}$, they are now conditioned upon $\textbf{x}$, i.e. the experimenter calculates $p(1|\textbf{x})=p(\bigoplus_{j=1}^{n}m_{j}=1|\textbf{x})$. The experimenter then relates their data $\textbf{x}$ to the experimental data $\textbf{m}$ and $\textbf{s}$.

If we return to the GHZ paradox, we have three parties but the bit-string $\textbf{x}\in\{0,1\}^{2}$. The experiment now accepts, or post-selects on data $\textbf{m}$ and $\textbf{s}$ when $s_1=x_1$, $s_2=x_2$ and $s_3=x_1\oplus x_2$ is satisfied. Then the experimenter calculates the correlator 
\begin{equation}
p(1|\textbf{x})=p(m_{1}\oplus m_{2}\oplus m_{3}=1|s_1=x_1,s_2=x_2,s_3=x_1\oplus x_2).
\end{equation}
A relationship between data $\textbf{s}$ and $\textbf{x}$ is established by the experimenter's post-selection. We now generalise this approach of relating $\textbf{s}$ to $\textbf{x}$ with the following form of post-selection.

\begin{defn}\label{defn421}
If an experimenter accepts, or post-selects on data $\textbf{m}$ and $\textbf{s}$ for every input $s_{j}$ satisfying $s_{j}=g_{j}(\textbf{x})$ where $g_{j}(\textbf{x})$ is some Boolean function on $\textbf{x}$, this is \textbf{input post-selection}. The experimenter fixes this relationship between $\textbf{x}$ and $\textbf{s}$ for all runs of the experiment. After post-selection, the experimenter calculates $p(1|\textbf{x})$ for their value of $\textbf{x}$.
\end{defn}

After the post-selection there are now $2^{|\textbf{x}|}$ correlators $p(1|\textbf{x})$ for all values of $\textbf{x}$. Just as with the correlators $p(1|\textbf{s})$, the correlators $p(1|\textbf{x})$ are elements of vectors $\vec{k}_{\textbf{x}}\in\mathbb{R}^{2^{|\textbf{x}|}}$. The deterministic correlators are $p(1|\textbf{x})=\delta^{1}_{f(\textbf{x})}$ for all Boolean functions $f(\textbf{x})$. Therefore, vectors $\vec{k}_{\textbf{x}}$ are contained in a convex polytope $\mathcal{P}_{\textbf{x}}$ with these extreme points being these deterministic correlators. There will also be the space of LHV correlators $\mathcal{L}_{\textbf{x}}$ in analogy to $\mathcal{L}$. If $n=|\textbf{x}|$ and the functions $g_{j}(\textbf{x})$ in input post-selection are $s_{j}=x_{j}$, then we recover the original $(n,2,2)$ Bell test. For this example, $\mathcal{L}_{\textbf{x}}$ is the convex hull of linear Boolean functions on $\textbf{x}$. In analogy with the detection loophole defined in the previous section, we now define a loophole for input post-selection.

\begin{defn}\label{defn422}
A \textbf{loophole} is introduced by an experimenter into a Bell test if after input post-selection, the space $\mathcal{L}_{\textbf{x}}$ is larger than the convex hull of linear Boolean functions on $\textbf{x}$.
\end{defn}

In the next subsection we will show how loopholes are avoided if the experimenter utilises input post-selection. We will then develop input post-selection in subsection \ref{sec42c}, now to encompass a relationship between $\textbf{x}$ and \emph{both} $\textbf{m}$ and $\textbf{s}$. This new form of post-selection will be called ``output-input post-selection''. In this case, we can still find a way to avoid loopholes in the sense that $\mathcal{L}_{\textbf{x}}$ remains the convex hull of linear Boolean functions. We now address loopholes in input post-selection.

\subsection{Linear Input Post-selection in $(n,2,2)$ tests}\label{sec42b}

\noindent
We begin our discussion with a key result for setting post-selection. This result informs us of how to avoid loopholes and will lead to us describing a particular class of input post-selections. 

\begin{thm}\label{thm421}
The space $\mathcal{L}_{\textbf{x}}$ of LHV correlators is the convex hull of linear Boolean functions on $\textbf{x}$ for input post-selections with $s_j=g_{j}(\textbf{x})$ if and only if every $g_{j}(\textbf{x})$ is a linear Boolean function on $\textbf{x}$.
\end{thm}

\textit{Proof}: First we prove the if statement. We only need to consider the extreme points of $\mathcal{L}$ corresponding to the deterministic linear Boolean functions $f(\textbf{s})$ on $\textbf{s}$, i.e. $f(\textbf{s})=\bigoplus_{j=1}^{n}a_{j}s_{j}\oplus b$ with $a_{j}$, $b\in\{0,1\}$. If we post-select on $s_j=g_{j}(\textbf{x})$ being a linear Boolean function, then $f(\textbf{s})\rightarrow f(\textbf{x})=\bigoplus_{j=1}^{n}a_{j}g_{j}(\textbf{x})\oplus b$, which is again a linear Boolean function now on $\textbf{x}$. To prove the only if statement, if $g_{j}(\textbf{x})$ is a non-linear Boolean function, then extreme points of $\mathcal{L}$ producing $f(\textbf{x})=\bigoplus_{j=1}^{n}a_{j}g_{j}(\textbf{x})\oplus b$ will in general be a non-linear Boolean function for all $a_{j}$ and $b$. $\square$
\newline

From this result, $\mathcal{L}_{\textbf{x}}$ will be defined by the facet Bell inequalities for the $(|x|,2,2)$ setting replacing $p(1|\textbf{s})$ with $p(1|\textbf{x})$. Returning to our example, the Mermin inequality (with replacing $\textbf{s}$ with $\textbf{x}$) is the CHSH-like inequality defining a facet of $\mathcal{L}_{\textbf{x}}$ with $|\textbf{x}|=2$. This all occurs only if the input post-selection consists of $g_{j}(\textbf{x})$ being a linear Boolean function. We now formally define this particular class of post-selections:

\begin{defn}\label{dfn423}
\textbf{Linear Input Post-selection} (LI) is input post-selection but where all of the functions $g_{j}(\textbf{x})$ are linear Boolean functions on $\textbf{x}$.
\end{defn}

This post-selection can be seen to simulate nMBQC as described in section \ref{sec34} of chapter \ref{chap3}. Instead of pre-processing on inputs which are then distributed to $n$ parties, we post-select on inputs satisfying the expressions that are described by the $\textbf{P}$ matrices. Since LHV resources can only produce linear Boolean functions in nMBQC, then our post-selection simulates a model with the same computational power. Crucially both the pre-processing and post-selection is restricted to the linear Boolean functions. 

The connection to MBQC can now be extended by considering adaptivity. In adaptive MBQC, inputs, or measurement settings at each site are influenced by previous measurement outcomes. Translating this into a Bell test, the input $s_{j}$ is now a function $h(m_{j'}|\forall j'\neq j)$ of measurements outcomes $m_{j'}$ from other sites. Directly, this would assume communication between sites. However, if we post-select on inputs $s_{j}$ satisfying this function $h(m_{j'}|\forall j'\neq j)$, then we can simulate this communication. We now discuss this form of post-selection and show, remarkably, that we can avoid loopholes. 

\subsection{Linear Output-Input Post-selection in $(n,2,2)$ tests}\label{sec42c}

We now extend input post-selection to consider functional relationships induced between the experimenter's variable $\textbf{x}$ and $\textbf{m}$ and $\textbf{s}$. In particular, for the $j$th site, $s_{j}$ can be related to outcomes $m_{j'}$ for $j\neq j'$. We introduce the notation $\textbf{m}^{\setminus j}$ to describe a $(n-1)$-length bit-string which is $\textbf{m}$ but \emph{without} the bit-value $m_{j}$. For example, if $j=1$, then $\textbf{m}^{\setminus j}=\{m_{2},m_{3},...,m_{n}\}$. With this new piece of notation we now introduce a new form of post-selection, first studied by Hoban and Browne \cite{hoban2}. ``Output-input post-selection'' is now the same as input post-selection but the experimenter now accepts data when $s_{j}=g_{j}(\textbf{x},\textbf{m}^{\setminus j})$ instead of $s_{j}=g_{j}(\textbf{x})$. Again, after the post-selection, the experimenter again calculates $p(1|\textbf{x})$ for each $\textbf{x}$.

For this output-input post-selection, the space of all possible correlators is $\mathcal{P}_{\textbf{x}}$, the same as input post-selection. For LHV correlators, we describe the space of correlators after output-input post-selection as $\mathcal{L}_{\textbf{x},\textbf{m}^{\setminus j}}$. As an extension of the definition of a loophole for input post-selection, a loophole emerges if $\mathcal{L}_{\textbf{x},\textbf{m}^{\setminus j}}$ is larger than the convex hull of linear Boolean functions on $\textbf{x}$. We now show when loopholes in output-input post-selection can be avoided.

\begin{thm}\label{thm422}
The space $\mathcal{L}_{\textbf{x},\textbf{m}^{\setminus j}}$ of LHV correlators is the convex hull of linear Boolean functions on $\textbf{x}$ for output-input post-selections $s_j=g_{j}(\textbf{x},\textbf{m}^{\setminus j})$ if and only if every $g_{j}(\textbf{x},\textbf{m}^{\setminus j})$ is a linear Boolean function on $\textbf{x}$ and $\textbf{m}^{\setminus j}$.
\end{thm}

\textit{Proof}: First we prove the if statement. We recall that all deterministic LHV single-site maps can be written as $m_{j}=\alpha_{j}s_{j}\oplus \beta_{j}$ and we can take their convex combination. We assume that $\alpha_{j}$ and $\beta_{j}$ is dependent on an LHV $\lambda$ but these variables are in no way correlated with the inputs $\textbf{s}$. Therefore, all extreme points of $\mathcal{L}$ from these deterministic maps result in $\bigoplus_{j=1}^{n}m_{j}$ being a linear Boolean function on $\textbf{x}$ and $\textbf{m}^{\setminus j}$. 

If we do not assume that the values $\alpha_{j}$ and $\beta_{j}$ for all $j$ are not correlated to $\textbf{s}$, there is a way in which this post-selection can allow correlations between bits from the LHV, $\alpha_{j}$, $\beta_{j}$ and inputs $\textbf{s}$. We now demonstrate this method. We can decompose a linear function $g_{j}(\textbf{x},\textbf{m}^{\setminus j})$ as $g^{(1)}_{j}(\textbf{m}^{\setminus j})\oplus g^{(2)}_{j}(\textbf{x})$, i.e. in terms of the linear functions $g^{(1)}_{j}(\textbf{m}^{\setminus j})$ and $g^{(2)}_{j}(\textbf{x})$ on $\textbf{m}^{\setminus j}$ and $\textbf{x}$ respectively. The outcomes in $\textbf{m}^{\setminus j}$ contain information about $\lambda$, but $s_{j}$ is random and uncorrelated to $\lambda$, $\textbf{m}$ and $\textbf{x}$. Therefore $g^{(1)}_{j}(\textbf{m}^{\setminus j})=g^{(2)}_{j}(\textbf{x})\oplus s_{j}$ means that $g^{(1)}_{j}(\textbf{m}^{\setminus j})$ is random and uncorrelated to $g^{(2)}_{j}(\textbf{x})$\footnote{If $g^{(2)}_{j}(\textbf{x})=0$, the bit $s_j$ does become correlated with other sites' measurements $m_k$ and hence $\lambda$ but $s_j$ will be uncorrelated to $\textbf{x}$. If $g^{(1)}_{j}(\textbf{m}^{\setminus j})=0$, we recover LI post-selection.}. These random bits $s_j$ play the role of the pad-bit in one-time pad cryptography which Shannon \cite{shannon} proved is perfectly secure for encrypting messages.

We finally prove the only if statement. If $g_{j}(\textbf{m}^{\setminus j},\textbf{x})$ becomes non-linear then we can always produce this function $f(\textbf{x})=g_{j}(\textbf{m}^{\setminus j},\textbf{x})$ as an output. Since values of $\textbf{m}^{\setminus j}$ can be made to be equal to values of $\textbf{x}$, there always exists a non-linear function in $\textbf{x}$ if $g_{j}(\textbf{m}^{\setminus j},\textbf{x})$ is non-linear. $\square$
\newline

We now call output-input post-selection where $g_{j}(\textbf{x},\textbf{m}^{\setminus j})$ is a linear Boolean function on $\textbf{x}$ and $\textbf{m}^{\setminus j}$, \textbf{Linear Output-Input Post-selection} (LOI). With LOI, we can simulate signalling processes by making inputs dependent on outputs at other sites. But, we can also keep the space of correlators confined to the linear Boolean functions on $\textbf{x}$. This means that for all $n\geq |\textbf{x}|$, the space $\mathcal{L}_{\textbf{x},\textbf{m}^{\setminus j}}$ for LOI is $\mathcal{L}_{\textbf{x}}$, the convex hull of linear Boolean functions. The $n$-independence in the space of correlators is unusual given that in traditional Bell tests, the role of the number of parties is important. In some way, by considering $|\textbf{x}|$, we unify all possible multi-party Bell settings for $n\geq|\textbf{x}|$.

With regards to quantum correlators, we have already indicated that there is an $n$-dependence in the example of the Mermin inequality. For $n=|\textbf{x}|=2$, the maximal violation of the CHSH inequality is $2\sqrt{2}$. After LI, for $n=3$ and $|\textbf{x}|=2$, the same CHSH inequality in terms of $p(1|\textbf{x})$ has the maximal violation of $4$. In the next subsection we will discuss the effect of LI and LOI upon the space of quantum correlators.

\subsection{Bipartite Quantum correlators under post-selection}\label{sec42d}

The space $\mathcal{Q}_{\textbf{x}}$ of quantum correlators under LI needs to be specified for a particular value of $n$, i.e. $\mathcal{Q}^{n}_{\textbf{x}}$. For LOI, the corresponding space of quantum correlators is $\mathcal{Q}^{n}_{\textbf{x},\textbf{m}^{\setminus j}}$ for $n$ number of parties. Since LOI includes all possible post-selections in LI, then necessarily $\mathcal{Q}^{n}_{\textbf{x}}\subseteq\mathcal{Q}^{n}_{\textbf{x},\textbf{m}^{\setminus j}}$. 

We now focus on $|\textbf{x}|=2$ as the smallest example of non-trivial behaviour of $\mathcal{Q}^{n}_{\textbf{x}}$ and $\mathcal{Q}^{n}_{\textbf{x},\textbf{m}^{\setminus j}}$. Since $\mathcal{Q}^{2}_{\textbf{x}}$ is strictly smaller than $\mathcal{P}_{\textbf{x}}$ for $|\textbf{x}|=2$, we can initially ask whether $\mathcal{Q}^{2}_{\textbf{x},\textbf{m}^{\setminus j}}$ is larger than $\mathcal{Q}^{2}_{\textbf{x}}$? This turns out not to be the case as we now demonstrate. For this situation, the most general LOI  possible involves post-selecting on the following relations being satisfied: $s_{1}=x_{1}\oplus \alpha m_{2}$ and $s_{2}=x_{2}\oplus \beta m_{1}$ with $\alpha,\beta\in\{0,1\}$. When $\alpha=\beta=0$, we retrieve the standard, well-studied scenario. The two scenarios where $\alpha\neq\beta$ are equivalent up to changing of labels. If we consider the scenario where $\{\alpha,\beta\}=\{0,1\}$ then the probabilities $p(1|\textbf{x})$ can be rewritten in terms of probabilities $p(m_{1},m_{2}|s_{1},s_{2})$:
\begin{eqnarray}\label{mxcorr}
p(1|\textbf{x})&=&\sum_{m_{1},m_{2}}\delta^{m_{1}\oplus m_{2}}_{1}p(m_{1},m_{2}|s_{1}=x_{1},s_{2}=x_{2}\oplus m_{1})\nonumber \\
&=&p(0,1|s_{1}=x_{1},s_{2}=x_{2})+p(1,0|s_{1}=x_{1},s_{2}=x_{2}\oplus 1).
\end{eqnarray}
The correlator can be written in this way as $p(\textbf{m}|\textbf{s})$ is a non-signalling distribution. Any non-signalling probability distribution can be written as a convex combination of the vertices of $\mathcal{NS}$. For the bipartite scenario there are two types of vertices: 1) local vertices where $p(m_{1},m_{2}|s_{1},s_{2})=\prod_{j=1}^{2}\delta^{m_{j}}_{a_{j}s_{j}\oplus b_{j}}$ for $a_{j}$, $b_{j}\in\{0,1\}$; and 2) ``non-local'' vertices for $p(m_{1},m_{2}|s_{1},s_{2})=\frac{1}{2}$ for $m_{1}\oplus m_{2}=(s_{1}\oplus a)(s_{2}\oplus b)\oplus c$, and $0$ otherwise where $a$, $b$, $c\in\{0,1\}$. In the former case, when $s_{1}=x_{1}$, $m_{1}$ is either $0$ or $1$ deterministically and so \eqref{mxcorr} must be $1$ for a local vertex. For a non-local vertex, \eqref{mxcorr} takes any of the values $\{0,\frac{1}{2},1\}$, so for non-signalling distributions, every correlator of the form \eqref{mxcorr} is at most $1$.

To see if any correlators are outside of $\mathcal{L}_{\textbf{x}}$, we put the correlators in \eqref{mxcorr} into the CHSH inequality (and any in its symmetry group) to obtain
\begin{eqnarray}
p(0,1|0,0)+p(1,0|0,0)+p(0,1|0,1)+p(1,0|0,1)+& & \nonumber \\
p(0,1|1,0)-p(1,0|1,0)-p(0,1|1,1)+p(1,0|1,1)&\leq& 2.
\end{eqnarray}
All bipartite non-local vertices satisfy
\begin{equation}
p(0,1|1,0)-p(1,0|1,0)=-p(0,1|1,1)+p(1,0|1,1)=0. 
\end{equation}
Therefore, all non-signalling probability distributions do not violate the CHSH inequality with LOI for $\alpha\neq\beta$. As a corollary, quantum correlators satisfy the CHSH inequality\footnote{The inequalities in the CHSH inequality symmetry group are also not violated as we can map to all inequalities in this group via local re-labellings or an overall sign change, and we can also map from every non-local vertex of $\mathcal{NS}$ via the same operations}.

Finally, for the scenario of LOI  with $\alpha=\beta=1$, then $m_{1}\oplus m_{2}=s_{1}\oplus s_{2}\oplus x_{1}\oplus x_{2}$. The correlators $p(1|\textbf{x})$ are calculated when $s_{1}\oplus s_{2}=x_1 \oplus x_2\oplus 1$, so $p(1|\textbf{x})=p(1|\textbf{x}')$ for $\textbf{x}=\textbf{x}'$ if $x_{1}\oplus x_{2}=x_{1}'\oplus x_{2}'$. Substituting these values of the correlators into all of the CHSH inequalities never yields a violation, as two of the correlators will cancel\footnote{After the terms that are equal but have opposite sign pre-factors in the inequality cancel, the inequalities reduce to either $2p(1|\textbf{x})\leq 2$ or $-2p(1|\textbf{x})\leq 0$ for a particular value of $\textbf{x}$}. To summarise then, for LOI for $\alpha\neq\beta$ or $\alpha=\beta=1$, quantum correlators do not exceed the LHV polytope. Therefore the space of quantum correlators $\mathcal{Q}^{2}_{\textbf{x},\textbf{m}^{\setminus j}}=\mathcal{Q}^{2}_{\textbf{x}}$, with $\mathcal{Q}^{2}_{\textbf{x}}$ being $\mathcal{Q}$ for $(2,2,2)$.

\subsection{Multipartite quantum correlators}\label{sec42e}

We have looked at the scenario when $n=2$, we will now consider the space of quantum correlators $\mathcal{Q}^{n}_{\textbf{x},\textbf{m}^{\setminus j}}$ for general $n$ and $|\textbf{x}|$. Having shown that LIO has no impact on quantum correlators, we now show the opposite in the multipartite setting. That is, the space of quantum correlators under LIO can be \emph{larger} than the space of quantum correlators under LI. We begin by considering the $n=3$ scenario and then use it to consider larger $n$ for a particular $|\textbf{x}|$.

Firstly, we observe that $\mathcal{Q}^{3}_{\textbf{x},\textbf{m}^{\setminus j}}=\mathcal{P}_{\textbf{x}}$ for $|\textbf{x}|=2$. From the GHZ paradox, $p(m_1\oplus m_2\oplus m_3=1|\textbf{x})=\delta^{x_{1}x_{2}\oplus 1}_{1}=x_{1}x_{2}\oplus 1$. We can map from the function $x_{1}x_{2}\oplus 1$, to the other non-linear Boolean functions $(x_{1}\oplus a)(x_{2}\oplus b)\oplus c$ (with $a$, $b$, $c\in\{0,1\}$) with relabelling of bit-values $x_{j}$. All vertices of $\mathcal{P}_{\textbf{x}}$ can be achieved by quantum correlators for $n=3$ and $|\textbf{x}|=2$.

We might ask whether quantum correlators can saturate the whole of $\mathcal{P}_{\textbf{x}}$ for particular values of $n$ and $|\textbf{x}|$? In the following lemma, quantum correlators for a given $n$ and $|\textbf{x}|$ can saturate the whole space $\mathcal{P}_{\textbf{x}}$. In particular, any vertex of $\mathcal{P}_{\textbf{x}}$ corresponding to non-linear Boolean functions can be attained with quantum correlators for a particular $n$. We use the $n=3$, $|\textbf{x}|=2$ case to demonstrate this fact.

\begin{lem}\label{lem421}
For all $|\textbf{x}|$, $\mathcal{Q}^{n}_{\textbf{x},\textbf{m}^{\setminus j}}$ contains the vertex $p(1|\textbf{x})=\delta^{1}_{f(\textbf{x})}$ for $f(\textbf{x})=\prod_{j=1}^{|\textbf{x}|}x_{j}$ if $n=3(|\textbf{x}|-1)$.
\end{lem}

\textit{Proof}: We prove this with an explicit LIO  protocol. If we have $n=3y$ parties for $y$ as some non-zero positive integer, and divide them into $y$ sets of three neighbouring parties in the following way $\{\{1,2,3\},...,\{3y-2,3y-1,3y\}\}$. For each of these $y$ sets $\{j,(j+1),(j+2)\}$, if inputs are $s_{j}$, $s_{(j+1)}$ and $s_{(j+2)}=s_{j}\oplus s_{(j+1)}$, then we can have $m_{j}\oplus m_{(j+1)}\oplus m_{(j+2)}=s_{j}s_{(j+1)}$ with quantum correlators\footnote{Corresponding to the maximal quantum violation $1$ of the Mermin inequality $-p(1|0,0,0)-p(1|0,1,1)-p(1|1,0,1)+p(1|1,1,0)\leq 0$, equivalent to the original Mermin inequality with the symmetry operation being adding $1$ (modulo $2$) to the joint outcome $m_{1}\oplus m_{2}\oplus m_{3}$.}. For $j=1$, if the experimenter post-selects upon $s_{1}=x_{1}$, $s_{2}=x_{2}$ and $s_{3}=x_{1}\oplus x_{2}$, then we have the situation for $n=3$ and $|\textbf{x}|=2$ discussed above. 

For $j=3k+1$ with $k\in\{1,2,...,(y-1)\}$, the experimenter post-selects data if $s_{j}=\bigoplus_{l=1}^{j-1}m_{l}$, $s_{(j+1)}=x_{k+2}\oplus 1$ and $s_{(j+2)}=\bigoplus_{l=1}^{j-1}m_{l}\oplus x_{k+2}\oplus 1$. For $j=4$, this results in $s_{4}=x_{1}x_{2}$, $s_{5}=x_{3}\oplus 1$ and $s_{6}=x_{1}x_{2}\oplus x_{3}\oplus 1$, and so $m_{4}\oplus m_{5}\oplus m_{6}=x_{1}x_{2}(x_{3}\oplus 1)$, resulting in $\bigoplus_{l=1}^{6}m_{l}=x_{1}x_{2}x_{3}$. Then by iteration, for $k\geq 2$, the above protocol results in $\bigoplus_{l=1}^{3y}m_{l}=\prod_{l=1}^{y+1}x_{l}$. Therefore, if $y=|\textbf{x}|-1$, the function $f(\textbf{x})=\prod_{l=1}^{|\textbf{x}|}x_{l}$ can be achieved deterministically with $n=3(|\textbf{x}|-1)$. $\square$
\newline

If we consider LI, we know that for $n=2^{|\textbf{x}|}-1$, every Boolean function can be achieved deterministically. This is because LI can simulate nMBQC directly, and in nMBQC we need at most this number of parties to achieve all Boolean functions. Translated into the language of post-selection, $\mathcal{Q}^{n}_{\textbf{x},\textbf{m}^{\setminus j}}=\mathcal{P}_{\textbf{x}}$ for $n=2^{|\textbf{x}|}-1$. What is more, we showed in Theorem \ref{thm342}, that to achieve $p(1|\textbf{x})=\delta^{1}_{f(\textbf{x})}$ for $f(\textbf{x})=\prod_{j=1}^{|\textbf{x}|}x_{j}$ with nMBQC, we require no fewer than $n=2^{|\textbf{x}|}-1$ parties. For $|\textbf{x}|\geq 3$, $2^{|\textbf{x}|}-1>3(|\textbf{x}|-1)$. This then gives us the following result.

\begin{thm}\label{thm423}
$\mathcal{Q}^{n}_{\textbf{x},\textbf{m}^{\setminus j}}$ can be be larger than $\mathcal{Q}^{n}_{\textbf{x}}$ for a fixed $n$ and $|\textbf{x}|$.
\end{thm}

If we utilise LOI for a particular number $n$ of parties, then we can get a larger violation of a Bell inequality with this LOI than with LI. Quantum correlators can be perceived to be ``more non-local'' if we process our measurement data in a particular way. Then the action of discarding data can not only allow classical, or LHV correlators to simulate quantum correlators (as in the detection loophole), but used to emphasize the non-classical aspect of quantum physics.

The LOI can simulate a circuit where some outputs can affect some inputs. Traditionally, Boolean circuits have sequential gates so there is a temporal order of processes. In our post-selection, Boolean functions result from resources that are without temporal order or space-like separated. There has been a great deal of research into a field called Boolean circuit complexity and there is a natural overlap with discussion of this field to our discussion of LOI. Boolean circuit complexity asks how many fundamental operations or \emph{gates} are required to perform any Boolean function (e.g. the AND and NOT gates) \cite{pap}. The application of Boolean circuit complexity results to LOI would be an interesting avenue of research. 

In this section we have shown that post-selection can be used to conceptually change Bell tests. The post-selection described also establishes a link between Bell tests and the full MBQC model described by Briegel and Raussendorf \cite{mbqc1}. Whilst LI simulates nMBQC, the adaptivity in MBQC can be simulated by LOI. With this post-selection, processing on measurement data utilises addition modulo $2$, as with the classical computer in MBQC. Heuristically, the Bell test with LOI is akin to a single round of measurements in MBQC, but we only accept the circuit if it corresponds to an adaptive circuit in MBQC; we discard the circuit otherwise. This is analogous to post-selected quantum state teleportation where we accept, post-select our system on the ``correct'' measurement outcome resulting in teleportation \cite{lloyd}. 

Central to our discussion in this section has been the computational description of correlators. We then used this computational description to consider input and output-input post-selection. We can limit the computational implications of this post-selection if we restrict ourselves to measurement data being related by linear Boolean functions. Linear Boolean functions are associated with LHV correlators. If all processing on data consists of linear Boolean functions, the computational power of LHV correlators remains linear. However, for general $(n,c,d)$ scenarios, LHV correlators are associated with $n$-partite linear functions. We show in the next section, that generalising LI and LOI to these more general scenarios can be very problematic. 

\section{General settings and Input Post-selection}\label{sec43}

\noindent
The discussion in the previous two sections \ref{sec41} and \ref{sec42} of this chapter have been in the $(n,2,2)$ scenario. We now consider generalisations of input post-selection to $(n,d,d)$ scenarios where $d$ is prime. We have demonstrated that we can avoid loopholes in the $(n,2,2)$ scenario, is this true for all $d$? In order to address this question we need to generalise the approach developed in the previous section. We will introduce two natural generalisations of the post-selection in LI, and we show that it is not loophole free. Despite this, we will again give an indication that the region of quantum correlators can be \emph{enlarged} by post-selection. We now proceed to introduce the framework for input post-selection in the $(n,d,d)$ scenario.

As before, the experimenter has some $|\textbf{x}|$-length digit string $\textbf{x}\in\mathbb{Z}_{d}^{|\textbf{x}|}$ which they have chosen. He receives data $\textbf{m}$ and $\textbf{s}$ from all $n$ parties and then accepts this data if inputs $s_{j}$ are equal to some function $g_{j}(\textbf{x})$ on $\textbf{x}$. If this function is not satisfied by all $s_{j}$ then the experimenter discards this data. Once the data has been accepted by the experimenter they calculate $k=\left[\sum_{j=1}^{n}m_{j}\right]_{d}$ and produce the correlator $p(k|\textbf{x})$. 

The space of all possible correlators is $\mathcal{P}_{\textbf{x}}$, the convex polytope of correlators $p(k|\textbf{x})=\delta^{k}_{f(\textbf{x})}$ for any function $f(\textbf{x}):\mathbb{Z}_{d}^{n}\rightarrow\mathbb{Z}_{d}$. For the trivial post-selection $s_{j}=x_{j}$ where $n=|\textbf{x}|$, then the space of LHV correlators is $\mathcal{L}_{\textbf{x}}$: the convex hull of $n$-partite linear functions on $\textbf{x}$. This space might be dependent on $n$, but this is implicitly assumed in our notation. As in the $(n,2,2)$ case, we define a \textbf{loophole} as a form of input post-selection that results in $\mathcal{L}_{\textbf{x}}$ being larger than the convex hull of $n$-partite linear functions. 

So far, this framework for all $(n,d,d)$ scenarios for prime $d$ is almost identical to the $(n,2,2)$ case. What is the generalisation of LI for this more general case? For $(n,2,2)$, the linear Boolean functions on $\textbf{x}$ are both $n$-partite linear functions \emph{and} the addition modulo $2$ of variables $x_{j}$ (upto some additional constant). In the $(n,d,d)$ scenario, functions consisting of sums of elements $x_{j}$ modulo $d$ are a subclass of all $n$-partite linear functions; we call these functions \textbf{affine functions} on $\textbf{x}$. In the next subsection we will consider input post-selection for affine functions $g_{j}(\textbf{x})$. We will then consider the case where $g_{j}(\textbf{x})$ is any $n$-partite linear function. In both cases, loopholes are introduced by the input post-selection. For all $n$-partite linear functions $g_{j}(\textbf{x})$, the space of LHV correlators is equal to $\mathcal{P}_{\textbf{x}}$ for some $n$; this is not possible for the affine functions $g_{j}(\textbf{x})$. Finally we will briefly discuss the space of quantum correlators under input post-selection. 

\subsection{Input Post-selection with Affine Functions}\label{sec43a}

We now consider input post-selection where $g_{j}(\textbf{x})$ are the affine functions. The affine functions can be written as $h(\textbf{x})=\left[b+\sum_{j=1}^{|\textbf{x}|}a_{j}x_{j}\right]_{d}$ for $a_{j}$, $b\in\mathbb{Z}_{d}$. It can be readily seen that for $d=2$, these functions are the linear Boolean functions\footnote{Linear Boolean functions are often referred to as affine Boolean functions.}. We now define the class of input post-selections for the affine functions.

\begin{defn}\label{defn431}
\textbf{Affine Input Post-selection} (AI) is input post-selection where the experimenter accepts data when all $s_{j}$ satisfy $s_{j}=h(\textbf{x})$ where $h(\textbf{x})$ is an affine function on $\textbf{x}$.
\end{defn}

The space of LHV correlators under AI is written as $\mathcal{L}_{\textbf{x}}^{\textrm{AI}}$. We are now in a position to present the following result that shows that AI introduces loopholes. Whilst loopholes are introduced, $\mathcal{L}_{\textbf{x}}^{\textrm{AI}}$ is still smaller than the space of all possible correlators. This fact will be utilised in subsection \ref{sec43c} to highlight the space of quantum correlators for AI.

\begin{prop}\label{prop431}
The space $\mathcal{L}^{\textrm{AI}}_{\textbf{x}}$ is larger than the convex hull of $n$-partite linear functions but smaller than $\mathcal{P}_{\textbf{x}}$ for $n\geq|\textbf{x}|$.
\end{prop}

\textit{Proof}: We first use the results from chapter \ref{chap2} to describe $n$-partite linear functions for $c=d$ being prime:
\begin{eqnarray}
f(\textbf{x})&=&\left[\alpha+\sum_{j=1}^{n}\sum_{k=1}^{(d-1)}\beta_{j,k}\left(1-\sum_{l=0}^{(d-1)}(-1)^{l}{(d-1) \choose l}k^{l}(s_{j})^{d-(l+1)}\right)\right]_{d}\nonumber \\
&=&\left[\alpha'+\sum_{j=1}^{n}\sum_{q=1}^{(d-1)}\beta'_{j,q}(s_{j})^{q}\right]_{d},
\end{eqnarray}
with $\alpha\in\mathbb{Z}_{d}$ and $\beta_{j,k}\in\mathbb{Z}_{d}$ where
\begin{eqnarray}
 \alpha'&=&[\alpha+\sum_{j=1}^{n}\sum_{k=1}^{(d-1)}\beta_{j,k}(1-(-k)^{(d-1)})]_{d} \nonumber \\
&=&\alpha
\end{eqnarray}
and 
\begin{equation}
\beta'_{j,q}=\sum_{j=1}^{n}\sum_{k=1}^{(d-1)}\beta_{j,k}(-1)^{d-q}{(d-1) \choose d-(q+1)}k^{d-(q+1)}.
\end{equation}
Thus $n$-partite linear functions are the sum modulo $d$ of powers of $s_{j}$. In AI we calculate correlators $p(k|\textbf{x})$ after post-selecting on $s_{j}=\left[b+\sum_{j=1}^{|\textbf{x}|}a_{j}x_{j}\right]_{d}$ for $a_{j}$, $b\in\mathbb{Z}_{d}$. Therefore the extreme points of $\mathcal{L}$ corresponding to the $n$-partite linear functions get mapped to extreme points of $\mathcal{L}^{\textrm{AI}}_{\textbf{x}}$ with extreme points $p(k|\textbf{x})=\delta^{1}_{f(\textbf{x})}$ corresponding to the functions
\begin{equation}\label{asfunction}
f(\textbf{x})=\left[\alpha'+\sum_{j=1}^{n}\sum_{q=1}^{(d-1)}\beta'_{j,q}(b+\sum_{j=1}^{|\textbf{x}|}a_{j}x_{j})^{q}\right]_{d},
\end{equation}
which is a function consisting of multiplication between elements of $\textbf{x}$. This function is not an $n$-partite linear function on $\textbf{x}$. The space $\mathcal{L}^{\textrm{AI}}_{\textbf{x}}$ of LHV correlators under AI is then not confined to the convex hull of  $n$-partite linear functions on $\textbf{x}$. 

However, $\mathcal{L}_{\textbf{x}}^{\textrm{AI}}$ does not contain all vertices of $\mathcal{P}_{\textbf{x}}$. In other words, the function in \eqref{asfunction} is not equal to all functions $f:\mathbb{Z}_{d}^{|\textbf{x}|}\rightarrow\mathbb{Z}_{d}$. We can demonstrate this by the example of the function $f(\textbf{x})=\left[\prod_{j=1}^{|\textbf{x}|}(x_{j})^{(d-1)}\right]_{d}$ that cannot be produced by powers of $\left[b+\sum_{j=1}^{|\textbf{x}|}a_{j}x_{j}\right]_{d}$. Therefore $\mathcal{L}^{\textrm{AI}}_{\textbf{x}}\neq\mathcal{P}_{\textbf{x}}$.  $\square$
\newline

We have shown that AI is \emph{not} a loophole-free form of post-selection but LHV correlators cannot saturate the whole space $\mathcal{P}_{\textbf{x}}$. This is somewhat analogous to discussion of the detection loophole, where for detection efficiency above some threshold, LHV correlators do not saturate the space of all possible correlators. In the subsequent subsection we will consider a more general class of input post-selections where $g_{j}(\textbf{x})$ is now an $n$-partite linear function on $\textbf{x}$. As a corollary of the above result, these input post-selections are also not loophole-free. However, in this new class of input post-selections, LHV correlators have greater computational expressiveness.

\subsection{Input Post-selection with $n$-Partite Linear Functions}\label{sec43b}

We now define input post-selection for $n$-partite linear functions $g_{j}(\textbf{x})$. As can be seen from this definition, this post-selection includes AI, and therefore is not loophole-free.

\begin{defn}\label{defn432}
$n$\textbf{-Partite Linear Input Post-selection} (PI) is input post-selection where the experimenter accepts data when all $s_{j}$ satisfy $s_{j}=h(\textbf{x})$ where $h(\textbf{x})$ is an $n$-partite linear function on $\textbf{x}$.
\end{defn}

Again, we can define the space of LHV correlators under PI as $\mathcal{L}^{\textrm{PI}}_{\textbf{x}}$. This space is thus larger than the convex hull of $n$-partite functions on $\textbf{x}$. In the following result we show that the space $\mathcal{L}^{\textrm{PI}}_{\textbf{x}}$ can be equal to $\mathcal{P}_{\textbf{x}}$ for particular instances of $|\textbf{x}|$ and $n$.

\begin{prop}\label{prop432}
The space $\mathcal{L}^{\textrm{PI}}_{\textbf{x}}$ is $\mathcal{P}_{\textbf{x}}$ for a large enough $n$ if $|\textbf{x}|\leq (d-1)$.
\end{prop}

\textit{Proof}: First we point out that for $d$ being prime, any function $f(\textbf{x})$ can be written as a polynomial of elements $x_{j}$ in the following way:
\begin{equation}
f(\textbf{x})=\left[\sum_{\textbf{z}\in\mathbb{Z}_{d}^{|\textbf{x}|}}a_{\textbf{z}}\prod_{j=1}^{|\textbf{x}|}(x_{j})^{z_{j}}\right]_{d},
\end{equation}
with $a_{\textbf{z}}\in\mathbb{Z}_{d}$ where $\textbf{z}\in\mathbb{Z}_{d}^{|\textbf{x}|}$ are digit-strings. We now demonstrate that there are values of $n=n'$ when we can achieve any of the polynomials $\left[\prod_{j=1}^{|\textbf{x}|}(x_{j})^{z_{j}}\right]_{d}$, and then we can take $d^{|\textbf{x}|}$ sets of these $n'$ parties; each set outputs $\left[a_{\textbf{z}}\prod_{j=1}^{|\textbf{x}|}(x_{j})^{z_{j}}\right]_{d}$ and we take the sum modulo $d$ of all the sets outputs and as a result produce $f(\textbf{x})$.

Now we demonstrate that for $n=n'$ parties we can produce the outcome $\left[\sum_{j=1}^{n'}m_{j}\right]_{d}=\left[\prod_{j=1}^{|\textbf{x}|}(x_{j})^{z_{j}}\right]_{d}$ deterministically. First, we show that all polynomial terms of length $2$, i.e. $\left[\prod_{j=1}^{|\textbf{x}|}(x_{j})^{z_{j}}\right]_{d}$ with only $2$ non-zero terms in $\textbf{y}$, can be produced and proceed by induction. The length $2$ polynomials can be achieved if a party outputs $m_{j}=\left[(s_{j})^{2}\right]_{d}$ which is an $n$-partite linear function on $\textbf{s}$. We then post-select on $s_{j}$ satisfying the $n$-partite linear function on $\textbf{x}$ in the following way $s_{j}=\left[(x_{l})^{y_{l}}+(x_{l'})^{y_{l'}}\right]_{d}$ where $l$ and $l'$ labels the $2$ elements of $\textbf{y}$ which are non-zero. After this post-selection $m_{j}=\left[(x_{l})^{2y_{l}}+(x_{l'})^{2y_{l'}}+2(x_{l})^{y_{l}}(x_{l'})^{y_{l'}}\right]_{d}$ and if we have two other parties that each outputs $m_{j+1}=\left[-s_{j+1}\right]_{d}$ and $m_{j+2}=\left[-s_{j+2}\right]_{d}$ and post-select on $s_{j+1}=(x_{l})^{2y_{l}}$ and $s_{j+2}=(x_{l'})^{2y_{l'}}$. Then if we take the sum modulo $d$ of these three outcomes we obtain $m_{j}\oplus m_{j+1}\oplus m_{j+2}=\left[2(x_{l})^{y_{l}}(x_{l'})^{y_{l'}}\right]_{d}$, which is a length $2$ polynomial. We can repeat this process with $q$ sets of three parties and take the sum modulo $d$ of the joint outcomes of all sets to obtain $\left[2q(x_{l})^{y_{l}}(x_{l'})^{y_{l'}}\right]_{d}=\left[(x_{l})^{y_{l}}(x_{l'})^{y_{l'}}\right]_{d}$ such that $\left[2q\right]_{d}=1$ as $d$ is prime. 

For $|\textbf{x}|=3$, we have another party outputting $m_{j'}=[(s_{j})^{3}]_{d}$ and post-selecting on the $n$-partite linear function  on $\textbf{x}$, $s_{j'}=\left[(x_{1})^{y_{1}}+(x_{2})^{y_{2}}+(x_{3})^{y_{3}}\right]_{d}$. Thus we produce $m_{j'}=\left[((x_{1})^{y_{1}}+(x_{2})^{y_{2}}+(x_{3})^{y_{3}})^{3}\right]_{d}=\left[3!(x_{1})^{y_{1}}(x_{2})^{y_{2}}(x_{3})^{y_{3}}+...\right]_{d}$ where ``...'' represents length $2$ polynomials of $\textbf{x}$. The length $2$ and $1$ polynomials can be subtracted from this output from the $j'$th site as they can be produced by other parties as shown above, so that the joint outcome can produce $\left[3!(x_{1})^{y_{1}}(x_{2})^{y_{2}}(x_{3})^{y_{3}}\right]_{d}$. Again by taking $q$ sets of parties that output this in total and taking the joint outcome of all $q$ sets produces 
\begin{equation}
\left[q3!(x_{1})^{y_{1}}(x_{2})^{y_{2}}(x_{3})^{y_{3}}\right]_{d}=\left[(x_{1})^{y_{1}}(x_{2})^{y_{2}}(x_{3})^{y_{3}}\right]_{d}
\end{equation}
for $\left[6q\right]_{d}=1$ as $d$ is prime.

We can repeat this process for $|\textbf{x}|>3$, where a party outputs $m_{j''}=\left[(s_{j''})^{|\textbf{x}|}\right]_{d}$ and we post-select upon $s_{j''}=\left[\sum_{k=1}^{|\textbf{x}|}(x_{k})^{y_{k}}\right]_{d}$. This results in 
\begin{equation}
m_{j''}=\left[(\sum_{k=1}^{|\textbf{x}|}(x_{k})^{y_{k}})^{|\textbf{x}|}\right]_{d}=\left[|\textbf{x}|!\prod_{k=1}^{|\textbf{x}|}(x_{k})^{y_{k}}+...\right]_{d}
\end{equation}
 where ``...'' represents length $(|\textbf{x}|-1)$ polynomials of $\textbf{x}$ which can be subtracted. Finally, again we can taking an arbitrary number of parties and the sum modulo $d$ of the parties outputs will be $\left[\prod_{k=1}^{|\textbf{x}|}(x_{k})^{y_{k}}\right]_{d}$. This all applies when $|\textbf{x}|\leq (d-1)$, and so when this is satisfied, all functions on $\textbf{x}$ can be achieved with large enough $n$. $\square$
\newline

Therefore in the presence of data post-selection that is a natural generalisation of LI post-selection, not only do we avoid loopholes, but we can completely saturate the space of all possible correlators $\mathcal{P}_{\textbf{x}}$. This truly highlights the uniqueness of the scenario with binary inputs and outputs at each site. We now discuss the effect of input post-selection upon quantum correlators.

\subsection{Quantum Correlators and Input Post-selection}\label{sec43c}

For LI and LIO, the space of LHV correlators was unaffected, but the space of quantum correlators was $n$-dependent and could completely saturate $\mathcal{P}_{\textbf{x}}$. Since LHV correlators can also saturate the whole correlator space with PI, we briefly consider the effect of AI on quantum correlators. The space of quantum correlators under AI post-selection is $\mathcal{Q}^{\textrm{\textrm{AI}}}_{\textbf{x}}$. As with $\mathcal{L}_{\textbf{x}}$, there may be an $n$-dependence on the size of $\mathcal{Q}^{\textrm{AI}}_{\textbf{x}}$, but for brevity we will not make this explicit in our notation. The main result of this subsection is that for $|\textbf{x}|=2$ and $n=3$, $\mathcal{Q}^{\textrm{AI}}_{\textbf{x}}$ is larger than $\mathcal{L}_{\textbf{x}}^{\textrm{AI}}$. We demonstrate this by an example for $d=3$.

We have already shown in Proposition \ref{prop431} that the vertex of $\mathcal{P}_{\textbf{x}}$ corresponding to the function $f(\textbf{x})=\left[(x_{1}x_{2})^{2}+1\right]_{3}$ is not in $\mathcal{L}_{\textbf{x}}^{\textrm{AI}}$. Therefore, we can adapt the non-trivial Bell inequality \eqref{nontrivialthree} from subsection \ref{sec33c} in chapter \ref{chap3} for correlators $p(1|\textbf{x})$:
\begin{eqnarray}\label{nontrivialai}
\frac{1}{9}\left(p(1|00)+p(1|01)+p(1|02)+p(110)+p(2|11)\right)&\nonumber \\
+\frac{1}{9}\left(p(2|12)+p(1|20)+p(2|21)+p(2|22)\right)\leq& \frac{8}{9}.
\end{eqnarray}
The right-hand-side is exactly the same as \eqref{nontrivialthree}, as all of the $n$-partite linear functions coincide with $f(\textbf{x})=\left[(x_{1}x_{2})^{2}+1\right]_{3}$ for $8$ out of $9$ values of $\textbf{x}$. Therefore, for all functions not equal to $f(\textbf{x})$, this is the maximum overlap between functions. $\mathcal{L}_{\textbf{x}}^{\textrm{AI}}$ will be a convex polytope of functions not including $f(\textbf{x})=\left[(x_{1}x_{2})^{2}+1\right]_{3}$, thus giving at most $\frac{8}{9}$ (for the Bell expression) for each of its extreme points. As discussed in chapter \ref{chap3}, this inequality is not violated by quantum correlators for $n=2$. However, this inequality can be violated by quantum correlators for $n=3$ with AI if $s_{1}=x_{1}$, $s_{2}=x_{2}$ and $s_{3}=[x_{1}+x_{2}]_{3}$. We used the MBS approach to find a lower bound of $\approx 0.9314>\frac{8}{9}$ on the quantum violation of \eqref{nontrivialai}. 

Even in the presence of post-selection that introduces loopholes, the space of quantum correlators can be larger than the space of LHV correlators. Whilst not as dramatic as the effect that LI and LIO has on the quantum region, it is never-the-less interesting how ``tactile'' quantum correlators can be. That is, even if we imbue LHV correlators with more computational power (as with AI), quantum correlators can still have more computational expressiveness. It would be an interesting avenue of research to consider how quantum correlators are affected by non-loophole-free post-selection and whether their power can always be ``boosted'' by this post-selection. 

\section{Chapter Summary}\label{sec44}

\noindent
The practical motivations of implementing Bell tests in the laboratory have motivated the study of loopholes and how they emerge when we have to reject ``imperfect'' measurement data \cite{pearle}. In this chapter we have used the insight from considering Bell tests from a computational point-of-view to say how and why loopholes emerge. By post-selecting on measurement data only when we have successful detection, we establish a relationship between the inputs and local hidden variables. This relationship allows other parties to indirectly learn the inputs of other sites via this shared data. By modelling this behaviour we retrieved the GM \cite{garg} bound on the necessary detection efficiency required to establish a loophole-free violation of a Bell inequality. We then subsequently improved upon their bound by considering more parties. 

Our improvement on the GM bound is not as impressive as the improvement attained by Eberhard \cite{eberhard} in the Clauser-Horne inequality setting. Eberhard's bound of $\frac{2}{3}$ has been improved upon further \cite{vertesi}, this was a result of considering more measurement settings at each site. It would be interesting to consider the Bell inequalities on the full probability distribution for $(n,2,2)$ for $n>2$, and whether the detection efficiency can be lowered further in analogy to our results. 

Despite the issues associated with post-selection, there is a scenario where if we have perfect detections but the experimenter post-selects on measurement data by choice, we do not introduce loopholes. We associate LHV correlators with a limited computational expressiveness in the $(n,2,2)$ scenario: only linear Boolean functions can be achieved. If we post-select on data but only in a way that does not introduce non-linear Boolean functions, we avoid loopholes and still allow the possibility for a violation of a Bell inequality. In fact, we can increase the amount of quantum violation for particular Bell inequalities if we utilise post-selection. 

However, we have also shown that the $(n,2,2)$ scenario is unique in the respect of not introducing loopholes; if we allow a greater number of inputs and outputs at each site, loopholes can again emerge. For $c=d>2$, LHV correlators can produce powers of its input, and this inherent multiplication can be used to simulate all possible correlators. The ability to produce addition and multiplication modulo $d$ for $d$ being prime can be enough to produce any function $f:\mathbb{Z}^{n}_{d}\rightarrow\mathbb{Z}_{d}$. This has highlighted both how fragile Bell tests are in establishing a distinction between quantum and LHV correlators, and also how much descriptive power is accumulated by considering Bell tests from the computational point-of-view. Since we have shown the intimate link between correlators and functions on digit-strings, discussing functions has allowed to capture part of the picture of loopholes in Bell tests.

Interestingly, the models of post-selection we have discussed for $(n,2,2)$ involve the same level of data processing involved in MBQC. With LIO post-selection, we can simulate time-like separated processes such as adaptive MBQC circuits without introducing loopholes. Modelling signalling processes within Bell tests could lead to an insight into why we obtain improvements in information processing for quantum resources. We will summarise and consider some of these ideas in the final chapter.

\chapter{Summary and Outlook}\label{chap5}

The Bell inequalities have dictated and continue to dictate much of the discussion about the nature of quantum mechanics. In this thesis we have suggested that a general framework for Bell tests has a computational aspect. This both allows us to use methods and ideas in computer science to say something about Bell tests and methods developed in Bell tests to say something about computation. This collaboration between applied and fundamental science is what drives a large part of quantum information science \cite{hardy4}. The diversity of connections addressed in this thesis have been made between the CGLMP-type Bell tests and basic number theory (in the form of functions on cyclic groups); loopholes, post-selection and quantum computing; we also connected quantum computing to non-local games and WW Bell tests.

In chapter \ref{chap2}, we outlined our approach to Bell tests, in particular looking at correlators: the expectation value of joint measurements. We showed that correlators can be associated with a notion of computation, that is functions on inputs. The calculation of a correlator maps raw statistical data into a stochastic map from inputs to a single output. This operational description allows to then think about information processing. This framework and description also has something to say about non-signalling theories and Svetlichny's model of correlations.

The discussion of correlators in chapter \ref{chap2} was mostly in terms of the vertex description of convex polytopes. In chapter \ref{chap3}, we shifted to discussing the Bell inequality as defining the convex polytope of LHV correlators. We used the vertex description from chapter \ref{chap2} to numerically calculate the linear inequalities, or facet Bell inequalities that define this polytope. However, it was only computationally feasible to find these inequalities for a relatively small number of settings. Given the hardness of the computational problem, we then just discussed non-trivial Bell inequalities, relaxing the need for the inequality to be facet-defining, but still potentially be violated. These non-trivial expressions then necessarily bound the space of LHV correlators to be smaller than the space of all possible correlators.

Non-trivial Bell inequalities are not only useful for bounding classical correlations, they have a natural interpretation in terms of non-local games. We looked at these non-local games in the many party, two-input, two-output scenario and showed that they have a concrete connection to Measurement-based Quantum Computing (MBQC). In particular, nMBQC, the class of non-adaptive circuits in the Raussendorf and Briegel model of MBQC can be shown to be inequivalent to a full quantum computer. However, within this nMBQC structure we still obtain natural generalisations of both the GHZ paradox and PR non-local box. 

An interesting aspect of the nMBQC model is that data processing by a classical computer does not imbue LHV theories with any more computational power. In chapter \ref{chap4}, we applied this insight to data post-selection in Bell tests. We showed that post-selection in Bell tests, such as post-selecting on detecting outcomes in imperfect experiments, is problematic and introduces ``loopholes''. We used the computational insight from the rest of the thesis to show how the detection loophole can emerge and then rederived the Garg-Mermin bound on detection efficiency \cite{garg}. Throughout this discussion our computational perspective drove the understanding of loopholes.

After showing how in imperfect experimental Bell tests, loopholes can emerge, we turned to a different framework for data post-selection. We assume that we have perfect detection and data collection, but we post-select on inputs satisfying certain constraints. We showed that LHV correlators are unaffected in their computational expressiveness by this post-selection. We associate this conservation of computational power with the post-selection being ``loophole-free''. The notion of a loophole in both frameworks for post-selection is heuristically connected as allowing LHV correlators to have more computational expressiveness than just the linear Boolean functions.

The post-selection in the second framework is loophole-free if we constrain its form. These constraints however can still allow the post-selection to simulate the processing a classical computer imposes on data sent to measurement sites in MBQC. Also we can simulate signalling processes with this post-selection. This offers a potentially fruitful way of viewing quantum protocols and processes that have time-like separated elements into a framework where processes are now in the context of space-like separated parties. All of these results were developed in the $(n,2,2)$ scenario, and we showed that generalisations of these methods to other scenarios is problematic, thus highlighting the uniqueness of LHV expressed in terms of linear Boolean functions.


The work in this thesis is by no means a complete analysis of the role of computation in Bell tests, but perhaps strengthens the study of the relationship between the two. There is much work to be done still in understanding quantum correlations and whilst we have discovered new phenomena, the characterisation of quantum correlations remains broadly ill-understood. We have conjectured that all quantum correlators for the bipartite scenario can be captured by a particular set of quantum operators in the Navacu\'{e}s-Pironio-Ac\'{i}n hierarchy. It would be of great interest if this were true and if a similar behaviour occurred in the multipartite setting. This is an immediate problem raised by work in this thesis and worth pursuing as a continuation. 

In recent years, a significant amount of effort into classifying the geometric nature of non-signalling correlations (see e.g. \cite{pironio3}). We have shown that some of the extremal structure of the polytope of non-signalling theories can be revealed by the extremal structure of correlators. It would be interesting to see if there is a connection between correlators and the rest of the vertices of the non-signalling polytope. This picture is not clear as some of the non-LHV vertices of the non-signalling polytope for $(3,2,2)$ do not violate any of the facet Bell inequalities for correlators in this setting. However, the generality of the correlator description in terms of computations could give a handle on some of these ideas.

Continuing with the theme of characterising the full probability distribution instead of correlators, it would be interesting to study the effect of data post-selection on non-signalling resources. The difficulty in relating the inputs of parties to each other as we have done can allow LHV resources to achieve correlations that violate locality. Since in the correlator framework all single-site maps get mapped to a single output, this violation of locality has little or no effect. It would be interesting to allow resources that exploit ``non-locality'' in this way but still cannot produce something that quantum mechanics can produce. This is akin to the detector loophole where the LHV region is enlarged by post-selection, but below a threshold detection efficiency, still is not large enough to simulate quantum correlators. 

Can our approach to correlators in terms of functions to applied to other issues in the study of Bell tests? An interesting potential avenue for further research could be the ``monogamy of Bell correlations'' \cite{pawlowski,toner}. This is similar to the ``monogamy of entanglement''\footnote{This expression is thought to originate with Charlie Bennett \cite{toner}.} where we have three parties and if two parties are maximally entangled then the third party cannot be entangled with either of these two parties. It has been shown that Bell correlations behave in an analogous fashion where if two parties out of three violate a bipartite Bell inequality, then the correlations between either of these two parties and a third party cannot achieve a violation of the same inequality. This has been generalised to many parties with these parties divided into two overlapping sets \cite{pawlowski}. Can the language of functions, or computations explain that if one set of parties is trying to perform a computation, then by a satisfiability argument, the other set cannot produce this same function?

Finally, since we have established a connection between MBQC and Bell tests, it would be interesting to simulate quantum computations such as, say, Shor's algorithm \cite{shor} and see if it violates a Bell inequality. In some sense then it could be seen that this computation cannot be resolved with a classical picture of the world, or it would highlight the non-classical aspects of this algorithm. Post-selection and quantum computation have been studied before by Scott Aaronson \cite{aaronson2}, in a different format to our own framework. It was shown by Aaronson that quantum computation with post-selection of a different kind to ours is incredibly powerful. Speculatively, there may be some connection between our work and ideas in computational complexity. We have already made the connection to the class IQP \cite{shepherd}, this class may be amenable to the study of our Bell tests with post-selection.

We hope to address the issues raised by this thesis in further research. We also hope that the work presented has produced the motivation to consider ``device-independent'' computing. This would be the ability to confirm that we have built something that uses quantum mechanics to compute but without knowing anything about the device. We have shown that the Bell inequality is a useful metric for quantum behaviour, in particular with regards to computation. More importantly, we hope that the work in this thesis can lead to new approaches of thinking about Bell tests, perhaps motivated by computation. 

N. David Mermin once quoted a ``distinguished Princeton physicist'' as saying \cite{mermin3},``Anybody who's not bothered by Bell's theorem has to have rocks in his head.'' The Bell inequality has been a profound addition to science and we hope that the work in this thesis contributes to new aspects of its study.


\end{document}